\documentclass[a4paper, twoside, 12pt]{book}
\usepackage[Lenny]{fncychap}
\usepackage[utf8]{inputenc}
\usepackage[T1]{fontenc}
\usepackage{geometry}
\usepackage{graphicx}
\usepackage{amsmath}
\usepackage{array}
\usepackage{subcaption}
\usepackage{multirow}
\usepackage{adjustbox}
\usepackage{lmodern} 
\usepackage[]{algorithm2e}
\RestyleAlgo{boxruled}
\usepackage{amssymb}
\usepackage{rotating}
\usepackage{xcolor}
\usepackage{comment}
\usepackage[square,numbers]{natbib}
\usepackage{tensor}
\usepackage{accents}
\usepackage{appendix}
\usepackage{tikz}
\usepackage{doi}
\usepackage{pgfplotstable}
\usepackage{pdfpages}

\usepackage{hyperref}
\hypersetup{
	colorlinks=true,
	linkcolor=black,
	urlcolor=blue,
	citecolor=blue,
	linktoc=all
}

\definecolor{UnivBlue}{RGB}{71, 165, 158}
 
\newcommand\numberthis{\addtocounter{equation}{1}\tag{\theequation}}
\DeclareMathOperator{\e}{e}

\newcommand\subtitle[1]{\begin{flushright}\small\textit{#1}\end{flushright}}

\include{pdg.tex}

\author{Romain GERVALLE}
\date{13 D{\'e}cembre 2023}
\title{Trous noirs chevelus et autres objets compacts dans des théories de la gravité}
\specialite{Physique Théorique}
\directeur{Mikhail S. VOLKOV}
\encadrant{Loïc VILLAIN}

\president{\textbf{Julien GARAUD}}{Professeur, Université de Tours}
\rapporteurb{\textbf{Eugen RADU}}{Chercheur associé, Université d'Aveiro -- Portugal}
\rapporteura{\textbf{Éric GOURGOULHON}}{DR CNRS, Laboratoire Univers et Théories}
\juryb{\textbf{Jutta KUNZ-DROLSHAGEN}}{Professeure, Université d'Oldenburg -- Allemagne}
\jurya{\textbf{Claudia DE RHAM}}{Professeure, Imperial College London -- Royaume-Uni}
\juryc{\textbf{Loïc VILLAIN}}{Maître de conférence, Université de Tours}
\juryd{\textbf{Mikhail S. VOLKOV}}{Professeur, Université de Tours}
\jurye{\textbf{Eugen RADU}}{Chercheur associé, Université d'Aveiro -- Portugal}
\juryf{\textbf{Éric GOURGOULHON}}{DR CNRS, Laboratoire Univers et Théories}

\pgfplotsset{compat=1.18}

\begin{document}
	
    \pagedegarde
    
    \frontmatter
 
	\newpage
	
	\thispagestyle{empty}

    \chapter*{Remerciements}

    \vspace{-1.3cm}
	
	Après le long et fastidieux travail de rédaction vient le temps des remerciements. Je tiens tout d'abord à remercier les membres du jury, et tout particulièrement les rapporteurs, {\'E}ric Gourgoulhon et Eugen Radu. Avoir un regard extérieur sur son travail avant de soutenir est très important. La soutenance ne pourrait tout simplement pas avoir lieu sans eux. Merci aussi à Claudia de Rham, Jutta Kunz et Julien Garaud pour avoir accepté de venir compléter le jury en tant qu'examinateur et examinatrices.

    \smallskip
    Mes pensées se dirigent ensuite naturellement vers mes deux directeurs de thèse, Michael et Loïc. Avant la thèse, ils ont été mes enseignants à la fac. Pour Loïc, la rencontre s'est faite dès ma première année de licence. Il ne se souvient probablement pas de moi en ce temps là, mais j'ai encore l'image de ce professeur qui tentait de nous illustrer le phénomène d'électrisation par frottement en parlant de ses cheveux longs. Il a grandement contribué à mon intérêt pour la relativité générale grâce à ses nombreuses conférences de vulgarisation scientifique sur des sujets qui faisaient rêver tous les étudiants curieux. J'en faisais partie. J'ai également connu Michael en licence, mais j'ai surtout été marqué par ses enseignements en master, en particulier son cours de cosmologie. Je me souviens d'un professeur tout aussi passionné que nous. Lorsque j'ai envisagé une thèse dans le domaine de la gravitation relativiste, c'est vers Loïc que je me suis tourné en premier. Il a par la suite contacté Michael, qui avait une idée de projet en tête, et c'est comme ça que tout a commencé. 

    \smallskip
    Me voilà donc parti, il y a un peu plus de quatre ans, pour une thèse avec deux directeurs aux cheveux longs. Les miens aussi ont bien poussé depuis. Il n'y pas que nos trous noirs qui sont chevelus... Je suis reconnaissant envers Michael pour son encadrement que je considère comme très équilibré. C'est lui qui m'a mis sur la piste des deux principaux projets de cette thèse, me laissant parfois en autonomie, mettant aussi parfois les mains dans le cambouis. C'est au travers de lui que j'ai appris, je le pense, à écrire des articles scientifiques. De nos premiers rendez-vous au début de la thèse, jusqu'à ceux de ces derniers mois, je pense avoir beaucoup gagné en assurance. C'est aussi grâce à lui. Loïc a suivi l'avancement de ces projets d'un peu plus loin, mais bien souvent, lorsque j'avais trop la tête dans le guidon, il m'a permis de prendre du recul sur mon travail. C'est aussi lui qui a assisté à presque tous les entraînements lorsque je devais présenter mes résultats lors de conférences. Il a également effectué un travail de relecture considérable sur mon manuscrit. Merci pour tous les cafés-debrief à Velpeau !
    
    \smallskip
    Je me dois aussi de remercier les autres personnes qui ont contribué à mon travail de recherche. Je pense notamment à Julien Garaud, ancien doctorant de Michael, qui m'a introduit à la méthode des éléments finis et à la librairie FreeFem. Je n'aurai jamais pu apprendre à résoudre des équations aux dérivées partielles aussi rapidement sans son aide. Merci aussi à Eugen Radu, avec qui j'ai pu échanger lors de mon projet annexe sur les étoiles à bosons. Il m'a invité plusieurs fois au Portugal pour présenter mes travaux au sein de l'équipe de recherche dirigée par Carlos Herdeiro, que je remercie également. Une pensée pour {\'E}ric Gourgoulhon que j'ai pu rencontrer au tout début de mon doctorat. Il m'a introduit au code qu'il a développé avec ses collaborateurs et qui permet de produire des images de trous noirs. Je ne me suis finalement pas servi de cet outil pendant ma thèse, mais je suis certain que cela pourra m'être utile dans le futur. 
    
    \smallskip
    Un laboratoire de recherche ne fonctionne pas qu'avec des chercheurs, nous devons tous beaucoup à nos très chères secrétaires : Laetitia Portier, Anouchka Lépine et {\'E}lodie Demoussis. Merci à elles qui ont notamment participé à la logistique pour la venue des membres du jury.

    \bigskip
    Venons en maintenant à tous ceux qui ont permis de rendre ces quatre années de thèse plus plaisantes : mes collègues et amis doctorants. Le premier que j'ai côtoyé au sein de l'Institut Denis Poisson fut Thomas Helpin. On a le même directeur, forcément ça créé des liens. Je le considère comme mon grand frère de thèse. Le second a été Guillaume Lemarthe, arrivé un an avant moi à l'IDP. Avec lui, je suis devenu un expert aux fléchettes (même si je perds souvent contre lui, je dois l'admettre). Yégor Goncharov est venu compléter notre bureau un peu plus tard. Le moins qu'on puisse dire, c'est qu'il a particulièrement contribué à la bonne ambiance : Guillaume et lui ont fondé le \textit{Club Amande}. Il s'agit d'un club de viennoiseries et de pâtisseries (aux règles d'admission très strictes) qui nous permet à tous de profiter d'un bon goûter à quatre heures. Je salue également Andrii Naidiuk, dernier arrivant en date de notre bureau. 

    \smallskip
    Les personnes de mon bureau ne sont pas les seules que je tiens à remercier. Merci à Léa Gohier, excellente partenaire de karaoké, Antonin Jacquet, au style vestimentaire toujours irréprochable, un modèle, Théo Girard, jamais à cours d'idées pour trouver des choses rigolotes à faire en soirée, Marion Meutelet, dont la bonne humeur est particulèrement contagieuse, Igor Haladjian, pour tous les after chez lui jusqu'à 6h du matin, Romane Coutard, pour les gâteaux maisons incroyables au Club Amande et Yohan Potaux, vous l'avez déjà vu avec des lunettes de soleil ? Une star. Je remercie aussi Jad Abou Yassin, Maxime Ligonniere et Julien Verges. Une pensée aussi pour les anciens doctorants que j'ai pu croiser au cours de mon séjour à l'IDP. J'ai beaucoup de souvenirs avec vous tous, pas seulement au labo, mais aussi à notre QG : le \textit{Shamrock} (dédicace à Annie et Fafa).

    \smallskip
    Mon cercle d'amis s'étend au-delà des doctorants de l'IDP. Je remercie Thibault Doré que j'ai rencontré en licence et avec qui j'ai toujours gardé un contact proche. Je salue aussi sa compagne, Cécile Ginestet, qui vient tout juste de devenir docteure au moment où j'écris ces lignes. Thibault m'a également permis de devenir ami avec Gisèle Gruener. Merci à elle d'avoir été, au travail, une directrice du département d'enseignement toujours très arrangeante, et en dehors du travail, la meilleure hôte qui soit pour organiser des apéros ou des soirées pizzas ! Par l'intermédiaire de Gisèle, j'ai aussi croisé le chemin de Patrick Blanchedeau, également collègue d'enseignement, et de Anne Thibault. Merci à vous pour tous les bons moments partagés. 
    
    \smallskip
    Comment ne pas remercier Floriane Picolo ? Elle a lancé avec moi la mode des soirées pétanque à Velpeau avant de rapidement devenir une amie hors pair et une camarade de soirée indéfectible. Le chemin entre le centre ville et la rue du Docteur Fournier, on pourrait le faire les yeux fermés. Merci pour toutes les fois où tu nous as accueilli chez toi, notre QG de quartier. Merci le sang ! Tant qu'on est à Velpeau, j'en profite pour saluer Pierre Vendé qui a été parmi les premiers à rejoindre notre fine équipe de boulistes. Merci aussi à Samara Benkel, toi qui est toujours partante pour une virée en boîte jusqu'au bout de la nuit ou pour aller en festival. Je signe sans hésitation pour retourner aux Terres du Son avec toi et Flo l'été prochain !

    \bigskip
    Après les amis, vient la famille. Merci à vous d'être venus assister à ma soutenance. Quelle drôle d'idée que de venir écouter un truc incompréhensible en anglais de bon matin. Je suis très touché par votre présence à tous. Je remercie en particulier ma mère, qui m'a laissé choisir le cursus que je voulais, à Tours, alors que j'aurai pu rester quelques années de plus à Orléans. Ça en valait la peine non ? D'ailleurs, il y a sûrement dans l'audience la personne qui a été la première à m'enseigner les mathématiques : Valérie Levallois, mon institutrice de CP. Je suis sûr que je ne suis pas le seul de ses élèves à finir en doctorat, mais elle voulait venir voir ça de ses propres yeux. 

    \smallskip
    Je sais qu'il manque une personne pour compléter ce tableau familial, il s'agit de mon père. Tout comme il l'a été, je suis plutôt cartésien, mais une partie de moi espère tout de même qu'il ait entrevu cette soutenance de là où il est. En tout cas, j'aurai beaucoup aimé te raconter ce que je fais.

    \bigskip
    Il reste une dernière personne à remercier, Lami Suleiman, ma compagne depuis maintenant quatre ans. On s'est connus pendant nos études avant de partir chacun faire une thèse. Elle travaille sur les étoiles à neutrons, moi sur les trous noirs. J'espère que l'on pourra bientôt collaborer ensemble. Merci pour ton précieux soutient, ta douceur sincère, et ton amour que je perçois toujours même quand tu es à 9100 kilomètres de moi. 
    
	\newpage
	
	\mbox{}
	
	\vspace{220 mm}
	
	{\hfill \large \textit{«~Les machines un jour pourront résoudre tous les problèmes, mais jamais aucune d'entre elles ne pourra en poser un !~»}}
    \begin{flushright}
        Prophétie attribuée à Albert Einstein.
    \end{flushright}
	
	\newpage
	
	\normalsize

    \section*{R\'esum\'e}
	
	Dans le cadre d’un espace-temps décrit par la relativité générale d’Einstein, sur lequel évolue uniquement le champ électromagnétique de Maxwell, les trous noirs stationnaires sont complètement caractérisés par leur masse, leur charge électrique ou magnétique, et leur moment cinétique : il s'agit de l'une des versions du théorème de calvitie. Pour autant, lorsque certaines des hypothèses de ce théorème sont omises, il a été établi qu'il cesse de s’appliquer. Cela conduit à l’émergence de trous noirs dits chevelus. Jusqu'à présent, les observations astronomiques ne permettent pas de détecter les «~cheveux~» des trous noirs. Cependant, avec le développement de détecteurs d'ondes gravitationnelles toujours plus précis, les trous noirs chevelus restent un sujet d'étude important en physique théorique. Dans cette thèse, nous considérons deux options permettant de s'affranchir du théorème de calvitie.

    La première option consiste à décrire la métrique de l’espace-temps par une théorie de gravitation alternative. Nous étudierons la stabilité de trous noirs chevelus dans un espace-temps vide décrit par la théorie de la bigravité massive. Cette théorie est connue pour ses solutions cosmologiques permettant de décrire un Univers en expansion accélérée sans avoir besoin de recourir à la constante cosmologique. Nous montrerons que les trous noirs chevelus en bigravité, obtenus à l’aide de méthodes numériques, sont capables de représenter tant des trous noirs stellaires que des trous noirs supermassifs.

    Une autre possibilité est de garder les équations d’Einstein, mais de considérer un contenu matériel autre que le champ électromagnétique de Maxwell. Nous choisirons pour cela les champs de la théorie électrofaible. Lorsque la gravitation est omise, cette théorie décrit des monopôles magnétiques de masse infinie. La relativité générale permet de les régulariser en masquant leur singularité coulombienne derrière un horizon des évènements. Les monopôles deviennent alors des trous noirs chargés magnétiquement, pouvant être chevelus. Ces trous noirs électrofaibles pourraient s'être formés lors de fluctuations primordiales, aux tout premiers instants de l’Univers. Après avoir étudié en détail la structure interne des monopôles en espace-temps plat, nous verrons comment leurs propriétés se généralisent au cas gravitationnel.

    Lorsqu’un trou noir chevelu voit le rayon de son horizon se réduire à zéro, les champs externes restent, et la configuration ainsi obtenue est appelée un soliton. Nous étudierons pour terminer un cas particulier de soliton obtenu lorsqu’un champ scalaire complexe est couplé à la relativité générale. Nous construirons des chaînes de ces solitons, appelés étoiles à bosons. Les équations aux dérivées partielles sous-jacentes seront résolues à l’aide de la méthode des éléments finis : une approche originale et peu utilisée par la communauté de la relativité numérique.

    \newpage

    \section*{Abstract}
	
	In the realm of spacetimes governed by Einstein's general relativity and containing only Maxwell's electromagnetic field, stationary black holes are fully characterized by their mass, electric or magnetic charge, and angular momentum -- a property encapsulated in a version of the no-hair theorem. However, the validity of this theorem is contingent on certain assumptions, and when these are relaxed, it has been established that the theorem does not always apply. This gives rise to the so-called hairy black holes. To date, astronomical observations have not provided concrete evidence of any type of black hole "hair". Nevertheless, the development of increasingly precise gravitational wave detectors has sparked renewed interest in hairy black holes. In this thesis, we delve into two approaches to circumvent the no-hair theorem.

    The first option consists in describing the spacetime metric by an alternative theory of gravitation. We investigate the dynamical stability of hairy black holes in a vacuum spacetime described by the theory of massive bigravity. This theory is known for its cosmological solutions which can account for a self-accelerating expansion of the Universe without requiring the use of the cosmological constant. We show that hairy black holes in bigravity, which are obtained using numerical methods, can describe both stellar black holes and supermassive black holes.

    Another approach is to keep Einstein's equations but to consider a different material content than Maxwell's electromagnetic field. For this we choose the fields of the electroweak theory. In the absence of gravitation, this theory describes magnetic monopoles with infinite mass. General relativity allows for their regularization by concealing their Coulombian singularity within an event horizon. As a result, monopoles become magnetically charged black holes which can exhibit a non-Abelian hair. These electroweak black holes might have formed during primordial fluctuations in the early Universe. After providing a detailed analysis of the internal structure of monopoles in flat space, we investigate how their properties generalize to the black hole case.

    When the horizon radius of a hairy black hole shrinks to zero, only its external fields remain, giving rise to a configuration known as a soliton. Lastly, we study a particular example of soliton that arises when a complex scalar field is coupled to general relativity. We construct chains of these solitons, which are referred to as boson stars, by solving the underlying partial differential equations using the finite element method. This technique is not very common in the numerical relativity community and provides an alternative to the finite difference method.

	\tableofcontents

    \renewcommand{\chaptermark}[1]{\MakeUppercase{\markboth{}{#1}}}
	
	\listoffigures
	
	\addcontentsline{toc}{chapter}{List of Figures}
	
	\listoftables
	
	\addcontentsline{toc}{chapter}{List of Tables}
	
	\chapter*{List of Acronyms}\chaptermark{List of Acronyms}
	
	\addcontentsline{toc}{chapter}{List of Acronyms}
	
	\textbf{ADM} \centerline{Arnowitt-Deser-Misner}\\
	\textbf{(A)dS} \centerline{(anti)-de Sitter}\\
    \textbf{BS} \centerline{Boson star} \\
	\textbf{dRGT} \centerline{de Rham-Gabadadze-Tolley}\\
	\textbf{FLRW} \centerline{Friedmann-Lemaître-Robertson-Walker} \\
	\textbf{FP} \centerline{Fierz-Pauli}\\
	\textbf{GL} \centerline{Gregory-Laflamme} \\
    \textbf{GR} \centerline{General Relativity}\\
    \textbf{GUT} \centerline{Grand Unified Theory} \\
	\textbf{ODE} \centerline{Ordinary Differential Equation}\\
    \textbf{PDE} \centerline{Partial Differential Equation}\\
    \textbf{RN} \centerline{Reissner-Nordström}\\
	\textbf{vDVZ} \centerline{van Dam-Veltman-Zakharov}\\
    \textbf{WS} \centerline{Weinberg-Salam}\\

    \mainmatter

\chapter*{Introduction}
\addcontentsline{toc}{chapter}{Introduction}\chaptermark{Introduction}

En 1915, le physicien Albert Einstein énonce les principes de la relativité générale~: une nouvelle théorie de la gravitation~\cite{Einstein1915}, plus de 200 ans après la loi de l'attraction universelle dévelopée par Newton. Cette dernière n'était pas compatible avec les principes de la relativité restreinte \cite{Einstein1905}, une autre théorie élaborée par Einstein en 1905. L'attraction gravitationnelle, telle qu'elle est décrite par Newton, agit à distance entre deux corps de façon instantanée. Or, selon la relativité restreinte, rien ne peut se déplacer plus vite que la lumière, y compris les interactions entre deux objets distants. Il fallut donc concilier ces deux propositions contradictoires, tout en gardant à l'esprit que les lois de Newton, en leur temps, n'étaient contredites par quasiment\footnote{Le seul échec de la théorie newtonienne concernait alors la précession du périhélie de Mercure, un problème qui fut résolu par Einstein lui-même en 1915 grâce à la relativité générale.} aucune expérience. Revenons un instant sur les principes fondateurs de la relativité restreinte.

Le principal postulat d'Einstein est le principe de relativité selon lequel les lois de la physique s'expriment de la même manière dans tous les référentiels \textit{inertiels}. On rappelle que ces derniers ont pour caractéristique que tous les objets isolés qui s'y trouvent sont, soit en mouvement rectiligne et uniforme, soit immobiles. Pour comprendre les raisons qui poussèrent Einstein à introduire ce célèbre principe, nous devons faire un détour par les équations de Maxwell. Ces dernières furent proposées en 1865 afin d'unifier les phénomènes électriques et magnétiques. Les équations de Maxwell permettent de décrire la lumière comme une onde dite, électromagnétique, c'est-à-dire une oscillation couplée du champ électrique et du champ magnétique se propageant dans l'espace. Ces équations prédisent une vitesse \textit{finie} pour la lumière, communément dénotée $c$, et valant approximativement $3\times 10^8\,\text{m}.\text{s}^{-1}$. En observant un rayon lumineux depuis le même référentiel que sa source, on mesure effectivement qu'il se déplace à la vitesse $c$, mais les choses se compliquent lorsque l'on change de référentiel. Par exemple, imaginons qu'un appareil permettant de mesurer la vitesse de la lumière soit en mouvement rectiligne et uniforme par rapport au référentiel dans lequel est émis le rayon lumineux. D'après la loi de composition des vitesses de Galilée, on devrait mesurer une vitesse plus petite que $c$, si l'appareil se déplace dans le même sens que le rayon lumineux, ou plus grande que $c$, s'il se déplace dans la direction opposée. Or, plusieurs expériences (dont celles de Michelson et Morley en 1887) démontrent que cette prédiction est erronée. L'expérience montre en effet que la lumière se déplace à la vitesse $c$, quel que soit le référentiel dans lequel on effectue la mesure.

{\`A} l'époque des expériences de Michelson et Morley, la communauté scientifique voyait la lumière comme une perturbation d'un milieu appelé éther. La constance de la vitesse $c$, quel que soit le référentiel de la mesure, était incompatible avec le premier modèle simple de l'éther proposé par Maxwell dans le cadre de la cinématique galiléenne. Plusieurs physiciens, dont Lorentz et Poincaré développèrent une modélisation plus sophistiquée de l'éther qui permettait d'expliquer les expériences sur la vitesse de la lumière et qui culmina, en 1905, avec l'obtention des transformations de Lorentz. La même année, Einstein aborda le problème à l'envers, postulant que les lois de la physique doivent être les mêmes dans tous les référentiels inertiels et remettant en cause la cinématique galiléenne. En particulier, la vitesse de la lumière, qui est prédite par les équations de Maxwell, doit donc être invariante sous les changements de référentiels. Einstein en déduit alors les transformations de Lorentz, mais l'interprétation qu'il en fait est radicalement différente~: selon lui, l'éther devient complètement inutile pour la description des ondes électromagnétiques et les notions d'espace et de temps \textit{absolus} s'effondrent. Les transformations de Lorentz, qui sont au coeur de la relativité restreinte, prédisent notamment la dilatation du temps et la contraction des longueurs entre deux observateurs se déplaçant à vitesse constante l'un par rapport à l'autre. 

La relativité restreinte donne un cadre robuste à l'étude des phénomènes électromagnétiques et plus encore, puisqu'elle permet de formaliser correctement toutes les théories classiques\footnote{La qualification de « classique » fait référence ici, et dans le reste de ce manuscrit, à tout ce qui n'est pas de nature « quantique ». L'unification de la relativité restreinte avec la mécanique quantique est également possible, on parle dans ce cas de théorie quantique des champs.} des champs. Ces théories décrivent la dynamique de champs physiques qui peuvent être d'une autre nature que le champ électromagnétique de Maxwell. En revanche, la théorie newtonienne de la gravitation n'est pas compatible avec la relativité restreinte. Ce constat poussa Einstein à développer une nouvelle théorie relativiste de la gravitation.

Mathématiquement, l'espace-temps de la relativité restreinte est caractérisé par la métrique de Minkwoski. Cette dernière permet la mesure de distances spatio-temporelles, plus communément appelés intervalles d'espace-temps. Tout objet isolé verra sa trajectoire suivre une ligne droite dans l'espace-temps de la relativité restreinte, et la distance entre deux objets isolés se déplaçant dans la même direction à la même vitesse restera toujours la même : on dit que cet espace est de courbure nulle. L'idée qui mena Einstein à la théorie de la relativité générale fut alors de décrire la gravitation, non pas par une interaction à distance instantanée\footnote{En relativité, la notion de simultanéité est d'ailleurs dépendante du référentiel.}, mais par une courbure non-nulle de l'espace-temps. De ce point de vue, les objets se déplaçant librement dans un champ gravitationnel ont des trajectoires qui suivent ce qui s'apparente à des lignes droites dans un espace-temps courbe : des géodésiques. Ces dernières peuvent être déterminées à l'aide de l'équation des géodésiques, une équation faisant intervenir la métrique, qui dans le cas général sera différente de celle de Minkowski. Le développement de la relativité générale requiert l'utilisation d'outils mathématiques associés à la géométrie différentielle, des outils qu'Einstein ne maîtrisait pas. Il fut donc fut assisté dans ses travaux par son ami mathématicien Marcel Grossmann~\cite{E.1915}. David Hilbert figure également parmi les noms associés au développement de la relativité générale. Il fut le premier à proposer une action de laquelle peut être déduite l'équation d'Einstein au moyen d'un principe variationnel. 

Bien que la relativité générale se réduise aux lois de Newton dans la limite des champs faibles, il convient de souligner encore une fois que cette théorie relativiste de la gravitation est conceptuellement très différente de la théorie newtonienne. L'équation proposée par Einstein est une relation entre la courbure de l'espace-temps et son contenu matériel ou énergétique. La courbure est encodée dans le tenseur d'Einstein, membre de gauche de l'équation, tandis que le contenu matériel est décrit par le tenseur énergie-impulsion apparaissant dans le membre de droite. Ce dernier peut être issu d'une théorie des champs, auquel cas les équations de cette théorie se retrouvent couplées à l'équation d'Einstein. La résolution de cette équation est en général un problème très complexe, puisqu'il s'agit d'un système d'équations aux dérivées partielles non-linéaires qui déterminent les composantes de la métrique de l'espace-temps. Une fois la métrique connue, il est possible de calculer les trajectoires de particules tests à l'aide de l'équation des géodésiques. Une autre propriété remarquable de la relativité générale est qu'elle permet d'étendre le principe de relativité à tous les référentiels, même non-inertiels : les lois de la physique s'expriment ainsi de la même façon pour tous les observateurs, à condition de remplacer les dérivées usuelles par des dérivées covariantes.

La première validation expérimentale de la relativité générale fut fournie par la trajectoire de Mercure, en particulier le mouvemement de précession de son périhélie. Einstein mis en évidence dès 1915 que les prédictions de sa théorie concernant l'obite de Mercure étaient meilleures que celles de la théorie newtonienne. Les succès de la relativité générale furent par la suite nombreux. On peut citer par exemple la déviation des rayons lumineux qui fut vérifiée expérimentalement par Eddington \cite{1920a} en 1919. Un autre effet remarquable est la dilatation gravitationnelle du temps : le temps s'écoule plus lentement à la surface de la Terre qu'en altitude. Cet effet a pu être vérifié expérimentalement à l'aide d'horloges atomiques \cite{Hafele1972,Hafele1972a} et doit être pris en compte par les satellites du système GPS. Parmi les autres prédictions les plus importantes de la théorie d'Einstein figurent l'expansion de l'Univers, les ondes gravitationnelles et les trous noirs. Notre attention se portera désormais sur ces derniers, qui occupent une place centrale dans le domaine d'étude de cette thèse.

Historiquement, la première solution décrivant un trou noir en relativité générale est aussi la première solution non triviale aux équations d'Einstein à avoir été découverte. Le physicien Karl Schwarzschild cherchait une solution exacte décrivant le champ gravitationnel autour d'un corps sphérique statique. La solution qu'il trouva en 1916, appelée métrique de Schwarzschild \cite{Schwarzschild1916}, possède une propriété étrange : l'espace-temps qu'elle décrit possède une frontière qui ne peut être traversée que dans un seul sens. On appelle cette frontière l'horizon des évènements. Il convient de noter que rien, pas même la lumière, ne peut s'échapper de l'horizon, si bien qu'un observateur situé à l'intérieur ne peut communiquer d'aucune façon avec le monde extérieur. La région à l'intérieur de l'horizon est ce que l'on appelle de nos jours un trou noir. Pour un astre de rayon suffisamment important, l'horizon des évènements est absent car l'espace-temps à l'intérieur de l'objet est décrit par une métrique qui n'est pas celle de Schwarzschild. Pendant longtemps il était alors admis que l'horizon décrit par la métrique de Schwarzschild n'était qu'un artefact mathématique, et que tous les objets physiques de l'Univers avaient des rayons suffisamment grands pour éviter l'apparition d'un horizon. Néanmoins, les travaux successifs de Chandrasekhar \cite{Chandrasekhar1931}, Tolman, Oppenheimer et Volkoff \cite{Tolman1939,Oppenheimer1939} dans les années 30 sur les naines blanches et les étoiles à neutrons -- des étoiles essentiellement constituées de fermions -- mirent en évidence l'existence d'une masse limite au-delà de laquelle l'effondrement gravitationnel ne pouvait être entravé par la pression de dégénérescence des fermions. Oppenheimer et Synder \cite{Oppenheimer1939a} effectuèrent notamment le premier calcul d'effondrement gravitationnel conduisant à la formation d'un trou noir. Bien que ce résultat soit basé sur un modèle de la composition des étoiles à neutrons particulièrement simpliste, il est désormais largement admis que l'effondrement d'étoiles extrêmement massive mène à la formation d'un trou noir. 

Après les travaux de Schwarzschild, des solutions décrivant des trous noirs plus généraux furent découvertes : la métrique de Reissner-Nordström \cite{Reissner1916,Weyl1917,Nordstrom1918} décrivant des trous noirs portant une charge électrique ou magnétique, la métrique de Kerr \cite{Kerr1963} qui décrit des trous noirs en rotation, et enfin la métrique de Kerr-Newman \cite{Newman1965} pour des trous noirs chargés en rotation. 

Au-delà de la théorie, la première observation indirecte d'un trou noir remonte aux années 70 avec la détection du système binaire Cygnus X-1 dans la constellation du Cygne \cite{WEBSTER1972,BOLTON1972}. Ce système est constitué d'une étoile d'environ 20 à 30 masses solaires ainsi que d'un compagnon invisible dont la masse est estimée autour de 7 à 13 masses solaires. Le fait que ce dernier soit invisible suggère qu'il s'agit d'un objet compact, autrement dit, une étoile à neutron ou un trou noir. Cependant, sa masse, estimée à l'aide des paramètres orbitaux, permet de trancher en faveur du trou noir, puisqu'on considère qu'une étoile à neutron ne peut avoir une masse supérieure à environ 3 fois la masse du soleil. Ce trou noir serait d'origine stellaire, c'est-à-dire issu de l'effondrement gravitationnel d'une étoile en fin de vie. Durant la même décennie furent détectés des objets extrêmement massifs, notamment au centre de la Voie Lactée \cite{Balick1974}. Les étoiles les plus proches du centre de notre galaxie semblent en effet orbiter autour d'un objet dont la masse est estimée à environ 4 millions de fois la masse du soleil \cite{Backer1982}. Il est aujourd'hui admis qu'il s'agit d'un trou noir dit supermassif, dont la formation diffère de celle des trous noirs stellaires. Il pourrait s'agir de l'effondrement de nuages de gaz peu de temps après le Big Bang. On dispose depuis 2022 d'une image issue de la collaboration Event Horizon Telescope qui révèle le disque d'accrétion entourant ce trou noir, nous permettant ainsi de discerner ce qui semble être l'horizon des évènements\footnote{Il faut noter que d'autres candidats alternatifs à un trou noir supermassif sont parfois envisagés. Voir par exemple les travaux de Vincent et al. \cite{Vincent2016}, ou de Herdeiro et al. \cite{Herdeiro2021c}.} \cite{EHT2022}. Il faut souligner que la résolution angulaire impressionnante de cette image, 50 microsecondes d'arc, est obtenue grâce à un réseau connecté de radiotélescopes autour du monde. Un autre moyen technique très populaire permettant la détection de trous noirs sont les observatoires d'ondes gravitationnelles. Il s'agit d'interféromètres de Michelson dotés de bras extrêmement longs (quelques kilomètres). Ces dispositifs sont capables de détecter des variations de longueurs infimes, de l'ordre du diamètre d'un proton, qui sont associées au passage d'une onde gravitationnelle. Ces dernières sont émises lors de processus astrophysiques très violents, comme par exemple la fusion de deux trous noirs. Les deux principales collaborations travaillant sur la détection de ces ondes sont LIGO aux États-Unis, et Virgo en Italie. En 2015, elles ont réalisé la première détection d'un signal d'onde gravitationnelle, dont l'analyse a confirmé qu'il résultait de la fusion de deux trous noirs \cite{Abbott2016}.

Les propriétés intrigantes des trous noirs peuvent faire penser que ces objets sont parmi les systèmes physiques les plus complexes qui puissent exister dans l'Univers. D'aucuns considèrent pourtant les trous noirs comme des « particules fondamentales gravitationnelles », les entités les plus élémentaires que l'on puisse constituer à l'aide de l'interaction gravitationnelle. En effet, en supposant que la théorie qui décrit le mieux la matière à grande échelle soit l'électromagnétisme de Maxwell, les trous noirs, en relativité générale, sont complètement caractérisés par leur masse, leur charge électrique ou magnétique, ainsi que leur moment cinétique. Ces grandeurs sont analogues à la masse, la charge et le spin des particules fondamentales du modèle standard. On dit que les trous noirs les plus généraux dans le modèle Einstein-Maxwell sont décrits par la métrique de Kerr-Newman~\cite{Mazur1982}. Il s'agit d'un théorème d'unicité, communément appelé théorème de calvitie\footnote{Nous parlons ici du théorème de calvitie s'appliquant lorsque seul le champ électromagnétique de Maxwell est couplé à la relativité générale et que l'espace-temps, contenant quatre dimensions, est asymptotiquement plat. Ce théorème repose sur d'autres hypothèses plus techniques qui ne seront pas abordées ici.}, en référence à l'aphorisme employé par John A. Wheeler~\cite{Ruffini1971} en 1971 : « les trous noirs n’ont pas de cheveux ». Historiquement, la conjecture de ce théorème (et de ses variantes) avait été proposée par des physiciens soviétiques dans les années 60 (Ginzburg, Zeldovitch et Novikov) \cite{Gourgoulhon2014}. Leur formulation de la conjecture était la suivante : les trous noirs résultant de l'effondrement gravitationnel d'étoiles massives dépendent uniquement de paramètres associés aux symétries du modèle considéré. Cette conjecture de la calvitie des trous noirs est plus générale que le théorème mentionné plus haut, puisqu'elle ne suppose pas une théorie particulière pour décrire la matière. Elle suppose en revanche implicitement la stabilité des trous noirs. Aucune preuve de la conjecture de calvitie n'existe à ce jour, c'est pourquoi nous préférons parler du théorème s'appliquant dans le modèle Einstein-Maxwell. Si certaines des hypothèses de ce théorème sont omises, on peut alors obtenir ce que l'on appelle des trous noirs chevelus. Ces solutions ont le même comportement asymptotique qu'un trou noir de Kerr-Newman, mais le voisinage de leur horizon est décrit par une configuration des champs non triviale : c'est le cheveu du trou noir. La découverte théorique des premier exemples de trous noirs chevelus est attribuée à Volkov et Galt'sov \cite{Volkov1989} en 1989. Leurs solutions numériques ont été obtenues dans le cadre de la relativité générale couplée à un champ de jauge non-Abélien plus complexe que le champ électromagnétique de Maxwell. 

Nous considérerons dans cette thèse deux possibilités permettant de s'affranchir du théorème de calvitie. La première option est de remplacer la relativité générale par une théorie de la gravitation alternative. Cela nous amène à la question de savoir pourquoi il faudrait modifier les équations d'Einstein qui semblent si bien décrire l'Univers à grande échelle. Une première limitation inhérente à la relativité générale est son incompatibilité avec la physique quantique. Il n'existe actuellement aucun consensus sur une théorie qui permettrait de décrire la gravitation dans un cadre quantique. Des candidats pour une telle théorie existent (gravitation quantique à boucles, théorie des supercordes, théories supersymétriques, $\dots$) mais se pose alors le problème de leur vérification expérimentale. En effet, la gravitation quantique intervient à l'échelle de Planck, une échelle pour le moment largement inaccessible à nos technologies. Le développement d'une théorie quantique de la gravitation pourrait permettre de résoudre les problèmes de singularités qui surviennent en relativité générale, notamment au sein des trous noirs ou lors du Big Bang. Cette thèse ne traite pas des aspects quantiques de la gravitation, c'est pourquoi nous ne nous étendrons pas davantage sur la question de la quantification de la relativité.

À l'opposé du problème de la quantification qui se pose lorsque l'on s'intéresse à des échelles de longueur très petites, on trouve aussi d'autres types de problèmes à très grande échelle. En 1917, l'astronome américain Vesto M. Slipher découvre que la lumière émise par les galaxies nous parvient, en moyenne, décalée vers le rouge \cite{Slipher1917}. Plus tard, en 1929, Edwin Hubble observe que ce décalage vers le rouge est environ proportionnel à la distance de la source par rapport à la Terre \cite{Hubble1929}. En interprétant ce phénomène à l'aide de l'effet Doppler relativiste, on en vient à la conclusion que les galaxies s'éloignent de nous à une vitesse proportionnelle à leur distance par rapport à la Terre. Cette interprétation naïve est toutefois erronée car elle impliquerait, notamment, des vitesses supérieures à celle de la lumière pour les galaxies trop éloignées. Il s'agit en réalité d'une manifestation de l'expansion de l'Univers. Pour comprendre cette idée, on peut se représenter l'espace-temps comme la membrane en latex d'un ballon de baudruche, et les galaxies comme des points dessinés sur cette membrane. Lorsque l'on gonfle le ballon, sa surface s'étire uniformément et tous les points s'éloignent les uns par rapport aux autres. Il convient de noter les limites de cette analogie : elle représente l'espace-temps comme un tissu matériel à deux dimensions qui est plongé dans un espace ambiant de dimension supérieure et elle donne un volume fini\footnote{La question de savoir si la courbure spatiale de l'Univers est positive, nulle ou négative n'est pas tranchée à ce jour. En fonction de ces trois possibilités, on obtient un Univers de volume fini (courbure positive) ou infini (courbure nulle ou négative).} à l'Univers. Elle permet néanmoins de comprendre une idée essentielle : l'éloignement apparent des galaxies s'explique par le fait que leurs positions relatives sont dépendantes de la dynamique même de l'espace-temps. L'expansion de l'Univers peut être décrite par la relativité générale dans le cadre du modèle de Friedmann-Lemaître-Robertson-Walker (FLRW) \cite{Friedman1922,Friedmann1924,Lemaitre1927,Robertson1935,Robertson1936,Robertson1936a,Walker1937}, du nom des quatre physiciens ayant travaillé sur la question de l'expansion. Leurs travaux sont le fondement théorique de ce qui constitue de nos jours le modèle standard de la cosmologie (aussi connu sous le nom de modèle de concordance). Néanmoins, en 1998, les observations astronomiques menées par Adam G. Riess \cite{Riess1998} et Saul Perlmutter~\cite{Perlmutter1999} démontrèrent que l'expansion de l'Univers est en fait accélérée. Pour expliquer ce phénomène dans le cadre théorique fourni par la relativité générale, on doit supposer que l'Univers est rempli à 68\% d'un ingrédient appelé énergie sombre : il s'agit d'une forme d'énergie aux propriétés très étranges puisque sa pression est \textit{négative}, ce qui lui confère un caractère répulsif. Mathématiquement, l'énergie sombre peut être modélisée en ajoutant à l'équation d'Einstein la fameuse constante cosmologique. Il est cependant très difficile d'admettre qu'une telle forme d'énergie puisse réellement exister. Une alternative à l'énergie sombre est alors de considérer que l'équation d'Einstein doit être modifiée pour décrire les phénomènes de très grande échelle, tels que l'expansion de l'Univers. Cette idée est l'une des motivations qui amena les physiciens théoriciens à développer de nouvelles théories (classiques) de la gravitation. Citons par exemple les théories tenseur-scalaire, qui décrivent l'interaction gravitationnelle par la combinaison du tenseur métrique ainsi que d'un champ scalaire, les théories métrique-affine qui introduisent un cadre géométrique plus général que la théorie d'Einstein, ou la gravité massive qui attribue une masse finie au graviton afin de donner à l'interaction gravitationnelle une portée finie. Dans certaines de ces théories, il est possible d'obtenir des trous noirs chevelus sans même devoir invoquer de champ matériel (espace-temps vide). Un exemple récent en théorie tenseur-scalaire est discuté dans l'article \cite{Aelst2020}. Dans le chapitre~\ref{chap_bigrav} de cette thèse, nous étudierons un autre exemple de trou noir chevelu dans une théorie de gravité massive à deux métriques. Nous renvoyons le lecteur à l'ouvrage \cite{Saridakis2021} pour plus de détails sur les théories alternatives de la gravitation et leurs applications en cosmologie. 

Un autre problème ouvert est celui de la matière noire. En mesurant les vitesses de rotation des étoiles autour de leur centre galactique, on observe que les étoiles les plus externes tournent plus vite que ce que prévoit la théorie \cite{Rubin1980}. La prédiction théorique peut se faire dans le cadre de la théorie newtonienne, elle se base sur la « masse lumineuse » des galaxies -- c'est-à-dire la masse que l'on peut estimer en fonction de leur luminosité. Cela suppose que toute la masse des galaxies est constituée d'étoiles. En réalité, tout se passe comme si les galaxies contenaient une forme de matière invisible, c'est-à-dire, n'interagissant pas par interaction électromagnétique, que l'on nomme matière noire. La densité de cette matière doit être suffisante pour contribuer significativement à la masse totale des galaxies. La présence de matière noire a par la suite été mise en évidence dans le milieu intergalactique par des effets de lentille gravitationnelle\footnote{Une lentille gravitationnelle, ou mirage gravitationnel, est un effet de déviation de la lumière qui se produit lorsque qu'une distribution de masse se trouve entre un observateur et une source lumineuse lointaine.} \cite{Taylor1998}. On estime que la matière noire doit représenter 27\% du contenu total de l'Univers, c'est-à-dire cinq fois plus que la matière baryonique ordinaire. Comme pour l'énergie sombre, on peut choisir d'aborder ce problème en tentant de modifier la théorie de la gravitation. Cependant, compte tenu du nombre important d'observations indépendantes indiquant la présence de matière noire, la communauté astrophysique admet communément l'existence de cette forme de matière invisible \cite{Hossenfelder2018}, bien que sa nature exacte demeure encore inconnue à ce jour.

Revenons maintenant aux trous noirs chevelus. Nous avons vu qu'il était possible d'en obtenir en considérant une théorie de la gravitation alternative. La seconde option que nous considérerons dans cette thèse est de garder les équations d'Einstein, mais de choisir un contenu matériel autre que le champ électromagnétique. Les équations de Maxwell qui décrivent la dynamique de ce champ s'appliquent à l'échelle macroscopique; nous savons toutefois qu'à plus hautes énergies, cette théorie doit être remplacée par la théorie électrofaible de Weinberg \cite{Weinberg1967} et Salam \cite{Salam1959,Salam1964} qui unifie l'interaction nucléaire faible et l'électromagnétisme. Les symétries de cette théorie sont plus complexes, ce qui implique un plus grand nombre de degrés de libertés physiques, et donc une plus grande richesse dans le spectre des solutions. En espace-temps plat, la théorie électrofaible admet des solutions de type « monopôles magnétiques » \cite{Cho1996,Gervalle2022a,Gervalle2023}. Ces configurations présentent un champ magnétique non-nul, qui se comporte asymptotiquement comme un champ électrique coulombien. Malheureusement, la théorie électrofaible n'est capable d'aucune prédiction expérimentale sur ces monopôles puisqu'elle leur attribue une masse infinie, associée à une singularité coulombienne à l'origine. La prise en compte de la gravitation permet une régularisation des monopôles électrofaibles grâce à la présence d'un horizon des évènements pouvant masquer la singularité centrale de tout observateur externe. On obtient alors des trous noirs porteurs d'une charge magnétique qui peuvent être chevelus. De tels trous noirs ont été considéré en 1994 par Lee et Weinberg dans le cadre d'une théorie des champs proche de la théorie électrofaible \cite{Lee1994}. Le physicien sud-américain Juan M. Maldacena évoque à nouveau ce type de trou noir bien plus tard, en 2021, sans pour autant fournir de solution explicite \cite{Maldacena2021}. Il mentionne que, dans leur version chevelue, les trous noirs électrofaibles sont entourés d'une région dans laquelle la symétrie complète de la théorie est restaurée. Ce phénomène physique est associé à de très hautes énergies, potentiellement inaccessibles dans les accélérateurs de particules. Asymptotiquement, la symétrie électrofaible est brisée et la théorie se réduit à l'électromagnétisme de Maxwell. Des versions à symétrie sphérique de ces trous noirs électrofaibles chevelus furent construites numériquement par un groupe américain \cite{Bai2021}. Néanmoins, les solutions à symétrie sphériques ont une taille de l'ordre de la longueur de Planck, ce qui compromet la validité de la théorie classique. Afin d'éviter de devoir prendre en compte des effets quantiques, il faut donc obtenir des solutions de plus grande taille. Maldacena estime que les trous noirs électrofaible chevelus peuvent avoir une taille de l'ordre du centimètre. Il reste néanmoins à clarifier le mécanisme permettant leur production, mais on peut raisonnablement considérer que ces trous noirs, s'ils existent, sont d'origine primordiale, c'est-à-dire issus de fluctuations quantiques des champs aux tout premiers instants de l'Univers \cite{Stojkovic2005,Bai2020a}. 

\newpage

\section*{Contenu de la thèse}
\addcontentsline{toc}{section}{Contenu de la thèse}

Dans cette thèse, nous proposons l'étude des trous noirs chevelus dans deux modèles différents. Notre objectif est de rendre nos résultats accessibles à un large public, notamment à des étudiants en master curieux ou qui envisagent une thèse dans des domaines proches (trous noirs chevelus, gravité modifiée, théorie des champs non-Abéliens, $\dots$). Pour cela, nous alternons entre des chapitres présentant de nouveaux résultats publiés dans des revues scientifiques, et des chapitres rappelant certaines notions théoriques essentielles pour la compréhension de nos travaux. Des connaissances de base en relativité générale et en théorie classique des champs sont toutefois recommandées afin d'appréhender au mieux le contenu de cette thèse. L'organisation du manuscrit est la suivante.

Le \hyperref[bh_gr]{premier chapitre} est une introduction courte aux trous noirs en relativité générale. Nous commençons par rappeler l'équation d'Einstein ainsi que la formulation variationnelle de la relativité générale qui repose sur l'action de Einstein-Hilbert. Cette formulation sera très utile lorsque nous présenterons d'autres modèles. Le tenseur énergie-impulsion est introduit comme la dérivée variationnelle du Lagrangien de la matière par rapport à la métrique. Nous présentons ensuite la solution non triviale la plus simple de l'équation d'Einstein dans le vide : la métrique de Schwarzschild. Cet exemple simple nous permet d'aborder des notions clefs associées aux trous noirs : l'horizon des évènements, la gravité de surface et la température de l'horizon. Nous aborderons brièvement le processus d'évaporation des trous noirs par le rayonnement de Hawking. Avant de passer à des trous noirs plus généraux, nous expliquons comment les quantités telles que la masse, la charge et le moment cinétique sont définies pour un espace-temps asymptotiquement plat. Nous présentons alors la métrique de Reissner-Nordstr{\"o}m (trous noirs chargés) puis la métrique de Kerr (trous noirs en rotation). Nous terminons ce chapitre par une discussion autour du théorème de calvitie.

Dans le \hyperref[chap_bigrav]{deuxième chapitre}, nous étudions des trous noirs chevelus dans un espace-temps vide décrit par la théorie de la bigravité massive. Cette théorie a été proposée par Hassan et Rosen \cite{Hassan2012} en 2012, et elle est connue pour ses solutions cosmologiques qui sont capables de décrire un Univers en expansion accélérée sans devoir recourir à la constante cosmologique. La bigravité massive décrit la dynamique de deux champs tensoriels qui sont associés à deux gravitons : l'un est sans masse, tandis que le second est massif. C'est l'interaction entre ces deux champs qui produit naturellement une constante cosmologique \textit{effective}. La théorie est aussi capable de décrire des trous noirs, on y retrouve notamment le trou noir le plus simple de la relativité générale, celui de Schwarzschild. Malheureusement, ce dernier est instable en dessous d'une certaine valeur critique pour son rayon d'horizon \cite{Babichev2013}. En parallèle du trou noir de Schwarzschild, des solutions chevelues asymptotiquement plates ont été découvertes par Brito, Cardoso et Pani \cite{Brito2013a} en 2013. Cette découverte est le point de départ de nos travaux. Tout d'abord, notons que l'existence de ces trous noirs chevelus a été remise en question ultérieurement par un groupe suédois~\cite{Torsello2017}. Nos résultats confirment l'existence des solutions chevelues. Pour résoudre les équations du champ de la théorie, nous avons utilisé un schéma numérique plus sophistiqué que celui employé par Brito et al. En particulier, les équations sont intégrées en partant exactement de l'horizon des évènements -- un point singulier du système d'équations différentielles. De plus, nous avons pris en compte des corrections non-linéaires dans la région asymptotique, afin de nous assurer que les solutions construites sont bien asymptotiquement plates. Dans un second temps, nous avons voulu analyser la stabilité des trous noirs chevelus dans le cadre de la théorie des perturbations linéaires. Nous montrons que ces solutions peuvent être stables ou instables, en fonction des paramètres de la théorie. Nous mettons en évidence une région de l'espace des paramètres dans laquelle les solutions chevelues sont stables alors que les trous noirs de Schwarzschild de mêmes masses ne le sont pas. Pour cette région de l'espace des paramètres, la théorie est capable de décrire des solutions cosmologiques en accord avec les observations astronomiques. Ainsi, si la dynamique de l'espace-temps est effectivement régie par la théorie de la bigravité, l'instabilité des trous noirs de Schwarzschild pourrait conduire à la formation de cheveux. La masse des trous noirs ainsi formés pourrait aller de quelques masses solaires (trous noirs stellaires) à environ un million de fois la masse du Soleil. Des trous noirs encore plus massifs peuvent toujours être décrits en bigravité par la solution de Schwarzschild, qui est stable pour de très grandes masses. 

Le \hyperref[chap_bigrav]{troisième chapitre} est un interlude proposant une courte introduction à des notions apparaissant en théorie de jauge. Nous commençons par décrire le mécanisme de brisure spontanée de symétrie, en partant du cas d'une symétrie discrète, jusqu'au cas le plus pertinent pour nos travaux : la brisure d'une symétrie de jauge. Les théories de jauge dans lesquelles la symétrie est spontanément brisée donnent lieu à des solutions de type «~monopôles magnétiques~». Nous décrivons le monopôle le plus simple qui apparaît dans le contexte de l'électromagnétisme de Maxwell, puis nous abordons ses généralisations dans les théories de jauge non-Abéliennes. Ce sera également l'occasion de faire quelques rappels sur la notion de champ de jauge non-Abélien. 

Dans le \hyperref[chap_mon]{quatrième chapitre}, nous étudions des trous noirs chevelus qui portent une charge magnétique dans le cadre du secteur bosonique de la théorie électrofaible de Weinberg et Salam couplé à la relativité générale. La théorie électrofaible est une partie du modèle standard des particules, un modèle ayant largement fait ses preuves expérimentalement, notamment avec la découverte du boson de Higgs en 2012 \cite{Aad2012,Chatrchyan2012}. Notre point de départ est la découverte par Bai et Korwar \cite{Bai2021} de trous noirs chevelus à symétrie sphérique chargés magnétiquement. Il s'agit d'une généralisation gravitationnelle du monopôle non-Abélien de Cho et Maison \cite{Cho1996}. Les solutions de Bai et Korwar correspondent à la plus petite charge magnétique qui peut être supportée par des champs non-Abéliens. Malheureusement, la taille de leur horizon, de l'ordre d'une dizaine de fois la longueur de Planck, en fait des objets très spéculatifs, puisqu'ils sont construits dans le cadre de la théorie classique des champs, sans tenir compte de corrections quantiques. Pour remédier à ce problème, nous proposons l'étude d'un cas plus général, celui d'un espace-temps à symétrie axiale. Cela nous permet d'obtenir des trous noirs chevelus de charge magnétique plus élevée, et donc, de plus grande taille. Avant de construire de telles solutions, nous étudions la stabilité des trous noirs de Reissner-Nordstr{\"o}m magnétiques dans la théorie. Nous constatons que pour la même charge magnétique que les solutions chevelues de Bai et Korwar, les trous noirs de Reissner-Nordstr{\"o}m présentent une instabilité dans le secteur à symétrie sphérique. L'évolution de cette instabilité pourrait naturellement conduire à la formation de cheveux. Pour des charges magnétiques plus élevées, l'instabilité de Reissner-Nordstr{\"o}m persiste, mais n'est plus dans le secteur à symétrie sphérique. C'est pourquoi les cheveux qui pourraient se former dans ce cas ne peuvent pas non plus être invariants sous toutes les rotations spatiales. Nous revenons ensuite aux trous noirs chevelus de Bai et Korwar, en proposant une étude plus approfondie de leurs propriétés. Nous passons enfin au cas axisymétrique, en commençant par la construction des monopôles en espace-temps plat. Ces derniers généralisent le monopôle de Cho-Maison à des charges magnétiques plus élevées et n'avaient jamais été construits auparavant. Le cas gravitationnel est finalement abordé, ce qui nous conduit à des trous noirs dont les cheveux sont constitués d'un anneau porteur de charge non-Abélienne et de deux boucles de courant électrique de sens opposés. La construction de ces trous noirs électrofaibles  nécessite la résolution d'un système de dix équations aux dérivées partielles elliptiques. Nous utilisons pour cela la méthode des éléments finis.

Le \hyperref[chap_mon]{cinquième chapitre} présente des résultats annexes qui ont été obtenues à l'aide de la méthode des éléments finis. Cette méthode n'a pratiquement jamais été utilisée pour des problèmes gravitationnels. Avant de l'appliquer à la construction de trous noirs électrofaibles, nous avons souhaité l'utiliser dans un cadre plus simple. On considère pour cela un champ scalaire complexe couplé à la relativité générale. Si l'on admet une dépendance temporelle harmonique pour le champ scalaire, il existe des solutions stationnaires appelées étoiles à bosons \cite{Lee1992a,Friedberg1987,Mielke1998}. Le potentiel du champ scalaire peut contenir uniquement un terme de masse, mais dans ce chapitre, on considère un potentiel incluant des auto-interactions. Dans ce contexte, les solutions présentent une limite en espace-temps plat connue sous le nom de $Q$-balls. Il s'agit d'un exemple particulier de soliton non-topologique : des solutions localisées dans l'espace qui ont une énergie finie et dont l'existence est garantie par la conservation d'une charge de Noether. Notre point de départ est l'article de Herdeiro, Kunz, Perapechka, Radu et Shnir \cite{Herdeiro2021} dans lequel des chaînes d'étoiles à bosons statiques sont construites à l'aide de la méthode des différences finies. Nous reproduisons leurs résultats en utilisant la méthode des éléments finis, et proposons ensuite une généralisation naturelle de ces chaînes en ajoutant de la rotation. Des tests de convergence numérique sont effectués et un argument qualitatif quant à la stabilité de ces structures est proposé. Enfin, nous étudions la limite en espace-temps plat des chaînes d'étoiles à bosons, qui n'existe que dans le cas avec rotation.

Nous concluons cette thèse par une synthèse générale et proposons des perspectives pour l'extension de nos divers travaux. Ce manuscrit comprend également une \hyperref[app_num]{annexe} décrivant les différentes méthodes numériques que nous avons utilisées pour la résolution des équations différentielles.

\newpage

\section*{Notations et conventions}
\addcontentsline{toc}{section}{Notations et conventions}

Nous représentons l'espace-temps par une variété différentielle de dimension 4 -- une dimension temporelle, trois dimensions spatiales. L'espace-temps est doté d'une métrique lorentzienne $g$ de signature $(-,+,+,+)$. Nous utiliserons toujours un système de coordonnées $(x^0,x^1,x^2,x^3)$ qui sera adapté aux symétries du problème considéré. Les indices spatio-temporels seront dénotés par des lettres grecques telles que $\mu,\nu,\dots\in\{0,1,2,3\}$. Les indices prenant uniquement des valeurs de $1$ à $3$ seront dénotés par des lettres latines telles que $i,j,a,b,\dots$ Des exceptions à ces règles peuvent survenir au cours du manuscrit~; le cas échéant, elles seront explicitement signalées afin d'éviter toute ambiguïté. Nous écrivons,
\begin{center}
\begin{tabular}{ll}
    \hline
    $\partial_\mu=\partial/\partial x^\mu$ & le champ de vecteurs associé à $x^\mu$,  \\
    $dx^\mu$ & le champ de 1-formes associé à $x^\mu$, \\
    $g_{\mu\nu}$ & les composantes de la métrique $g$, \\
    $\delta^\mu_\nu=\tensor{g}{^\mu_\nu}$ & le symbole de Kronecker, \\
    $\epsilon_{abc},\;\epsilon_{\mu\nu\alpha\beta}$ & le symbole de Levi-Civita, \\
    \hline
    $\sqrt{-g}$ & la racine carrée de la valeur absolue du déterminant de $(g_{\mu\nu})$,\\
    $\nabla_\mu$ & la dérivée covariante géométrique associée à $g$, \\
    $\Gamma^\sigma_{\mu\nu}$ & le symbole de Christoffel de $g$, \\
    $R_{\mu\nu\rho\sigma}$ & le tenseur de Riemann de $g$, \\
    $R_{\mu\nu}=\tensor{R}{^\alpha_\mu_\alpha_\nu}$ & le tenseur de Ricci de $g$, \\
    $R=\tensor{R}{^\mu_\mu}$ & le scalaire de Ricci de $g$, \\
    $G_{\mu\nu}$ & le tenseur d'Einstein de $g$. \\
    \hline
\end{tabular}
\end{center}
Nous définissons les symboles de Christoffel par,
\begin{equation*}
    \Gamma^\sigma_{\mu\nu}=\frac{1}{2}g^{\sigma\rho}\left(\partial_\mu g_{\rho\nu}+\partial_\nu g_{\mu\rho}-\partial_\rho g_{\mu\nu}\right),
\end{equation*}
et le tenseur de Riemann par,
\begin{equation*}
    \tensor{R}{^\mu_\nu_\rho_\sigma}=\partial_\rho\Gamma^\mu_{\nu\sigma}-\partial_\sigma\Gamma^\mu_{\nu\rho}+\Gamma^\mu_{\lambda\rho}\Gamma^\lambda_{\nu\sigma}-\Gamma^\mu_{\lambda\sigma}\Gamma^\lambda_{\nu\rho}.
\end{equation*}

Dans le chapitre~\ref{chap_bigrav}, une deuxième métrique $f$ est introduite. Nous noterons alors $G(g)_{\mu\nu}$ (resp. $G(f)_{\mu\nu}$) le tenseur d'Einstein associé à la métrique $g$ (resp. $f$), et de même pour les autres quantités géométriques. Nous utilisons la convention d'Einstein de sommation des indices répétés. Les tenseurs seront souvent désignés par leurs composantes, par exemple, $\tensor{T}{^\mu_\nu}$ se réfère au tenseur $T$ défini comme suit,
\begin{equation*}
    T=\tensor{T}{^\mu_\nu}\partial_\mu\otimes dx^\nu.
\end{equation*}
Le symbole $T$ désignera alors plutôt la trace du tenseur.

Lorsque nous introduisons des grandeurs sans dimension, nous noterons leurs homologues avec dimension en utilisant des symboles en gras. Nous notons,
\begin{center}
\begin{tabular}{ll}
    \hline
    $c$ & la vitesse de la lumière dans le vide, \\
    $\hbar$ & la constante de Planck réduite, \\
    $G$ & la constante gravitationnelle de Newton, \\
    $\kappa=8\pi G/c^4$ & la constante gravitationnelle d'Einstein, \\
    $k_B$ & la constante de Boltzmann. \\
    \hline
\end{tabular}
\end{center}
Sauf mention explicite du contraire, nous choisissons la convention $c=\hbar=k_B=1$. Ainsi, le temps et l'espace ont la même dimension physique, la masse et la température ont la dimension de l'inverse d'une longueur, et les constante $G$ et $\kappa$ ont la dimension d'une longueur au carré.

Nous utilisons les unités de Gauss pour les quantités électromagnétiques. Ainsi, les équations de Maxwell s'écrivent, $\nabla_\nu F^{\mu\nu}=4\pi\,J_e^\mu$, où $F_{\mu\nu}$ est le tenseur électromagnétique et $J^\mu$ est le quadrivecteur densité de courant électrique.

    \renewcommand{\chaptermark}[1]{\markboth{\MakeUppercase{\chaptername\ \thechapter. {#1}}}{}}

\chapter{Black holes in general relativity}
\label{bh_gr}

The notion of black holes was first proposed by the British physicist John Michell in 1783, long before the theory of General Relativity was developed. Indeed, in the Newtonian theory of gravitation, one can imagine an object so massive and small that its escape velocity is greater than the speed of light. This object would be invisible since no light ray could escape from its gravitational field. The same idea was rediscovered by Pierre-Simon de Laplace in 1796. However the concept of Newtonian black holes relies on the particle nature of light and requires light particles to be massive. The particle description of light was challenged in 1801 by Young's interference experiment which demonstrated the wave-like behavior of light. Of course, from a modern viewpoint, the wave-particle duality in quantum mechanics specifies that light can behave as a wave, or as a particle, but in the latter case the light particles are massless which is not compatible with the concept of Newtonian black hole. 

In 1915, Albert Einstein \cite{Einstein1915} came up with his famous theory of General Relativity (GR), a geometric theory of gravitation which generalizes the principles of special relativity and refines Newton's law of universal gravitation. The basic idea behind Einstein's theory has been summarized by John Wheeler as follows: matter tells spacetime how to curve, and curved spacetime tells matter how to move. In other words, the gravitational interaction is a consequence of spacetime curvature, and the concept of gravitational force has to be abandoned.

In 1916, Karl Schwarzschild found the first non-trivial exact solution to the Einstein equations \cite{Schwarzschild1916}. His solution describes a spherically symmetric gravitational field -- the spacetime metric -- in vacuum. It can describe the external field around a spherical object such as, for example, a star. However, the Schwarzschild solution contains the so-called \textit{event horizon} whose circumference depends linearly on the mass of the object. It corresponds to a spacetime boundary beyond which gravity is so strong that nothing, not even light, can escape. If the object radius is larger than the horizon radius then nothing special happens since the interior spacetime is not described by the Schwarzschild metric. For a long time, it was considered that all objects in the Universe were large enough to avoid the appearance of an event horizon. However, the successive works of Chandrasekhar \cite{Chandrasekhar1931} in 1931 and of Tolman, Oppenheimer, Volkoff \cite{Tolman1939,Oppenheimer1939,Oppenheimer1939a} in 1939 demonstrated that the gravitational collapse of relativistic stars, when they exceed certain maximal masses, cannot be counterbalanced by the degeneracy pressure of fermions. As a very massive star collapses, its radius shrinks, eventually becoming smaller than the horizon radius of the Schwarzschild metric. The spacetime region inside the horizon is what we call nowadays a black hole.

In this chapter, we review the different black hole solutions existing in GR and their properties. For a more detailed discussion about GR and black holes, we refer the reader to the textbooks \cite{Wald1984,Misner2017,Carroll2019}. In Section~\ref{ein_eq_sch}, we introduce the Einstein equation and present the Schwarzschild solution. In Section~\ref{global_quant}, we explain how to compute classical quantities such as mass, charge, and angular momentum within an asymptotically flat spacetime. This leads us to Sections~\ref{intro_rn} and \ref{intro_kerr} where we introduce the spacetime metrics describing respectively charged and rotating black holes. Finally, Section~\ref{more_gen_bh} presents the no-hair theorem in GR, its implications, and how to circumvent it.

\section{Einstein equation and the Schwarzschild solution}
\label{ein_eq_sch}

The Einstein field equation relates the spacetime curvature with the energy, momentum and stress contained in that spacetime. It can be written in the following form,
\begin{equation}
\label{ein_eq_intro}
    G_{\mu\nu}=\kappa\,T_{\mu\nu},
\end{equation}
where $G_{\mu\nu}$ is the Einstein tensor describing the curvature of spacetime, $T_{\mu\nu}$ is the stress-energy tensor of the non-gravitational fields and $\kappa=8\pi G$ with $G$, the Newtonian constant of gravitation. The Einstein tensor is constructed from the spacetime metric denoted $g_{\mu\nu}$ as 
\begin{equation}
    G_{\mu\nu}=R_{\mu\nu}-\frac{1}{2}g_{\mu\nu}R,
\end{equation}
where $R_{\mu\nu}$ is the Ricci tensor and $R=g_{\mu\nu}R^{\mu\nu}$ is the Ricci scalar. The Einstein equation must respect the conservation of energy, which can be expressed in curved spacetime by 
\begin{equation}
    \nabla^\mu T_{\mu\nu}=0.
\end{equation}
This is compatible with the Einstein equation by virtue of the (contracted) Bianchi identity,
\begin{equation}
    \nabla^\mu R_{\mu\nu}=\frac{1}{2}\nabla_\nu R.
\end{equation}

The Einstein equation \eqref{ein_eq_intro} can be derived by means of a variational principle applied to the action,
\begin{equation}
\label{einstein-hilbert}
    S=\int{\left(\frac{1}{2\kappa}R+\mathcal{L}_\text{M}\right)\sqrt{-g}\,d^4 x},
\end{equation}
where $\mathcal{L}_\text{M}$ is the Lagrangian density describing non-gravitational fields and $g=\det(g_{\mu\nu})$ is the determinant of the metric tensor. The first part of this action containing the Ricci scalar $R$ is called the \textit{Einstein-Hilbert action}. Imposing that the variations of \eqref{einstein-hilbert} with respect to any small metric fluctuations $\delta g_{\mu\nu}$ vanish implies the Einstein field equation \eqref{ein_eq_intro}. The stress-energy tensor is derived from the variations of $\mathcal{L}_\text{M}$ with respect to the metric,
\begin{equation}
    T_{\mu\nu}=-\frac{2}{\sqrt{-g}}\frac{\delta(\sqrt{-g}\,\mathcal{L}_\text{M})}{\delta g^{\mu\nu}}.
\end{equation}
The description of GR in terms of the action \eqref{einstein-hilbert} allows for easy unification with any other classical field theory. For example the electromagnetism is coupled to GR by taking $\mathcal{L}_\text{M}=-(1/4)F_{\mu\nu}F^{\mu\nu}$ where $F_{\mu\nu}$ is the electromagnetic tensor. The Einstein-Hilbert action is also the starting point for modifying GR. In practice, modified gravity theories are often defined by their action, not by their field equations.

Let us now consider vacuum solutions in GR. We may first rewrite the Einstein equation in the following equivalent form,
\begin{equation}
\label{ein_alt}
    R_{\mu\nu}=\kappa\left(T_{\mu\nu}-\frac{1}{2}g_{\mu\nu}T\right),
\end{equation}
where $T=g_{\mu\nu}T^{\mu\nu}$ is the trace of the stress-energy tensor. In vacuum, one has $T_{\mu\nu}=0$ so that the Einstein equation reduces to,
\begin{equation}
\label{vac_ein}
    R_{\mu\nu}=0.
\end{equation}
Even in this case, the equation is far from simple to solve. It consists of 10 nonlinear Partial Differential Equations (PDEs) which determine the components of the metric. Taking into account the freedom to choose the coordinate system, this can be reduced to 6 PDEs. The solution established by Schwarzschild in the special case of a spherically symmetric and static spacetime can be written in the following form,
\begin{equation}
\label{sch_intro}
    ds^2=g_{\mu\nu}dx^\mu dx^\nu=-\left(1-\frac{2GM}{r}\right)dt^2+\left(1-\frac{2GM}{r}\right)^{-1}dr^2+r^2d\Omega^2,
\end{equation}
where $d\Omega^2=d\vartheta^2+\sin^2\vartheta\,d\varphi^2$ is the metric on a unit 2-sphere in spherical coordinates and $M$ is a free parameter with the dimension of a mass. To interpret correctly the mass $M$, one can for example consider the geodesic equation,
\begin{equation}
    \frac{d^2 x^\alpha}{d\tau^2}+\Gamma^\alpha_{\mu\nu}\frac{dx^\mu}{d\tau}\frac{dx^\nu}{d\tau}=0,
\end{equation}
where $\tau$ is the proper time measured by a test particle\footnote{A test particle is an idealized physical system which, subjected to the action of an external field, does not influence it in return.} and $\Gamma^\alpha_{\mu\nu}$ are the Christoffel symbols associated with the spacetime metric. This equation determines the trajectory $x^\mu(\tau)$ of the test particle, hence, it can be seen as the equivalent of Newton's second law in GR. If the particle moves only in the $r$-direction in the Schwarzschild spacetime \eqref{sch_intro}, its acceleration is given by $d^2 r/d\tau^2=-GM/r^2$, in accordance with the Newtonian result if we identify $\tau$ with the absolute time of classical mechanics and $M$ with the mass of the gravitational source. This correspondence holds in the asymptotic region where GR should reduces to Newton's law of gravitation. We can further identify $r$ with the spherical radial coordinate in this region.

In GR, one should be careful with the physical interpretation of the coordinates used to describe a given solution. First, by taking the limit $r\to\infty$, the Schwarzschild solution reduces to the Minkowski metric,
\begin{equation}
    ds^2=-dt^2+dr^2+r^2\left(d\vartheta^2+\sin^2\vartheta\,d\varphi^2\right).
\end{equation}
This property is referred to as asymptotic flatness\footnote{Of course, this gives only a coordinate-dependent meaning of what is an asymptotically flat spacetime. The notion of asymptotic flatness can be defined independently of the system of coordinates, see for example the section 4.3.2. of Ref.~\cite{Gourgoulhon2023}.}. The timelike coordinate $t$ can be interpreted as the time measured by the clock of an observer at spatial infinity. Then, let us consider the metric induced by Eq.~\eqref{sch_intro} on a hypersurface with  $t$ and $r$ constant. This is equivalent to taking $dt=dr=0$ in Eq.~\eqref{sch_intro} and we obtain the metric on a 2-sphere of radius $r$. Thus the $(\vartheta,\varphi)$ coordinates are just the same spherical coordinates as we are used to in flat spacetime. One can also identify $r$ as a quantity related to the area $A$ of the 2-sphere by $A=4\pi r^2$. But can we identify $r$ as the distance between the 2-sphere and its center? The answer is negative. Indeed, from Eq.~\eqref{sch_intro}, one can see that an infinitesimal change $dr$ of the $r$ coordinate is related to an infinitesimal proper distance $ds$ by
\begin{equation}
\label{prop_dist}
    ds=\left(1-\frac{2GM}{r}\right)^{-1/2}dr,
\end{equation}
hence the proper distance between two events A and B located at $r=r_\text{A}$ and $r=r_\text{B}$ is greater than the difference $|r_\text{A}-r_\text{B}|$. Moreover, if $r\leq 2GM$, then the quantity in the right hand side of Eq.~\eqref{prop_dist} is not well-defined. The special hypersurface with $r=r_H\equiv 2GM$ is actually an event horizon.

The metric components $g_{rr}$ is not defined at $r=r_H$. However, the components of the metric are coordinate-dependent quantities. A direct computation of curvature invariants such as the Ricci scalar $R$ or the Kretschmann scalar $R_{\mu\nu\alpha\beta}R^{\mu\nu\alpha\beta}$ reveals that the curvature is not divergent at $r=r_H$. Therefore the apparent singularity here is only a coordinate singularity: something that has nothing to do with an actual divergence of any physical quantity. On the other hand the metric components and the curvature invariants both diverge for $r=0$. This means that $r=0$ is a true curvature singularity where the theory breaks down.

If $r<r_H$, the component $g^{rr}$ is negative which means that $r$ is a timelike coordinate. This explains why Eq.~\eqref{prop_dist} is not valid if $r<r_H$: a change of radial coordinate $r$ in this case is not associated with a proper distance but with a proper time. At the same time, if  $r<r_H$, $g_{tt}$ is positive so that $t$ is a spacelike coordinate. Similarly, hypersurfaces with $r=\text{const}.$ are timelike if $r>r_H$, spacelike if $r<r_H$ and in the limiting case $r=r_H$, the hypersurface is null. This is a first important property: an event horizon is a null hypersurface. It is worth noting however that not all null hypersurfaces are event horizons. For example the light cone is a null hypersurface in Minkowski spacetime but it is not an event horizon. The change of nature of the hypersurfaces with $r=\text{const.}$ has an important consequence. Indeed, spacelike hypersurfaces are one-way membranes: the trajectories of physical particles (\textit{a.k.a.} causal curves) can only pass through them in one direction. Therefore, if a particle crosses the event horizon, all its possible trajectories are necessarily oriented in the direction of decreasing $r$. In other words, it cannot escape back to spatial infinity. This key feature is what defines the event horizon. It is worth noting that the event horizon is a \textit{global} concept: one must consider the possibility of reaching spatial infinity to conclude whether or not a null hypersurface is an event horizon. For a more rigorous and technical definition of event horizons, see for example \cite{Gourgoulhon2023}.

Another possibility to characterize the event horizon is to consider Killing vector fields. These are associated with the symmetries of spacetime. A Killing vector $K$ satisfies the so-called Killing equation,
\begin{equation}
\label{killing_eq}
    \nabla_\mu K_\nu+\nabla_\nu K_\mu=0.
\end{equation}
Roughly speaking, the Killing vectors indicate the directions along which the geometry does not change. For example the Schwarzschild geometry \eqref{sch_intro} has three rotational Killing vectors which are associated with the spherical symmetry plus one Killing vector associated with the time-translation symmetry. The latter is of particular interest in the context of our discussion about the event horizon of the Schwarzschild metric. Up to an overall constant, this Killing vector is $\partial_t$ and it goes from being timelike to spacelike at the event horizon, just as the hypersurfaces with constant $r$. We say that the event horizon is a \textit{Killing horizon} of $\partial_t$: it is a null hypersurface with the normal vector being the Killing vector $\partial_t$. In GR, every event horizon in a stationary and asymptotically flat spacetime is a Killing horizon for some Killing vector $K$. This property is known as Hawking's rigidity theorem (see for example Sec.~5.5 of Ref.~\cite{Gourgoulhon2023}). Then, to every Killing horizon $\Sigma$ we can associate a quantity called the surface gravity $\kappa_g$ which is defined by
\begin{equation}
\label{surf_grav_intro}
    \kappa_g^2=-\left.\frac{1}{2}\nabla_\mu K_\nu\,\nabla^\mu K^\nu\right|_\Sigma.
\end{equation}
If $\kappa_g=0$, the horizon is said to be \textit{degenerate}. If the Killing horizon is an event horizon then we define the temperature of the black hole as
\begin{equation}
    T_H=\frac{\kappa_g}{2\pi}.
\end{equation}
According to the zeroth law of black hole thermodynamics\footnote{We refer to the Chap.~16 of \cite{Gourgoulhon2023} for a discussion on the other laws of black hole thermodynamics and the assumptions on which they are based.}, this quantity is constant on the horizon. For a Schwarzschild black hole described by the metric \eqref{sch_intro}, one has $\kappa_g=1/(4GM)$ and $T_H=1/(8\pi GM)$. The notion of black hole temperature has been introduced by Stephen Hawking in 1974 \cite{Hawking1974}. It is based on the idea that black holes should emit thermal radiation due to quantum effects\footnote{Quantum effects here are taken into account by a semi-classical approach. One considers quantum fields on a black hole background but one neglects the back-reaction on the spacetime geometry.} near the event horizon. This process is known as the Hawking radiation and it causes the \textit{evaporation} of the black hole as it loses mass over time. For stellar or larger black holes, Hawking radiation is negligible due to its extremely low intensity. However, for primordial black holes formed during the early Universe, the role of Hawking radiation becomes significant. Neutral and spherically symmetric primordial black holes could evaporate completely through this process. We shall see in Secs.~\ref{intro_rn} and \ref{intro_kerr} that this picture is different for charged or rotating black holes.

As mentioned in the introduction of this section, the Schwarzschild metric does not only describes black holes. It also describes the exterior geometry surrounding any spherical object with mass $M$. If the surface of the object is located at $r>2GM$, then there is no event horizon since the metric of the interior will be different from Schwarzschild. More importantly, the Schwarzschild solution still describes the exterior region if the massive object is collapsing into a black hole. Indeed, the Birkhoff's theorem \cite{Jebsen1921,Birkhoff1923} states that spherically symmetric solutions to the Einstein equation in vacuum must be static and are described by the Schwarzschild metric \eqref{sch_intro}.

Finally, let us discuss the central curvature singularity $r=0$ inside Schwarzschild black holes. If we assume these object to be the end state of gravitational collapse of spherically symmetric mass distributions, the appearance of a singularity is pathological. This actually reveals the limit of the classical theory: singularities are expected to be regularized by taking into account quantum effects.

\section{Mass, charge and angular momentum}

\label{global_quant}

We have seen that a Schwarzschild black hole is characterized by a unique parameter, its mass $M$. Before moving on to different black hole solutions, we shall define the so-called global quantities in curved spacetime. These are physical quantities that can be measured by an observer far from the gravitational source such as for example, the mass, the electric (or magnetic) charge and the angular momentum.

Let us consider first the case of the electric charge. Maxwell's equations in a curved spacetime read 
\begin{equation}
    \nabla_\nu F^{\mu\nu}=4\pi J_e^\mu,
\end{equation}
where $F_{\mu\nu}$ is the electromagnetic tensor and $J_e^\mu$ is the 4-current density. The charge $Q$ inside a spacelike hypersurface $\Sigma$ is then given by,
\begin{equation}
\label{charge_vol}
    Q=-\int_\Sigma{d^3 x\sqrt{\gamma}\,n_\mu J_e^\mu}=-\frac{1}{4\pi}\int_\Sigma{d^3 x\sqrt{\gamma}\,n_\mu\nabla_\nu F^{\mu\nu}},
\end{equation}
where $\gamma$ is the determinant of the induced metric on $\Sigma$ and $n^\mu$ is the unit normal vector to $\Sigma$. The minus sign ensures that a positive charge distribution and a future-directed normal vector yield a positive total charge. We know from Gauss's theorem that the volume integral in Eq.~\eqref{charge_vol} can be transformed into a surface integral over the boundary $\partial\Sigma$. This means that the knowledge of the electromagnetic field outside the charge distribution is sufficient to compute the total charge. From a practical point of view, this is of particular interest for black holes because it can be difficult to compute the integral \eqref{charge_vol} in the presence of a singularity inside $\Sigma$. Applying Gauss's theorem to the last expression in Eq.~\eqref{charge_vol}, we obtain 
\begin{equation}
\label{charge_surf}
    Q=-\frac{1}{4\pi}\oint_{\partial\Sigma}{d^2 x\sqrt{\gamma^{(2)}}\,n_\mu \sigma_\nu F^{\mu\nu}},
\end{equation}
where $\gamma^{(2)}$ is the determinant of the induced metric on $\partial\Sigma$ and $\sigma^\mu$ is the outgoing unit normal vector to $\partial\Sigma$. This formula gives the total charge enclosed by the surface $\partial\Sigma$. Thus to compute the electric charge in the whole spacetime, we can choose this surface to be a 2-sphere at spatial infinity. The total magnetic charge is computed by replacing in the above formula the electromagnetic tensor by its Hodge dual,
\begin{equation}
    *F^{\mu\nu}=\frac{1}{2\sqrt{-g}}\epsilon^{\mu\nu\alpha\beta}F_{\alpha\beta},
\end{equation}
where $\epsilon^{\mu\nu\alpha\beta}$ is the antisymmetric Levi-Civita symbol.

We turn now to the total mass of an asymptotically flat spacetime. The concept of mass (or energy) is more subtle to define in GR. In classical field theory, the energy density is typically described by the $(t,t)$ component of the stress-energy tensor. However in GR, this tensor describes only the properties of the matter fields and not those of the gravitational field (the spacetime curvature). It turns out that different definitions of the mass exist in GR and they agree only under some specific assumptions \cite{Wald1984}.

By analogy with the definition of the charge, we would like to construct a conserved 4-current to be integrated as in Eq.~\eqref{charge_vol}. If the spacetime is stationary, a first candidate could be 
\begin{equation}
    J_T^\mu=\xi_\nu T^{\mu\nu},
\end{equation}
where $\xi^\mu$ is the asymptotically timelike Killing vector associated with the time-translation symmetry. The divergence of $J_T^\mu$ is
\begin{equation}
    \nabla_\mu J_T^\mu=T^{\mu\nu}\nabla_\mu\xi_\nu+\xi_\nu\nabla_\mu T^{\mu\nu}=0,
\end{equation}
where the first term vanishes since one has $\nabla_\mu\xi_\nu=-\nabla_\nu\xi_\mu$ (Killing equation) and the second one is also zero by virtue of the conservation of energy. One can then define a conserved quantity by integrating $J_T^\mu$ over a spacelike hypersurface. However the corresponding volume integral cannot be transformed into a surface integral of the form~\eqref{charge_surf}. 

A second candidate for a conserved 4-current is,
\begin{equation}
    J_R^\mu=\xi_\nu R^{\mu\nu}.
\end{equation}
One can check that it is indeed conserved by computing its divergence,
\begin{equation}
    \nabla_\mu J_R^\mu=R^{\mu\nu}\nabla_\mu\xi_\nu+\xi_\nu\nabla_\mu R^{\mu\nu}=\frac{1}{2}\xi_\nu\nabla^\nu R=0,
\end{equation}
where the last expression vanishes because the directional derivative of the Ricci scalar along a Killing vector field is zero (this is related to the fact that the geometry does not change in the direction of a Killing vector). Then we define the total mass by integrating this 4-current, 
\begin{equation}
\label{komar_vol}
    M_\text{Komar}=\frac{1}{4\pi G}\int_\Sigma{d^3 x\sqrt{\gamma}\,n_\mu\xi_\nu R^{\mu\nu}}=2\int_\Sigma{d^3 x\sqrt{\gamma}\,n_\mu\xi_\nu\left(T^{\mu\nu}-\frac{1}{2}g^{\mu\nu}T\right)},
\end{equation}
where we have used the Einstein equation \eqref{ein_alt} to obtain the second expression and the normalization factor has been introduced for future convenience. This definition of the mass is due to the physicist Arthur Komar \cite{Komar1963}. It can be rewritten as a surface integral by using another identity which holds for any Killing vector field,
\begin{equation}
    \nabla_\nu\nabla_\mu\xi^\nu=\xi^\nu R_{\mu\nu}.
\end{equation}
Substituting this into Eq.~\eqref{komar_vol} and using Gauss's theorem we obtain,
\begin{equation}
\label{komar_surf}
    M_\text{Komar}=\frac{1}{4\pi G}\oint_{\partial\Sigma}{d^2 x\sqrt{\gamma^{(2)}}\,n_\mu \sigma_\nu\nabla^\mu\xi^\nu}.
\end{equation}
One can check that this definition for the mass agrees with the expected result\footnote{More rigorously, the Komar mass in the form \eqref{komar_vol} yields a vanishing mass for the Schwarzschild metric. This is because the maximal extension of Schwarzschild spacetime actually contains two asymptotic boundaries and the corresponding surface integral of the form \eqref{komar_surf} should contain two contributions which cancel each other.} for the Schwarzschild metric \eqref{sch_intro}, $M_\text{Komar}=M$. This formula is sometimes called the \textit{Komar integral} associated with the timelike Killing vector $\xi^\mu$.

Another definition of the mass is due to Arnowitt, Deser and Misner (ADM). Their Hamiltonian formulation of GR \cite{Arnowitt1959} allows for a natural definition of the conserved energy for an asymptotically flat spacetime. At spatial infinity, one can introduce small metric fluctuations $h_{\mu\nu}$ around Minkowski, 
\begin{equation}
    g_{\mu\nu}=\eta_{\mu\nu}+h_{\mu\nu},
\end{equation}
where $\eta_{\mu\nu}$ is the Minkowski metric. The ADM energy (or ADM mass) is then defined as
\begin{equation}
    M_\text{ADM}=\frac{1}{16\pi G}\oint_{\partial\Sigma}{d^2 x\sqrt{\gamma^{(2)}}\,\sigma^i\left(\partial_j\tensor{h}{^j_i}-\partial_i\tensor{h}{^j_j}\right)}.
\end{equation}
It turns out that the ADM mass coincides with the Komar mass if $h_{\mu\nu}$ is time-independent at infinity. In GR, the ADM definition of the mass is associated with a positive energy theorem which ensures that it is non-negative under reasonable assumptions on the matter fields\footnote{See for example the section 6.4 in Ref.~\cite{Carroll2019} for more details.}.

Let us finally consider the case of the angular momentum. It is defined by a Komar integral associated with the rotational Killing vector $\chi^\mu$ of spacetime,
\begin{equation}
    J_\text{Komar}=-\frac{1}{8\pi G}\oint_{\partial\Sigma}{d^2 x\sqrt{\gamma^{(2)}}\,n_\mu \sigma_\nu\nabla^\mu\chi^\nu},
\end{equation}
where only the normalization factor differs from the Komar integral for the mass \eqref{komar_surf}. Note that in an adapted system of coordinates such that the rotation is along the azimuthal direction $\varphi$, the rotational Killing vector is $\chi=\partial_\varphi$. 

\section{Charged black holes}

\label{intro_rn}

Non-rotating black holes carrying an electric charge and/or a magnetic charge are described in GR by the \textit{Reissner-Nordström} (RN) solution. It has been discovered between 1916 and 1921 by Hans Reissner \cite{Reissner1916}, Hermann Weyl \cite{Weyl1917}, Gunnar Nordström \cite{Nordstrom1918} and George Barker Jeffery \cite{Jeffery1921} independently. 

The RN metric is a solution to the Einstein equation \eqref{ein_eq_intro} coupled to the vacuum Maxwell's equation,
\begin{equation}
\label{maxwell_rn}
    \nabla_\mu F^{\mu\nu}=0,
\end{equation}
where $F_{\mu\nu}=\partial_\mu A_\nu-\partial_\nu A_\mu$ is the electromagnetic tensor, which is also referred to as the field strength tensor or Faraday tensor. The stress-energy tensor entering the Einstein equation in this case is given by
\begin{equation}
    T_{\mu\nu}=F_{\mu\lambda}\tensor{F}{_\nu^\lambda}-\frac{1}{4}g_{\mu\nu}F_{\alpha\beta}F^{\alpha\beta}.
\end{equation}
The field equations can be recovered by varying the following action,
\begin{equation}
    S=\int{d^4 x\sqrt{-g}\left(\frac{R}{2\kappa}-\frac{1}{4}F_{\mu\nu}F^{\mu\nu}\right)}.
\end{equation}
It is sometimes referred to as the Einstein-Maxwell action. The gauge potential describing a pointlike electric charge and a pointlike magnetic charge\footnote{The aim this chapter is not to provide a precise description of magnetic charges, for this topic, we refer to Sec.~\ref{monop}.} is
\begin{equation}
\label{gauge_rn}
    A_\mu dx^\mu=\frac{Q}{r}dt+P(\cos\vartheta-1)\,d\varphi,
\end{equation}
where $P$ and $Q$ are constants. The radial components of the electric and magnetic fields are respectively $E^r=F^{tr}=-Q/r^2$ and $B^r=*F^{tr}=-P/r^2$ so that the potential \eqref{gauge_rn} describes an electric charge $-Q$ and a magnetic charge $-P$. The RN metric can then be written as
\begin{equation}
\label{rn_intro}
    ds^2=-N(r)\,dt^2+\frac{dr^2}{N(r)}+r^2 d\Omega^2,
\end{equation}
with 
\begin{equation}
    N(r)=1-\frac{2GM}{r}+\frac{4\pi G}{r^2}\left(P^2+Q^2\right).
\end{equation}
One can check that the fields \eqref{gauge_rn} and \eqref{rn_intro} solve the equations \eqref{ein_eq_intro} and \eqref{maxwell_rn}. The system of coordinates here has the same interpretation as for the Schwarzschild metric \eqref{sch_intro}. The latter is recovered if the charges are set to zero, $P=Q=0$. Note that the $1/r^2$ dependence of the electric and magnetic fields is what we are used to in flat space. However it should be emphasized that the $r$ coordinate here approaches that of Minkowski spacetime only in the limit $r\to\infty$.

\begin{figure}
    \centering
    \includegraphics[scale=1.0]{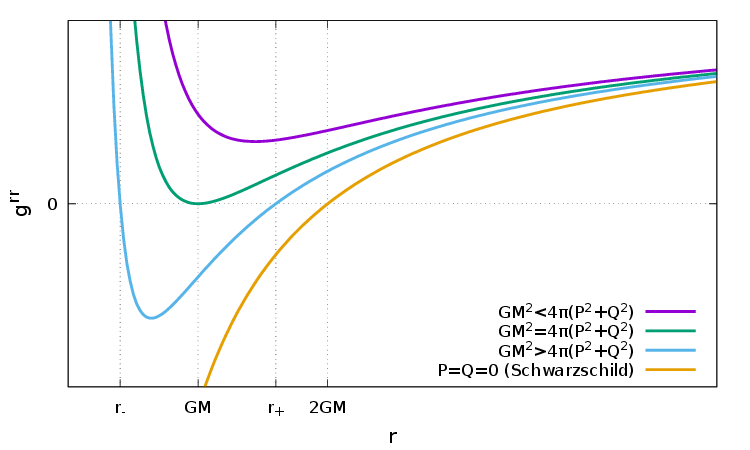}
    \caption[Profiles of the metric component $g^{rr}$ for a Reissner-Nordström black hole in the three typical cases]{Profiles of component $g^{rr}$ of the RN metric against $r$ for three typical cases. The zeroes indicate the location of an event horizon. The Schwarzschild case with $P=Q=0$ is also shown for comparison.}
    \label{fig_rn}
\end{figure}

The RN spacetime has a true curvature singularity at $r=0$ where the curvature invariants diverge. The event horizon can be located by determining where the hypersurfaces with $r=\text{const.}$ become null. Since $\partial_\mu r$ is a one-form normal to these hypersurfaces with squared norm
\begin{equation}
    g^{\mu\nu}(\partial_\mu r)(\partial_\nu r)=g^{rr},
\end{equation}
it suffices to determine at which $r$ the component $g^{rr}=N(r)$ of the inverse metric vanishes. Of course this simple condition for locating the event horizon is coordinate-dependent. Here one has $g^{rr}(r)=0$ for
\begin{equation}
\label{rpm}
    r=r_\pm\equiv GM\pm \sqrt{G^2M^2-4\pi G(P^2+Q^2)}.
\end{equation}
Hence the horizon structure of RN spacetime is more complicated than for Schwarzschild as there can be two, one or zero event horizons, see the Fig.~\ref{fig_rn}. We shall therefore distinguish the three following cases.
\begin{itemize}
    \item $GM^2<4\pi(P^2+Q^2)$ (no horizon).\\
    In this case there is no event horizon, hypersurfaces with constant $r$ are always timelike. This means that there is no obstruction to an observer to approach the central singularity and return to report what he observed. This is called a \textit{naked singularity}, a curvature singularity without event horizon. In 1969 the physicist Roger Penrose proposed the so-called \textit{cosmic censorship conjecture} \cite{Penrose1969} which states that any singularity that forms in the Universe as a result of gravitational collapse must be hidden behind an event horizon, thus preventing it from being seen by external observers. It turns out that there are actually good reasons to consider the RN solution with $GM^2<4\pi(P^2+Q^2)$ as physically unacceptable. Roughly speaking, this condition means that the total energy of the solution is less than the energetic contribution of the electromagnetic field alone. In other words, this would correspond to charges with negative masses. 
    \item $GM^2>4\pi(P^2+Q^2)$ (two horizons).\\
    This situation is considered to be more realistic as the energy of the electromagnetic field is less than the total energy. The spacetime in this case has two event horizons which are located at $r=r_\pm$ as in Eq.~\eqref{rpm}. As seen from from the outside, the outer horizon $r_{+}$ is similar to that of a Schwarzschild black hole but what happens inside is very different. When going through the first horizon, hypersurfaces with constant $r$ become spacelike so that an infalling observer necessarily moves in the direction of decreasing $r$. Then he reaches the inner horizon $r_{-}$ where the hypersurfaces switch back to being timelike, hence the observer can stop his motion at some finite value $r=r_0<r_{-}$. He is now free to choose to move in the direction of increasing $r$. At $r=r_{-}$, the hypersurfaces once again become spacelike so that the observer is now forced to move in the direction of increasing $r$. He will eventually be ejected through the outer horizon $r_{+}$, thus reaching the exterior region. One may think that this observer has been able to get into and out of a black hole. However he is now in a different spacetime region, causally disconnected from where he was before entering the black hole. This can be seen by maximally extending the RN metric \eqref{rn_intro} and constructing the corresponding conformal diagram (see the figure 6.4 in Ref.~\cite{Carroll2019}). The spacetime in this case consists of infinitely many asymptotically flat region connected to each other via wormholes. One must keep in mind that this picture is actually unphysical as classical GR is certainly not appropriate to describe what happens close to the central singularity, in the region $r\leq r_{-}$.
    \item $GM^2=4\pi(P^2+Q^2)$ (one horizon).\\
    This case is known as the \textit{extremal} RN solution. The metric describes a single event horizon located at $r=GM$ where the corresponding hypersurface is null. However the hypersurfaces with constant $r$ are timelike on either side. What happens inside the horizon is similar to the previous case without the region $r_{-}<r<r_{+}$. It is worth noting that the temperature of extremal RN black holes is zero hence the horizon is degenerate. A careful computation reveals that if a non-extremal RN black hole evaporates through the Hawking process, it can only approach the extremal limit but it will never reach it. 
\end{itemize}

\section{Rotating black holes}

\label{intro_kerr}

While the solutions describing static black holes presented above were found soon after GR was developed, the solution for rotating black holes was found by Roy Kerr only in 1963 \cite{Kerr1963}. The equation to be solved in this case is that of vacuum GR, $R_{\mu\nu}=0$, but the spherical symmetry has to be abandoned. The spacetime metric of a Kerr black hole is only stationary and axially symmetric, thus it has two Killing vectors which can be expressed (in adapted coordinates) as,
\begin{equation}
    \xi=\partial_t,\quad\quad\chi=\partial_\varphi.
\end{equation}
The \textit{Kerr metric} can be written in the following (complicated) form,
\begin{align*}
    ds^2=&-\left(1-\frac{2GMr}{\rho^2}\right)dt^2-\frac{4GMa}{\rho^2}r\sin^2\vartheta\,dt\,d\varphi\\
\label{kerr}
    &+\frac{\rho^2}{\Delta}dr^2+\rho^2 d\vartheta^2+\frac{\sin^2\vartheta}{\rho^2}\left((r^2+a^2)^2-a^2\Delta\sin^2\vartheta\right)d\varphi^2,\numberthis{}
\end{align*}
where
\begin{equation}
\label{q_kerr}
    \Delta\equiv r^2-2GMr+a^2,\quad\quad\rho^2\equiv r^2+a^2\cos^2\vartheta.
\end{equation}
The Kerr solution is thus parameterized by two free parameters, $M$ and $a$, which correspond respectively to the mass and the angular momentum per unit mass. Note that it is also possible to include electric and magnetic charges $Q$ and $P$, the result is called the \textit{Kerr-Newman metric} \cite{Newman1965}. However the inclusion of charges does not change significantly the phenomena that occur in a Kerr spacetime so we will consider here only the Kerr metric \eqref{kerr}.

In the limit $a\to 0$, the Schwarzschild metric \eqref{sch_intro} is recovered. However when $M$ vanishes, the line element \eqref{kerr} reduces to the Minkowki metric but not in usual spherical coordinates. The coordinates $(t,r,\vartheta,\varphi)$ which are used here are known as Boyer-Lindquist coordinates \cite{Boyer1967} and in the flat space limit one has, 
\begin{align*}
    x&=\sqrt{r^2+a^2}\sin\vartheta\cos\varphi,\\
    y&=\sqrt{r^2+a^2}\sin\vartheta\sin\varphi,\\
    z&=r\cos\vartheta,\numberthis{}
\end{align*}
where $(x,y,z)$ are the usual Cartesian coordinates of the Euclidean 3-space. Hence the parameter $a$ is not anymore related to an angular momentum in this limit.

A direct inspection of the Kretschmann scalar reveals that the curvature singularity of the Kerr metric is located where $\rho(r,\vartheta)=0$. From Eq.~\eqref{q_kerr}, this occurs at
\begin{equation}
\label{ring_sing}
    (r,\vartheta)=(0,\pi/2).
\end{equation}
This should not seems surprising as $r=0$ is not a single point in space but a disk, and the set of points which satisfy \eqref{ring_sing} is actually the boundary of this disk. Hence the singularity inside a Kerr black hole has the topology of a ring.

As before, the event horizons are located where the hypersurfaces with constant $r$ are null. This occurs when $g^{rr}=\Delta/\rho^2=0$ and since $\rho^2\geq 0$, we obtain the quadratic equation,
\begin{equation}
    \Delta(r)=r^2-2GMr+a^2=0.
\end{equation}
Thus, as for the RN geometry, there are three possibilities: $GM>a$ (two horizons), $GM=a$ (one horizon) and $GM<a$ (no horizon). The latter case corresponds to a naked singularity and must be excluded while for $GM\geq a$ the event horizons are located at
\begin{equation}
    r_\pm=GM\pm\sqrt{G^2M^2-a^2}.
\end{equation}
Therefore, the horizon structure of Kerr spacetime is similar to that of RN and the conformal diagram consists of infinitely many asymptotically flat exterior regions connected via various wormholes (see the figure 6.8 in Ref.~\cite{Carroll2019}). However the event horizons of Kerr black holes are not Killing horizons for the time-translation Killing vector $\xi$. Instead, they are Killing horizons for a linear combination of $\xi$ and $\chi$. For example, for the outer horizon, the Killing vector which is null at $r=r_{+}$ is
\begin{equation}
    K_{+}\equiv\xi+\Omega_H\,\chi,
\end{equation}
where 
\begin{equation}
    \Omega_H\equiv\frac{a}{2GMr_{+}}=\frac{a}{2GM(GM+\sqrt{G^2M^2-a^2})}.
\end{equation}
Hence the surface gravity and the temperature of Kerr black holes are computed by using $K_{+}$ in Eq.~\eqref{surf_grav_intro}. 

The norm of the time-translation Killing vector $\xi$ is
\begin{equation}
    \xi^\mu\xi_\mu=-(1-\frac{2GMr}{\rho^2})=-\frac{1}{\rho^2}(\Delta-a^2\sin^2\vartheta),
\end{equation}
and in particular at $r=r_{+}$ (where $\Delta=0$) one has $\xi^\mu\xi_\mu=a^2\sin^2\vartheta/\rho^2\geq 0$ so that $\xi$ is already spacelike at the outer horizon. The norm of $\xi$ is null on the surface such that
\begin{equation}
    (r-GM)^2=G^2M^2-a^2\cos^2\vartheta,
\end{equation}
which is called the \textit{stationary limit surface}. The region between this surface and the outer horizon is referred to as the \textit{ergosphere} (or \textit{ergoregion}). In this region, the trajectories with constant $(r,\vartheta,\varphi)$ are not allowed since they are tangent to $\xi$ which is spacelike there. However it is possible for an observer to keep constant his coordinates $r$ and $\vartheta$ while moving in the $\varphi$ direction. As a result, an observer inside the ergoregion necessarily moves in the direction of the black hole rotation. Of course this observer is able to escape the ergoregion as he is still outside the event horizon. 

The presence of an ergoregion have an important physical consequence known as the \textit{Penrose process} \cite{Penrose1971}. The energy of a particle is defined as
\begin{equation}
\label{energy_kerr}
    E=-\xi_\mu p^\mu,
\end{equation}
where $p$ is the particle's four-momentum. Outside the ergoregion, both $\xi$ and $p$ are timelike so that the energy \eqref{energy_kerr} is positive-definite. However inside the ergoregion, $\xi$ is spacelike which means that it is possible for particles to have a negative energy. Now imagine an observer entering the ergoregion from the outside with energy $E^{(0)}>0$. Once he enters the ergoregion, he throws an object inside the black hole in such a way that the energy of the object is negative. If we call $E^{(1)}$ the energy of the observer after leaving the ergoregion and $E^{(2)}$ the energy of the object thrown into the black hole, we have
\begin{equation}
    E^{(0)}=E^{(1)}+E^{(2)},
\end{equation}
and since $E^{(2)}<0$, it follows that $E^{(0)}<E^{(1)}$. Thus, the observer has left the ergoregion with more energy than he entered with. Of course this energy does not come from nowhere and it can be shown that the negative energy object falling into the black hole causes the angular momentum of the hole to decrease. Therefore, energy has been extracted from the black hole rotation.

Finally, let us mention black holes are not the only objects that can have an ergoregion, see for example the rotating boson stars in Chap.~\ref{chap_bs}.

\section{More general black holes?}
\label{more_gen_bh}

We have seen in the above sections black hole solutions that are characterized by their mass $M$, their electric/magnetic charges $Q$, $P$ and their angular momentum (per unit mass) $a$. Are there more general solutions? In the spherically symmetric case, the Birkhoff's theorem ensures that the Schwarzschild metric \eqref{sch_intro} is the only solution of vacuum GR. Including charges, the Birkhoff's theorem can be generalized to prove that the only spherically symmetric and asymptotically flat solution to the Einstein-Maxwell field equations is the RN metric \eqref{rn_intro}. These results can be seen as gravitational counterparts of the situation in classical electromagnetism. Indeed, in a region free of charges, the only spherically symmetric field configuration is the Coulombian field described by the gauge potential \eqref{gauge_rn}.

What happens beyond spherical symmetry? For a planet, the external field depends on the specific topography. If one decomposes the metric into multipole moments, then an infinite number of coefficients is required to describe the external field exactly. One might imagine the situation to be similar for black holes. However, it turns out that the number of parameters needed to describe the metric of a stationary black hole is always very small. If the only non-gravitational field is the electromagnetic one, we have a uniqueness theorem which states that stationary and asymptotically flat black hole solutions that are non-singular outside the event horizon are fully characterized by their mass, their electric/magnetic charges and their angular momentum. This theorem is often referred to as a \textit{no-hair theorem}, named after the famous statement by Wheeler: \textit{"black holes have no hair"} \cite{Ruffini1971}. This means that black holes are completely characterized by a small amount of parameters. Specifically, the Kerr-Newmann metric describes the most general black hole solution in GR coupled to electromagnetism. The no-hair theorem considered here follows from different results from Israel \cite{Israel1967}, Carter \cite{Carter1971}, Hawking \cite{Hawking1972}, Robinson \cite{Robinson1975} and Mazur \cite{Mazur1982}. The Ref.~\cite{Heusler1996} provides proofs of the different black hole uniqueness theorems in GR.

The no-hair theorem states nothing about the non-stationary case. However, stationary black holes are of particular interest as they are expected to be the end state of gravitational collapse. For example, if we consider oscillating configurations, then they will lose energy through the emission of gravitational waves so that the oscillations are necessarily damped. 

An important consequence of the no-hair theorem is the \textit{information loss paradox} \cite{Hawking1976}. Consider a very complicated collection of matter fields (for example, a star) and collapse it into a black hole. According to the no-hair theorem, the final configuration is completely determined by its mass, electric/magnetic charges and angular momentum, implying that information about the initial configuration seems to have been lost during the collapse. In classical GR, one usually states that the missing information is somehow "hidden" behind the event horizon. However, black holes should also evaporate through the emission of Hawking radiation and they may eventually disappear thus leaving no trace of the information hidden inside. It is often speculated that the Hawking radiation itself may encode the information about the original configuration. In any case, a theory of quantum gravity would certainly cure this issue.

In this thesis, we investigate \textit{hairy} black holes. These are black holes that arise when some assumptions of the no-hair theorem are relaxed. Their asymptotic behavior is the same as for a Kerr-Newman black hole, but the vicinity of their horizon is described by a non-trivial field configuration: this is the black hole "hair". We shall consider two possibilities which lead to hairy black holes. The first option is to consider a different theory of gravitation than GR. Within the so-called modified gravity theories, examples of hairy black holes are numerous. For instance, Ref.~\cite{Aelst2020} presents rotating hairy black hole solutions in the context of scalar-tensor theories, in which the gravitational interaction is described not only by the metric tensor, but also by a scalar field. In Chap.~\ref{chap_bigrav} we will present asymptotically flat hairy black holes in the massive bigravity theory. The second possibility to evade the no-hair theorem is to keep GR but to couple it to a different field than the electromagnetic one. The very first examples of hairy black holes were reported in 1989 by Volkov and Galt'sov in this context \cite{Volkov1989,Volkov1990} (see also Ref.~\cite{Volkov1999} for a review). In their work, they considered non-Abelian Yang-Mills fields minimally coupled to GR. It turns out that hairy black holes arise within even simpler models. For instance, in Ref.~\cite{Herdeiro2014}, Herdeiro and Radu constructed hairy rotating black hole solutions in the simple framework of a massive complex scalar field minimally coupled to GR. In Chap.~\ref{chap_mon}, we will construct black holes with magnetically charged hair by coupling GR to the non-Abelian fields of the electroweak theory. For a recent review of hairy black holes that arise in different contexts, we refer the reader to Ref.~\cite{Volkov2017}.

\chapter[Hairy black holes in massive bigravity]{Asymptotically flat hairy black holes in massive bigravity}
\label{chap_bigrav}
\subtitle{This chapter is based on \cite{Gervalle2020}.}

\section{Introduction}

Among the several options for modifying GR, theories with massive gravitons have aroused strong interest since the pioneering work of Fierz and Pauli (FP) \cite{FP1939a} in 1939. In their paper, they present a linear theory describing a massive spin-2 field. The kinetic part of the FP Lagrangian is the same as for GR linearized around flat spacetime, but their theory also contains a mass term that is suitably chosen to avoid the presence of nonphysical degrees of freedom. In this sense, the fundamental field $h_{\mu\nu}$ of the FP theory can be viewed as a massive field (a massive graviton) evolving in the flat Minkowskian spacetime whereas the metric fluctuation in linearized GR corresponds to a massless field (a massless graviton).

To understand the motivation of giving a mass to the graviton, one can have the following qualitative argument. A massless interaction as gravitation in GR is known to be of \textit{infinite} range. Therefore, if the Universe is filled only with ordinary matter or energy\footnote{Ordinary matter/energy is characterized by positive pressure and energy density.}, the corresponding (attractive) gravitational interactions forbid the acceleration of the expansion. On the other hand, a massive graviton is associated to a \textit{finite} range interaction that can presumably allow accelerated expansion of the Universe without invoking exotic types of energy such as Dark Energy. The FP theory is thus a natural candidate to describe such a massive gravitational interaction at the linear level.

Many decades after the original paper describing the FP theory, van Dam, Veltman and Zakharov discovered that this linear theory suffers from a pathology. In presence of matter fields, the predictions of the FP massive gravity differ from those of linearized GR, even in the limit of vanishing graviton mass \cite{Dam1970,Zakharov1970}. This so-called vDVZ \textit{discontinuity} ruled out massive graviton theories since modifications of GR must agree with actual observations, for example, at the level of the Solar system. However, two years later, in 1972, Vainshtein showed that this issue can be cured by taking into account nonlinear corrections to the FP theory \cite{Vainshtein1972}. The nonlinearities are expected to restore GR at short scales by screening the massive interactions, while in the far field region where the linear theory applies, one should recover the FP massive gravity. This is refered to as the \textit{Vainshtein screening mechanism}.

Unfortunately, shortly after this finding, the seek for a nonlinear theory of massive gravity encountered a new issue. Boulware and Deser shown that massive gravity theories with a mass term reducing to that of FP upon linearization generically propagate an additional degree of freedom with negative energy \cite{Boulware1972}. The latter is referred to as the \textit{Boulware-Deser ghost} and renders the theory unstable: any small fluctuation around a given field configuration would grow without bound.

For almost forty year the idea of a massive graviton was abandoned until 2010 when de Rham, Gabadadze and Tolley finally constructed an explicit example of a nonlinear theory of massive gravity free of the Boulware-Deser ghost \cite{Rham2011}. Their theory, commonly called the dRGT theory, is described by the action
\begin{equation}
\label{dRGT}
    \boldsymbol{S}_\text{dRGT}=\frac{1}{\boldsymbol{\kappa}}\int\left(\frac{1}{2}\boldsymbol{R}(g)-\boldsymbol{m}^2\mathcal{U}(g,f)\right)\sqrt{-g}\,d^4\boldsymbol{x},
\end{equation}
where $\boldsymbol{\kappa}=8\pi\boldsymbol{G}$ is the gravitational coupling, $\boldsymbol{R}(g)$ is the (dimensionful) scalar curvature of the spacetime metric $g_{\mu\nu}$, $\boldsymbol{m}$ is the graviton mass, and $\mathcal{U}(g,f)$ is a potential that depends on $g_{\mu\nu}$, but also on a second fixed metric $f_{\mu\nu}$, which is usually called the \textit{reference} metric. The latter is a necessary ingredient for the construction of a FP-like mass term. In the original dRGT theory, the reference metric is chosen to be Minkowski $f_{\mu\nu}=\eta_{\mu\nu}$. The Boulware-Deser ghost is avoided by a specific choice of the potential $\mathcal{U}$ -- see Eq.~\eqref{pot_hr} below. 

Unfortunately, it turns out that flat and closed FLRW cosmologies do not exist in the original dRGT theory with a flat reference metric \cite{DAmico2011}. Although such cosmological solutions can be constructed by choosing more a general reference metric, it was shown that these cosmologies suffer from instabilities \cite{Felice2013}. Moreover, the introduction of a reference metric that is completely fixed seems to violate Einstein's "no prior geometry" requirement. A natural way to solve this issue is to render the reference metric dynamical by adding a Einstein-Hilbert term for $f_{\mu\nu}$ in the dRGT action \eqref{dRGT}. Remarkably, Hassan and Rosen showed that this straightforward generalization of the dRGT massive gravity was possible, \textit{i.e.} the resulting theory is free from any ghosts \cite{Hassan2012}. The action of the Hassan-Rosen massive bigravity is given in Eq.~\eqref{action_hr} below. This theory describes at the linear level two gravitons interacting together, one of which is massive and the other is massless.

A crucial question must be answered in order to interpret physically the massive bigravity theory: which metric is the one that describes the actual geometry of spacetime? Or, in other words, to what metric should the matter fields be coupled? Hassan and Rosen has considered different possibilities \cite{Hassan2012}. First, if the two metrics are coupled to the same matter fields,
\begin{equation}
    \sqrt{-g}\,\mathcal{L}_{M,1}(g_{\mu\nu},\phi),\quad\sqrt{-f}\,\mathcal{L}_{M,2}(f_{\mu\nu},\phi),
\end{equation}
then the geodesic equation would be modified, which violates the equivalence principle. Second, one can imagine a coupling to an effective metric $\hat{g}_{\mu\nu}$ constructed out of $g_{\mu\nu}$ and $f_{\mu\nu}$,
\begin{equation}
    \sqrt{-\hat{g}}\,\mathcal{L}_M(\hat{g}_{\mu\nu},\phi).
\end{equation}
Unfortunately, this would generically bring back the Boulware-Deser ghost. The remaining possibility is to consider two independent types of matter fields, say $A$ and $B$, that are each coupled to one of the metrics,
\begin{equation}
    \sqrt{-g}\,\mathcal{L}_{M,A}(g_{\mu\nu},\phi^A),\quad\sqrt{-f}\,\mathcal{L}_{M,B}(f_{\mu\nu},\phi^B).
\end{equation}
This choice preserves the ghost-free feature of the massive bigravity theory. In this case, one has two spacetime geometries: the particles of type $A$ follow the geodesics of the $g$-metric while the trajectories of type $B$ particles obey the geodesic equation of the $f$-metric. In what follows, we consider that only one type of matter exists and that it is coupled to one of the two metrics, say $g_{\mu\nu}$. The spacetime geometry is thus described by the $g$-metric and we interpret $f_{\mu\nu}$ simply as an additional tensor field coupled to the spacetime metric $g_{\mu\nu}$.

The massive bigravity theory admits viable self-accelerating cosmological solutions \cite{Volkov2012b,Strauss2012,Comelli2012,Akrami2015,Moertsell2015,Aoki2015,Lueben2020,Hoegaas2020} with the cosmological constant mimicked by the graviton mass. The theory also admits solutions describing black holes and stars \cite{Volkov2012a} or even wormholes \cite{Sushkov2015}. We refer the reader to Ref.~\cite{Volkov2013} for a review about cosmological and black hole solutions. In this chapter we shall be discussing black holes.

First, the black hole solutions of GR also exist in massive bigravity. One usually call them \textit{bald} black holes, as opposed to the \textit{hairy} ones that circumvent the no-hair theorem. The bald black holes were reported long ago \cite{Isham1978,Guerses1979,Gurses1981} in the context of a different theory with two metrics inspired by the physics of strong interaction \cite{Isham1971}. In the spherically symmetric case, both of their metrics describe a Schwarzschild-(anti) de Sitter~((A)dS) geometry. Choosing the Eddington-Finkelstein coordinates, one can represent them as \cite{Babichev2014}
\begin{align*}
    g_{\mu\nu}dx^\mu dx^\nu&=-\Sigma_g\,dv^2+2\,dvdr+r^2d\Omega^2,\\
    \label{SadS-EF}
    f_{\mu\nu}dx^\mu dx^\nu&=C^2\left(-\Sigma_f\,dv^2+2\,dvdr+r^2d\Omega^2\right),\numberthis{}
\end{align*}
where $\Sigma_g=1-2M_g/r+\Lambda_g/(3r^2)$, $\Sigma_f=1-2M_f/r+\Lambda_f/(3r^2)$ and $C$, $\Lambda_g$, $\Lambda_f$ are constants whose values are determined by the field equations. It is possible to pass to the Schwarzschild coordinates but only one of the two metrics can be diagonal in this coordinate system. These solutions have been shown to exist also in the ghost-free bigravity of Hassan and Rosen \cite{Comelli2012a}, and they admit charged \cite{Babichev2014a} and spinning \cite{Babichev2014b} generalizations. In the special case of $M_f=\Lambda_f=0$ one obtain the limit of the dRGT theory where the $f$-metric is flat whereas the $g$-metric is non-trivial and describes a bald black hole geometry. It is worth noting that this covers all possible static and spherically symmetric black holes in the dRGT massive gravity theory (other time-dependent black holes were reported in Ref.~\cite{Rosen2018}).

It is also possible to construct bald black holes in massive bigravity whose metrics are both diagonal in the Schwarzschild coordinates. Indeed, as it was noticed in \cite{Volkov2012a}, for special values of the parameters entering the potential, the theory reduces to vacuum GR when the two metrics coincide $g_{\mu\nu}=f_{\mu\nu}$. Then it follows immediately that all black holes of vacuum GR can be embedded into the ghost-free bigravity. For example, one has the bi-Schwarzschild solution:
\begin{equation}
\label{bi_sch}
    g_{\mu\nu}dx^\mu dx^\nu=f_{\mu\nu}dx^\mu dx^\nu=-\left(1-\frac{2M}{r}\right)dt^2+\left(1-\frac{2M}{r}\right)^{-1}dr^2+r^2 d\Omega^2.
\end{equation}
An effective cosmological constant $\Lambda$ can be included by assuming the two metrics to be proportional instead of being equal \cite{Volkov2012a,Volkov2013}. However it has to be emphasized that these solutions are different from those described by Eq.~\eqref{SadS-EF}. For example, they do not admit the dRGT theory limit (the mass $M$ in Eq.~\eqref{bi_sch} is the same for both metrics) and also the solution \eqref{SadS-EF} is linearly stable \cite{Babichev2016}, whereas \eqref{bi_sch} is unstable for small masses with respect to fluctuations which do not satisfy the condition\footnote{There is no reason for small fluctuations around $f_{\mu\nu}$ and $g_{\mu\nu}$ to be equal.} $g_{\mu\nu}=f_{\mu\nu}$ \cite{Babichev2013}. 

Second, the massive bigravity theory also admits hairy black hole solutions. These solutions cannot be constructed analytically because it requires to solve a very involved system of Ordinary Differential Equations (ODEs). The first numerical construction of hairy black holes was carried out in Ref.~\cite{Volkov2012a}, but none of the solutions found were asymptotically flat. The hairy black holes exist in the case where both metrics are simultaneously diagonal and they must share the same horizon \cite{Banados2011,Deffayet2012}. Its radius $r_H^g$ measured by the $g$-metric is generically different from the radius $r_H^f$ measured by the $f$-metric. One can always set $r_H$ to unity via rescaling but the ratio $u=r_H^g/r_H^f$ is a scale-invariant quantity. The choice of a value for $u$ completely determines the boundary conditions at the horizon. Therefore, the set of all black hole solutions (with bi-diagonal metrics) can be labeled by the parameter $u$.

The special case $u=1$ corresponds to the bi-Schwarzschild solution \eqref{bi_sch}. Choosing $u\neq 1$ yields more general black holes supporting massive \textit{hair} outside the horizon. Generically, the two metrics do not approach that of Minkowski at spatial infinity but one rather finds either a curvature singularity at a finite radius outside the horizon, or an asymptotic behavior close to that of an AdS spacetime \cite{Volkov2012a}. 

However it cannot be excluded that asymptotically flat hairy black holes exist for a special, fine-tuned, value of the parameter $u$. Moreover, the stability analysis of the bi-Schwarzschild solution \eqref{bi_sch} within the linear perturbation theory provides a good evidence in favor of their existence. Indeed, bi-Schwarzschild black holes are stable for $r_H>0.86$ (in units of the graviton mass) but unstable for $r_H\leq 0.86$ \cite{Babichev2013,Brito2013}. This suggests that the unstable bi-Schwarzschild black holes should decay to something else while the asymptotic structure of spacetime must be preserved. Consequently, asymptotically flat hairy black holes should exist as the final state of this instability. It is worth noting that the mathematical structure of the pertubation equations is identical to that of black strings in $d=5$ GR. The instability of these black strings was studied by Gregory and Laflamme (GL) \cite{Gregory1993} and therefore we shall refer to the bi-Schwarzschild black hole with $r_H=0.86$ as the GL point.

Finding the asymptotically flat hairy black hole solutions requires to use an appropriate numerical scheme to solve the underlying boundary value problem such as, for example, the \textit{shooting method} presented in Appendix \ref{num_bvp}. This was accomplished by Brito, Cardoso and Pani \cite{Brito2013a}. However, a few years later, a different group analyzed the asymptotic structure of the spherically symmetric solutions in massive bigravity \cite{Torsello2017}, and it was claimed that the bi-Schwarzschild solution \eqref{bi_sch} was the only asymptotically flat black hole in the theory. Therefore, a controversy emerged: it became unclear if asymptotically flat black holes with hair exist or not.

In this chapter, we reconsider this issue by ourselves and we are able to construct the asymptotically flat hairy black holes \cite{Gervalle2020}, thereby confirming the finding in Ref.~\cite{Brito2013a}. Our numerical scheme, the \textit{multishooting method}, is similar to the one employed by the authors in Ref.~\cite{Brito2013a}, but we refine it to be able to integrate the field equations starting exactly from the horizon radius (a singular point of the differential equations). We also take into account nonlinear corrections in the far field region by using a technique involving integral equations similar to that presented in, for example, Refs.~\cite{Breitenlohner1992,Breitenlohner1994,Breitenlohner1995}. From the methodological viewpoint, this chapter provides an example of how one should properly tackle a nonlinear boundary value problem with singular endpoints. We will discuss the contradictory conclusion of Ref.~\cite{Torsello2017} at the end of this chapter.

Apart from confirming the existence of asymptotically flat hairy black holes with strong numerical evidence, we explore in more detail the parameter space of the theory and discover many new features of these solutions. We also perform the stability analysis of the hairy black holes within the linear perturbation theory and search for a region in the parameter space where the hairy solutions are stable and have masses relevant to describe astrophysical black holes.

The rest of the chapter is organized as follows. In Section~\ref{setup_bigravity} we present the massive bigravity theory of Hassan and Rosen \cite{Hassan2012}. The field equations in the spherically symmetric case and the simplest analytical solutions are described in Secs.~\ref{spher_spym_bigravity} and \ref{anal_bigravity}. In Secs.~\ref{bc1_bigravity} and \ref{bc2_bigravity} we present the boundary conditions at the horizon and at infinity, and then summarize our numerical procedure in Section~\ref{num_bigravity}. The asymptotically flat hairy black holes and their new features are presented in Section~\ref{hairy_bh_bigravity}. After that, we describe the stability analysis of hairy solutions in Section~\ref{stab_bigravity}. Finally our discussion culminates in Section~\ref{param_bigravity} where we explore the parameter space and its various limits. We also identify in this section a region in the parameter space where the hairy solutions can consistently describe astrophysical black holes. Section~\ref{conclusion_bigravity} gives our conclusion and we discuss the arguments of Ref.~\cite{Torsello2017}. The two appendices of this chapter contain the description of the desingularization of the equations at the horizon and the time-dependent ansatz which is used for the stability analysis.

\section{The ghost-free massive bigravity}
\label{setup_bigravity}

The massive bigravity theory of Hassan and Rosen is defined on a four-dimensional spacetime manifold equipped with two Lorentzian metrics $g_{\mu\nu}$ and $f_{\mu\nu}$ with the signature $(-,+,+,+)$. It is described by the action \cite{Hassan2012}
\begin{equation}
\label{action_hr}
    \boldsymbol{S}_\text{HR}=\frac{1}{2\boldsymbol{\kappa}_1}\int{\boldsymbol{R}(g)\sqrt{-g}\,d^4\boldsymbol{x}}+\frac{1}{2\boldsymbol{\kappa}_2}\int{\boldsymbol{R}(f)\sqrt{-f}\,d^4\boldsymbol{x}}-\frac{\boldsymbol{m}^2}{\boldsymbol{\kappa}}\int{\mathcal{U}(g,f)\sqrt{-g}\,d^4\boldsymbol{x}},
\end{equation}
where $\boldsymbol{\kappa}_1$ and $\boldsymbol{\kappa}_2$ are the gravitational couplings, $\boldsymbol{\kappa}$ is a parameter with the same dimension, and $\boldsymbol{m}$ is the mass parameter. The interaction between the two metrics is described by the potential $\mathcal{U}$ which is a function of the scalar invariants of the tensor
\begin{equation}
\label{gamma_tens}
    \hat{\gamma}=\sqrt{\hat{g}^{-1}\hat{f}},
\end{equation}
where the hats denote matrices and the square root is understood in the sense that $\hat{\gamma}^2=\hat{g}^{-1}\hat{f}$. One can express this relation in components as
\begin{equation}
    \tensor{(\gamma^2)}{^\mu_\nu}=\tensor{\gamma}{^\mu_\alpha}\tensor{\gamma}{^\alpha_\nu}=g^{\mu\alpha}f_{\alpha\nu}.
\end{equation}
Denoting by $\lambda_a$ with $a\in\{1,2,3,4\}$ the eigenvalues of $\hat{\gamma}$, the interaction potential is
\begin{equation}
\label{pot_hr}
    \mathcal{U}=\sum_{k=0}^4{b_k\,\mathcal{U}_k},
\end{equation}
where $b_k$ are dimensionless parameters and $\mathcal{U}_k$ are defined by
\begin{align*}
    \mathcal{U}_0&=1,\quad\quad\mathcal{U}_1=\sum_a\lambda_a=[\gamma],\\
    \mathcal{U}_2&=\sum_{a<b}\lambda_a\lambda_b=\frac{1}{2!}\left([\gamma]^2-[\gamma^2]\right),\\
    \mathcal{U}_3&=\sum_{a<b<c}\lambda_a\lambda_b\lambda_c=\frac{1}{3!}\left([\gamma]^3-3[\gamma][\gamma^2]+2[\gamma^3]\right),\\
    \mathcal{U}_4&=\lambda_1\lambda_2\lambda_3\lambda_4=\text{det}(\hat{\gamma}).\numberthis{}
\end{align*}
Here $[\gamma]=\text{Tr}(\hat{\gamma})=\tensor{\gamma}{^\mu_\mu}$ and $[\gamma^k]=\text{Tr}(\hat{\gamma}^k)=\tensor{(\gamma^k)}{^\mu_\mu}$.

In vacuum, the two metrics actually enter the action in a symmetric way, since Eq.~\eqref{action_hr} is invariant under the interchange
\begin{equation}
\label{inter_sym}
    g_{\mu\nu}\leftrightarrow f_{\mu\nu},\quad\quad\boldsymbol{\kappa}_1\leftrightarrow\boldsymbol{\kappa}_2,\quad\quad b_k\leftrightarrow b_{4-k}.
\end{equation}
As a result, it is completely equivalent to couple the matter fields to either $g_{\mu\nu}$ or $f_{\mu\nu}$. Therefore we will assume without loss of generality that the metric to be minimally coupled to matter is $g_{\mu\nu}$ and we shall call it the \textit{physical} metric (the test particles follow its geodesics). The theory is also invariant under rescalings $\boldsymbol{\kappa}\rightarrow\pm\lambda^2\boldsymbol{\kappa}$, $b_k\rightarrow\pm b_k$, $\boldsymbol{m}\rightarrow\lambda\boldsymbol{m}$. This allows one to impose the normalization condition $\boldsymbol{\kappa}=\boldsymbol{\kappa}_1+\boldsymbol{\kappa}_2$. The variations of the action with respect to each metric give two sets of Einstein equations,
\begin{equation}
\label{ein_eq_dim}
    \boldsymbol{G}(g)_{\mu\nu}=\boldsymbol{m}^2\kappa_1 T_{\mu\nu},\quad\quad \boldsymbol{G}(f)_{\mu\nu}=\boldsymbol{m}^2\kappa_2 \mathcal{T}_{\mu\nu},
\end{equation}
where $\kappa_1=\boldsymbol{\kappa}_1/\boldsymbol{\kappa}$ and $\kappa_2=\boldsymbol{\kappa}_2/\boldsymbol{\kappa}$ so that $\kappa_1+\kappa_2=1$. The source terms in the right-hand sides of \eqref{ein_eq_dim} come from the variations of the potential $\mathcal{U}$,
\begin{equation}
\label{eff_stress_tensors}
    \tensor{T}{^\mu_\nu}=g^{\mu\alpha}T_{\alpha\nu}=\tensor{\tau}{^\mu_\nu}-\mathcal{U}\,\delta^\mu_\nu,\quad\tensor{\mathcal{T}}{^\mu_\nu}=f^{\mu\alpha}\mathcal{T}_{\alpha\nu}=-\frac{\sqrt{-g}}{\sqrt{-f}}\tensor{\tau}{^\mu_\nu},
\end{equation}
where $f^{\mu\alpha}$ is the inverse of $f_{\mu\alpha}$ and
\begin{align*}
    \tensor{\tau}{^\mu_\nu}=&(b_1\,\mathcal{U}_0+b_2\,\mathcal{U}_1+b_3\,\mathcal{U}_2+b_4\,\mathcal{U}_3)\tensor{\gamma}{^\mu_\nu}-(b_2\,\mathcal{U}_0+b_3\,\mathcal{U}_1+b_4\,\mathcal{U}_2)\tensor{(\gamma^2)}{^\mu_\nu}\\
    &+(b_3\,\mathcal{U}_0+b_4\,\mathcal{U}_1)\tensor{(\gamma^3)}{^\mu_\nu}-b_4\,\mathcal{U}_0\tensor{(\gamma^4)}{^\mu_\nu}.\numberthis{}
\end{align*}
Finally, there is an identity following from the diffeomorphism invariance of the potential term in the action,
\begin{equation}
\label{diffeo_rel}
    \sqrt{-g}\accentset{(g)}{\nabla}_\mu\tensor{T}{^\mu_\nu}+\sqrt{-f}\accentset{(f)}{\nabla}_\mu\tensor{\mathcal{T}}{^\mu_\nu}=0,
\end{equation}
where $\accentset{(g)}{\nabla}_\mu$ and $\accentset{(f)}{\nabla}_\mu$ denote the covariant derivatives with respect to each metric.

The massive bigravity theory describes two interacting gravitons, one massive and one massless. To make it apparent, one can consider the flat space limit. Setting $g_{\mu\nu}$ and $f_{\mu\nu}$ to be equal to the Minkowski metric $\eta_{\mu\nu}$, Eqs.~\eqref{ein_eq_dim} reduce to
\begin{equation}
    0=-\boldsymbol{m}^2\kappa_1(P_0+P_1)\eta_{\mu\nu},\quad\quad 0=-\boldsymbol{m}^2\kappa_2(P_1+P_2)\eta_{\mu\nu},
\end{equation}
with $P_m=b_m+2b_{m+1}+b_{m+2}$. It follows that the flat space will be solution only if the parameters $b_k$ satisfy the conditions $P_1=-P_0=-P_2$. Assuming this to be the case, one can consider small deviations $\delta g_{\mu\nu}$ and $\delta f_{\mu\nu}$ around the flat space solution, $g_{\mu\nu}=\eta_{\mu\nu}+\delta g_{\mu\nu}$ and $f_{\mu\nu}=\eta_{\mu\nu}+\delta f_{\mu\nu}$. Then, the linearization of Eqs.~\eqref{ein_eq_dim} with respect to the deviations yields
\begin{equation}
    \hat{\mathcal{E}}^{\alpha\beta}_{\mu\nu}h^{(0)}_{\alpha\beta}=0,\quad\quad\hat{\mathcal{E}}^{\alpha\beta}_{\mu\nu}h_{\alpha\beta}+\frac{\boldsymbol{m}_\text{FP}^2}{2}(h_{\mu\nu}-\eta_{\mu\nu} h)=0,
\end{equation}
where $h^{(0)}_{\mu\nu}=\kappa_1\,\delta f_{\mu\nu}+\kappa_2\,\delta g_{\mu\nu}$, $h_{\mu\nu}=\delta f_{\mu\nu}-\delta g_{\mu\nu}$, $h=\eta^{\mu\nu}h_{\mu\nu}$ and $\hat{\mathcal{E}}^{\alpha\beta}_{\mu\nu}$ denotes the linear part of the Einstein tensor,
\begin{equation}
    \hat{\mathcal{E}}^{\alpha\beta}_{\mu\nu}=-\delta^\alpha_\mu\delta^\beta_\nu\Box+\eta^{\alpha\lambda}\delta^\beta_\nu\partial_\lambda\partial_\mu+\eta^{\alpha\lambda}\delta^\beta_\mu\partial_\lambda\partial_\nu-\eta^{\alpha\beta}\partial_\mu\partial_\nu-\eta_{\mu\nu}\partial^\alpha\partial^\beta+\eta_{\mu\nu}\eta^{\alpha\beta}\Box.
\end{equation}
The equation for $h^{(0)}_{\mu\nu}$ is exactly the same as GR linearized around flat spacetime therefore, it describes a massless graviton with two dynamical degrees of freedom. The second effective metric fluctuation $h_{\mu\nu}$ fulfills a field equation equivalent to that of the FP massive gravity, it corresponds to a massive graviton with five degrees of freedom and a mass
\begin{equation}
\label{fp_mass}
    \boldsymbol{m}_\text{FP}^2=P_1\,\boldsymbol{m}^2.
\end{equation}
Therefore it will be convenient to impose $P_1=1$ so that the mass parameter $\boldsymbol{m}$ entering the action \eqref{action_hr} will coincide with the graviton mass $\boldsymbol{m}_\text{FP}$ in flat space. Then, the condition to have Minkowski as a solution becomes $P_0=P_2=-1$ and this can be used to express the five $b_k$ in terms of only two independent parameters $c_3$ and $c_4$,
\begin{align*}
\label{bk}
    b_0&=4c_3+c_4-6,\quad b_1=3-3c_3-c_4,\quad b_2=2c_3+c_4-1,\\
    b_3&=-(c_3+c_4),\quad b_4=c_4.\numberthis{}
\end{align*}
At the same time, it has been shown that the theory propagates exactly $2+5=7$ degrees of freedom also away from the flat space limit and for arbitrary $b_k$ (see \cite{Hassan2012a,Alexandrov2013,Soloviev2020} for the Hamiltonian formulation of massive bigravity).

Finally, we introduce dimensionless coordinates $x^\mu$ via
\begin{equation}
    x^\mu=\boldsymbol{m}\boldsymbol{x}^\mu.
\end{equation}
This is equivalent to the conformal rescaling of the metrics,
\begin{equation}
    g_{\mu\nu}\rightarrow\frac{1}{\boldsymbol{m}^2}g_{\mu\nu},\quad\quad  f_{\mu\nu}\rightarrow\frac{1}{\boldsymbol{m}^2}f_{\mu\nu}.
\end{equation}
The mass parameter $\boldsymbol{m}$ disappears from the field equations so that Eqs.~\eqref{ein_eq_dim} become
\begin{equation}
\label{ein_eq}
    \tensor{G(g)}{^\mu_\nu}=\kappa_1\tensor{T}{^\mu_\nu},\quad\quad\tensor{G(f)}{^\mu_\nu}=\kappa_1\tensor{\mathcal{T}}{^\mu_\nu}.
\end{equation}
The Bianchi identities for these two Einstein equations imply that
\begin{equation}
    \accentset{(g)}{\nabla}_\mu\tensor{T}{^\mu_\nu}=0,\quad\quad\accentset{(f)}{\nabla}_\mu\tensor{\mathcal{T}}{^\mu_\nu}=0,
\end{equation}
which is consistent with \eqref{diffeo_rel}. In this dimensionless setting, the unit of length is $1/\boldsymbol{m}$, the Compton wavelength of the massive graviton.

In what follows, we shall be analyzing the field equations \eqref{ein_eq} without any assumptions about the values of $\kappa_1$, $\kappa_2$ and $b_k$. However, when integrating the equations numerically, we shall take into account that $\kappa_1+\kappa_2=1$ by introducing a mixing angle $\eta$,
\begin{equation}
    \kappa_1=\cos^2\eta,\quad\quad\kappa_2=\sin^2\eta,
\end{equation}
and we shall choose the $b_k$ according to \eqref{bk}. Therefore, our numerical solutions depend on three parameters of the theory, $c_3$, $c_4$ and $\eta$.

\section{Spherically symmetric field equations}
\label{spher_spym_bigravity}

Let us introduce spherical-like coordinates $(x^0,x^1,x^2,x^3)=(t,r,\vartheta,\varphi)$ where $t$ is a timelike coordinate, $r$ is a radial coordinate and $\vartheta$, $\varphi$ are respectively the polar and azimuthal angles. We assume that both metrics are static, spherically symmetric and diagonal. Therefore, following for example Ref.~\cite{Volkov2012a}, one has,
\begin{align*}
    ds^2_g&=g_{\mu\nu}dx^\mu dx^\nu=-Q^2 dt^2+\frac{dr^2}{\Delta^2}+R^2 d\Omega^2,\\
\label{ansatz_hr}
    ds^2_f&=f_{\mu\nu}dx^\mu dx^\nu=-q^2 dt^2+\frac{dr^2}{W^2}+U^2 d\Omega^2,\numberthis{}
\end{align*}
where $d\Omega^2=d\vartheta^2+\sin^2\vartheta\,d\varphi^2$ and $Q$, $q$, $\Delta$, $W$, $R$, $U$ are functions depending on the radial coordinate $r=\boldsymbol{m}\boldsymbol{r}$ only. We emphasize that this is actually not the most general ansatz for spherically symmetric fields, since one could also include an off-diagonal metric element $f_{01}$ as shown in Appendix \ref{time_dep_ansatz}. However, in the \textit{static} case, the only possible black hole solution with $f_{01}\neq 0$ corresponds to Eqs.~\eqref{SadS-EF}. Therefore the bi-diagonal ansatz \eqref{ansatz_hr} is the most general one for obtaining non-Schwarzschild solutions.

The tensor $\tensor{\gamma}{^\mu_\nu}$ encoding the coupling between the two metrics is then given according to \eqref{gamma_tens} by
\begin{equation}
    \tensor{\gamma}{^\mu_\nu}=\text{diag}\left(\frac{q}{Q},\frac{\Delta}{W},\frac{U}{R},\frac{U}{R}\right),
\end{equation}
and the effective stress-energy tensors in Eq.~\eqref{eff_stress_tensors} read
\begin{equation}
    \tensor{T}{^\mu_\nu}=\text{diag}\left(\tensor{T}{^0_0},\tensor{T}{^1_1},\tensor{T}{^2_2},\tensor{T}{^2_2}\right),\quad\quad\tensor{\mathcal{T}}{^\mu_\nu}=\text{diag}\left(\tensor{\mathcal{T}}{^0_0},\tensor{\mathcal{T}}{^1_1},\tensor{\mathcal{T}}{^2_2},\tensor{\mathcal{T}}{^2_2}\right),
\end{equation}
where
\begin{align*}
    \tensor{T}{^0_0}&=-\mathcal{P}_0-\mathcal{P}_1\frac{\Delta}{W},\\
    \tensor{T}{^1_1}&=-\mathcal{P}_0-\mathcal{P}_1\frac{q}{Q},\\
    \tensor{T}{^2_2}&=-\mathcal{D}_0-\mathcal{D}_1\left(\frac{q}{Q}+\frac{\Delta}{W}\right)-\mathcal{D}_2\frac{q\Delta}{QW},\\
    \left(\frac{U}{R}\right)^2\tensor{\mathcal{T}}{^0_0}&=-\mathcal{P}_2-\mathcal{P}_1\frac{W}{\Delta},\\
    \left(\frac{U}{R}\right)^2\tensor{\mathcal{T}}{^1_1}&=-\mathcal{P}_2-\mathcal{P}_1\frac{Q}{q},\\
    \left(\frac{U}{R}\right)^2\tensor{\mathcal{T}}{^2_2}&=-\mathcal{D}_3-\mathcal{D}_2\left(\frac{Q}{q}+\frac{W}{\Delta}\right)-\mathcal{D}_1\frac{QW}{q\Delta}.
\end{align*}
Here we have introduced the shortened notations
\begin{equation}
\label{Pk}
    \mathcal{P}_k=b_k+2b_{k+1}\frac{U}{R}+b_{k+2}\left(\frac{U}{R}\right)^2,\quad\quad\mathcal{D}_k=b_k+b_{k+1}\frac{U}{R}.
\end{equation}
The independent field equations are 
\begin{align*}
\label{indep_eqs}
    \tensor{G(g)}{^0_0}&=\kappa_1\tensor{T}{^0_0},\quad\quad\tensor{G(f)}{^0_0}=\kappa_2\tensor{\mathcal{T}}{^0_0},\\
    \tensor{G(g)}{^1_1}&=\kappa_1\tensor{T}{^1_1},\quad\quad\tensor{G(f)}{^1_1}=\kappa_2\tensor{\mathcal{T}}{^1_1},\numberthis{}
\end{align*}
plus the conservation condition $\accentset{(g)}{\nabla}_\mu\tensor{T}{^\mu_\nu}=0$, which has one non-vanishing component,
\begin{equation}
\label{cons_eq}
    \accentset{(g)}{\nabla}_\mu\tensor{T}{^\mu_1}=\left(\tensor{T}{^1_1}\right)'+\frac{Q'}{Q}\left(\tensor{T}{^1_1}-\tensor{T}{^0_0}\right)+2\frac{R'}{R}\left(\tensor{T}{^1_1}-\tensor{T}{^2_2}\right)=0,
\end{equation}
where the primes denote the differentiation with respect to $r$. The second stress-energy tensor also fulfills a conservation condition with one non-trivial component,
\begin{equation}
\label{cons_eq_2}
    \accentset{(f)}{\nabla}_\mu\tensor{\mathcal{T}}{^\mu_1}=\left(\tensor{\mathcal{T}}{^1_1}\right)'+\frac{q'}{q}\left(\tensor{\mathcal{T}}{^1_1}-\tensor{\mathcal{T}}{^0_0}\right)+2\frac{U'}{U}\left(\tensor{\mathcal{T}}{^1_1}-\tensor{\mathcal{T}}{^2_2}\right)=0,
\end{equation}
but this follows from Eq.~\eqref{cons_eq} by virtue of the relation \eqref{diffeo_rel}. As a result, we have 5 independent differential equations \eqref{indep_eqs}, \eqref{cons_eq}, which is enough to determine the 6 metric functions $Q$, $q$, $\Delta$, $W$, $R$, $U$, because the freedom of reparametrization of the radial coordinate $r\rightarrow\tilde{r}(r)$ allows us to fix one of the unknown functions.

We shall now introduce new functions
\begin{equation}
\label{def_NY}
    N=\Delta R',\quad\quad Y=W U',
\end{equation}
in terms of which the two metrics can be written as
\begin{align*}
    ds^2_g&=-Q^2 dt^2+\frac{dR^2}{N^2}+R^2 d\Omega^2,\\
\label{ansatz_hr_2}
    ds^2_f&=-q^2 dt^2+\frac{dU^2}{Y^2}+U^2 d\Omega^2.\numberthis{}
\end{align*}
This parametrization has the advantage of showing only first derivatives in the Einstein tensor. The four equations \eqref{indep_eqs} become 
\begin{align}
    \label{eqN}
    N'&=-\frac{\kappa_1}{2}\frac{R}{NY}\left(R'Y\,\mathcal{P}_0+U'N\,\mathcal{P}_1\right)+\frac{(1-N^2)R'}{2RN},\\
    \label{eqY}
    Y'&=-\frac{\kappa_2}{2}\frac{R^2}{UNY}\left(R'Y\,\mathcal{P}_1+U'N\,\mathcal{P}_2\right)+\frac{(1-Y^2)U'}{2UY},\\
    \label{eqQ}
    Q'&=-\left(\kappa_1(Q\,\mathcal{P}_0+q\,\mathcal{P}_1)+\frac{Q(N^2-1)}{R^2}\right)\frac{RR'}{2N^2},\\
    \label{eqq}
    q'&=-\left(\kappa_2(Q\,\mathcal{P}_1+q\,\mathcal{P}_2)+\frac{q(Y^2-1)}{R^2}\right)\frac{R^2U'}{2Y^2U}.
\end{align}
The conservation condition \eqref{cons_eq} reads
\begin{equation}
\label{cons_eq_bis}
    \accentset{(g)}{\nabla}_\mu\tensor{T}{^\mu_1}=\frac{U'}{R}\left(1-\frac{N}{Y}\right)\left(d\mathcal{P}_0
+\frac{q}{Q}d\mathcal{P}_1\right)+\left(\frac{q'}{Q}-\frac{NQ'U'}{YQR'}\right)\mathcal{P}_1=0,
\end{equation}
with
\begin{equation}
    d\mathcal{P}_k=2\left(b_{k+1}+b_{k+2}\frac{U}{R}\right).
\end{equation}
The first derivatives of $Q$ and $q$ in Eq.~\eqref{cons_eq_bis} can be substituted by using Eqs.~\eqref{eqQ} and \eqref{eqq}, giving
\begin{equation}
\label{cons_eq_exp}
    R^2Q\,\accentset{(g)}{\nabla}_\mu\tensor{T}{^\mu_1}=\frac{U'}{Y}\boldsymbol{C}=0,
\end{equation}
where
\begin{align*}
    \boldsymbol{C}=&\left(\kappa_2\frac{R^4\mathcal{P}_1^2}{2UY}-\kappa_1\frac{R^3\mathcal{P}_0\mathcal{P}_1}{2N}-\frac{(N^2-1)R\,\mathcal{P}_1}{2N}+(N-Y)R\,d\mathcal{P}_0\right)Q\\
\label{defC}
    &+\left(\kappa_2\frac{R^4\mathcal{P}_1\mathcal{P}_2}{2UY}-\kappa_1\frac{R^3\mathcal{P}_1^2}{2N}+\frac{(Y^2-1)R^2\mathcal{P}_1}{2UY}+(N-Y)R\,d\mathcal{P}_1\right)q.\numberthis{}
\end{align*}
The second conservation condition \eqref{cons_eq_2} becomes
\begin{equation}
\label{cons_eq_2_exp}
    -U^2q\,\accentset{(f)}{\nabla}_\mu\tensor{\mathcal{T}}{^\mu_1}=\frac{R'}{N}\boldsymbol{C}=0.
\end{equation}
A first possibility to satisfy the two conditions \eqref{cons_eq_exp} and \eqref{cons_eq_2_exp} is $U'=R'=0$. However, a direct inspection of the line elements \eqref{ansatz_hr_2} shows that in this case, both metrics are degenerate (one has $g_{11}=R'^2/N^2$, $f_{11}=U'^2/Y^2$). Because physically relevant metrics are not degenerate, one should consider the second possibility to satisfy \eqref{cons_eq_exp}, \eqref{cons_eq_2_exp}, which is
\begin{equation}
\label{eqC}
    \boldsymbol{C}=0.
\end{equation}
This is an algebraic equation that can be resolved with respect to $q$ to give
\begin{equation}
\label{alg_q} 
     q=-\Sigma(R,U,N,Y)\,Q,
\end{equation}
where $\Sigma(R,U,N,Y)$ is the ratio of the coefficients in front of $Q$ and $q$ in Eq.~\eqref{defC}.

As a result, we obtain four differential equations of first order \eqref{eqN}-\eqref{eqq} plus one algebraic equation \eqref{eqC}. As a consistency check, one can insert the ansatz \eqref{ansatz_hr_2} directly to the action \eqref{action_hr}, which gives
\begin{equation}
    \boldsymbol{S}_\text{HR}=\frac{4\pi}{\boldsymbol{m}^2\boldsymbol{\kappa}}\int{\mathcal{L}\,dtdr},
\end{equation}
where, dropping a total derivative,
\begin{align*}
    \mathcal{L}=&\frac{1}{\kappa_1}\left(\frac{(1-N^2)R'}{N}-2RN'\right)Q+\frac{1}{\kappa_2}\left(\frac{(1-Y^2)U'}{Y}-2UY'\right)q\\
    \label{effL}
    &-\frac{QR^2R'}{N}\mathcal{P}_0-\left(\frac{QR^2U'}{Y}+\frac{qR^2R'}{N}\right)\mathcal{P}_1-\frac{qR^2U'}{Y}\mathcal{P}_2.\numberthis{}
\end{align*}
The variations of \eqref{effL} with respect to $N$, $Y$, $Q$, $q$ reproduce Eqs.~\eqref{eqN}-\eqref{eqq}, while varying it with respect to $R$, $U$ gives the conditions \eqref{cons_eq_exp} and \eqref{cons_eq_2_exp}. All the equations and the reduced Lagrangian $\mathcal{L}$ are left invariant under the interchange \eqref{inter_sym}, which now reads
\begin{equation}
    \label{inter_sym_exp}
    \kappa_1\leftrightarrow\kappa_2,\quad Q\leftrightarrow q,\quad N\leftrightarrow Y,\quad R\leftrightarrow U,\quad b_k\leftrightarrow b_{4-k}.
\end{equation}

At this stage, one cannot solve the system of field equations since \eqref{eqN}-\eqref{eqq} contain the first derivatives of $R$ and $U$ which are not yet known. One of these two radial functions can be fixed by imposing a gauge condition to completely specify the radial coordinate, but the other one should be determined dynamically. We need therefore one more differential equation. Since the algebraic constraint \eqref{eqC} should be stable, one can differentiate it to obtain the secondary constraint
\begin{equation}
    \boldsymbol{C}'=\frac{\partial\boldsymbol{C}}{\partial N}N'+\frac{\partial\boldsymbol{C}}{\partial Y}Y'+\frac{\partial\boldsymbol{C}}{\partial Q}Q'+\frac{\partial\boldsymbol{C}}{\partial q}q'+\frac{\partial\boldsymbol{C}}{\partial R}R'+\frac{\partial\boldsymbol{C}}{\partial U}U'=0.
\end{equation}
The derivatives $N'$, $Y'$, $Q'$, $q'$ can be replaced by using Eqs.~\eqref{eqN}-\eqref{eqq} and $q$ can be substituted by virtue of Eq.~\eqref{alg_q}. This yields
\begin{equation}
\label{second_cons}
    \boldsymbol{C}'=\mathcal{A}(R,U,N,Y)\,R'+\mathcal{B}(R,U,N,Y)\,U'=0,
\end{equation}
where the explicit expressions of $\mathcal{A}(R,U,N,Y)$ and $\mathcal{B}(R,U,N,Y)$ are rather complicated and we do not show them. It is worth noting that Eq.~\eqref{second_cons} is invariant under a change of radial coordinate,
\begin{equation}
    r\rightarrow\tilde{r}(r)\quad\Rightarrow\quad R'\rightarrow\tilde{R}'=\frac{dr}{d\tilde{r}}R',\quad U'\rightarrow\tilde{U}'=\frac{dr}{d\tilde{r}}U'.
\end{equation}
We can express $U'$ by using the secondary constraint,
\begin{equation}
\label{proto_eqU}
    U'=-\frac{\mathcal{A}(R,U,N,Y)}{\mathcal{B}(R,U,N,Y)}R'\equiv\mathcal{D}_U(R,U,N,Y)\,R'.
\end{equation}
Finally we specify completely the radial coordinate by setting 
\begin{equation}
    R'=1\quad\Rightarrow\quad R=r.
\end{equation}
This choice is very natural since it corresponds to a Schwarzschild-like radial coordinate\footnote{Surfaces with $r=\text{const.}$ are 2-spheres whose surface is $4\pi r^2$.}. Then, Eq.~\eqref{proto_eqU} reduces to
\begin{equation}
\label{eqU}
    U'=\mathcal{D}_U(r,U,N,Y).
\end{equation}
Now, one can substitute $U'$ in the right-hand sides of Eqs.~\eqref{eqN} and \eqref{eqY}, these two equations together with \eqref{eqU} form a closed system of three differential equations
\begin{align*}
    N'=\mathcal{D}_N(r,U,N,Y),\\
    Y'=\mathcal{D}_Y(r,U,N,Y),\\
    \label{ode_hr}
    U'=\mathcal{D}_U(r,U,N,Y).\numberthis{}
\end{align*}
To determine $Q$ and $q$, we inject \eqref{alg_q} to \eqref{eqQ} to obtain,
\begin{equation}
\label{eqQ_fin}
    Q'=-\frac{r}{2N^2}\left(\kappa_1(\mathcal{P}_0+\Sigma(r,U,N,Y)\mathcal{P}_1)+\frac{N^2-1}{r^2}\right)Q\equiv\mathcal{F}(r,U,N,Y)\,Q.
\end{equation}
This differential equation determines $Q$ whereas the function $q$ can be obtained by using its algebraic expression \eqref{alg_q}.

Summarizing, spherically symmetric spacetimes with bi-diagonal metrics are determined by the three coupled equations \eqref{ode_hr} for $N$, $Y$ and $U$. As soon as their solution is obtained, the remaining functions $Q$, $q$ are determined from \eqref{eqQ_fin} and \eqref{alg_q}. Notice that the system of equations \eqref{ode_hr} coincides with the one examined in previous works, such as Refs.~\cite{Volkov2012a,Brito2013a}.

\section{Analytical solutions}
\label{anal_bigravity}

We shall now describe the simplest solutions which can be found analytically \cite{Volkov2012a,Hassan2013}. For this, it will be convenient to use the equations in the form \eqref{eqN}-\eqref{eqq}.

\subsection{Proportional backgrounds}

One can recover GR black holes by choosing the two metrics to be proportional \cite{Volkov2012a,Hassan2013},
\begin{equation}
\label{prop_met}
    ds^2_f=C^2 ds^2_g,
\end{equation}
where $C$ is a constant that is determined by the field equations (see Eq.~\eqref{eff_lambda} below). Then the equations \eqref{eqN}-\eqref{eqq} are solved by
\begin{equation}
\label{prop_bg}
    Q^2=N^2=Y^2=1-\frac{2M}{r}-\frac{\Lambda(C)}{3}r^2,\quad R=r,\quad q=CQ,\quad U=CR.
\end{equation}
It describes two proportional Schwarzschild-(A)dS geometries. We emphasize that the cosmological constant $\Lambda(C)$ here is \textit{effective}, in the sense that it is produced by the massive graviton, and not added by hand as in GR. The constants $C$ and $\Lambda(C)$ are determined by 
\begin{equation}
\label{eff_lambda}
    \kappa_1(\mathcal{P}_0+C\,\mathcal{P}_1)=\frac{\kappa_2}{C}(\mathcal{P}_1+C\,\mathcal{P}_2)\equiv\Lambda(C).
\end{equation}
The $\mathcal{P}_k$ terms are polynomials of order 2 in $U/R=C$ so that the above equation is algebraic and can have up to four real roots which determine the possible values of $C$. Of course these values depend on the theory parameter. If the parameters $b_k$ are choosen according to \eqref{bk}, then $C=1$ will automatically be one of the roots and $\Lambda(C=1)=0$. It corresponds to a bald black hole with asymptotically flat geometry, or in other words, a (bi-)Schwarzschild black hole with mass $M$.

At the same time, the dimensionful cosmological constant $\boldsymbol{\Lambda}$ should agree with the observations of the accelerating expansion of the Universe. Therefore, one should have
\begin{equation}
    \boldsymbol{\Lambda}=\boldsymbol{m^2}\Lambda\sim 1/(\boldsymbol{R_\text{Hub}})^2,
\end{equation}
where $\boldsymbol{R_\text{Hub}}$ is the Hubble radius. One possible way to fulfill this condition is to assume the graviton mass to be such that its Compton wavelength is of the order of the Hubble radius,
\begin{equation}
\label{tiny_m}
    1/\boldsymbol{m}\sim\boldsymbol{R_\text{Hub}}.
\end{equation}
This is actually the historical choice in massive gravity theories. A second possibility is to assume that the dimensionless $\Lambda$ is very small. This is possible if one of the gravitational coupling is negligible in front of the other : $\kappa_1 \ll \kappa_2=1-\kappa_1\sim 1$. Indeed, Eq.~\eqref{eff_lambda} then implies that $\Lambda\sim\kappa_1$ and that $C$ should be very close to a root of $\mathcal{P}_1+C\,\mathcal{P}_2$. The hierarchy between the two couplings is in fact necessary to reconcile with the observations the perturbation spectrum of cosmological solutions in massive bigravity since the latter contains an instability in the scalar sector \cite{Comelli2012b,Koennig2014,Lagos2014}. This instability can be shifted toward early times of the Universe if \cite{Akrami2015,Moertsell2015,Aoki2015,Lueben2020,Hoegaas2020}
\begin{equation}
    \frac{\kappa_1}{\kappa_2}\sim\kappa_1\leq\left(\frac{\boldsymbol{M_\text{ew}}}{\boldsymbol{M_\text{Pl}}}\right)^2\sim 10^{-34}\ll 1,
\end{equation}
where $\boldsymbol{M_\text{ew}}\sim 100\,\text{GeV}$ is the electroweak energy scale and $\boldsymbol{M_\text{Pl}}\sim 10^{19}\,\text{GeV}$ is the Planck scale. Here $10^{-34}$ is the \textit{upper bound} for $\kappa_1$ in order to make the instability unobservable. Therefore, one can define a range of physcially acceptable values for $\kappa_1$ by setting
\begin{equation}
\label{kappa_phys_choice}
    \kappa_1=\gamma^2\times 10^{-34}\quad\text{with}\quad\gamma\in[0,1].
\end{equation}
Then it follows that
\begin{equation}
\label{m_phys_choice}
    1/\boldsymbol{m}\sim\sqrt{\Lambda}\,\boldsymbol{R_\text{Hub}}=\sqrt{\kappa_1}\,\boldsymbol{R_\text{Hub}}=\gamma\times\left(\frac{\boldsymbol{M_\text{ew}}}{\boldsymbol{M_\text{Pl}}}\right)\boldsymbol{R_\text{Hub}}\sim\gamma\times 10^6\,\text{km},
\end{equation}
which is of the order of the solar size if $\gamma\sim 1$. However, in what follows we shall not be assuming $\kappa_1$ to be small and shall present our results for arbitrary $\kappa_1\in[0,1]$.

\subsection{Deformed AdS background}

Another simple analytical solution can be obtained by choosing $U$, $q$ to be constant,
\begin{equation}
\label{deformedAdS}
    U=U_0,\quad\quad q=q_0.
\end{equation}
This solves Eqs.~\eqref{eqq} and \eqref{cons_eq_exp} and the remaining equations \eqref{eqN}-\eqref{eqQ} then can be integrated in quadratures \cite{Volkov2012a}. However, this solution is unphysical, since the $f$-metric is degenerate if $U'=0$. It turns out that other more general solutions approach \eqref{deformedAdS} at spatial infinity. For these solutions, the leading behavior of the functions at large $r$ is given by
\begin{align*}
    N^2&=-\kappa_1\frac{b_0}{3}r^2-\kappa_1 b_1 U_0\,r+\mathcal{O}(1),\\
    Y&=-\frac{\sqrt{3}\,\kappa_2 b_1}{4U_0\sqrt{-\kappa_1 b_0}}r^2+\mathcal{O}(r),\\
    Q&=\frac{q_0}{4U_0}r+\mathcal{O}(1),\\
    \label{asymp_deformedAdS}
    U&=U_0+\mathcal{O}\left(\frac{1}{r}\right),\quad q=q_0+\mathcal{O}\left(\frac{1}{r}\right),\numberthis{}
\end{align*}
where $b_0$, $b_1$ are integration constants. The $g$-metric approaches the AdS geometry in the leading $\mathcal{O}(r^2)$ order, but the subleading terms are different from AdS. This is why we refer to this asymptotic behavior of the metrics functions as \textit{deformed} AdS.

It has been shown numerically that black hole solutions of Eqs.~\eqref{eqN}-\eqref{eqq} generically approach for $r\rightarrow\infty$ either \eqref{prop_bg} or \eqref{asymp_deformedAdS} (or they exhibit a curvature singularity at a finite $r$ outside the horizon), hence they are not asymptotically flat \cite{Volkov2012a}.

\section{Boundary conditions at the horizon}
\label{bc1_bigravity}

Let us now require the $g$-metric to have a regular event horizon located at $r=r_H$. In terms of the metric functions, this means that $g_{00}=Q^2$ and $g^{11}=N^2$ have simple zeroes at this point. In other words, close to the horizon radius, one has $Q^2\sim N^2\sim r-r_H$ and we shall consider only the exterior region $r\geq r_H$ where $Q^2\geq 0$ and $N^2\geq 0$. This behavior is compatible with the field equations only if the $f$ metric also has a regular horizon at the same place, hence $q^2\sim Y^2\sim r-r_H$. As a result, both metrics share a horizon located at the radial coordinate $r=r_H$, in agreement with \cite{Deffayet2012,Banados2011}. However, it should be emphasized that the horizon radius measured by the $g$-metric, $r_H$, can be different from that measured by the second metric, $U(r_H)$. We therefore introduce the parameter $u\equiv U(r_H)/r_H$.

As a result, close to the horizon we can construct a local solution close by expanding the metrics functions in powers of $(r-r_H)$,
\begin{equation}
\label{loc1}
    N^2=\sum_{n\geq 1}{a_n(r-r_H)^n},\quad Y^2=\sum_{n\geq 1}{b_n(r-r_H)^n},\quad U=u\,r_H+\sum_{n\geq 1}{c_n(r-r_H)^n},
\end{equation}
and 
\begin{equation}
\label{loc2}
    Q^2=\sum_{n\geq 1}{d_n(r-r_H)^n},\quad q^2=\sum_{n\geq 1}{e_n(r-r_H)^n}.
\end{equation}
The coefficient $a_n$, $b_n$, $c_n$, $d_n$, $e_n$ can then be determined by the field equations. It turns out that they all can be expressed in terms of $a_1$ and the latter should fulfill a quadratic equation,
\begin{equation}
\label{biquad}
    \mathcal{A}a_1^2+\mathcal{B}a_1+\mathcal{C}=0\;\Rightarrow\; a_1=\frac{1}{2\mathcal{A}}\left(-\mathcal{B}+\sigma\sqrt{\mathcal{B}^2-4\mathcal{A}\mathcal{C}}\right),\;\;\sigma=\pm 1,
\end{equation}
where $\mathcal{A}$, $\mathcal{B}$, $\mathcal{C}$ are functions of $u$, $r_H$ and of the theory parameters $b_k$, $\kappa_1$, $\kappa_2$. The two possible signs for $\sigma$ describe two branches of solutions. The negative sign, $\sigma=-1$, always yields solutions with a curvature singularity at a finite $r>r_H$. Therefore, we choose $\sigma=1$ and then, for a given value of the horizon radius $r_H$, the local solutions \eqref{loc1}, \eqref{loc2} comprise a set labeled by a continuous parameter $u$. 

The surface gravity for each metric is \cite{Volkov2012a}
\begin{equation}
\label{surf_grav}
    \kappa_g^2=\lim_{r\rightarrow r_H}Q^2N'^2=\frac{1}{4}d_1 a_1,\quad\quad\kappa_f^2=\lim_{r\rightarrow r_H}Q^2\left(\frac{Y}{U'}\right)^2=\frac{e_1 b_1}{4c_1^2},
\end{equation}
and using the values of the expansion coefficients determined by the equations yields the identity $\kappa_g=\kappa_f$. Hence the two surface gravities coincide as well as the Hawking temperatures 
\begin{equation}
\label{temp}
    T_H=\frac{\kappa_g}{2\pi}=\frac{\kappa_f}{2\pi}.
\end{equation}

Close to the horizon, one has $N(r)\sim Y(r)\sim\sqrt{r-r_H}$ hence the derivatives $N'$ and $Y'$ are not defined at the horizon. In the Refs.~\cite{Volkov2012a,Brito2013a}, the authors start the numerical integration at a point $r=r_H+\epsilon$ with $\epsilon\ll 1$. The boundary conditions for the integration are set using the local solutions \eqref{loc1} and \eqref{loc2} evaluated at this initial point. However, although the dependence of the numerical solutions on $\epsilon$ is expected to be small, still its presence in the procedure may lead to numerical instabilities. This point was emphasized in \cite{Torsello2017}. To eliminate this arbitrary small parameter of the numerical procedure, we introduce new functions 
\begin{equation}
    \nu(r)=\frac{N(r)}{S(r)},\quad y(r)=\frac{Y(r)}{S(r)}\quad\text{with}\quad S(r)=\sqrt{1-\frac{r_H}{r}}.
\end{equation}
The functions $\nu$, $y$ and all their derivatives assume finite values at $r=r_H$. Making this change of variables in Eqs.~\eqref{ode_hr} gives a \textit{desingularized} version of the equations that allows us to start the numerical integration exactly at $r=r_H$. This is described in Appendix \ref{desing_hor_bigravity}.

To recapitulate, all black hole solutions with a given $r_H$ can be labeled by only one parameter $u$. If $u=1$ then the two metrics coincide and they describe the bi-Schwarzschild solution \eqref{bi_sch}. If $u=C$ where $C$ is a root of the algebraic equation \eqref{eff_lambda}, then the solution corresponds to a Schwarzschild-(A)dS black hole \eqref{prop_bg}. For other values of $u$, more general solutions can be obtained, they correspond to hairy black holes and they can be of the following three qualitative types, depending on their asymptotic behavior \cite{Volkov2012a}.
\begin{itemize}
    \item Solutions extending up to arbitrarily large values of $r$ and asymptotically approaching a proportional AdS background \eqref{prop_met}, \eqref{prop_bg}. At large $r$ one has $N=N_0(1+\delta N)$, $Y=Y_0(1+\delta Y)$, $U=U_0(1+\delta U)$ where $N_0$, $Y_0$, $U_0$ are given by Eq.~\eqref{prop_bg}, while the deviations $\delta N$, $\delta Y$, $\delta U$ approach zero. Linearizing the field equations with respect to these deviations, one finds that
    \begin{equation}
        \delta N=\frac{A}{r^3},\quad \delta U=B_1\text{e}^{\lambda_1 r}+B_2\text{e}^{\lambda_2 r},\quad\delta Y =\mathcal{O}(\delta U),
    \end{equation}
    where $A$, $B_1$, $B_2$ are integration constants and the real parts of $\lambda_1$ and $\lambda_2$ are \textit{negative}. As a result, all these pertubative modes vanish for $r\rightarrow\infty$, and since the number of integration constants is the same as the number of equations \eqref{ode_hr}, it follows that the AdS background is an \textit{attractor} at large $r$.
    \item Solutions extending up to arbitrarily large values of $r$ and asymptotically approaching a deformed AdS background \eqref{asymp_deformedAdS}. The latter is also an attractor at large $r$.
    \item Solutions extending only up to $r=r_S<\infty$ where derivatives of some metric functions diverge. Computation of curvature invariants shows that this is associated to a curvature singularity.
\end{itemize}

This exhausts the possible types of generic hairy black holes. We shall see in the next section that the flat space is not an attractor. For example, choosing $u=1+\epsilon$ yields solutions which are close to Schwarzschild in the vicinity of the horizon, but for larger values of $r$ the metrics deviate from that of Schwarzschild \cite{Volkov2012a}. This means the bi-Schwarzschild solution is Lyapunov unstable \cite{Torsello2017}. However, it does not exclude the existence of other asymptotically flat solutions which may exist for specific discrete values of $u$. 

\section{Boundary conditions at infinity}
\label{bc2_bigravity}

We are looking for solutions approaching the flat space at infinity: $g_{\mu\nu}=f_{\mu\nu}=\eta_{\mu\nu}$ as $r\rightarrow\infty$. We thus set 
\begin{equation}
    N=1+\delta N,\quad\quad Y=1+\delta N,\quad\quad U=r+\delta U,
\end{equation}
where $\delta N$, $\delta Y$, $\delta U$ are small deviations. Injecting this into the field equations \eqref{ode_hr} yields
\begin{align*}
    \delta N'&=-\frac{1}{r}(\kappa_2\,\delta N+\kappa_1\,\delta Y)-\kappa_1\,\delta U+\mathcal{N}_N,\\
    \delta Y'&=-\frac{1}{r}(\kappa_2\,\delta N+\kappa_1\,\delta Y)+\kappa_2\,\delta U+\mathcal{N}_Y,\\
    \label{flat_space_eq}
    \delta U'&=\left(1+\frac{2}{r^2}\right)(\delta Y-\delta N)+\mathcal{N}_U,\numberthis{}
\end{align*}
where $\mathcal{N}_N$, $\mathcal{N}_Y$, $\mathcal{N}_U$ represent all the terms that are nonlinear in $\delta N$, $\delta Y$, $\delta U$. Keeping only the linear terms, the solution to these equations is
\begin{align*}
    \delta N&=\frac{A}{r}+B\,\kappa_1\frac{1+r}{r}\e^{-r}+\,C\,\kappa_1\frac{1-r}{r}\e^{+r},\\
    \delta Y&=\frac{A}{r}-B\,\kappa_2\frac{1+r}{r}\e^{-r}-\,C\,\kappa_2\frac{1-r}{r}\e^{+r},\\
    \label{full_lin_sol}
    \delta U&=B\frac{1+r+r^2}{r^2}\e^{-r}+\,C\frac{1-r+r^2}{r^2}\e^{+r},\numberthis{}
\end{align*}
where $A$, $B$, $C$ are integration constants. The terms proportional to $A$ are the Newtonian modes describing the massless graviton. The other two modes proportional to $B$ and $C$ correspond to the massive graviton contributions, they contain the Yukawa exponents $\e^{\pm r}$ (remember that $r=\boldsymbol{m}\boldsymbol{r}$). The terms proportional to $C$ correspond to unstable\footnote{Here we are implicitly referring to the instability in the Lyapunov sense: the small deviations are time independent. We will study the dynamical stability with time-dependent perturbation in Sec.~\ref{stab_bigravity}.} modes: they diverge exponentially in the limit $r\rightarrow\infty$. As a result, \textit{flat space is not an attractor}. If one numerically integrates Eqs.~\eqref{ode_hr} starting from the horizon, trying to approach flat space in this way, the modes proportional to $\e^{+r}$ would grow rapidly and would drive the solution away from flat space. The only way to proceed is to suppress the unstable mode from the very beginning by requiring $C=0$, thus the solution at large $r$ should be 
\begin{align*}
    \delta N&=\frac{A}{r}+B\,\kappa_1\frac{1+r}{r}\e^{-r}+\dots,\\
    \delta Y&=\frac{A}{r}-B\,\kappa_2\frac{1+r}{r}\e^{-r}+\dots,\\
    \label{asymp_flat_space}
    \delta U&=B\frac{1+r+r^2}{r^2}\e^{-r}+\dots,\numberthis{}
\end{align*}
where the dots denote nonlinear corrections. The usual practice would be to neglect these corrections and assume that the linear terms approximate the solution correctly for $r>r_\star\gg r_H$. However, we have noticed that already the quadratic corrections contain an additional factor $\log(r)$ and hence dominate the linear terms for $r\rightarrow\infty$. Therefore, nonlinear corrections can be important and should not be excluded from the procedure.

Fortunately, problems of this kind have already been studied in the literature, see for example the Ref.~\cite{Breitenlohner1995}. Let us first express $\delta N$, $\delta Y$, $\delta U$ in terms of three new functions $Z_0$, $Z_{+}$, $Z_{-}$:
\begin{align*}
    \delta N&=Z_0+\kappa_1\frac{1+r}{r}Z_{+}+\kappa_1\frac{1-r}{r}Z_{-},\\
    \delta Y&=Z_0-\kappa_2\frac{1+r}{r}Z_{+}-\kappa_2\frac{1-r}{r}Z_{-},\\
    \label{def_Z}
    \delta U&=\frac{1+r+r^2}{r^2}Z_{+}+\frac{1-r+r^2}{r^2}Z_{-}.\numberthis{}
\end{align*}
Equations \eqref{flat_space_eq} then become
\begin{align*}
    Z_0'+\frac{Z_0}{r}&=\mathcal{S}_0(r,Z_0,Z_\pm)\equiv\kappa_1\,\mathcal{N}_Y+\kappa_2\,\mathcal{N}_N,\\
    Z_{+}'+Z_{+}&=\mathcal{S}_{+}(r,Z_0,Z_\pm)\equiv\frac{1-r+r^2}{2r^2}(\mathcal{N}_N
-\mathcal{N}_Y)-\frac{1-r}{2r}\mathcal{N}_U,\\
\label{eq_Z}
    Z_{-}'-Z_{-}&=\mathcal{S}_{-}(r,Z_0,Z_\pm)\equiv\frac{1+r+r^2}{2r^2}(\mathcal{N}_Y
-\mathcal{N}_N)+\frac{1+r}{2r}\mathcal{N}_U.\numberthis{}
\end{align*}
We emphasize that all the terms on the left in these equations are linear in $Z_0$, $Z_\pm$, while those on the right are nonlinear. Keeping only the linear terms, the solution is $Z_0\propto 1/r$, $Z_{+}\propto\e^{-r}$ and $Z_{-}\propto\e^{+r}$. Hence the redefinition of the functions \eqref{def_Z} allows one to separate the different modes present in Eq.~\eqref{full_lin_sol}. If we set
\begin{equation}
\label{Z_sol_lin}
    Z_0=\frac{A}{r},\quad\quad Z_{+}=B\,\e^{-r},\quad\quad Z_{-}=0,
\end{equation}
this reproduces the linear part of \eqref{asymp_flat_space}. Now, to take nonlinear corrections into account, we convert Eqs.~\eqref{eq_Z} into the equivalent set of integral equations,
\begin{align*}
    Z_0(r)&=\frac{A}{r}-\int_r^\infty{\frac{\bar{r}}{r}\mathcal{S}_0(\bar{r},Z_0(\bar{r}),Z_\pm(\bar{r}))\,d\bar{r}},\\
    Z_{+}(r)&=B\e^{-r}+\int_{r_\star}^r{\e^{\bar{r}-r}\mathcal{S}_+(\bar{r},Z_0(\bar{r}),Z_\pm(\bar{r}))\,d\bar{r}},\\
    \label{integ_Z}
    Z_{-}(r)&=-\int_r^\infty{\e^{r-\bar{r}}\mathcal{S}_-(\bar{r},Z_0(\bar{r}),Z_\pm(\bar{r}))\,d\bar{r}},\numberthis{}
\end{align*}
where $r_\star$ is some large value. These equations determine the solution in the far field region $r>r_\star$, and they are solved by iterations. We start the iterations by injecting the configuration \eqref{Z_sol_lin} to the integrals, which gives a corrected configuration, and so on. In practice, we introduce compactified variables $x=r_\star/r$ and $\bar{x}=r_\star/\bar{r}$ which map the infinite interval $[r_\star,\infty)$ to the finite range $[0,1]$ and we discretize this interval to compute the integrals by the trapezoid rule.

\begin{figure}
    \centering
    \includegraphics[width=7.5cm]{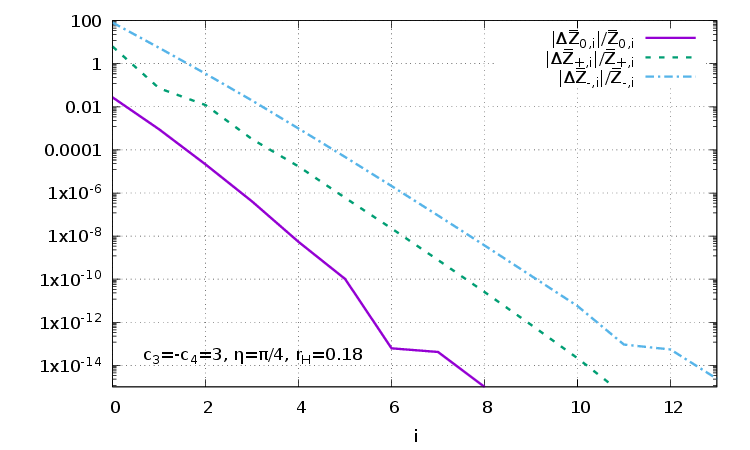}
    \includegraphics[width=7.5cm]{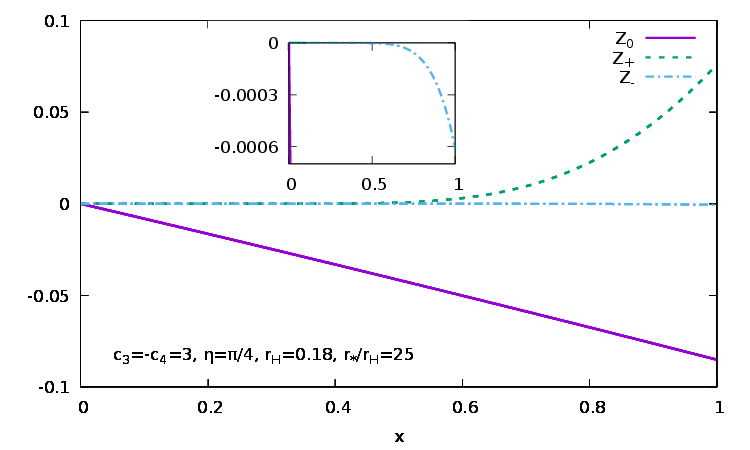}
    \caption[Convergence of the interations of the integral equations \eqref{integ_Z} and the corresponding solution profiles.]{Left: convergence of the iterations of the integral equations \eqref{integ_Z}. Right: the functions $Z$, $Z_\pm$ against $x=r_\star/r$. The insertion shows a closeup of $Z_{-}$.}
    \label{plot_Z}
\end{figure}

To see if the iterations converge, we compute for each function $Z\in\{Z_0,Z_{\pm}\}$ and for each discretization node the difference $\Delta Z_i=Z_{i+1}-Z_{i}$ between each successive iteration, and then we take the average $\overline{\Delta Z}_i$ over all discretization nodes. Computing similarly the average $\overline{Z}_i$ of $Z_i$, we see that the ratios $\overline{\Delta Z}_i/\overline{Z}_i$ decrease with $i$ exponentially, see the left panel of Fig.~\ref{plot_Z}. An example of solution of the integral equations is shown on the right panel in Fig.~\ref{plot_Z}. In the insertion of the plot, we can see that the amplitude $Z_{-}$ is always small but nonvanishing. All the three amplitudes vanish at spatial infinity ($x=0$), hence the solution is indeed asymptotically flat.

To recapitulate, the procedure described above yields an asymptotically flat solution in the region $r>r_\star$ and takes nonlinear corrections into account. To extend this solution to the region $r_H<r<r_\star$, one only needs its value at $r=r_\star$,
\begin{align*}
    Z_0(r_\star)&=\frac{A}{r_\star}-\int_{r_\star}^\infty{\frac{\bar{r}}{r_\star}\mathcal{S}_0(\bar{r},Z_0(\bar{r}),Z_\pm(\bar{r}))\,d\bar{r}},\\
    Z_{+}(r_\star)&=B\e^{-r_\star},\\
    \label{Z_rstar}
    Z_{-}(r_\star)&=-\int_{r_\star}^\infty{\e^{r_\star-\bar{r}}\mathcal{S}_-(\bar{r},Z_0(\bar{r}),Z_\pm(\bar{r}))\,d\bar{r}}.\numberthis{}
\end{align*}
It is worth noting that the parameter $A$ is related to the ADM mass,
\begin{equation}
    M=-A.
\end{equation}

\section{Numerical procedure}
\label{num_bigravity}

Summarizing the above discussion, the asymptotically flat black holes in the massive bigravity theory are described by solutions of the three coupled first order ODEs \eqref{ode_hr} which determine the functions $N(r)$, $Y(r)$ and $U(r)$. At the horizon $r=r_H$ one has
\begin{equation}
    N(r)=\nu(r)\sqrt{1-\frac{r_H}{r}},\quad\quad Y(r)=y(r)\sqrt{1-\frac{r_H}{r}},
\end{equation}
where the horizon value $\nu(r_H)\equiv\nu_H$ and $y(r_H)\equiv y_H$ are finite and determined by the Eqs.~\eqref{biquad_nu}, \eqref{yh} in Appendix~\ref{desing_hor_bigravity}. On the other hand, the value $U(r_H)\equiv u r_H$ can be arbitrary. Therefore, the boundary conditions at the horizon are labeled by just one free parameter $u$, and choosing some value for it completely determines the solution for $r>r_H$. However, the solutions are not necessarily asymptotically flat. One has to consider the appropriate boundary conditions in the far field region.

Far from the horizon, at $r=r_\star\gg r_H$, one has
\begin{align*}
    N(r_\star)&=1+Z_0(r_\star)+\kappa_1 B\frac{1+r_\star}{r_\star}\e^{-r_\star}+\kappa_1\frac{1-r_\star}{r_\star}Z_{-}(r_\star),\\
    Y(r_\star)&=1+Z_0(r_\star)-\kappa_2 B\frac{1+r_\star}{r_\star}\e^{-r_\star}-\kappa_2\frac{1-r_\star}{r_\star}Z_{-}(r_\star),\\
    U(r_\star)&=r_\star+B\frac{1+r_\star+r_\star^2}{r_\star^2}\e^{-r_\star}+\frac{1-r_\star+r_\star^2}{r_\star^2}Z_{-}(r_\star),\numberthis{}
\end{align*}
where $Z_0(r_\star)$ and $Z_{-}(r_\star)$ are functions of $A$, $B$ determined by \eqref{Z_rstar}.

As a result, we have the boundary conditions at $r=r_H$ labeled by $u$ and the boundary conditions at $r=r_\star$ labeled by $A$, $B$. We use them to construct solutions in the region $r_H\leq r\leq r_\star$ by using the \textit{multishooting} method presented in Appendix \ref{num_bvp}. The three shooting parameters to be adjusted are $u$, $A$ and $B$.

As a consistency check, one can verify that the bi-Schwarzschild solution is obtained for the values of the shooting parameters
\begin{equation}
\label{uab_sch}
    u=1,\quad\quad A=-\frac{r_H}{2},\quad\quad B=0.
\end{equation}
Are there other solutions? Since the bi-Schwarzschild solution is Lyapunov unstable \cite{Torsello2017}, the different asymptotically flat black holes are \textit{parametrically isolated} from each other. This creates a practical problem. Indeed, the initial guess for $u$, $A$, $B$ must be close to the "true" values for the Newton's iterations \eqref{newton} to converge. Hence, some additional information is needed to specify an appropriate initial guess.

This additional information is provided by the stability analysis of the bi-Schwarschild solution \eqref{bi_sch} \cite{Babichev2013,Brito2013}. In this analysis, one finds that for $r_H=0.86$ the time-dependent perturbation equations admit a \textit{static} solution (a zero mode) for which the small fluctuations around the bi-Schwarzschild background depend only on $r$ and are bounded everywhere in the exterior region $r\geq r_H$. This can be viewed as a perturbative approximation of a new solution that merges with the bi-Schwarzschild one for $r_H=0.86$.

This suggests that these new solutions should be obtained by choosing the event horizon radius to be close to $r_H=0.86$ and choose the initial guess for $u$, $A$, $B$ to be close to \eqref{uab_sch}. Then the numerical iterations should converge to values $u$, $A$, $B$ which are slightly different from \eqref{uab_sch} and correspond to an almost bi-Schwarzschild black hole distorted by massive hairs. Changing then iteratively the value of $r_H$ yields solutions which can deviate considerably from the bi-Schwarzschild metrics in the vicinity of the horizon, but always approach flat spacetime in the far field region.

\section{Asymptotically flat hairy black holes}
\label{hairy_bh_bigravity}

Applying the procedure outlined above, we have been able to construct asymptotically flat hairy black hole solutions. We confirm the results of Ref.~\cite{Brito2013a} and obtain also new results.

\subsection{General properties}

\begin{figure}
    \centering
    \includegraphics[width=7.8cm]{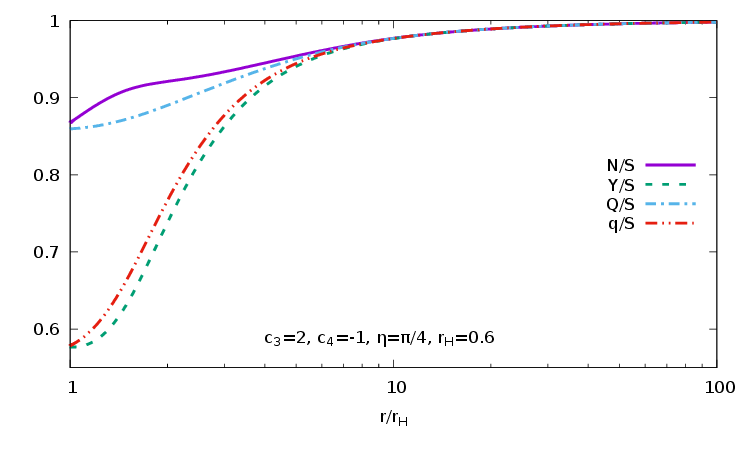}
    \includegraphics[width=7.8cm]{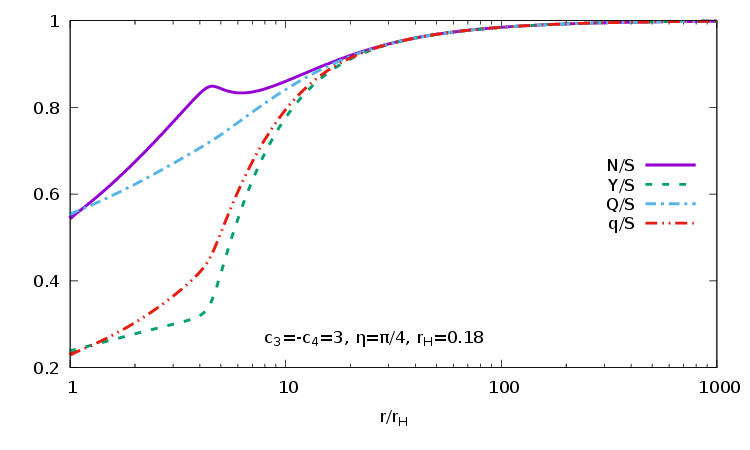}
    \includegraphics[width=7.8cm]{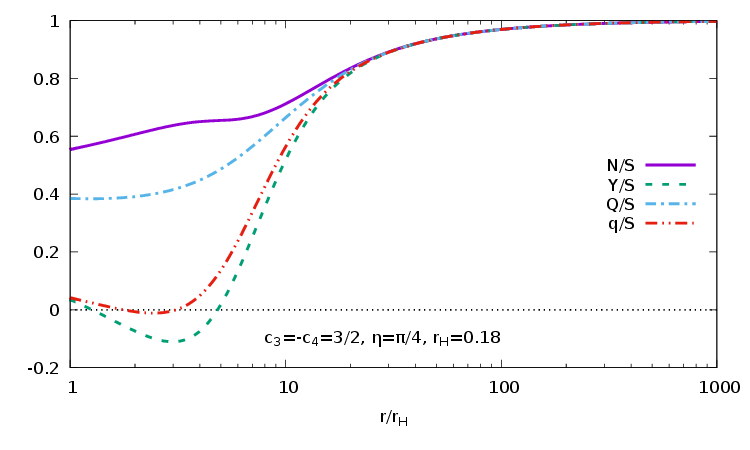}
    \includegraphics[width=7.8cm]{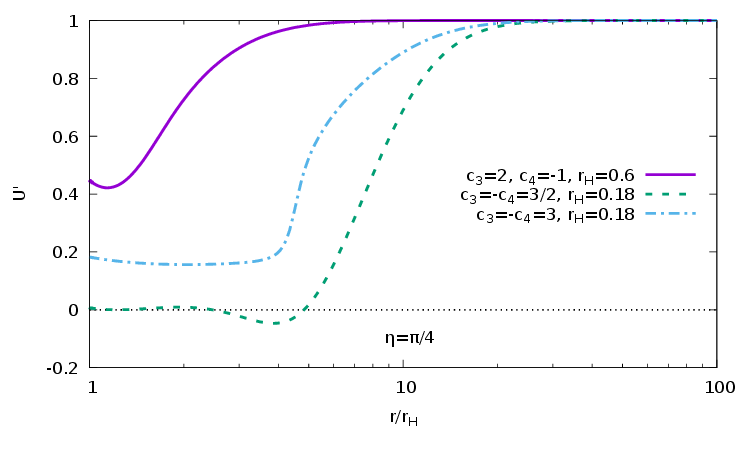}
    \caption[Profiles of hairy black hole solutions with $\eta=\pi/4$ and various values of $r_H$, $c_3$, $c_4$.]{Profiles of $N/S$, $Q/S$, $Y/S$, $q/S$ with $S=\sqrt{1-r_H/r}$ and that of $U'$ for solutions with $\eta=\pi/4$ but for various values of $r_H$, $c_3$, $c_4$. The solution with $c_3=-c_4=3/2$ shown on the two lower panels have a singular $f$-metric because the functions $q$, $Y$, $U'$ develop zeroes outside the horizon.}
    \label{amp1_hr}
\end{figure}

First, we find that for $r_H$ approaching from below the GL value, $r_H\approx 0.86$, there are asymptotically flat hairy solutions for any $c_3$, $c_4$, $\eta$. As expected, they are very close to the bi-Schwarzschild solution: one has $u=U_H/r_H\approx 1$ and the ADM mass $M\approx r_H/2$. However for smaller values of $r_H$ the solutions deviate more and more from Schwarzschild. To illustrate this, we present in Fig.~\ref{amp1_hr} the ratios $N/S$, $Q/S$, $Y/S$, $q/S$ and the derivative $U'$ for $\eta=\pi/4$ and different values of $c_3$, $c_4$. If these functions are equal to one, then the solution corresponds to the bi-Schwarzschild metrics \eqref{bi_sch}. As one can see, they deviate considerably from unity in the vicinity of the horizon. Hence the massive graviton hair is concentrated in this region.

\begin{figure}
    \centering
    \includegraphics[width=7.8cm]{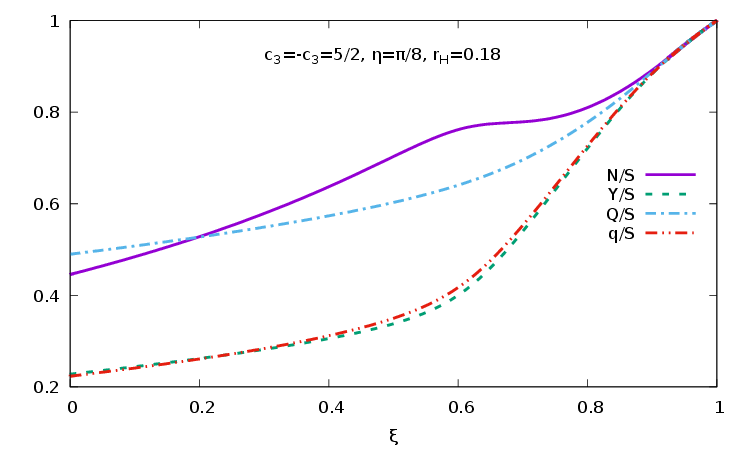}
    \includegraphics[width=7.8cm]{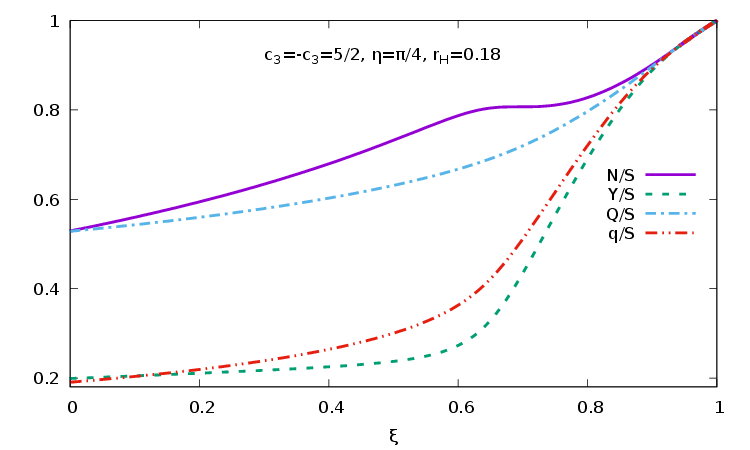}
    \includegraphics[width=7.8cm]{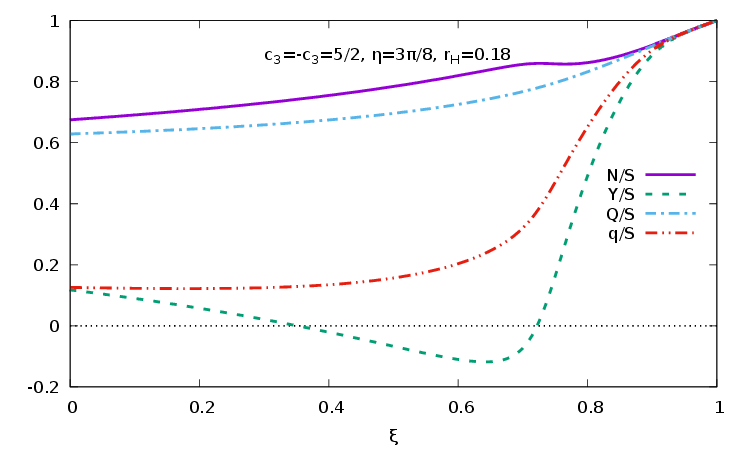}
    \includegraphics[width=7.8cm]{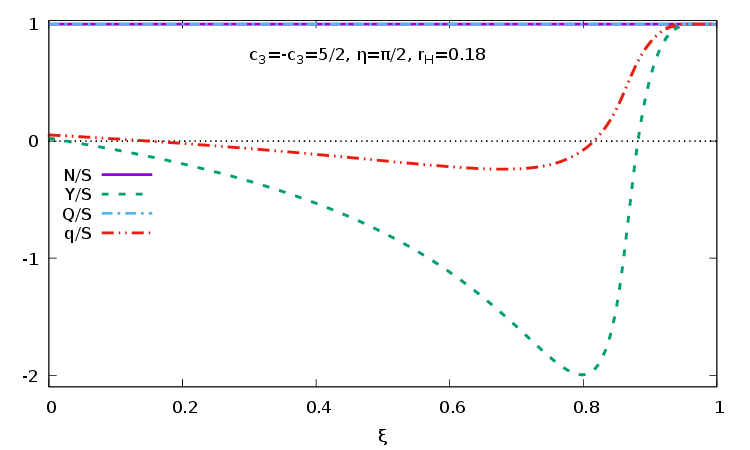}
    \caption[Profiles of hairy black hole solutions with $c_3=-c_4=5/2$, $r_H=0.18$ and various values of $\eta$.]{Profiles of $N/S$, $Q/S$, $Y/S$, $q/S$ with $S=\sqrt{1-r_H/r}$ against $\xi=(r-r_H)/(r+r_H)$ for solutions with the same $c_3$, $c_4$, $r_h$ but different values of $\eta$. When $\eta$ approaches $\pi/2$ then the functions $Y$, $q$ start showing zeroes. For $\eta=\pi/2$ the $g$-metric is Schwarzschild with $N/S=Q/S=1$.}
    \label{amp2_hr}
\end{figure}

Solutions are regular for $r_H$ close to $0.86$. However for smaller $r_H$ and depending on the values of $c_3$, $c_4$, $\eta$, the functions $Y$, $q$, $U'$ may show negative values outside the horizon, whereas $Q$, $N$ always remain positive. This implies that the $f$-metric is singular, because the curvature invariants of its Riemann tensor diverge where $Y$, $q$ or $U'$ vanish. An example of this is shown on the lower two panels in Fig.~\ref{amp1_hr}, and also on the lower two panels in Fig.~\ref{amp2_hr} where one can see that the phenomenon occurs when $\eta$ approaches $\pi/2$. Nevertheless, a curvature singularity of the $f$-metric does not invalidate the solution because the $f$-geometry is not directly measurable by test particles. The latter follow the geodesics of the $g$-metric which is always regular when $f_{\mu\nu}$ is not. We shall therefore keep such solutions in our considerations.

The solutions in Fig.~\ref{amp1_hr} are shown up to large but still finite values of the radial coordinate ($r/r_H\leq 100$ or $r/r_H\leq 1000$). What is shown is the combination of the solutions of the differential equations \eqref{ode_hr} in the region $r_H\leq r\leq r_\star$ and of the integral equations \eqref{integ_Z} for $r>r_\star$ where $r_\star/r_H=25$. At the same time, our procedure yields solutions in the whole region $r\in[r_H,\infty)$. Introducing the compactified radial variable
\begin{equation}
    \xi=\frac{r-r_H}{r+r_H}\in[0,1],
\end{equation}
we plot in Fig.~\ref{amp2_hr} the ratios $N/S$, $Y/S$, $Q/S$, $q/S$ against $\xi$. As one can see, they approach unity as $\xi\rightarrow 1$ (and the same is true for $U'$), hence the solutions are indeed asymptotically flat. 

\begin{figure}
    \centering
    \includegraphics[width=7.8cm]{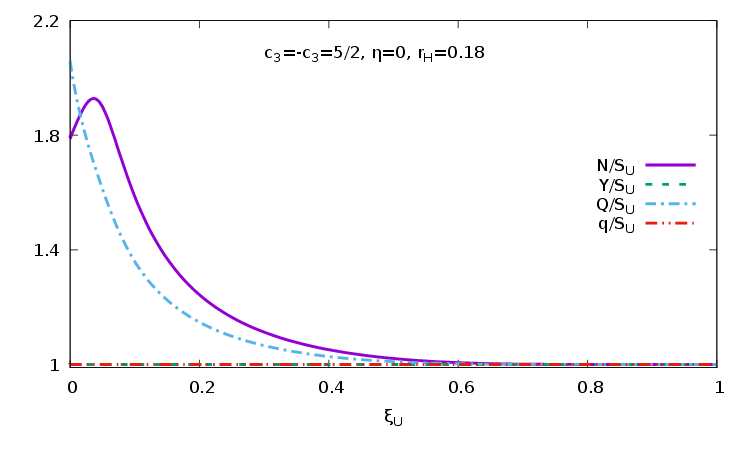}
    \includegraphics[width=7.8cm]{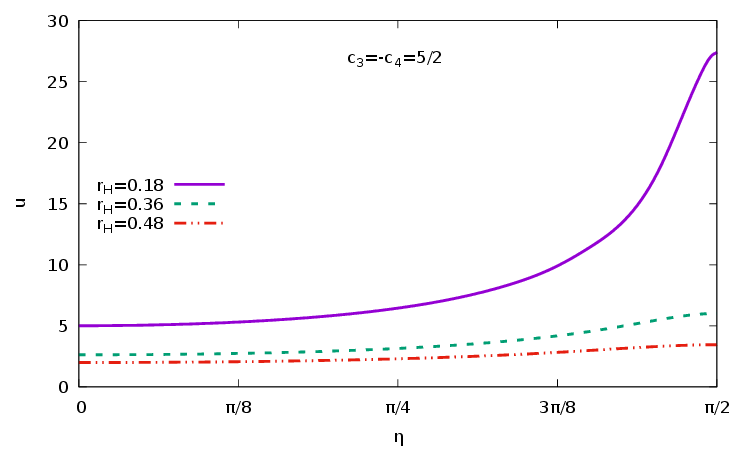}
    \includegraphics[width=7.8cm]{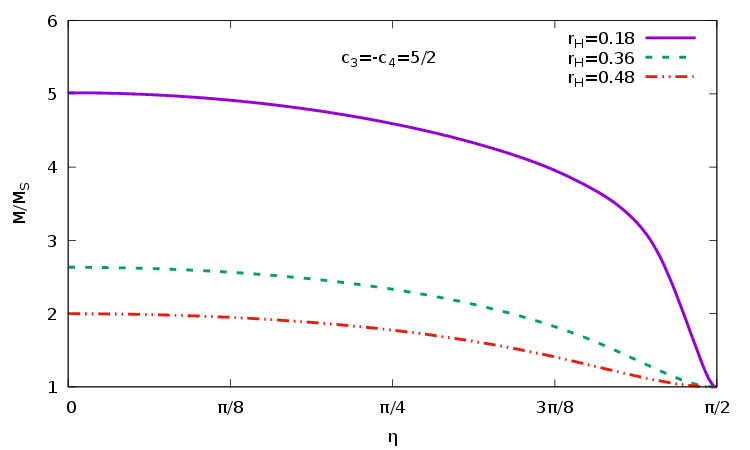}
    \includegraphics[width=7.8cm]{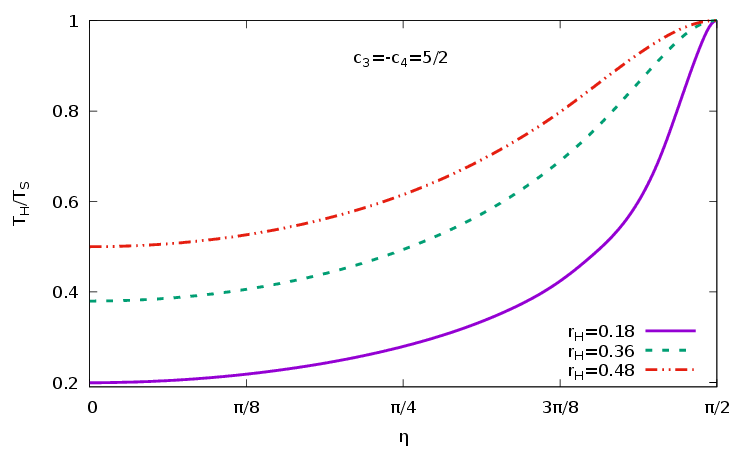}
    \caption[Profiles of the hairy black hole solution with $c_3=-c_4=5/2$, $r_H=0.18$, $\eta=0$ and the values of the horizon parameter $u$, of the ADM mass $M$, of the temperature $T_H$ against $\eta$.]{Upper left: $N/S_U$, $Q/S_U$, $Y/S_U$, $q/S_U$ with $S_U=\sqrt{1-U_H/U}$ against $\xi_U=(U-U_H)/(U+U_H)$ for $\eta=0$. One has $Y/S_U=q/S_U=1$ hence the $f$-metric is Schwarzschild. The other three panels show $u=U_H/r_H$, the ADM mass $M$ and the temperature $T_H$ against $\eta$ for three different values of $r_H$ and $c_3=-c_4=5/2$.}
    \label{uMT_eta}
\end{figure}

If $\eta=\pi/2$ then $\kappa_1=0$ and the field equation for $g_{\mu\nu}$ becomes identical to that of vacuum GR so that the $g$-metric becomes Schwarzschild. In this limit, the theory reduces to the massive gravity for the dynamical $f$-metric on a fixed Schwarzschild background. The solution for the $f$-metric is shown on the lower right panel in Fig.~\ref{amp2_hr}. Similarly, for $\eta=0$ one has $\kappa_2=0$ and the $f$-metric becomes Schwarzschild while the $g$-metric is solution of the massive gravity on a Schwarzschild background. This is shown on the upper left panel in Fig.~\ref{uMT_eta}. It should be emphasized that the radii of the background Schwarzschild black holes for $\eta=0$ and $\eta=\pi/2$ are not the same. Indeed the event horizon radius measured by the $g$-metric is $r_H$ but that measured by the $f$-metric is $U(r_H)\equiv U_H$. Thus, for the solution shown in Fig.~\ref{amp2_hr} with $\eta=\pi/2$, the Schwarzschild black hole is described by the $g$-metric and its horizon size is $r_H=0.18$, while for $\eta=0$ the Schwarzschild background is described by the $f$-metric with the horizon size $U_H=ur_H$ where $u\approx 5$ (as seen in Fig.~\ref{uMT_eta}); hence in this case the Schwarzschild black hole is much larger. As a result, solutions on these different backgrounds look quite different: the solution for the $f$-metric on the lower right panel in Fig.~\ref{amp2_hr} shows zeroes hence it is singular, while the solution for the $g$-metric on the upper left panel in Fig.~\ref{uMT_eta} is regular.

Solutions for $\eta=\pi/2$ will play an important role in the Section~\ref{param_bigravity} below. We shall call them "hairy Schwarzschild" because their $g$-metric is Schwarzschild but their $f$-metric supports hairs.

The figure~\ref{uMT_eta} presents the dependence on $\eta$ of $u=U_H/r_H$, of the ADM mass $M$ expressed in units of the Schwarzschild mass $M_S=r_H/2$ and of the temperature $T$ expressed in units of the Schwarzschild temperature $T_S=1/(4\pi r_H)$. As one can see, the dependence is rather strong for small $r_H$, in particular for $u$. It should be emphasized that the mass as well as the temperature is the same with respect to each metric. If $\eta=\pi/2$ then the $g$-metric is Schwarzschild hence $M=M_S$ and $T=T_S$. If $\eta=0$ then the $f$-metric is Schwarzschild but with a larger radius $U_H=ur_H$, hence the mass is larger, $M=U_H/2=uM_S$, while the temperature is smaller, $T=T_S/u$. Therefore, if $\eta=0$ then $M/M_S=u$ so that, for example, $M/M_S\approx 5\approx u$ for $r_H=0.18$, as seen in Fig.~\ref{uMT_eta}. Between the two limiting cases $\eta=0,\pi/2$, the mass $M$, the temperature $T_H$ and $u$ are monotonic functions of $\eta$.

\begin{figure}
    \centering
    \includegraphics[width=7.8cm]{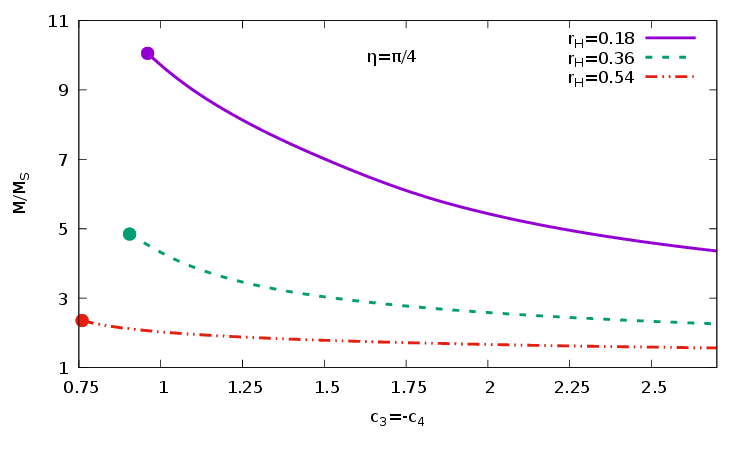}
    \includegraphics[width=7.8cm]{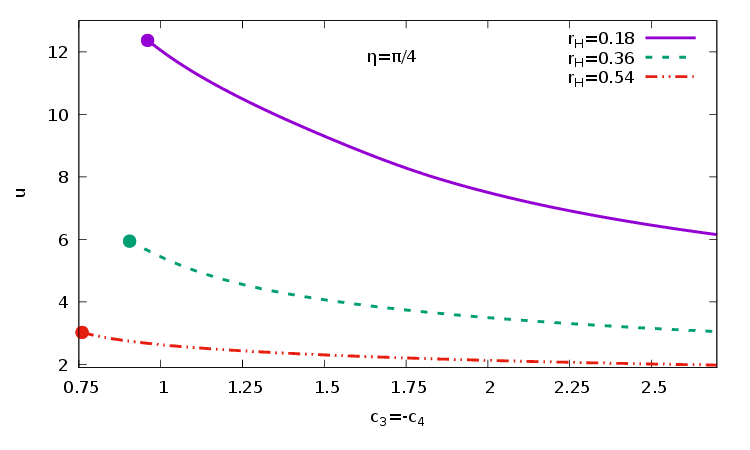}
    \caption[The ADM mass $M$ and the horizon parameter $u$ against $c_3=-c_4$ for $\eta=\pi/4$ and various values of $r_H$.]{The ADM mass $M$ (left) and $u=U_H/r_H$ (right) against $c_3=-c_4$ for three different values of $r_H$ and $\eta=\pi/4$.}
    \label{uM_c3c4}
\end{figure}

The figure~\ref{uM_c3c4} shows the dependence of $u$ and $M$ on $c_3$ in the special case $c_3=-c_4$. One can see that the solutions exist only if the value of $c_3$ is not too small. The points where the solutions cease to exist are represented in the figure by plain circles. Similarly, the hairy solutions generically do not exist for arbitrarily small values of $r_H$. As was noticed in Ref.~\cite{Brito2013a}, small $r_H$ black holes exist if the coefficient $b_3$ in the potential \eqref{pot_hr} vanishes so that the cubic part of the potential is absent. In view of \eqref{bk}, this requires that $c_3=-c_4$, but it is not the only condition. One can distinguish the following two cases:
\begin{equation}
\label{cases_hr}
    \text{I:}\;\;c_3\neq -c_4\quad\text{or}\quad c_3=-c_4<1,\quad\quad\text{II:}\;\;c_3=-c_4\geq 1.
\end{equation}
In case I, asymptotically flat hairy black holes do not exist if $r_H<r_H^\text{min}$ where the value of $r_H^\text{min}>0$ depends on the specific values of $c_3$, $c_4$, $\eta$. In case II, they exist for arbitrarily small $r_H$, although their $f$-metric may be singular. We shall see below in Sec.~\ref{param_bigravity} what happens when $r_H$ approaches the lower bound.

\subsection{Duality relation}

In Ref.~\cite{Brito2013a}, the authors report the existence of asymptotically flat hairy black holes only below the GL point, for $r_H\leq 0.86$. However the interchange symmetry \eqref{inter_sym} implies that hairy solutions should exist as well for $r_H>0.86$. Indeed, the theory is left invariant under the replacement 
\begin{equation}
\label{dual_expl}
    \eta\rightarrow\frac{\pi}{2}-\eta,\quad Q\leftrightarrow q,\quad N\leftrightarrow Y,\quad r\leftrightarrow U,\quad c_3\rightarrow 3-c_3,\quad c_4\rightarrow 4c_3+c_4-6.
\end{equation}
This means that if for some values of $\eta$, $c_3$, $c_4$ there is a solution
\begin{equation}
\label{sol_1}
    Q(r),\quad q(r),\quad N(r),\quad Y(r),\quad U(r),
\end{equation}
then for $\tilde{\eta}=\pi/2-\eta$, $\tilde{c}_3=3-c_3$, $\tilde{c}_4=4c_3+c_4-6$ there should be a "dual" solution described by
\begin{align*}
    \tilde{Q}(r)&=q(w(r)),\quad\tilde{q}(r)=Q(w(r)),\quad\tilde{N}(r)=Y(w(r)),\\
\label{sol_2}
    \tilde{Y}(r)&=N(w(r)),\quad\tilde{U}(r)=w(r),\numberthis
\end{align*}
where $w(r)$ is the inverse function of $U(r)$, such that $U(w(r))=r$. This duality correspondence relates between themselves black holes of different sizes: the solution \eqref{sol_1} has the horizon located at $r=r_H$ while that of \eqref{sol_2} is located at $r=\tilde{r}_H=U(r_H)$. Consequently, one has
\begin{equation}
    \tilde{u}=\frac{\tilde{U}(\tilde{r}_H)}{\tilde{r}_H}=\frac{r_H}{U(r_H)}=\frac{1}{u}.
\end{equation}
Now, for hairy black holes with $r_H<0.86$, one always has $U(r_H)>0.86$ and $u>1$. It follows that their duals are  characterized by $\tilde{r}_H>0.86$ and by $\tilde{u}<1$.

\begin{figure}
    \centering
    \includegraphics[width=7.8cm]{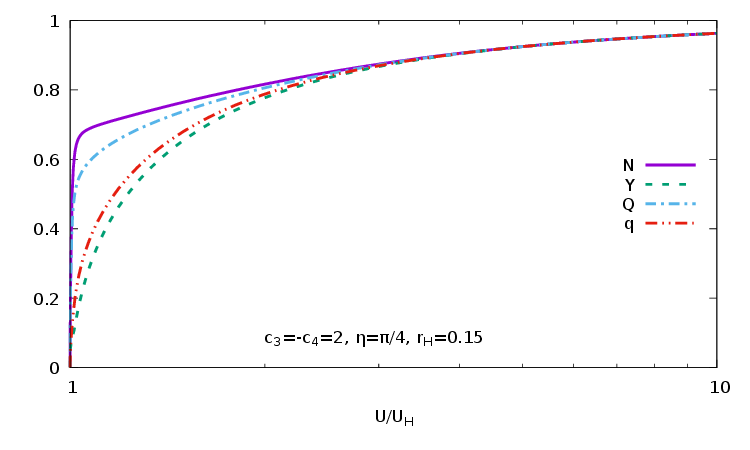}
    \includegraphics[width=7.8cm]{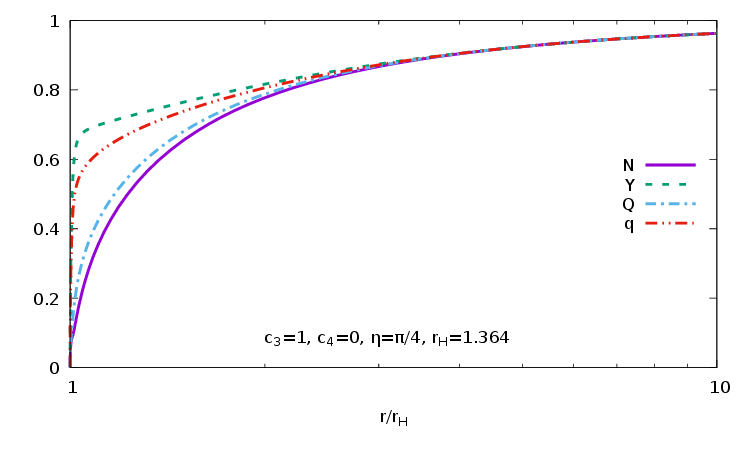}
    \caption[The hairy solution with $c_3=-c_4$, $\eta=\pi/4$, $r_H=0.15$ and its dual.]{The solution with $c_3=-c_4=2$, $\eta=\pi/4$, $r_H=0.15$ (left) and the dual solution with $c_3=1$, $c_4=0$, $\eta=\pi/4$, $r_H=1.364$ (right). The profiles on the two panels are exactly the same, up to the interchange $N\leftrightarrow Y$, $Q\leftrightarrow q$, $r/r_H\leftrightarrow U/U_H$.}
    \label{dual_bh}
\end{figure}

An explicit example of dual black holes is shown in Fig.~\ref{dual_bh}. The left panel presents the solution for $c_3=-c_4=2$, $\eta=\pi/4$, $r_H=0.15$ for which $U(r_H)=1.364$, hence $u=1.364/0.15=2.42$. The right panel presents the dual solution which exists for $c_3=1$, $c_4=0$, $\eta=\pi/4$. Its horizon radius is $r_H=1.364$ and $u=0.15/1.364=0.41$. Plotting the first solution against $U/U_H$ and the second against $r/r_H$ yields exactly the same profiles, up to the interchange $N\leftrightarrow Y$, $Q\leftrightarrow q$.

The reason why solutions with $r_H>0.86$ were not found in Ref.~\cite{Brito2013a} is unclear. A possible explanation is that the bi-Schwarzschild black hole \eqref{bi_sch} is unstable only for $r_H<0.86$. Therefore the standard argument for the existence of new solutions states that these should only exist for $r_H<0.86$ and they may be viewed as stable remnants of the decay of the bi-Schwarzschild black hole. However as we shall see in Sec.~\ref{param_bigravity} below, the situation for hairy black holes in Massive Bigravity is more complicated than this naive picture.

The duality is in fact a powerful tool for studying the solutions because, sometimes, their properties may look puzzling in one description, but become obvious within the dual description.

\section{Stability analysis}
\label{stab_bigravity}

\subsection{Derivation of the perturbation equation}

In this section we analyze the stability of the hairy solutions by studying their spherically symmetric perturbations within the ansatz described in Appendix \ref{time_dep_ansatz},
\begin{align*}
    ds^2_g&=-Q^2dt^2+\frac{dr^2}{N^2}+r^2d\Omega^2,\\
    ds^2_f&=-(q^2-\alpha^2 Q^2 N^2)dt^2-2\alpha\left(q+\frac{QNU'}{Y}\right)dtdr+\left(\frac{U'^2}{Y^2}-\alpha^2\right)dr^2+U^2 d\Omega^2,\numberthis{}
\end{align*}
where $Q$, $q$, $N$, $Y$, $\alpha$, $U$ are functions of $r$ and $t$. The static and bi-diagonal ansatz \eqref{ansatz_hr_2} is recovered when nothing depends on time and when $\alpha=0$. Therefore, we describe small fluctuations around the static solutions by setting
\begin{align*}
    Q(r,t)&=\accentset{(0)}{Q}(r)+\delta Q(r,t),\quad\quad q(r,t)=\accentset{(0)}{q}\,(r)+\delta q(r,t),\\
    N(r,t)&=\accentset{(0)}{N}(r)+\delta N(r,t),\quad\quad Y(r,t)=\accentset{(0)}{Y}(r)+\delta Y(r,t),\\
    \label{pert_hr}
    U(r,t)&=\accentset{(0)}{U}(r)+\delta U(r,t),\quad\quad \alpha(r,t)=\delta\alpha(r,t),\numberthis
\end{align*}
where the functions $\accentset{(0)}{Q}(r)$, $\accentset{(0)}{q}\,(r)$, $\accentset{(0)}{N}(r)$, $\accentset{(0)}{Y}(r)$, $\accentset{(0)}{U}(r)$ correspond to the background black hole and $\delta Q$, $\delta q$, $\delta N$, $\delta Y$, $\delta U$, $\delta\alpha$ are the small fluctuations.

We inject \eqref{pert_hr} to the field equations and linearize with respect to the perturbations. Linearizing the equation $\tensor{G(g)}{^0_1}=\kappa_1\tensor{T}{^0_1}$ yields
\begin{equation}
\label{pert_off_diag}
    \frac{2}{r\, NQ^2}\delta\dot{N}=\kappa_1\frac{\mathcal{P}_1}{Q}\delta\alpha,
\end{equation}
where $N$, $Q$, $\mathcal{P}_1$ relate to the static background (we do not write their over sign "$(0)$" for simplicity). In the linear perturbation theory, one can consistently separate the time variable by assuming the harmonic time dependence for all fluctuations:
\begin{equation}
    \delta N(r,t)=\e^{i\omega t}\delta N(r),\quad\quad\delta\alpha(r,t)=\e^{i\omega t}\delta\alpha(r),
\end{equation}
and similarly for $\delta Y$, $\delta Q$, $\delta q$, $\delta U$. Injecting to \eqref{pert_off_diag} yields the algebraic relation
\begin{equation}
    \delta\alpha(r)=\frac{2i\omega}{r\,N Q\mathcal{P}_1}\delta N(r).
\end{equation}
Then the linearization of $\tensor{G(f)}{^0_1}=\kappa_2\tensor{\mathcal{T}}{^0_1}$ yields a relation between $\delta\alpha$, $\delta Y$, $\delta U$. Using these relations, one finds that the three equations $\tensor{G(g)}{^0_0}=\kappa_1\tensor{T}{^0_0}$, $\tensor{G(f)}{^0_0}=\kappa_2\tensor{\mathcal{T}}{^0_0}$ and $\accentset{(g)}{\nabla}_\mu\tensor{T}{^\mu_0}=0$ are equivalent. As a result, among the 8 equations \eqref{pert_eq_hr_1}, \eqref{pert_eq_hr_2}, only 6 are independent at the linearized level.

Taking all of this into account, one finds that the 6 perturbation functions $\delta Q(r)$, $\delta q(r)$, $\delta N(r)$, $\delta Y(r)$, $\delta U(r)$ and $\delta\alpha(r)$ can be expressed in terms of a single master amplitude $\Psi(r)$ subject to the Schr\"odinger-type equation,
\begin{equation}
\label{master_pert_hr}
    \frac{d^2\Psi}{dr_\ast^2}+(\omega^2-V(r_\ast))\Psi=0.
\end{equation}
The function $\Psi(r)$ is a linear combination of $\delta N(r)$ and $\delta Y(r)$ with rather complicated coefficients involving the static background functions. The potential $V(r)$ is also a complicated function that we do not show. The tortoise radial coordinate $r_\ast\in(-\infty,+\infty)$ is defined by the relation
\begin{equation}
    dr_\ast=\frac{1}{a(r)}dr,
\end{equation}
where the function $a(r)$ (also complicated) varies monotonically from 0 to 1 when $r$ ranges from $r_H$ to $\infty$. The potential $V$ always tends to zero at the horizon, for $r_\ast\to -\infty$, and it approaches unit value at infinity, for $r_\ast\to +\infty$.

For the bi-Schwarzschild background with $Q=q=N=Y=\sqrt{1-r_H/r}$ and $U=r$, one has $a(r)=Q(r)$ and the potential admits a simple expression
\begin{equation}
\label{potsch}
    V(r)=\left(1-\frac{r_H}{r}\right)\left(1+\frac{r_H}{r^3}+6\frac{r_H(r_H-2r)+r^3(r-2r_H)}{(r_H+r^3)^2}\right),
\end{equation}
in agreement with Ref.~\cite{Brito2013}. In the flat space limit, $r_H\to 0$, this reduces to $V(r)=1+6/r^2$, which is the potential of a massive particle of unit mass with spin 2 (remember that we use dimensionless quantities so that masses are expressed in units of the graviton mass).

Equation \eqref{master_pert_hr} defines an eigenvalue problem on the line $r_\ast\in(-\infty,+\infty)$. Solutions of this problem with $\omega^2>0$ describe scattering states of gravitons. There may also be solutions with purely imaginary frequency $\omega=\pm i\sigma$ (here $\sigma>0)$ and hence with $\omega^2=-\sigma^2<0$. Such solutions are usually called bound states because the wave function $\Psi$ is everywhere bounded and square-integrable. One has
\begin{equation}
\label{asymp_psi_bigravity_tor}
    A\e^{+\sigma r_\ast}\leftarrow\Psi\rightarrow B\e^{-\sqrt{1+\sigma^2}\,r_\ast}\quad\text{as}\quad -\infty\leftarrow r_\ast\rightarrow +\infty,
\end{equation}
where $A$, $B$ are integration constants. The time dependence of the perturbations in this case is $\e^{i\omega t}=\e^{\pm\sigma t}$ so that they may grow in time exponentially. Therefore, the bound state solutions to the eigenvalue problem \eqref{master_pert_hr} correspond to unstable modes of the background black hole.

\subsection{Computing the eigenfrequencies}

We shall investigate the potential existence of bound state solutions with $\omega^2<0$ in the spectrum of the eigenvalue problem \eqref{master_pert_hr}. If such solutions exist, then the background black hole is unstable, otherwise, the black hole is stable with respect to spherically symmetric perturbations. In principle, it would then be necessary to consider more general perturbations but in view of the complexity of the problem even in the case of spherically symmetric perturbations, we shall not pursue the analysis further. Moreover, in many cases, the potential instabilities reside in the spherically symmetric sector (of course, this should be proven on a case-by-case basis).

\begin{figure}
    \centering
    \includegraphics[width=7.8cm]{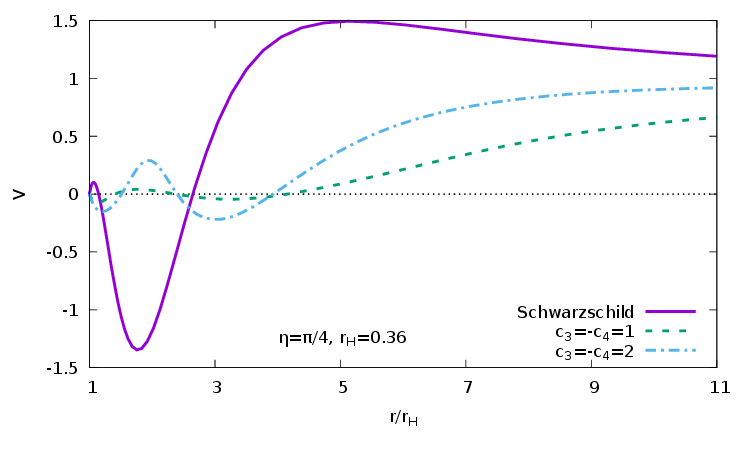}
    \includegraphics[width=7.8cm]{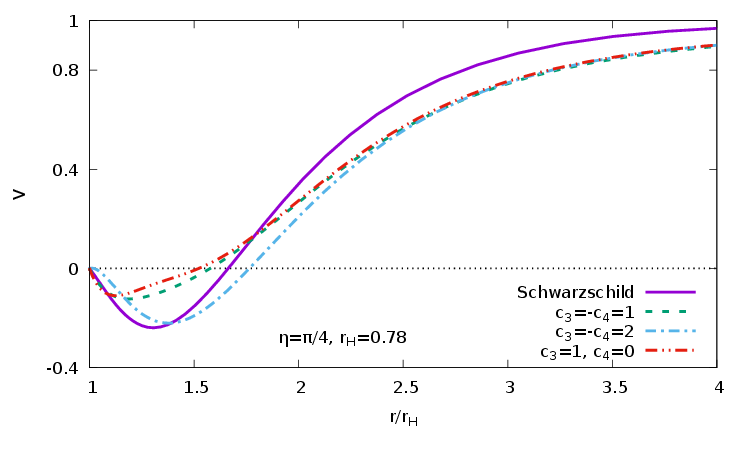}
    \caption[The potential entering the perturbation equation \eqref{master_pert_hr} for $\eta=\pi/4$ and various values of $c_3$, $c_4$, $r_H$.]{The potential $V(r)$ entering the perturbation equation \eqref{master_pert_hr} for $r_H=0.36$ (left) and for $r_H=0.78$ (right). Different values of $c_3$, $c_4$ are considered while $\eta=\pi/4$.}
    \label{pots}
\end{figure}

First of all, we shall check the profile of the potential $V(r)$. If it is positive definite, then there are no bound states. We thus show in Fig.~\ref{pots} typical examples of the potential $V(r)$ for hairy backgrounds with two different horizon sizes $r_H$ and different values of $c_3$, $c_4$. We also show for comparison the potential \eqref{potsch} of the bi-Schwarzschild solution with the same $r_H$. In each case the potential shows negative values in the vicinity of the horizon and therefore, bound states \textit{may} exist. However their existence is not yet guaranteed.

On the other hand, it is known that bound states certainly exist for bi-Schwarzschild black holes with $r_H<0.86$ \cite{Babichev2013,Brito2013}. Since the potentials for the hairy solutions with $r_H=0.78$ shown in Fig.~\ref{pots} are close to the Schwarzschild potential, bound states are likely to exist for these hairy backgrounds.

In order to know whether bound states exist or not, we use the well-known Jacobi criterion \cite{Amann1995} which requires to construct the solution of the perturbation equation \eqref{master_pert_hr} with $\omega=0$. We start the numerical integration in the asymptotic region where the tortoise coordinate $r_\ast$ becomes identical to the usual $r$. Here Eq.~\eqref{master_pert_hr} simply reduces to $\Psi''-\Psi=0$ so that the bounded solution is $\Psi=\e^{-r}$. Then we extend numerically this solution\footnote{At this stage, we treat the pertubation equation as an initial value problem so we simply use the \textit{Runge-Kutta 4} method presented in Appendix.~\ref{num_ivp}.} toward small values of $r$. The Jacobi criterion states that the number of bound states is equal to the number of nodes of the resulting solution for $\Psi(r)$. We find that, depending on the values of $r_H$, $\eta$, $c_3$, $c_4$, it may indeed show a zero as $r$ approaches $r_H$. Therefore, there exist a bound state.

The next step is to actually find the bound state by solving the eigenvalue problem \eqref{master_pert_hr} with the potential $V(r)$ obtained numerically from the hairy background configurations. For this we set $\omega^2=-\sigma^2$ (with $\sigma\in\mathbb{R}$) and use the local solutions at infinity and close to the horizon given by the Eq.~\eqref{asymp_psi_bigravity_tor}, which can be rewritten in terms of the usual $r$ coordinate as,
\begin{equation}
    A\times(r-r_H)^{\sigma r_H}\leftarrow\Psi(r)\rightarrow\e^{-\sqrt{1+\sigma^2}\,r}\quad\text{as}\quad r_H\leftarrow r\rightarrow +\infty.
\end{equation}
Here we have kept only the integration constant $A$ in the local solution close to the horizon while the integration constant in the local solution at infinity has been set to unity without loss of generality. Then we apply the \textit{multishooting} method to solve the perturbation equation by treating it as a boundary value problem (see Appendix.~\ref{num_bvp}). The second-order ODE \eqref{master_pert_hr} can be transformed into two first-order ODEs so that we need two shooting parameters. These parameters to be adjusted by the numerical method are $A$ and $\sigma$. This procedure yields the bound state solution $\Psi(r)$ and the corresponding eigenfrequency $\omega=\pm i\sigma$ (see \cite{Pani2013,Berti2009} for reviews on the black hole perturbation theory and the tools that can be used to solve the perturbation equation).

\begin{figure}
    \centering
    \includegraphics[width=7.8cm]{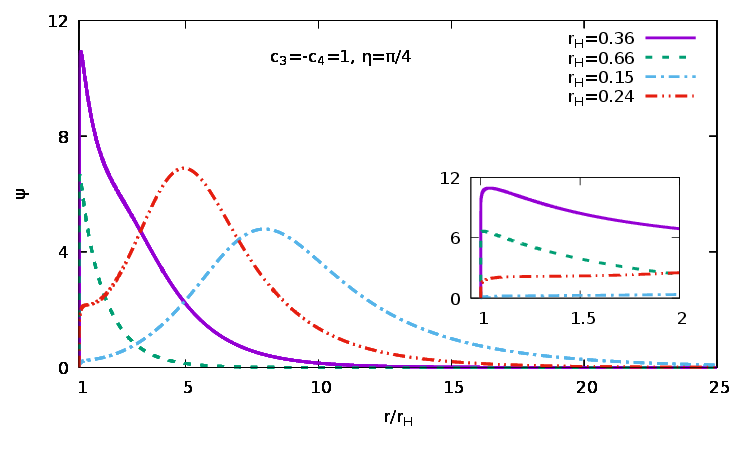}
    \includegraphics[width=7.8cm]{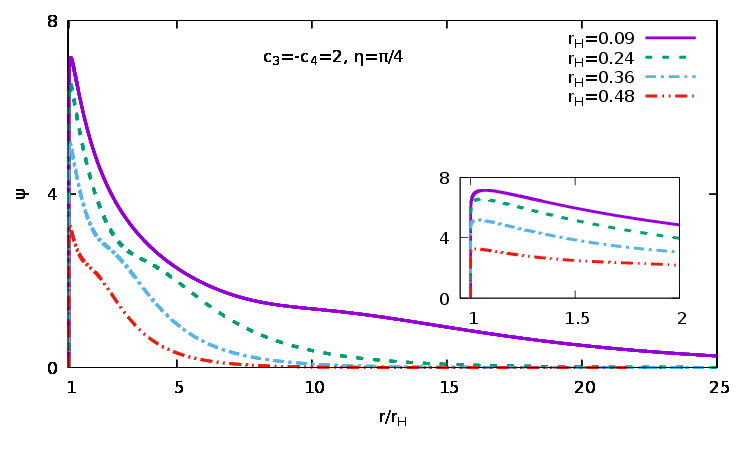}
    \caption[The bound state solutions of the perturbation equation \eqref{master_pert_hr} for $\eta=\pi/4$ and various values of $c_3$, $c_4$, $r_H$.]{The bound states $\Psi(r)$ for $\eta=\pi/4$ and different $r_H$ with $c_3=-c_4=1$ (left) and $c_3=-c_4=2$ (right). They vanish at the horizon and at infinity.}
    \label{psi}
\end{figure}

We consider first the hairy solutions with $\eta=\pi/2$ that have been reported previously in Ref.~\cite{Brito2013a}. Examples of profile of their wave functions $\Psi$ against the ordinary radial coordinate $r$ are shown in Fig.~\ref{psi}. They vanish at the horizon, then show a maximum very close to $r_H$, and approach zero for $r\to\infty$.

\begin{figure}
    \centering
    \includegraphics[width=7.8cm]{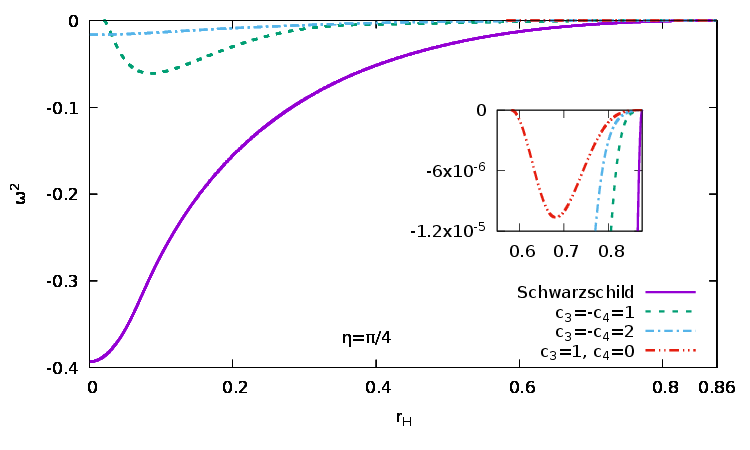}
    \includegraphics[width=7.8cm]{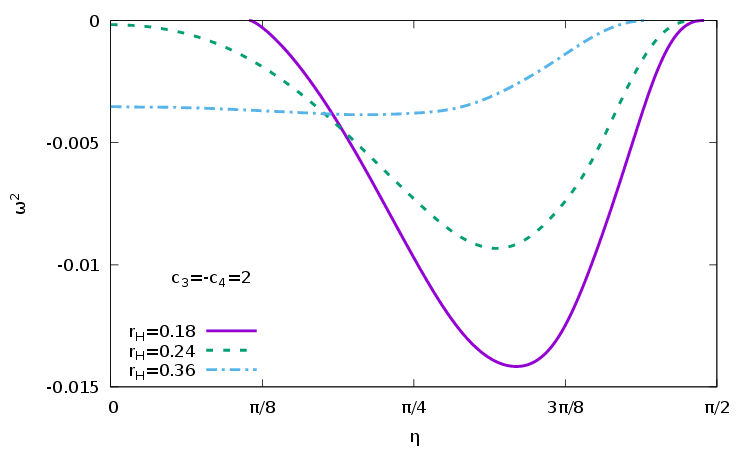}
    \caption[The eigenfrequency of the perturbation equation \eqref{master_pert_hr} against the horizon radius $r_H$ and against the mixing angle $\eta$ for various values of $c_3$, $c_4$.]{Left: the eigenfrequency $\omega^2$ for the hairy and for the bi-Schwarzschild black holes against the horizon radius $r_H$ for different values of $c_3$, $c_4$ and $\eta=\pi/4$. Right: the eigenfrequency $\omega^2$ for the hairy black holes against $\eta$ for different values of $r_H$ and $c_3=-c_4=2$.}
    \label{omega}
\end{figure}

We find bound states with $\omega^2<0$ for all the hairy black holes reported in Ref.~\cite{Brito2013a}, see the left panel on Fig.~\ref{omega}. Therefore, all these solutions are unstable. We emphasize that all of them correspond to the particular choice $\eta=\pi/4$, hence $\kappa_1=\kappa_2=1/2$. In order to test our method, we have also computed the eigenfrequencies for the bi-Schwarzschild solution as in Ref.~\cite{Brito2013}.

As seen in the left panel on Fig.~\ref{omega}, the absolute value of the negative $\omega^2$ for the bi-Schwarzschild solution is always larger than that of the hairy solutions. Therefore, the instability growth rate for the hairy black holes is not as large as for the bi-Schwarzschild black hole. Since one has $\omega=\boldsymbol{\omega}/\boldsymbol{m}$ where $\boldsymbol{\omega}$ is the dimensionful physical frequency, the instability growth time is $1/\boldsymbol{\omega}=1/(\omega\boldsymbol{m})$. If we assume the graviton mass $\boldsymbol{m}$ to be very small as in Eq.~\eqref{tiny_m}, then the instability growth time will be cosmologically large, hence the instability will not play an important role. However, as we shall see in the next section, the physical choice is rather to assume that $1/\boldsymbol{m}\leq 10^6\,\text{km}$ according to \eqref{m_phys_choice}, in which case the instability growth time will be less than $10^3$ seconds. Hence the instability must be avoided if we wish to describe the black holes of our Universe by the hairy solutions of the massive bigravity theory.

The eigenvalue $\omega^2(r_H)<0$ approaches zero as $r_H\to 0.86$. In this limit, all hairy black holes "lose their hair" and merge with the bi-Schwarzschild solution which admits a zero mode\footnote{A zero mode is a bound state with eigenfrequency $\omega=0$.} for $r_H=0.86$. Close to the value $r_H=0.86$, all solutions are close to each other and $\omega^2$ is close to zero for any $c_3$, $c_4$, $\eta$ while for smaller $r_H$, the backgrounds and $\omega^2$ become parameter dependent. The eigenfrequency may approach zero also for type I hairy black holes when the solutions cease to exist, \textit{i.e.} for $r_H=r_H^\text{min}>0$. For example, for $c_3=1$, $c_4=0$, the hairy solution disappears at $r_H\approx 0.58$, and at the same time $\omega^2$ approaches zero, as seen in the insertion of the left panel on Fig.~\ref{omega}.

The instability of hairy black holes is in fact a somewhat puzzling phenomenon, since it is unclear what they may decay to. Since the hairy solutions $r_H<0.86$ are more energetic than the bi-Schwarzschild solution (see next section), they may approach the latter via absorbing and/or radiating away their hair during the decay. However the bi-Schwarzschild black hole is also unstable for $r_H<0.86$ and should decay into something else. 

The perturbative instability of the bi-Schwarzschild solution in massive bigravity is mathematically equivalent \cite{Babichev2013} to the GL instability of black strings in $d=5$ vacuum GR~\cite{Gregory1993}. It is known that the nonlinear development of the latter leads to the formation of an infinite string of "black hole beads" but the event horizon topology does not change~\cite{Lehner2011}. This fact being established within the $d=5$ vacuum GR, a similar scenario is not possible in the $d=4$ massive bigravity theory. Hence the fate of the bigravity black holes should be different. For example, they may radiate away all their energy via spherically symmetric radiations (some radiative solutions are already known \cite{Kocic2017,Hoegaas2020a}), but it is unclear what happens to the horizon, whether it disappears or not. In GR, the horizon cannot evaporate via a classical process \cite{Hawking1973}, but in the massive bigravity theory the situation might be different.

Now, let us see how the $\eta$ parameter can affect the stability property of hairy black holes. In the right panel on Fig.~\ref{omega} we show $\omega^2$ as a function of $\eta$ for several values of $r_H$ and with $c_3=-c_4=2$. Remarkably, one can see that the eigenfrequcency $\omega^2(\eta)<0$ vanishes when $\eta$ approaches $\pi/2$ and thus the unstable bound state disappears in a region close to $\eta=\pi/2$. The same phenomenon occurs in a region close to $\eta=0$, but only for small values of the event horizon radius $r_H$. For example for $r_H=0.18$, the eigenfrequency vanishes for $\eta\approx\pi/8$, as seen in the figure. At the same time, the bi-Schwarzschild solution is certainly unstable for any values of $\eta$ since we are considering here horizon radii $r_H<0.86$. 

As a result, depending on the value of $\eta$, some hairy black hole solutions can be stable and thus be good candidates for the final state of unstable bi-Schwarzschild black hole decay. On the other hand, we have not analyzed in this section the stability property of hairy solutions with $r_H>0.86$. Their stability in this case is not a decisive criterion since the corresponding bi-Schwarzschild solution is stable when $r_H>0.86$. Below we shall describe in more detail the $(r_H,\eta)$-parameter space for a specific choice of $c_3$, $c_4$ which can lead to a large set of stable hairy solutions.

\section{Parameter space and the physical solutions}
\label{param_bigravity}

The full parameter space for asymptotically flat hairy black hole solutions in massive bigravity is a four dimensional space spanned by the parameters $(r_H,\eta,c_3,c_4)$. To simplify the study of this parameter space, we shall adopt the following strategy: we choose the particular values
\begin{equation}
\label{choice1}
    c_3=-c_4=5/2,
\end{equation}
which fulfill condition II in \eqref{cases_hr} and study the solutions for all possible $(r_H,\eta)$. Then, performing the duality transformation gives us all possible solutions for
\begin{equation}
\label{choice2}
    c_3=1/2,\quad\quad c_4=3/2,
\end{equation}
which values corresponds to the case I in \eqref{cases_hr}. This approach reveals interesting and rather complex features of the solutions which are presumably generic for any $c_3$, $c_4$.

\subsection{The ADM mass}

\begin{figure}
    \centering
    \includegraphics[height=5.5cm]{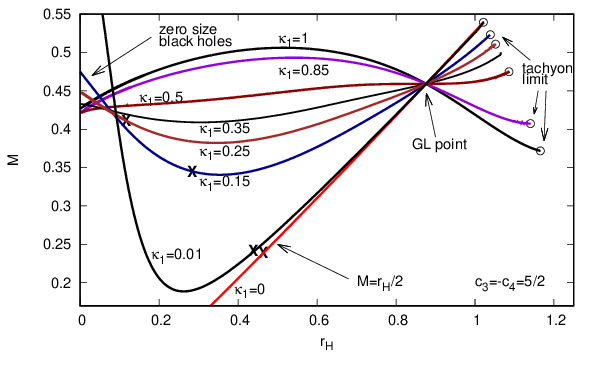}
    \includegraphics[height=5.5cm]{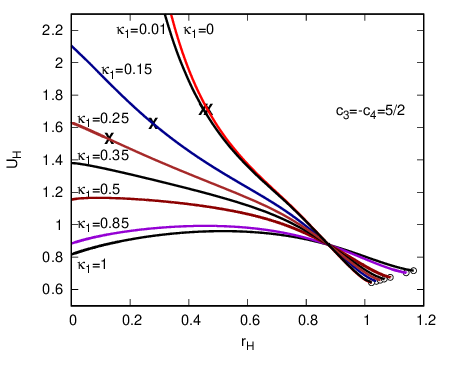}
    \caption[The ADM mass, $M$, and the horizon radius as measured by the $f$-metric, $U_H$, against $r_H$ for the hairy solutions with $c_3=-c_4=5/2$.]{The ADM mass $M(r_H)$ (left) and the function $U_H(r_H)$ (right) for the hairy solutions with $c_3=-c_4=5/2$. The crosses mark the points on the left of which the $f$-metric becomes singular. The circles mark the termination points beyond which the solutions would become complex-valued. When $\kappa_1=\cos^2\eta\to 0$, the mass $M(r_H)$ develops a more and more profound minimum, while the values of $M(0)$ and $U_H(0)$ grow without bounds.}
    \label{Mr_Ur}
\end{figure}

In the figure~\ref{Mr_Ur}, we show the ADM mass $M$ and the horizon radius as measured by the $f$-metric, $U_H$, as functions of $r_H$ for several values of $\eta\in[0,\pi/2]$. As one can see, all curves $M(r_H)$ intersect at the GL point $(r_H,M)=(0.86,0.43)$ where the solutions bifurcate with the bi-Schwarzschild solution,
\begin{equation}
    N(r)=Q(r)=Y(r)=q(r)=\sqrt{1-\frac{r_\text{GL}}{r}},\quad U(r)=r,
\end{equation}
where $r_\text{GL}=0.86$. At the same time, all curves $U_H(r_H)$ intersect at the point $(r_H,U_H)=(r_\text{GL},r_\text{GL})$. Away from the GL point, the $g$-metric still remains Schwarzschild if $\eta=\pi/2$, in which case $M(r_H)$ is a linear function,
\begin{equation}
    \eta=\frac{\pi}{2}\;\text{:}\quad N(r)=Q(r)=1-\frac{r_H}{r}\;\Rightarrow\; M=\frac{r_H}{2},
\end{equation}
but the $f$-metric for these solutions is not Schwarzschild. These are the solutions we call hairy Schwarzschild, as explainded in Sec.~\ref{hairy_bh_bigravity}. For $\eta\neq\pi/2$ the mass depends nonlinearly on $r_H$.

Let us now introduce the mass function $m(r)$ via $N^2(r)=1-2m(r)/r$. The equation.~\eqref{eqN} then assumes the form
\begin{equation}
\label{def_rho_hr}
    m'(r)=\kappa_1\frac{r^2}{2}\left(\mathcal{P}_0+U'\mathcal{P}_1\frac{N}{Y}\right)\equiv\kappa_1\rho(r),
\end{equation}
from which the ADM mass may be expressed as
\begin{equation}
\label{mass_hr}
    M=m(\infty)=\frac{r_H}{2}+\kappa_1\int_{r_H}^\infty{\rho(r)dr}\equiv M_\text{bare}+M_\text{hair}.
\end{equation}
Here the "bare" mass $M_\text{bare}=r_H/2$ is determined only by the horizon radius and coincides with the mass of a Schwarzschild black hole of radius $r_H$. The mass $M_\text{hair}$ expressed by the integral is the contribution of the massive hair distributed outside the horizon. As one can see in Fig.~\ref{Mr_Ur}, one has $M>r_H/2$ if $r_H<r_\text{GL}$, hence the hair mass is positive and the hairy solutions are more energetic than the bare bi-Schwarzschild black hole. On the contrary, the mass of the hair becomes negative above the GL point, when $r_H>r_\text{GL}$, and the hairy solutions are then less energetic than the bi-Schwarzschild. Therefore the energy density of the hair $\rho(r)$ can be negative. In fact, it is already known that the standard energy condition do not hold within the massive bigravity theory \cite{Baccetti2012}.

It is important to notice that, unless $\kappa_1=\cos^2\eta$ is very small, the ADM mass of all hairy solutions always varies within a finite range and remains close to the GL value,
\begin{equation}
\label{m_gl}
    M\approx \frac{r_\text{GL}}{2}=0.43,
\end{equation}
as seen in Fig.~\ref{Mr_Ur}. It seems that this fact was not reported in Ref.~\cite{Brito2013a}, which always shows only the ratio $M/r_H$. This ratio diverges as $r_H\to 0$, however the mass $M$ actually remains finite in this limit. Let us restore for the moment the speed of light $\boldsymbol{c}$ and Newton's constant $\boldsymbol{G}$, then the dimensionful ADM mass\footnote{We emphasize that what we call the graviton mass $\boldsymbol{m}$ is actually a quantity whose dimension is the inverse of a length: $1/\boldsymbol{m}$ is the Compton wavelength of the massive graviton.} is
\begin{equation}
    \boldsymbol{M}=\frac{\boldsymbol{c^2}M}{\boldsymbol{G}\boldsymbol{m}}.
\end{equation}
In view of Eq.~\eqref{m_gl}, this means that the dimensionful mass of the hairy solutions is always close to that of a Schwarzschild black hole of size $\boldsymbol{r}_H=r_\text{GL}/\boldsymbol{m}$, which is close to the Compton wavelength of the massive graviton. As a result, one cannot assume the graviton mass $\boldsymbol{m}$ to be of the order of the inverse Hubble radius as in Eq.~\eqref{tiny_m}. Indeed, this would imply the hairy black holes to be as heavy as a Schwarzschild black hole of the size of the Universe -- a physically meaningless result. One should rather assume that $1/\boldsymbol{m}=\gamma\times 10^6\,\text{km}$ with $\gamma\in[0,1]$ as in Eq.~\eqref{m_phys_choice}, which is consistent with the cosmological observations if $\kappa_1$ is very small, as expressed in Eq.~\eqref{kappa_phys_choice}. We shall see that this choice leads to physically acceptable values of the mass.

For $r_H\approx r_\text{GL}$, the masses of the hairy black holes are then close to that of a Schwarzschild black hole of radius $\gamma\times 10^6\,\text{km}$, that is $\boldsymbol{M}\sim 0.3\,\gamma\times 10^6\,\boldsymbol{M}_\odot$. If $\gamma\sim 1$ this gives the typical value for supermassive black holes observed in the center of many galaxies. However for smaller $r_H$ the ADM mass can deviate considerably from the GL value and becomes very small or very large. As seen in the left panel on Fig.~\ref{Mr_Ur}, for small $\kappa_1$, the mass $M(r_H)$ shows a minimum: first, it decreases with $r_H$, then reaches a minimal value $M_\text{min}$, and then increases up to some $M(r_H=0)$. For smaller values of $\kappa_1$, the minimum becomes more and more profound and the value $M_\text{min}$ approaches zero while $M(r_H=0)$ becomes larger and larger. Thus the curves $M(r_H)$ converge smoothly towards the limiting case $M(r_H)=r_H/2$ as $\kappa_1\to 0$, except for the point at $r_H=0$.

\begin{figure}
    \centering
    \includegraphics[width=7.8cm]{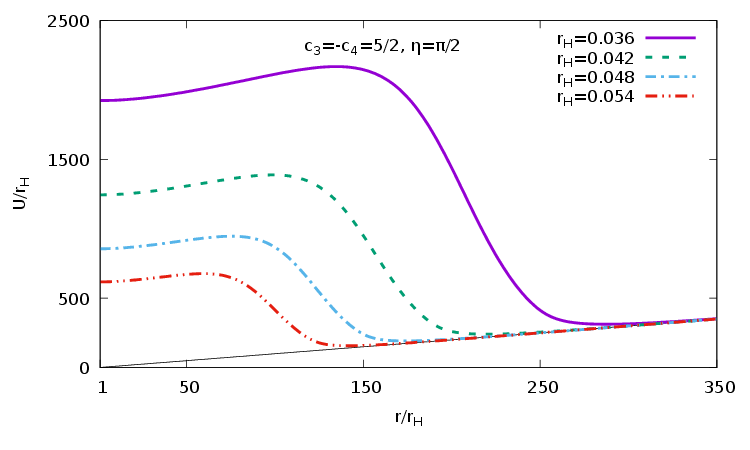}
    \includegraphics[width=7.8cm]{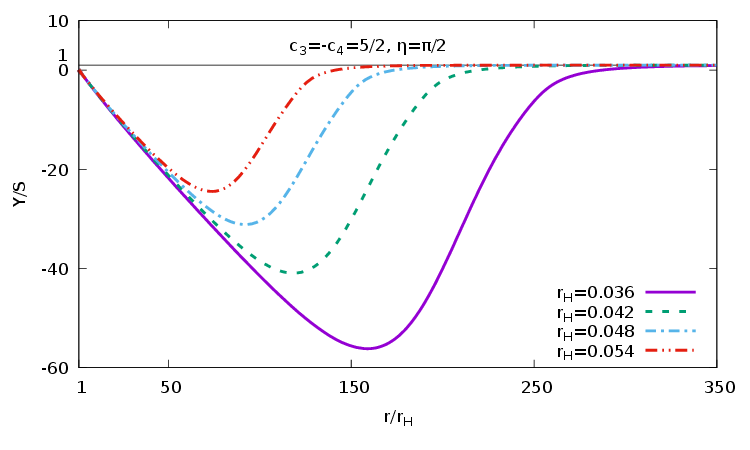}
    \includegraphics[width=7.8cm]{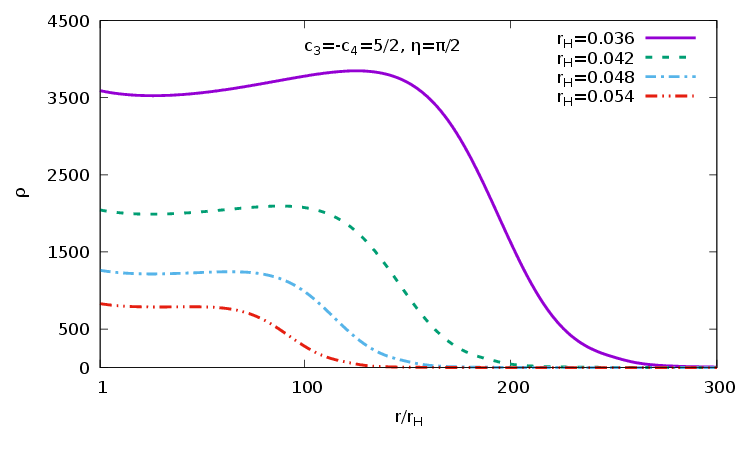}
    \includegraphics[width=7.8cm]{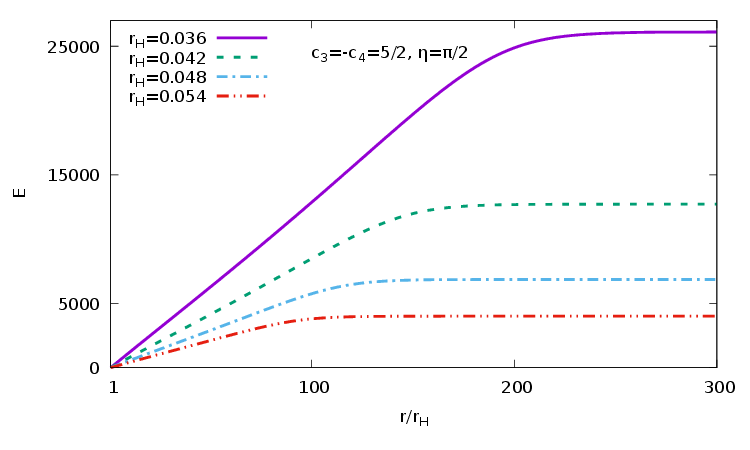}
    \caption[The metric functions $U(r)$, $Y(r)$, the hair energy density $\rho(r)$ and its integral $E(r)$ for the hairy solutions with $\kappa_1=0$.]{The functions $U(r)$, $Y(r)$, the hair energy density $\rho(r)$ and its integral $E(r)$ whose asymptotic value $E(\infty)$ is the "hair energy" for the hairy solutions with $\kappa_1=0$ and small values $r_H$.}
    \label{pio2}
\end{figure}

To get an approximation for $M(r_H)$ when $\kappa_1$ is very close to zero, we consider the hairy Schwarzschild solutions with $\kappa_1=0$. Their $g$-metric is Schwarzschild with all the hair contained in the $f$-metric. It turns out that the $Y$ and $U$ functions of the $f$-metric depend very strongly on $r_H$ when the latter is small, as seen in the top panels on Fig.~\ref{pio2}. We inject these solutions to Eq.~\eqref{def_rho_hr} to obtain the hair energy density $\rho(r)$ and $E(r)\equiv\int_{r_H}^r{\rho\,dr}$. As one can see in the bottom panels on Fig.~\ref{pio2}, these functions show very large values when $r_H$ is small. Thus the total "hair energy", $E(\infty)$, becomes larger and larger when $r_H$ decreases but it does not backreact since $\kappa_1=0$ so that the $g$-metric remains Schwarzschild. The hair energy starts to backreact if $\kappa\neq 0$ and if $\kappa_1\ll 1$ then one can deduce from Eq.~\eqref{mass_hr} that 
\begin{equation}
    M=\frac{r_H}{2}+\kappa_1 E(\infty)+\mathcal{O}(\kappa_1^2),
\end{equation}
where $E(\infty)$ is computed for $\kappa_1=0$. We evaluate numerically $E(\infty)$ for various values of $r_H$ and obtain the following best fit approximation:
\begin{equation}
\label{approx_M}
    M(r_H)\approx\frac{r_H}{2}+\kappa_1\frac{a}{(r_H)^s},
\end{equation}
with $a=0.0056$ and $s=4.61$. Assuming that $\kappa_1=\gamma^2\times 10^{-34}$, this function shows an absolute minimum at
\begin{equation}
    (r_H)_\text{min}\approx 5.2\,\gamma^{0.35}\times 10^{-7},\quad\quad M_\text{min}\approx 3.1\,\gamma^{0.35}\times 10^{-7},
\end{equation}
whose dimensionful values are obtained by multiplying by $1/\boldsymbol{m}=\gamma\times 10^{6}\,\text{km}$ (and restoring again the speed of light and Newton's constant):
\begin{equation}
    (\boldsymbol{r}_H)_\text{min}=\frac{(r_H)_\text{min}}{\boldsymbol{m}}\approx 0.52\,\gamma^{1.35}\,\text{km},\quad\boldsymbol{M}_\text{min}=\frac{\boldsymbol{c^2}M_\text{min}}{\boldsymbol{G}\boldsymbol{m}}\approx 0.2\,\gamma^{1.35}\,\boldsymbol{M}_\odot.
\end{equation}
This determines the minimum mass for the hairy black holes with $\kappa_1\ll 1$. When $r_H$ gets smaller then the mass starts to grow again, but it grows only up to a finite although very large value as $r_H\to 0$ because the approximation \eqref{approx_M} is not valid for however small $r_H$.

\subsection{The lower limit $r_H\to 0$: zero size black holes}
\label{zero_size_bh}

All the solutions extend down to arbitrarily small values of $r_H$. As seen in Fig.~\ref{Mr_Ur}, the ADM mass $M$ does not vanish when $r_H\to 0$ (except for $\kappa_1=\cos^2\eta=0$). It approaches finite values, even though the bare mass $M_\text{bare}=r_H/2\to 0$. Therefore, all mass comes from the hair contribution $M_\text{hair}$ in this limit. As a result, the configuration remains different from the flat space vacuum even when the horizon size $r_H$ shrinks to zero. This phenomenon is actually not very surprising, since in many nonlinear field theories there are solutions describing a small black hole inside a soliton\footnote{Solitons are nontrivial solutions which are localized in space and have no event horizon, see Chap.~\ref{chap_bs} for more details.} (for example, inside a magnetic monopole, see Ref.~\cite{Volkov1999} and Chap.~\ref{chap_mon}). Sending the horizon size to zero, the black hole disappears, but its external nonlinear matter fields remain and become a gravitating soliton containing a regular origin in its center instead of the horizon. Therefore, the $r_H\to 0$ limit of a hairy black hole may correspond to a regular soliton.

\begin{figure}
    \centering
    \includegraphics[width=7.8cm]{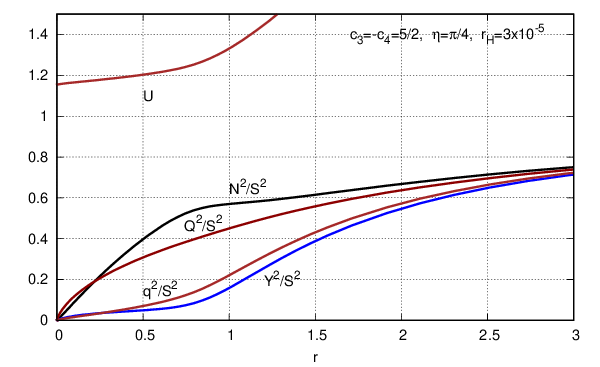}
    \includegraphics[width=7.8cm]{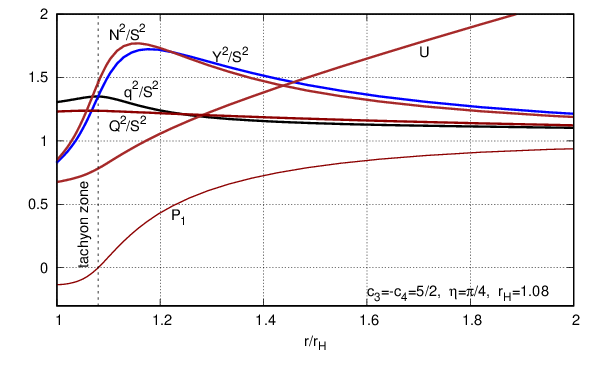}
    \caption[Profiles of a hairy solution close to the zero size black hole and of a hairy solution close to the tachyon limit.]{Profiles of the solution with $r_H\sim 10^{-5}$ that is close to the zero size black hole (left), and of that close to the tachyon limit, with $D\sim 10^{-6}$ (right). One has $S^2=1-r_H/r$. The function $\mathcal{P}_1$ determines the graviton mass via \eqref{grav_mass_generic} and the graviton bahaves as a tachyon if $\mathcal{P}_1<0$.}
    \label{limit_cases}
\end{figure}

However for the hairy solutions in massive bigravity, the limiting configuration exists, but it seems to be singular and not of the regular soliton type. First, as seen in the right panel on Fig.~\ref{Mr_Ur}, the value of $U_H$ which determines the size of the $f$-horizon remains finite when $r_H\to 0$, hence the $f$-geometry remains of black hole type even in this limit. Second, the left panel on Fig.~\ref{limit_cases} shows a solution with a very small $r_H$ and one can see that $N^2/S^2\sim r$ for $r\lesssim 0.5$. However, one has $S=\sqrt{1-r_H/r}\to 1$ as $r_H\to 0$, hence one has in this limit $N^2\sim r$. The numerical profiles shown on the figure actually suggests the following small $r$ behavior of the limiting configuration:
\begin{equation}
    N^2\sim Y^2\sim Q^2\sim q^2\sim r,\quad U=U_\text{lim}+\mathcal{O}(r).
\end{equation}
Therefore, the limiting form of the $g$-metric is something like a "zero size black hole" and it is singular since its Ricci scalar $R(g)=2/r^2+\mathcal{O}(1/r)$ diverges at $r=0$. On the contrary, the $f$-geometry remains of the regular black hole type because $U$ does not vanish. The temperature remains finite for $r_H\to 0$ and is always the same for both metrics. The limiting $g$-temperature can be formally computed by assuming $N^2=\alpha\,r$, $Q^2=\beta\,r$ with $\alpha\approx 0.7$ and $\beta\approx 6$, according to the profiles shown in the left panel on Fig.~\ref{limit_cases}. The equations \eqref{surf_grav} and \eqref{temp} then yield $T_H=\sqrt{\alpha\beta}/(4\pi)\approx 0.163$, which is very close to the value $T_H=0.16$ for the solution with $r_H\sim 10^{-5}$ shown in the figure. Of course these considerations are purely formal since the zero size black hole cannot evaporate and further reduce its size. Moreover, their geometry is singular at the "zero size" horizon. In other words, the physical sense of the black hole temperature for the limiting configuration is not well-defined and the configuration itself seems to be unphysical.

We finally emphasize again that the $f$-metric can become singular for small $r_H$ because the functions $q$ and $Y$ develop additional zeroes outside the horizon. This happens along the parts of the curves on the left of the points marked by crosses in Fig.~\ref{Mr_Ur}. As we have discussed before in Sec.~\ref{hairy_bh_bigravity}, we do not exclude such solutions because we consider the physical metric to be $g_{\mu\nu}$. The $f$-geometry is not probed by test particles and all the physical quantities of the solutions such as the ADM mass do not show anything special when the functions $q$ and $Y$ start to oscillate. The potential $V$ in the perturbation equation \eqref{master_pert_hr} also remains perfectly regular. As a result, we do not exclude that solutions with singular $f$-geometry are physically relevant. Only the zero size black holes with $r_H\to 0$ are considered to be non-physical.

\subsection{The upper "tachyon" limit $r_H\to r_H^\text{max}(\eta)$}

For any value of $\eta$, there is an upper bound $r_H^\text{max}(\eta)>r_\text{GL}$ beyond which hairy solutions cease to exist. At this limiting value of the horizon radius, the two roots of the algebraic equation \eqref{biquad_nu} (or, equivalently \eqref{biquad}) coincide. This equation determines the horizon values of the solutions. Its two roots correspond to two solution branches: the root with $\sigma=+1$ gives rise to asymptotically flat solutions whereas the second root with $\sigma=-1$ always yields solutions with a curvature singularity at a finite radius outside the horizon. The determinant of \eqref{biquad_nu} factorizes as
\begin{equation}
    \mathcal{D}\equiv\mathcal{B}^2-4\mathcal{A}\mathcal{C}=\mathcal{P}_1^2(r_H)\,D,
\end{equation}
where $\mathcal{P}_1(r_H)$ is defined by \eqref{Pk} and $D$ is a complicated function of $r_H$, $U_H$, $\eta$, $c_3$, $c_4$. When $r_H$ increases, then $\mathcal{P}_1(r_H)$ crosses zero at some $r_H=r_H^\text{tach}(\eta)$ while $D$ remains positive. When $r_H$ continues to increase, then $D$ approaches zero and vanishes as $r_H\to r_H^\text{max}(\eta)>r_H^\text{tach}(\eta)$. No further increase of $r_H$ is possible since $D$ would then be negative thus rendering the solutions complex-valued. It is the reason why solutions cease to exist in this upper limit.

Although the determinant $\mathcal{D}$ vanishes for $r_H=r_H^\text{tach}(\eta)$ when $\mathcal{P}_1(r_H)=0$ and also for $r_H=r_H^\text{max}(\eta)$ when $D=0$, the two solution branches never merge. Specifically, the two horizon values $\nu_H$ determined by \eqref{biquad_nu} coincide when $\mathcal{D}=0$, but a careful inspection reveals that $y_H$, $U'_H$ in Eqs.~\eqref{Uph} and \eqref{yh} remain different for the two branches when $\mathcal{P}_1(r_H)=0$. If $D=0$ then all horizon values $\nu_H$, $y_H$, $U'_H$ coincide for the two branches, but the derivatives $y'_H$ defined by \eqref{eq_nuyHbis} remain different. This is a consequence of the fact that the existence and uniqueness theorem applies only to regular points of the ODEs, whereas the event horizon $r=r_H$ is a singular point.

In the interval $r_H^\text{tach}(\eta)<r_h<r_H^\text{max}(\eta)$ one has $\mathcal{P}_1(r_H)<0$. This has an important consequence for the corresponding solutions. Let us remember that the relation \eqref{fp_mass} defining the graviton mass was obtained by linearizing the field equations around flat spacetime. Therefore this relation holds only for a Minkowski background and can be written as $\boldsymbol{m}_\text{FP}^2=\mathcal{P}_1(\infty)\,\boldsymbol{m}^2$. The graviton mass around an arbitrary spherically symmetric background has been derived in Ref.~\cite{Mazuet2018a} and reads
\begin{equation}
\label{grav_mass_generic}
    \boldsymbol{m}_\text{FP}^2=\mathcal{P}_1(r)\,\boldsymbol{m}^2.
\end{equation}
Therefore, if $\mathcal{P}_1(r)<0$ then the mass becomes purely imaginary: the graviton behaves as a tachyon. As a result, solutions for $r_H>r_H^\text{tach}(\eta)$ show unphysical features and we call $r_H\to r_H^\text{max}(\eta)$ the "tachyon limit". The horizon value $y'_H$ diverges in this limit, but it seems to be an integrable divergence similar to $y'(r)\sim1/\sqrt{r-r_H}$ and the limiting solution itself always stays regular. We were able to approach this solution rather closely, as shown in the right panel on Fig.~\ref{limit_cases} which presents a solution with the horizon value $D\sim 10^{-6}$.

\subsection{Parameter regions for solutions with $c_3=-c_4=5/2$}

\begin{figure}
    \centering
    \includegraphics[width=11cm]{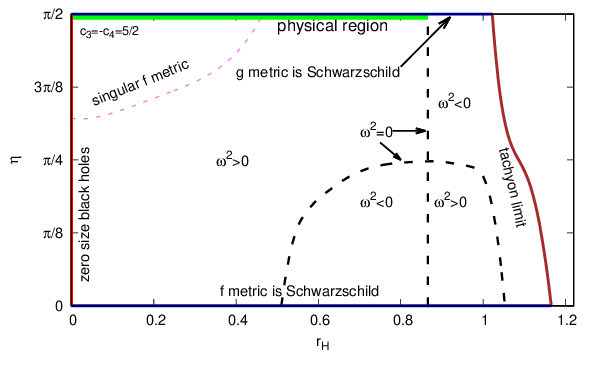}
    \caption[The $(r_H,\eta)$-parameter space for hairy black hole solutions with $c_3=-c_4=5/2$.]{The $(r_H,\eta)$-parameter space for hairy black hole solutions with $c_3=-c_4=5/2$. The dashed black lines corresponds to $\omega^2=0$ and separate stable and unstable sectors. The upper left corner contains solutions with a singular $f$-metric; however their $g$-geometry is regular.}
    \label{sector1}
\end{figure}

Let us now collect all the facts together. The diagram in Fig.~\ref{sector1} shows the region in the $(r_H,\eta)$ plane in which hairy black hole solutions exist. The low boundary at $\eta=0$ corresponds to solutions whose $f$-metric is Schwarzschild, while the upper boundary at $\eta=\pi/2$ corresponds to solutions whose $g$-metric is Schwarzschild. The zero size black holes are on the left boundary at $r_H=0$. The right boundary marks the tachyon limit beyound which the solutions would become complex-valued. The upper left corner of the diagram contains solutions with a singular $f$-metric, but their $g$-geometry, which is physically measurable, is regular.

The diagram also shows dashed lines corresponding to the zero modes, $\omega^2=0$, of the perturbative eigenvalue problem \eqref{master_pert_hr}. In particular, the vertical line corresponds to the GL value $r_H=0.86$. The eigenfrequency $\omega^2$ changes sign when crossing these lines. Therefore, the lines separate sectors where $\omega^2>0$ and hence the hairy solutions are stable, from sectors where $\omega^2<0$ and the solutions are unstable. There are altogether two stable and two unstable sectors. It is worth noting that the stable region is now much larger than for solutions with $c_3=-c_4=2$ considered in the previous section. One also notices that the tachyonic solutions are in an unstable sector.

Finally, the diagram shows the "physical region" corresponding to physically acceptable hairy solutions. As explained above, for such solutions the coupling $\kappa_1=\cos^2\eta$ should be very small for their mass not to be too large, hence $\eta$ should be very close to $\pi/2$. Therefore the physical region, represented by the thick green line, is at the top of the diagram, and it stops at the GL value, $r_H=0.86$, to the right of which hairy solutions are unstable.

Physical solutions are therefore described by a $g$-metric which is extremely close to Schwarzschid since
\begin{equation}
    G(g)_{\mu\nu}=\kappa_1\,T_{\mu\nu},\quad\text{with}\quad\kappa_1\leq 10^{-34}.
\end{equation}
The "hairy features" of the solutions hidden in the $f$-metric should be difficult to observe, unless in violent processes like black hole mergers producing large enough interaction $T_{\mu\nu}$ to overcome the $10^{-34}$ suppression. In summary, the physically acceptable hairy black holes in massive bigravity are extremely close to the GR black holes, but we expect their strong field dynamics to be different.

The physical region contains stable hairy black holes whose masses range from the minimal value $\sim 0.2\,\gamma^{1.35}\,\boldsymbol{M}_\odot$ up to the maximal value $\sim 0.3\,\gamma\times 10^6\,\boldsymbol{M}_\odot$ with $\gamma\in[0,1]$. Yet heavier black holes can be described within the massive bigravity theory but they cannot be hairy and should be described by the bi-Schwarzschild solution \eqref{bi_sch}, which is stable for $r_H>0.86$. Stable black holes with $\boldsymbol{M}<0.2\,\gamma^{1.35}\,\boldsymbol{M}_\odot$ can only be of the type \eqref{SadS-EF}.

\subsection{Parameter regions for dual solutions $c_3=1/2$, $c_4=3/2$}

\begin{figure}
    \centering
    \includegraphics[height=5.5cm]{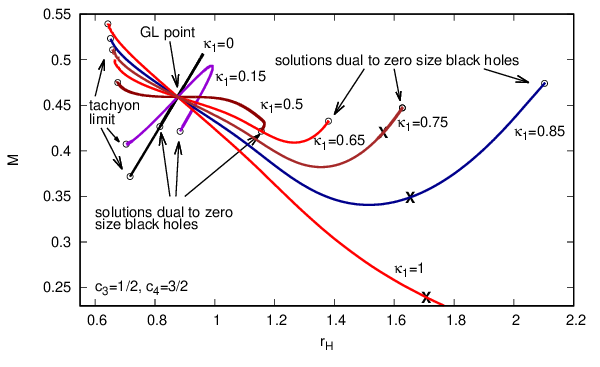}
    \includegraphics[height=5.5cm]{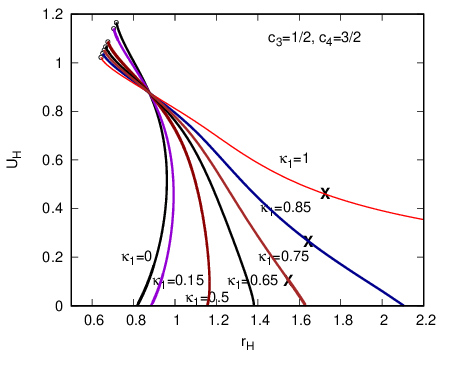}
    \caption[The ADM mass, $M$, and the horizon radius as measured by the $f$-metric, $U_H$, against $r_H$ for the hairy solutions with $c_3=1/2$, $c_4=3/2$.]{The ADM mass $M(r_H)$ (left) and the function $U_H(r_H)$ (right) for the hairy solutions with $c_3=1/2$, $c_4=3/2$. The crosses mark the points on the right of which the $g$-metric becomes singular, hence these parts of the curves correspond to unphysical solutions that should be excluded from considerations.}
    \label{MU_rU}
\end{figure}

Let us finally see how the solutions described above transform under the duality \eqref{dual_expl}. This transformation converts the parameter values \eqref{choice1} into \eqref{choice2}, flips the sign of $\eta-\pi/4$ and exchanges the $Q$, $N$, $r$ with $q$, $Y$, $U$. On the plots, this is equivalent to relabel the functions and plotting them against $U$ instead of $r$. The ADM mass and the temperature are invariant under duality, since these quantities do not depend on the metric from which they are computed. The stability property does not change neither since, for example, if a solution is unstable and admits growing in time perturbations, then its dual version will contain the same growing modes and hence will be unstable as well.

The figure \ref{MU_rU} shows the dual version of Fig.~\ref{Mr_Ur}. The mass curves $M(r_H)$ still intersect at the GL point but they look quite different as compared to those in Fig.~\ref{Mr_Ur}. In particular, not all of them are single-valued. The reason is that the functions $U_H(r_H)$ in the right panel on Fig.~\ref{Mr_Ur} are not always monotone. This means that solutions of different horizon radii $r_H$ and masses $M$ with the parameter choice $c_3=-c_4=5/2$ can have the same value for $U_H$. After the duality transformation, this yields two hairy solutions with the same $r_H$ but different $U_H$ and $M$.

The solutions now exist for $r_H\in[r_H^\text{min}(\eta),r_H^\text{max}(\eta)]$. The lower limit $r_H^\text{min}(\eta)$ corresponds to what used to be the upper limit before the duality -- tachyonic solutions with vanishing horizon determinant $\mathcal{D}$. The upper limit $r_H^\text{max}(\eta)$ corresponds for small $\eta$ to solutions dual to the zero size black holes presented above. After the duality, their $g$-metric describes a regular black hole while the $f$-metric has $U_H=0$, hence it describes the zero size black hole geometry. For larger values of $\eta$, the upper boundary $r_H^\text{max}(\eta)$ corresponds to points where two different solutions with the same $r_H$ but with different $M$ merge with each other. In this case, the zero size black hole limit is not located at $r_H^\text{max}(\eta)$, but at a smaller value of the horizon radius where the curves terminate, as seen in the Fig.~\ref{MU_rU}.

The solutions below the GL point, for $r_H<0.86$, are still more energetic than the hairy Schwarzschild with $\eta=\pi/2$, hence their hair mass $M_\text{hair}$ is positive, whereas above the GL point it becomes negative. 

\begin{figure}
    \centering
    \includegraphics[width=11cm]{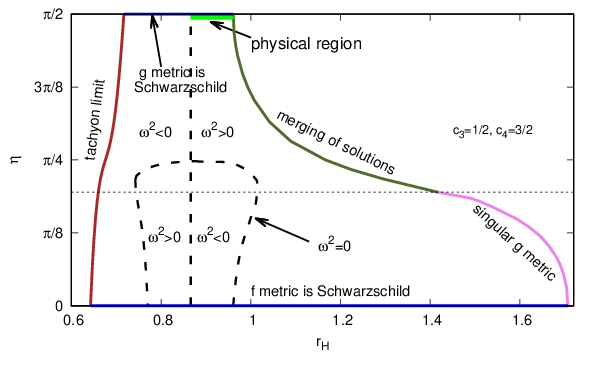}
    \caption[The $(r_H,\eta)$-parameter space for hairy black hole solutions with $c_3=1/2$, $c_4=3/2$.]{The $(r_H,\eta)$-parameter space for hairy black hole solutions with $c_3=1/2$, $c_4=3/2$. The dashed black lines corresponds to $\omega^2=0$ and separate stable and unstable sectors.}
    \label{sector2}
\end{figure}

Finally the figure \ref{sector2} shows the $(r_H,\eta)$-parameter space. The diagram now looks different as compared to that in Fig.~\ref{sector1}, although it corresponds to essentially the same solutions, up to the duality transformation. The solutions with a singular $g$-metric, which are dual to solutions with a singular $f$-metric in Fig.~\ref{sector1}, are now excluded since we are considering $g_{\mu\nu}$ as the one that actually describes the geometry of spacetime. The physical region corresponding to stable solutions with $\eta$ close to $\pi/2$ is now above the GL point, where the hair mass is negative. The physical solutions are again characterized by the $g$-metric that is extremely close to Schwarzschild, but the novel feature is that now for each value of $r_H$ in the physical region, there are two different solutions whose $g$-metrics are almost the same but the $f$-metrics are different. The physical region stops at $r_H^\text{max}(\eta)$ where the two different solutions merge with each other.

As a result, the physical region in Fig.~\ref{sector2} is rather short and corresponds only to supermassive black holes with $0.86<r_H<r_H^\text{max}(\eta)$. All black holes of smaller masses are unstable. Therefore, the parameter choice $c_3=1/2$, $c_4=3/2$ is not physically interesting.

\section{Conclusion}
\label{conclusion_bigravity}

We have presented above a detailed analysis of static and asymptotically flat hairy black holes in the ghost-free massive bigravity theory. Extending the earlier results of Ref.~\cite{Brito2013a}, we find that for given values of the theory parameters $c_3$, $c_4$, $\eta$, and for a given event horizon size $r_H\in[r_H^\text{min},r_H^\text{max}]$ there are one or sometimes two different black holes supporting a nonlinear massive graviton hair. These solutions coexist with the bi-Schwarzschild solution described by the Eq.~\eqref{bi_sch}. The hairy solutions are more energetic than the bi-Schwarzschild one if $r_H<0.86$ and they less energetic otherwise. When $r_H$ approaches the limiting values $r_H^\text{min}$ or $r_H^\text{max}$, the solutions either become complex-valued or merge between themselves. For some values of $c_3$, $c_4$, zero size black holes exist for which $r_H^\text{min}=0$ but the corresponding horizon size as measured by the $f$-metric, $U_H$, remains finite. Depending on values of $r_H$, $c_3$, $c_4$, $\eta$, the hairy solutions can be either stable or unstable.

To avoid the hairy black holes from being unphysically heavy, one is bound to assume the massive graviton Compton wavelength to be $1/\boldsymbol{m}=\gamma\times 10^6\,\text{km}$ where the parameter $\gamma$ may range in the interval $[0,1]$. The agreement with the cosmological data is then achieved by assuming that $\kappa_1=\cos^2\eta=\gamma^2\times(\boldsymbol{M}_\text{ew}/\boldsymbol{M}_\text{Pl})^2=\gamma^2\times 10^{-34}$. Therefore, in the $(r_H,\eta)$-parameter space, we identify the physical region to be close to the $\eta=\pi/2$ boundary. In this region, the Einstein equation for $g_{\mu\nu}$, $G(g)_{\mu\nu}=\kappa_1 T_{\mu\nu}$, is extremely close to that of vacuum GR since $\kappa_1<10^{-34}$. As a consequence, the physical hairy black holes are described by a $g$-metric extremely close to Schwarzschild while all hairy features are contained in the $f$-metric.

The stable hairy black holes have their masses ranging from $\sim 0.2\,\gamma^{1.35}\,\boldsymbol{M}_\odot$ to $\sim 0.3\,\gamma\times 10^6\,\boldsymbol{M}_\odot$. Yet heavier black holes can be described in the theory by the bald bi-Schwarzschild solution which is stable when the hairy solutions become unstable. Therefore, astrophysical black holes from stellar ones to supermassive ones can be described by stable solutions of the massive bigravity theory.

However, we expect the detection of the hairy features to be very complicated. Indeed, the $f$-metric is not coupled to matter and cannot be directly probed, while the deviation of the "visible" $g$-metric from Schwarzschild is suppressed by the factor $\kappa_1<10^{-34}$. Nevertheless, it is plausible that the interaction between the two metrics during violent processes such as black holes mergers may produce an energy-momentum tensor strong enough to overcome the $10^{-34}$ suppression. Therefore the signals detected by LIGO/VIRGO \cite{Abbott2016} may carry information about the hairy structure of the black holes. We expect the "hair imprint" in the signal to be stronger for \textit{small} black holes since we know that the functions $U$, $Y$ of the $f$-metric become very large when the horizon size is small, which should influence the effective energy-momentum tensor $T_{\mu\nu}$ of the merger. However, to actually determine the hair imprint in the signal would require dynamical simulations going beyond the scope this thesis. We therefore simply refer to the Ref.~\cite{Dong2021} where calculations of this type are carried out within the framework of the ghost-free massive gravity theory \cite{Rham2011} and also to the Ref.~\cite{Max2017} where the authors highlight the phenomenon of gravitational waves oscillations that arises in massive bigravity.

Finally, we should discuss the paper \cite{Torsello2017} where the authors conjecture that the bi-Schwarzschild solution \eqref{bi_sch} is the only asymptotically flat black hole solution in massive bigravity. First of all, it was emphasized in Ref.~\cite{Torsello2017} that the usual practice of starting the numerical integration not at the horizon $r=r_H$, a singular point of the ODEs, but at a regular nearby point $r=r_H+\epsilon$, as was done in Ref.~\cite{Brito2013a}, could in principle lead to numerical instabilities. We agree with this, and it is actually for this reason that we use a desingularization procedure (described in Appendix~\ref{desing_hor_bigravity}) which allows us to start the numerical integration exactly at $r=r_H$. Next, small initial deviations from the bi-Schwarzschild solution were considered in \cite{Torsello2017} via setting at the horizon $u=U_H/r_H=1+\epsilon$. Integrating the equations toward large $r$ then yields metrics whose components diverge as $r\to\infty$ instead of approaching finite values. This observation, made already in Ref.~\cite{Volkov2012a}, shows that the bi-Schwarzschild solution is \textit{Lyapunov} unstable: no regular and asymptotically flat solutions exists in a \textit{small vicinity} of the bi-Schwarzschild solution. However it does not exclude the existence of other asymptotically flat solutions for values of $u$ deviating considerably from unity. Ultimately, the authors in Ref.~\cite{Torsello2017} attempt to reproduce one of the asymptotically flat hairy solutions found in \cite{Brito2013a}. They find that some metrics components diverge as $r\to\infty$. The authors describe their numerical method in their Appendix D: a straightforward integration starting from the horizon with a standard routine of \textit{Mathematica}. In other words, they treat the ODEs as an initial value problem. To start the numerical integration, one must provide an initial value for the parameter $u$ at the horizon and the authors just take one of the numerical value given in \cite{Brito2013a}. However the asymptotically flat hairy black holes are also Lyapunov unstable, just as the bi-Schwarzschild solution. In fact, the exponentially growing mode $\e^{+\boldsymbol{m}\boldsymbol{r}}$ is generically present for small deviations around the flat space. To avoid this mode, one must treat the ODEs as a boundary value problem and use a suitable algorithm to solve the equations, such as for example the \textit{multishooting} method. This is the reason why the authors in Ref.~\cite{Torsello2017} were not able to obtain asymptotically flat hairy black holes. 

\begin{subappendices}
\section{Desingularization at the horizon}
\label{desing_hor_bigravity}

The horizon $r=r_H$ is a singular point of the differential equations \eqref{ode_hr} -- the derivatives $N'$ and $Y'$ are not defined at this point. In this appendix, we describe the procedure to desingularize the equations at the horizon. 

We first introduce new functions
\begin{equation}
    \nu(r)=\frac{N(r)}{S(r)},\quad\quad y(r)=\frac{N(r)}{S(r)}\quad\text{with}\quad S(r)=\sqrt{1-\frac{r_H}{r}}.
\end{equation}
Equations \eqref{eqN} and \eqref{eqY} then yield
\begin{equation}
\label{eq_nuy}
    \nu'=-\frac{\nu}{2r}+\frac{\mathcal{C}_1}{2\nu yr^2S^2},\quad y'=-\frac{yU'}{2U}+\frac{\mathcal{C}_2}{2\nu y r^2 U S^2},
\end{equation}
where
\begin{align*}
    \mathcal{C}_1&=(r^2-r_H\nu^2-\kappa_1 r^3\mathcal{P}_0)y-\kappa_1r^3\mathcal{P}_1U'\nu,\\
    \label{C1C2}
    \mathcal{C}_2&=\nu r^2(1-\kappa_2 r^2\mathcal{P}_2)U'-\kappa_2 r^4\mathcal{P}_1y-r_H U\nu y^2.\numberthis{}
\end{align*}
It follows directly from \eqref{eq_nuy} that the derivatives $\nu'$ and $y'$ will be finite at the horizon only if 
\begin{equation}
    \mathcal{C}_1\rvert_{r_H}=0,\quad\quad\mathcal{C}_2\rvert_{r_H}=0,
\end{equation}
from where one obtains the horizon values
\begin{align}
\label{Uph}
    U'_H&=\left.\frac{(1-\nu^2-\kappa_1 r^2\mathcal{P}_0)y}{\kappa_1 r^2\mathcal{P}_1\nu}\right|_{r_H},\\
\label{yh}
    y_H&=\left.\frac{1+(\kappa_2r^2\mathcal{P}_2-1)\nu^2+\kappa_1\kappa_2(\mathcal{P}_0\mathcal{P}_1-\mathcal{P}_1^2)r^4-(\kappa_1\mathcal{P}_0+\kappa_2\mathcal{P}_2)r^2}{\kappa_1 r\mathcal{P}_1U\nu }\right|_{r_H}.
\end{align}
At the same time, the horizon value of $U'$ can be obtained from \eqref{ode_hr},
\begin{equation}
\label{Uphbis}
    U'_H=\lim_{r\rightarrow r_H}\mathcal{D}_U(r,U,S\nu,Sy).
\end{equation}
This value must agree with the one given by Eq.~\eqref{Uph}, which yields a condition on $\nu_H$. If the parameters $b_k$ are chosen according to \eqref{bk}, this condition reduces to a biquadratic equation
\begin{equation}
\label{biquad_nu}
    \mathcal{A}\,(\nu_H^2)^2+\mathcal{B}\,\nu_H^2+\mathcal{C}=0,
\end{equation}
where Eq.~\eqref{yh} has been used to substitute $y_H$ and $\mathcal{A}$, $\mathcal{B}$, $\mathcal{C}$ are (rather complicated) functions of $r_H$, $U_H$. As a result, for given $r_H$, $U_H$, there are two possible horizon values $\nu_H^{(1)}$ and $\nu_H^{(2)}$. Injecting to Eqs.~\eqref{Uph} and \eqref{yh}, this determines the values $y_H$ and $U'_H$. Finally, the horizon values of $\nu'$ and $y'$ are obtained from \eqref{eq_nuy} by using l'Hopital's rule, which gives
\begin{equation}
\label{eq_nuyH}
    \nu_H'=-\frac{\nu_H}{2r_H}+\frac{\mathcal{C}'_1\rvert_{r_H}}{2r_H\nu_H y_H},\quad y'_H=-\frac{y_H U'_H}{2U_H}+\frac{\mathcal{C}'_2\rvert_{r_H}}{2r_H\nu_H y_H U_H}.
\end{equation}
There remains to compute the derivatives of $\mathcal{C}_1$ and $\mathcal{C}_2$. One has for example,
\begin{equation}
    \mathcal{C}'_1\rvert_{r_H}=\left.\left(\frac{\partial}{\partial r}+\nu_H'\frac{\partial}{\partial\nu}+y'_H\frac{\partial}{\partial y}+U'_H\frac{\partial}{\partial U}+U''_H\frac{\partial}{\partial U'}\right)\mathcal{C}_1(r,U,\nu,y,U')\right|_{r_H},
\end{equation}
where the second derivative of $U$ at $r=r_H$ is evaluated similarly by differentiating  Eq.~\eqref{Uphbis} and similarly for $\mathcal{C}'_2\rvert_{r_H}$. Injecting this to \eqref{eq_nuyH} yields a \textit{linear} algebraic system determining $\nu'_H$ and $y'_H$. Resolving it gives (we do not show explicit formulas in view of their complexity)
\begin{equation}
\label{eq_nuyHbis}
    \nu'_H=\nu'_H(r_H,U_H,\nu_H,y_H),\quad\quad y'_H=y'_H(r_H,U_H,\nu_H,y_H).
\end{equation}

Summarizing, the equations with the new functions $\nu$ and $y$ read
\begin{align*}
    \nu'&=-\frac{\nu}{2r}+\frac{\mathcal{C}_1}{2\nu yr^2S^2}\equiv\mathcal{F}_\nu(r,U,\nu,y),\\ y'&=-\frac{yU'}{2U}+\frac{\mathcal{C}_2}{2\nu y r^2 U S^2}\equiv\mathcal{F}_y(r,U,\nu,y),\\
    \label{ode_bis_hr}
    U'&=\mathcal{D}_U(r,U,S\nu,Sy)\equiv\mathcal{F}_U(r,U,\nu,y),\numberthis{}
\end{align*}
where $\mathcal{C}_1$ and $\mathcal{C}_2$ are defined by \eqref{C1C2} while $\mathcal{D}_U$ is the same as in \eqref{ode_hr}. These equations apply for $r>r_H$, while at $r=r_H$ they should be replaced by
\begin{align*}
    \nu'&=\nu'_H(r_H,U_H,\nu_H,y_H),\\
    y'&=y'_H(r_H,U_H,\nu_H,y_H),\\
    U'&=U'_H(r_H,U_H,\nu_H,y_H),\numberthis{}
\end{align*}
where $\nu'_H$, $y'_H$, $U'_H$ are given by Eqs.~\eqref{Uph} and \eqref{eq_nuyHbis}. These equations contain two free parameters, $r_H$ and $U_H\equiv ur_H$, while $\nu_H$ and $y_H$ are determined respectively by the equations \eqref{biquad_nu} and \eqref{yh}. This formulation allows us to start the numerical integration exactly at the horizon $r=r_H$. For $r>r_H$, we are free to choose either the original formulation \eqref{ode_hr} with the functions $N$ and $Y$ or the new formulation \eqref{ode_bis_hr} with the functions $\nu$ and $y$.

\section{Ansatz with time dependence}
\label{time_dep_ansatz}

Let us allow both metrics to depend on time, assuming that they are still spherically symmetric. The freedom of reparametrizations of the $t$, $r$ coordinates can be used to make the $g$-metric diagonal, but the $f$-metric will in general contain an off-diagonal term. The two metrics can be written as \cite{Volkov2012b}
\begin{align*}
    ds^2_g&=-Q^2dt^2+\frac{dr^2}{\Delta^2}+R^2d\Omega^2,\\
    ds^2_f&=-(q^2-\alpha^2Q^2\Delta^2)dt^2-2\alpha\left(q+\frac{Q\Delta}{W}\right)dtdr+\left(\frac{1}{W^2}-\alpha^2\right)dr^2+U^2d\Omega^2,\numberthis{}
\end{align*}
where $Q$, $q$, $\Delta$, $W$, $\alpha$, $U$, $R$ are functions of $r$ and $t$. The tensor $\tensor{\gamma}{^\mu_\nu}$ encoding the coupling between the two metrics is then given by
\begin{equation}
    \tensor{\gamma}{^\mu_\nu}=\begin{pmatrix}
        q/Q & \alpha/Q & 0 & 0 \\
        -\alpha Q\Delta^2 & \Delta/W & 0 & 0 \\
        0 & 0 & U/R & 0 \\
        0 & 0 & 0 & U/R
    \end{pmatrix}.
\end{equation}
This tensor is used to compute the effective stress-energy tensors $\tensor{T}{^\mu_\nu}$ and $\tensor{\mathcal{T}}{^\mu_\nu}$ in \eqref{eff_stress_tensors}. One can introduce new functions $N(r,t)$ and $Y(r,t)$ similarly as in \eqref{def_NY},
\begin{equation}
    N=\Delta R',\quad\quad Y=W U',
\end{equation}
where the primes denote the partial derivatives with respect to $r$, and one can specify completely the radial coordinate by setting
\begin{equation}
    R(r,t)=r.
\end{equation}
The independent field equations in \eqref{ein_eq} are
\begin{align*}
    \tensor{G(g)}{^0_0}&=\kappa_1\tensor{T}{^0_0},\quad\tensor{G(g)}{^1_1}=\kappa_1\tensor{T}{^1_1},\quad\tensor{G(g)}{^0_1}=\kappa_1\tensor{T}{^0_1},\\
\label{pert_eq_hr_1}
    \tensor{G(f)}{^0_0}&=\kappa_2\tensor{\mathcal{T}}{^0_0},\quad\tensor{G(f)}{^1_1}=\kappa_2\tensor{\mathcal{T}}{^1_1},\quad\tensor{G(f)}{^0_1}=\kappa_2\tensor{\mathcal{T}}{^0_1},\numberthis{}
\end{align*}
plus two nontrivial components of the conservation condition $\accentset{(g)}{\nabla}_\mu\tensor{T}{^\mu_\nu}=0$,
\begin{equation}
\label{pert_eq_hr_2}
    \accentset{(g)}{\nabla}_\mu\tensor{T}{^\mu_0}=0,\quad\quad\accentset{(g)}{\nabla}_\mu\tensor{T}{^\mu_1}=0.
\end{equation}
Here one has explicitly
\begin{equation}
    \tensor{G(g)}{^0_0}=\frac{N^2-1}{r^2}+\frac{2NN'}{r},\quad\tensor{G(g)}{^1_1}=\frac{N^2-1}{r^2}+\frac{2N^2Q'}{r\,Q},\quad\tensor{G(g)}{^0_1}=\frac{2\dot{N}}{r\,NQ^2},
\end{equation}
where the dot denotes the partial derivative with respect to $t$, and
\begin{equation}
    \tensor{T}{^0_0}=-\mathcal{P}_0-\mathcal{P}_1\frac{NU'}{Y},\quad\tensor{T}{^1_1}=-\mathcal{P}_0-\mathcal{P}_1\frac{q}{Q},\quad\tensor{T}{^0_1}=\mathcal{P}_1\frac{\alpha}{Q},
\end{equation}
where the $\mathcal{P}_k$ are defined in Eq.~\eqref{Pk}. The components of the second stress-energy tensor are
\begin{align*}
    \tensor{\mathcal{T}}{^0_0}&=-\frac{r^2}{NU^2\mathcal{A}}\left(\mathcal{P}_1 qY+\mathcal{P}_2\left(\alpha^2N^2QY+qNU'\right)\right),\\
    \tensor{\mathcal{T}}{^1_1}&=-\frac{r^2}{U^2\mathcal{A}}\left(\mathcal{P}_1 QU'+\mathcal{P}_2\left(\alpha^2 NQY+qU'\right)\right),\\
    \tensor{\mathcal{T}}{^0_1}&=-\frac{r^2}{NU^2\mathcal{A}}\mathcal{P}_1Y\alpha,\numberthis{}
\end{align*}
where $\mathcal{A}=NQY\alpha^2+qU'$. The components of the second Einstein tensor $\tensor{G(f)}{^\mu_\nu}$ are very complicated and their explicit expressions can be found in Ref.~\cite{Gervalle2020}.

\end{subappendices}

\chapter{Gauge theories and magnetic monopoles}
\label{chap_gauge_theory}

Having considered an explicit example of hairy black holes in modified gravity, we will now return to GR and study black hole with magnetically charged hair. However, we have seen in Sec.~\ref{more_gen_bh} that a no-hair theorem holds for the Einstein-Maxwell field equations. Therefore, we must consider a more sophisticated field theory than Maxwell electromagnetism. A very natural candidate is the electroweak theory of Weinberg \cite{Weinberg1967} and Salam \cite{Salam1959,Salam1964} which unifies electromagnetic and weak nuclear interactions\footnote{The weak nuclear interaction is responsible for $\beta$-decay of neutrons in atomic nuclei.}. This theory has been constructed based on the previous work of Glashow \cite{Glashow1959}, and it relies on the fundamental notions of gauge symmetries and spontaneous symmetry breaking.

While our primary focus is not on delving deeply into particle physics, it is essential to have a good understanding of gauge theories in flat space before moving on to the study of magnetic hairy black holes. Another important concept to grasp is magnetic monopoles, as they are flat space counterparts of magnetically charged black holes. The aim of this chapter is thus to provide an overview of all these theoretical concepts. In Section~\ref{gauge_sym_br}, we introduce the notion of spontaneously broken symmetries in gauge theory. Magnetic monopoles in electromagnetism and their generalizations in non-Abelian gauge theories are introduced in Section~\ref{monop}.

\section{Gauge theories and symmetry breaking}
\label{gauge_sym_br}

Gauge theories are a central topic in modern physics. In a nutshell, a gauge theory is a field theory whose Lagrangian is left invariant under a \textit{local} transformation of its fundamental fields. It follows that any measurable quantity is also invariant under the local transformation, which is often called the \textit{gauge} transformation. 


All the fundamental interactions are described by a gauge theory. What mainly distinguishes the different theories is the group $G$ to which the gauge transformations belong. For example, Maxwell electromagnetism is characterized by its $\text{U(1)}$ gauge group, the electroweak theory of Weinberg and Salam is invariant under $\text{SU(2)}\times\text{U(1)}$ gauge transformations and the strong interaction is described by a $\text{SU(3)}$ gauge theory. The combination of the electroweak theory with the strong interaction constitutes the so-called Standard Model of Particles.

For some theories the field configuration of minimal energy, the fundamental state, does not have the same symmetries as the theory itself. We say that the symmetry is \textit{spontaneously broken}. Understanding this phenomenon is of great interest as it is a crucial ingredient in the development of the Standard Model of Particles. We shall introduce spontaneously broken symmetries in three steps: first by considering discrete symmetries, then continuous global symmetries and finally local symmetries. 

This section will stick to the main steps addressed in the chapter 5 of Ref.~\cite{Quigg2013}. Interested readers are encouraged to refer to this source, and also to Refs.~\cite{Rubakov2002, Leng1988, Coleman1985a}.

\subsection{Breaking of a discrete symmetry}

Let us consider the following theory describing a real scalar field $\phi$ in Minkowski spacetime,
\begin{equation}
\label{th1}
    \mathcal{L}=-\frac{1}{2}\partial_\mu\phi\,\partial^\mu\phi-V(\phi),
\end{equation}
where $V(\phi)$ is the interaction potential defined by
\begin{equation}
\label{pot_broke}
    V(\phi)=\frac{1}{4}\phi^4+\frac{1}{2}\mu^2\phi^2.
\end{equation}
Here $\mu^2$ is a real parameter, which can be negative. Since the potential is an even function of $\phi$, the theory is invariant under the discrete symmetry $\phi\rightarrow -\phi$. The energy of a given field configuration is
\begin{equation}
    E=\int{d^3 x\left(\frac{1}{2}(\partial_t\phi)^2+\frac{1}{2}(\nabla\phi)^2+V(\phi)\right)},
\end{equation}
where $\partial_t$ denotes the time derivative and $\nabla$ is the gradient operator. The state of minimal energy should be stationary. Then, since the potential is assumed to depend only on $\phi$ and not on its gradient, it suffices to find the absolute minimum of $V$ to determine the fundamental state. By differentiating \eqref{pot_broke} with respect to $\phi$,
\begin{equation}
    \frac{\partial V}{\partial\phi}=\phi\left(\phi^2+\mu^2\right),
\end{equation}
we find two distinct cases:
\begin{enumerate}
    \item if $\mu^2>0$ then $\phi=0$ is the global minimum of the potential,
    \item if $\mu^2<0$ then $\phi=\pm\sqrt{-\mu^2}$ are two global minima of the potential while $\phi=0$ is a local maximum.
\end{enumerate}

In the case $\mu^2>0$, the fundamental state is $\phi=0$ and it is invariant under the parity transformation $\phi\rightarrow -\phi$ which is the discrete symmetry of the theory \eqref{th1}. The symmetry in this case is said to be \textit{manifest}. Introducing small perturbations around the fundamental state, $\phi=\delta\phi$, the Lagrangian describing the linearized theory is obtained by keeping only the quadratic terms in $\delta\phi$ in Eq.~\eqref{th1},
\begin{equation}
    \mathcal{L}_\text{lin}=-\frac{1}{2}\left(\partial_\mu\delta\phi\,\partial^\mu\delta\phi+\mu^2\delta\phi^2\right).
\end{equation}
This is precisely the Lagrangian describing a free massive field of mass $\mu$ whose dynamic is governed by the linear field equation
\begin{equation}
    (\partial_\mu\partial^\mu-\mu^2)\delta\phi=0.
\end{equation}

In the case $\mu^2<0$, there are two degenerate fundamental states, $\phi_\pm=\pm\sqrt{-\mu^2}$. The symmetry here is not manifest because $\phi_{+}$ and $\phi_{-}$ are not invariant under the parity transformation: we say that the symmetry is \textit{spontaneously broken}. However, because of the parity invariance of the theory, the two states are physically indistinguishable. We can therefore choose to study without loss of generality perturbations around $\phi_{+}$,
\begin{equation}
    \phi=\phi_{+}+\delta\phi=\sqrt{-\mu^2}+\delta\phi.
\end{equation}
Inserting this to the Lagrangian \eqref{th1} yields,
\begin{equation}
    \mathcal{L}=-\frac{1}{2}\partial_\mu\delta\phi\,\partial^\mu\delta\phi+\mu^2\left(\frac{\delta\phi^4}{4\phi_{+}^2}+\frac{\delta\phi^3}{\phi_{+}}+\delta\phi^2-\frac{\phi_{+}^2}{4}\right).
\end{equation}
We clearly see that the symmetry is not manifest because of the cubic term, although so far this Lagrangian describes exactly the same theory but written in terms of $\delta\phi$ instead of $\phi$. Keeping only the quadratic terms gives
\begin{equation}
    \mathcal{L}_\text{lin}=-\frac{1}{2}\left(\partial_\mu\delta\phi\,\partial^\mu\delta\phi-2\mu^2\delta\phi^2\right).
\end{equation}
Thus the linearized theory corresponds to a free massive field of mass $\sqrt{-2\mu^2}$. 

As a result, with this simple example we find that nothing special happens when the symmetry is spontaneoulsy broken: in any case the theory \eqref{th1} describes one single massive field whose mass depends on the parameter $\mu$. We shall move on to the more interesting case of continuous symmetries.

\subsection{Breaking of a continuous global symmetry}

\begin{figure}[b!]
    \centering
    \includegraphics[scale=1.2]{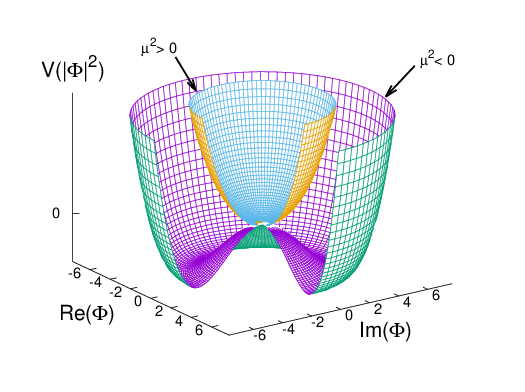}
    \caption[Profile of the scalar field potential for a complex scalar field theory whose global U(1) symmetry can be spontaneously broken.]{Profile of the potential $V(|\Phi|^2)$ for $\mu^2>0$ (blue/orange surface) and $\mu^2<0$ (purple/green surface). For $\mu^2>0$ there is an absolute minimum at $\Phi=0$ whereas for $\mu^2<0$ the minimum is not unique and corresponds to the circle $|\Phi|^2=-\mu^2$.}
    \label{fig_hat}
\end{figure}

The prototypical theory for studying the spontaneous breaking of a continuous symmetry is the Goldstone model \cite{Goldstone1961},
\begin{equation}
\label{th2}
    \mathcal{L}=-\partial_\mu\Phi^\ast\partial^\mu\Phi-V(|\Phi|^2),
\end{equation}
with $\Phi$ a complex scalar field, $\Phi^\ast$ its complex conjugate and
\begin{equation}
\label{mexican_hat}
    V(|\Phi|^2)=\frac{1}{2}|\Phi|^4+\mu^2|\Phi|^2.
\end{equation}
The theory is invariant under global phase transformations $\Phi\rightarrow e^{i\theta}\Phi$. Notice that these transformations belongs to the group $\text{U(1)}$ and the symmetry here is only \textit{global} because the transformations depend on a constant parameter $\theta$.

As before, we shall distinguish the two cases $\mu^2>0$ and $\mu^2<0$, see Fig.~\ref{fig_hat}. In the former case the fundamental state is $\Phi=0$ : the symmetry is manifest and the linearized theory describes two real scalar fields, $\phi_1=\text{Re}\,\Phi$ and $\phi_2=\text{Im}\,\Phi$, whose mass is $\mu$. The case $\mu^2<0$ is that of spontaneously broken symmetry. The fundamental state is not unique and is defined by 
\begin{equation}
\label{degen_fundam}
    |\Phi|^2=(\text{Re}\,\Phi)^2+(\text{Im}\,\Phi)^2=-\mu^2.
\end{equation}
The set of all fundamental states has the topology of a circle and all these states are related to each other by a global phase transformation $\Phi\rightarrow e^{i\theta}\Phi$. We shall consider without loss of generality $\Phi=\sqrt{-\mu^2}\equiv\Phi_0$ and introduce perturbations around this fundamental configuration by setting 
\begin{equation}
    \Phi=(\Phi_0+\eta) e^{i\xi},
\end{equation}
where $\eta$ and $\xi$ are two real fields. Inserting this to the Lagrangian \eqref{th2} and keeping only the quadratic terms yields
\begin{equation}
    \mathcal{L}_\text{lin}=-\partial_\mu\eta\,\partial^\mu\eta-2\Phi_0^2\eta^2-\partial_\mu\xi\,\partial^\mu\xi.
\end{equation}
This Lagrangian describes a massive field $\eta$ whose mass is $\sqrt{2}\,\Phi_0=\sqrt{-2\mu^2}$ and a \textit{massless} field $\xi$. The non-vanishing mass of the field $\eta$ can be viewed as a consequence of the restoring force of the potential against radial oscillations: any small deviation in the radial direction around the fundamental state increases the energy of the configuration. In contrast, the field $\xi$ corresponds to angular deviations around the fundamental state. These deviations do not change the energy, by virtue of the global U(1) invariance of the theory. Thus there is no restoring force against angular oscillations and the field $\xi$ is massless.

The appearance of a massless field when the global U(1) symmetry is spontaneously broken can be generalized to other internal symmetries\footnote{By \textit{internal} symmetries, we exclude spacetime symmetries such as, for example, the time invariance.}, this is the \textit{Goldstone theorem}. In the general case, one massless scalar field will occur for each broken generator of the original symmetry group or, in other words, for each generator that connects the different fundamental states. These massless fields are referred to as the Nambu-Goldstone bosons \cite{Goldstone1961,Nambu1960}, they are zero-energy excitations connecting the fundamental states together. 

Unfortunately, the Goldstone theorem does not apply to gauge theories. As we shall see, when the broken symmetry is local, the situation is different.

\subsection{Breaking of a continuous local symmetry}

Let us extend the Goldstone model \eqref{th2} by rendering local its U(1) symmetry. This can be achieved by replacing the usual derivative $\partial_\mu$ by a covariant derivative,
\begin{equation}
\label{cov_der}
    D_\mu\equiv\partial_\mu-iq\,A_\mu,
\end{equation}
where $q$ is a coupling constant and $A_\mu$ is the gauge vector field (also referred to as the gauge potential). The latter must change in a specific way when a local transformation is performed (see Eq.~\eqref{gauge_trans_intro} below). We can also add a kinetic term for the gauge field, which yields the Higgs-Abelian model \cite{Higgs1966},
\begin{equation}
\label{th3}
    \mathcal{L}=-(D_\mu\Phi)^\ast D^\mu\Phi-V(|\Phi|)^2-\frac{1}{4}F_{\mu\nu}F^{\mu\nu},
\end{equation}
where $F_{\mu\nu}$ is the field strength tensor defined by
\begin{equation}
    F_{\mu\nu}=\partial_\mu A_\nu-\partial_\nu A_\mu,
\end{equation}
and $V$ is the same potential as in Eq.~\eqref{mexican_hat}. This Lagrangian is left invariant under the local transformation
\begin{align*}
    \Phi&\rightarrow\Phi'= e^{iq\alpha(x)}\Phi,\\
\label{gauge_trans_intro}
    A_\mu&\rightarrow A'_\mu=A_\mu+\partial_\mu \alpha(x),\numberthis{}
\end{align*}
where $\alpha$ can be any function of the spacetime coordinates. This type of symmetry is also called \textit{gauge} symmetry. The usual derivative $\partial_\mu\Phi$ generates a new term proportional to $\partial_\mu\alpha$ when the transformation \eqref{gauge_trans_intro} is performed so that the quantity $\partial_\mu\Phi^\ast\partial^\mu\Phi$ is not invariant. However the introduction of the covariant derivative \eqref{cov_der} together with the transformation rule \eqref{gauge_trans_intro} for $A_\mu$ renders the kinetic term $(D_\mu\Phi)^\ast D^\mu\Phi$ invariant.

As before, there are two cases depending upon the sign of $\mu^2$ in the potential \eqref{mexican_hat}. If $\mu^2>0$ then the potential has a unique minimum at $\Phi=0$ which corresponds to the fundamental state. This case corresponds to an unbroken U(1) symmetry and the linearized theory describes a massless vector field $A^\mu$ plus two scalar fields, $\text{Re}\,\Phi$ and $\text{Im}\,\Phi$, with common mass $\mu$.

In the case $\mu^2<0$ the potential has a continuum of absolute minima corresponding to a continuum of degenerate fundamental states which are defined by the Eq.~\eqref{degen_fundam}. We shall study without loss of generality small perturbations around a specific fundamental configuration by setting 
\begin{equation}
    \Phi=(\Phi_0+\eta) e^{i\zeta/\Phi_0}=\Phi_0+\eta+i\zeta,
\end{equation}
where the second equality is valid at the linearized level and $\Phi_0=\sqrt{-\mu^2}$. Injecting this to the Lagrangian \eqref{th3} and keeping only the quadratic terms yields,
\begin{equation}
\label{th3_lin}
    \mathcal{L}_\text{lin}=-\partial_\mu\eta\,\partial^\mu\eta-2\Phi_0^2\eta^2-\frac{1}{4}F_{\mu\nu}F^{\mu\nu}-q^2\Phi_0^2A_\mu A^\mu-\partial_\mu\zeta\,\partial^\mu\zeta+2q\Phi_0 A^\mu\partial_\mu\zeta. 
\end{equation}
As it was the case for a broken global symmetry, the $\eta$-field associated with radial fluctuations around the fundamental state has a mass $\sqrt{2}\,\Phi_0=\sqrt{-2\mu^2}$. The gauge field $A_\mu$ seems to have acquired a mass too via the term $q^2\Phi_0^2 A_\mu A^\mu$. However the last term, which couples the gauge field with the $\zeta$-field, renders the interpretation of this Lagrangian complicated. 

The situation can be clarified by remarking that the last three terms in Eq.~\eqref{th3_lin} can be rewritten as
\begin{equation}
    -q^2\Phi_0^2A_\mu A^\mu-\partial_\mu\zeta\,\partial^\mu\zeta+2q\Phi_0 A^\mu\partial_\mu\zeta=-q^2\Phi_0^2\left(A_\mu-\frac{1}{q\Phi_0}\partial_\mu\zeta\right)\left(A^\mu-\frac{1}{q\Phi_0}\partial^\mu\zeta\right).
\end{equation}
This expression suggests that the field $\zeta$ disappear if we perform the following gauge transformation,
\begin{equation}
    A_\mu\rightarrow A'_\mu=A_\mu-\frac{1}{q\Phi_0}\partial_\mu\zeta(x).
\end{equation}
At the same time, the field $\Phi$ transforms as
\begin{equation}
    \Phi\rightarrow\Phi'= e^{-i\frac{\zeta(x)}{\Phi_0}}\Phi=\Phi_0+\eta.
\end{equation}
Returning to the original expression \eqref{th3} and using now the gauge transformed fields $A'_\mu$ and $\Phi'$, the quadratic Lagrangian reads
\begin{equation}
    \mathcal{L}_\text{lin}=-\partial_\mu\eta\,\partial^\mu\eta-2\Phi_0^2\eta^2-\frac{1}{4}F_{\mu\nu}F^{\mu\nu}-q^2\Phi_0^2A_\mu A^\mu.
\end{equation}
The interpretation of the field content in the linear theory is now completely clear:
\begin{itemize}
    \item one massive real scalar field $\eta$ with mass $\sqrt{2}\,\Phi_0=\sqrt{-2\mu^2}$,
    \item one massive real vector field $A_\mu$ with mass $\sqrt{2}\,q\Phi_0=q\sqrt{-2\mu^2}$,
    \item no $\zeta$-field!
\end{itemize}
The gauge field $A_\mu$ which was massless before spontaneous symmetry breaking is now massive. However the massless Nambu-Goldstone boson which was present in the case of a global symmetry breaking has disappeared. At first glance, this phenomenon can seem puzzling. Where did the degree of freedom of the $\zeta$-field go? A more careful analysis of the number of degrees of freedom in the theory actually reveals that none is missing. Before spontaneous symmetry breaking, the theory propagates four degrees of freedom: two modes are contained in the complex scalar field $\Phi$ and two others in the massless vector field $A_\mu$. After the symmetry breaking, we are left with one scalar mode $\eta$ plus three modes contained in the massive vector field $A_\mu$. As a result, the total number of degrees of freedom is preserved by the spontaneous symmetry breaking.

This mechanism applies for any gauge theory whose symmetry can be spontaneously broken (see for example \cite{Kibble1967}). It is called the Brout-Englert-Higgs-Hagan-Guralnik-Kibble mechanism \cite{Englert1964,Higgs1964,Higgs1964a,Guralnik1964} (often abbreviated the Higgs mechanism) and the remaining massive scalar is associated with the Higgs boson. It was the key ingredient for the construction of the electroweak theory of Weinberg and Salam. Indeed, the weak interaction involves massive bosons, but including a mass term in the Lagrangian would break the gauge invariance. Therefore, the Higgs boson and the associated Higgs mechanism provide an elegant way to get massive bosons within a gauge theory. 

The gauge group of the Weinberg-Salam model is $\text{SU(2)}\times\text{U(1)}$ but the vacuum configuration (fundamental state) exhibits only a U(1) symmetry. Thus the symmetry here is only partially broken: $\text{SU(2)}\times\text{U(1)}\rightarrow\text{U(1)}$. One has $\text{dim}\left(\text{SU(2)}\times\text{U(1)}\right)=4$ and $\text{dim}\left(\text{U(1)}\right)=1$. The residual U(1) symmetry is associated with the photon, a massless gauge boson, and the number of massive bosons is $4-1=3$ : these are the $W_\pm$ and $Z$ bosons whose masses were measured in colliders \cite{Arnison1983,LEP1992,Aad2012,Chatrchyan2012}. 

To summarize, the Higgs mechanism occurs in the electroweak part of the Standard Model. The remaining part, which corresponds to the strong nuclear interaction, is not affected by the spontaneous symmetry breaking. Consequently, the gauge bosons of the strong interaction -- the gluons -- remain massless. In their pursuit of unifying all the fundamental interactions, physicists have considered models beyond the Standard Model, where an additional Higgs mechanism takes place with the following symmetry breaking pattern,
\begin{equation}
    G\rightarrow SU(3)\times SU(2)\times U(1),
\end{equation}
where $G$ should be a group which contains the Standard Model group. In the simplest case, $G=\text{SU(5)}$. These models are known as Grand Unified Theories (GUTs). One of the key predictions of most GUTs is the possibility of proton decay. Numerous experiments had searched for this process but, to date, no conclusive evidence has been found \cite{Takenaka2020}.

\section{Magnetic monopoles}
\label{monop}

The development of spontaneously broken gauge theories has led to a great understanding of fundamental interactions and the particles that mediate them. If the gauge group is non-Abelian, meaning the group elements do not commute (as is the case with the SU(2) group), then the theory predicts the existence of finite-energy magnetic monopoles. In the simplest framework of unbroken Abelian gauge theories, such as Maxwell electromagnetism, magnetic monopoles also exist, but they have infinite energy. In this section, we review the construction of magnetic monopoles, from the simple Abelian monopole of Dirac, to its non-Abelian generalizations.

When examining Maxwell's equations, a classic statement often encountered in undergraduate studies is that isolated magnetic poles (\textit{a.k.a.} magnetic monopoles) do not exist. This is commonly illustrated in the following way: if a bar magnet is cut in half, one do not obtain one north pole and one south pole. Instead, each piece becomes a magnet. In other words, magnetic poles are always found in pairs of north and south poles. From the mathematical point of view, this property is encoded in the Maxwell-Thomson equation which has to be satisfied by the magnetic field $\vec{B}$,
\begin{equation}
\label{maxwell_thomson}
    \vec{\nabla}\cdot\vec{B}=0\quad\Leftrightarrow\quad\oint_\Sigma{\vec{B}\cdot\,\vec{n}\,dS}=0,
\end{equation}
where $\Sigma$ is a surface enclosing a volume $\Omega$ and $\vec{n}$ is the outgoing unit vector normal to $\Sigma$. The integral equation is obtained by integrating the first equation over $\Omega$ and then using Gauss's theorem.

However, in 1931 Dirac discovered that magnetic monopoles were actually compatible with Maxwell's equations \cite{Dirac1931a}. The construction of the monopole field configuration relies on the U(1) gauge invariance of classical electromagnetism and as we shall see below, it has a very important consequence for particle physics.

\subsection{The Dirac gauge potential}

Let us consider the Lagrangian describing electromagnetism in vacuum,
\begin{equation}
\label{lag_maxwell}
    \mathcal{L}=-\frac{1}{4}F_{\mu\nu}F^{\mu\nu},
\end{equation}
where $F_{\mu\nu}=\partial_\mu A_\nu-\partial_\nu A_\mu$ is the field strength tensor associated with the gauge potential $A_\mu$. The usual 3-dimensional magnetic field is then defined as $B^i=(1/2)\varepsilon^{ijk}F_{jk}$, where $\epsilon_{ijk}$ is the antisymmetric Levi-Civita tensor. The corresponding field equation is,
\begin{equation}
\label{max1}
    \nabla_\mu F^{\mu\nu}=0,
\end{equation}
where $\nabla_\mu$ is the geometrical covariant derivative (valid for any coordinate system, in any spacetime) and the definition of $F_{\mu\nu}$ also implies another equation,
\begin{equation}
\label{max2}
    \nabla_\alpha F_{\mu\nu}+\nabla_\mu F_{\nu\alpha}+\nabla_\nu F_{\alpha\mu}=0.
\end{equation}
Equation~\eqref{max1} reproduces the Maxwell-Gauss and Maxwell-Ampère equations while Eq.~\eqref{max2} is equivalent to the Maxwell-Thomson and Maxwell-Faraday equations. The theory is left invariant under the gauge transformations,
\begin{equation}
\label{gauge_trans_bis}
    A_\mu\rightarrow A'_\mu=A_\mu+\frac{i}{q}U\partial_\mu U^{-1}\quad\text{with}\quad U= e^{iq\alpha(x)}.
\end{equation}
This is exactly the same U(1) transformation defined by the Eq.~\eqref{gauge_trans_intro} above but written in more fashioned way. Note that the quantity $q$ which appears here is identified with the electron charge.

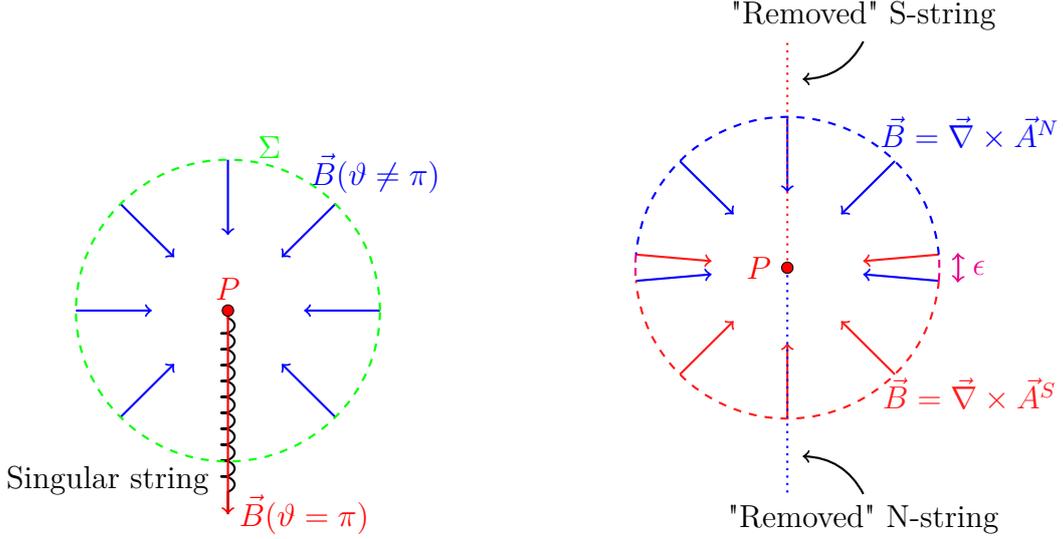
\begin{figure}
\centering
  \scalebox{1.0}{
\begin{tikzpicture}
    \tikzstyle{spring}=[thick,decorate,decoration={coil,pre length=0.1cm,post
  length=0.1cm,segment length=6}]
    \draw [thick,blue,->] (90:2)node[above=5pt,right=7pt,green]{$\Sigma$} -- (90:1) ;
    \draw [thick,blue,->] (45:2)node[right=15pt,above]{$\vec{B}(\vartheta\neq\pi)$} -- (45:1) ;
    \draw [thick,blue,->] (0:2) -- (0:1) ;
    \draw [thick,blue,->] (135:2) -- (135:1) ;
    \draw [thick,blue,->] (180:2) -- (180:1) ;
    \draw [thick,blue,->] (-135:2) -- (-135:1) ;
    \draw [thick,blue,->] (-45:2) -- (-45:1) ;
    \draw [spring] (0:0) -- (-90:2.7)node[left=45pt,above=3pt]{Singular string} ;
    \draw [thick,red,->] (0:0) -- (-90:2.7)node[right]{$\vec{B}(\vartheta=\pi)$} ;
    \draw [dashed,thick,green] (0,0) circle (2) ;
    \draw [fill=red] (0,0)node[above,red]{$P$} circle (0.075) ;
\end{tikzpicture}}
    \hspace{2cm}
  \scalebox{1.0}{
\begin{tikzpicture}
    \tikzstyle{spring}=[thick,decorate,decoration={coil,pre length=0.1cm,post
  length=0.1cm,segment length=6}]
    \draw [thick,blue,->] (90:2) -- (90:1) ;
    \draw [thick,blue,->] (45:2)node[right=28pt,above]{$\vec{B}=\vec{\nabla}\times\vec{A}^N$} -- (45:1) ;
    \draw [thick,blue,->] (-5:2) -- (-5:1) ;
    \draw [thick,blue,->] (135:2) -- (135:1) ;
    \draw [thick,blue,->] (-175:2) -- (-175:1) ;
    \draw [thick,color=red!90,->] (-135:2) -- (-135:1) ;
    \draw [thick,color=red!90,->] (-45:2)node[right=28pt,below=-3pt]{$\vec{B}=\vec{\nabla}\times\vec{A}^S$} -- (-45:1) ;
    \draw [thick,color=red!90,->] (-90:2) -- (-90:1) ;
    \draw [thick,color=red!90,->] (175:2) -- (175:1) ;
    \draw [thick,color=red!90,->] (5:2) -- (5:1) ;
    \draw [dotted,thick,blue] (0,0) -- (0,-3) ;
    \draw [dotted,thick,color=red!90] (0,0) -- (0,3) ;
    \draw [thick,->] (1,3.0)node[above]{"Removed" S-string} to[bend left] (0.2,2.5) ;
    \draw [thick,->] (1,-3.0)node[below]{"Removed" N-string} to[bend right] (0.2,-2.5) ;
    \draw [dashed,thick,blue] (5:2) arc (5:175:2) ;
    \draw [dashed,thick,color=red!90] (-5:2) arc (-5:-175:2) ;
    \draw [densely dashed,thick,magenta] (-5:2) arc (-5:5:2) ;
    \draw [densely dashed,thick,magenta] (175:2) arc (175:185:2) ;
    \draw [thick,magenta,<->] (-5:2.25) arc (-5:5:2.25)node[right=1pt,midway]{$\epsilon$} ;
    \draw [fill=red] (0,0)node[left=2pt,red]{$P$} circle (0.075) ;
\end{tikzpicture}}
  \caption[Schematic representation of the Dirac monopole with the singular string and its regularization.]{Left: schematic representation of the Dirac monopole with the singular string located at $\vartheta=\pi$. The ingoing magnetic flux through $\Sigma$ is compensated by the outgoing flux of the string. Right: schematic representation of the globally regular Dirac monopole described by two locally regular gauges.}
  \label{fig_dirac_mon}
\end{figure}

Let us describe the construction of the Dirac monopole by using the spherical coordinates $(r,\vartheta,\varphi)$ on Minkowski spacetime. The gauge potential found by Dirac can be expressed as,
\begin{equation}
\label{dirac_pot}
    A_\mu dx^\mu=P(\cos\vartheta-1)d\varphi,
\end{equation}
where $P$ is a constant. A direct computation reveals that this gauge field solves indeed the Eqs.~\eqref{max1}-\eqref{max2}. But does it actually describe a magnetic monopole ? Let us compute the 3-dimensional vector potential and the corresponding magnetic field:
\begin{equation}
\label{B_mon}
    \vec{A}=\frac{P(\cos\vartheta-1)}{r\sin\vartheta}\vec{e}_\varphi\quad\Rightarrow\quad\vec{B}=\vec{\nabla}\times\vec{A}=-\frac{P}{r^2}\vec{e}_r,
\end{equation}
where $\vec{e}_r$ and $\vec{e}_\varphi$ are respectively the radial and azimuthal unit vectors. As one can see, the magnetic field is radial and decays as $1/r^2$, just as the Coulombian electric field of a pointlike electric charge. Thus the gauge potential \eqref{dirac_pot} seems to effectively describes a magnetic monopole whose magnetic charge is $-P$. At the same time, it seems to be in contradiction with the integral form of the Maxwell-Thomson equation \eqref{maxwell_thomson}. To understand what happens here, one must inspect more carefully the naive computation that is performed in Eq.~\eqref{B_mon}. The vector potential $\vec{A}$ is actually singular at $\vartheta=\pi$ where the function $(\cos\theta-1)/\sin\vartheta$ diverges and the azimuthal vector $\vec{e}_\varphi$ is not defined there. As a result, the expression for $\vec{B}$ that is given above is valid everywhere except at $\vartheta=\pi$. A more accurate description of the Dirac monopole is presented on the left panel of Fig.~\ref{fig_dirac_mon}. For $\vartheta\neq\pi$, the magnetic field is described by the Eq.~\eqref{B_mon} and if $P>0$ then its contribution to the magnetic flux through $\Sigma$ is negative. For $\vartheta=\pi$, the magnetic field can be represented as a vector of "infinite" norm directed along the exterior. Thus its contribution to the flux is positive and the precise computation (which have to be carried out using distributions) shows that it exactly compensates the negative flux of $\vec{B}(\vartheta\neq\pi)$. Hence the gauge field \eqref{dirac_pot} does not describe an isolated magnetic pole, but rather a semi-infinite and infinitely thin solenoid lying on the negative $z$-axis: this is the singular \textit{Dirac string}.

\subsection{Removing the string singularity}
\label{removing_string_intro}

Dirac actually found that the singular string was nothing but a pure gauge artifact and that a completely regular description of the magnetic monopole was possible. The most rigorous description the string removal is due to Wu and Yang \cite{Wu1975,Wu1976} and we shall describe below their approach. 

First, we can show that the position of the Dirac string in space is defined up to a gauge transformation. Let us consider the gauge transformation \eqref{gauge_trans_bis} with
\begin{equation}
\label{move_string}
    U(\varphi)= e^{2iqP\varphi}.
\end{equation}
The gauge transformed potential is 
\begin{equation}
\label{dirac_pot_bis}
    A'_\mu dx^\mu=P(\cos\vartheta+1)d\varphi\quad\Rightarrow\quad\vec{A}'=\frac{P(\cos\vartheta+1)}{r\sin\vartheta}\vec{e}_\varphi.
\end{equation}
The gauge potential $A'_\mu$ is regular at $\vartheta=\pi$ but the string singularity is now located in the upper hemisphere at $\vartheta=0$. Thus the gauge transformation \eqref{move_string} acts as a rotation of the Dirac string by an angle of $\pi$. One can actually find a more general gauge transformation that rotates the singular string to any other direction. Thus the position of the string is completely arbitrary and the different configurations are all related to each other by some gauge transformation. This suggests that the magnetic field of the string should not be physically observable.

We have now all the ingredients for the description of a globally regular field configuration surrounding the monopole\footnote{Strictly speaking, the origin where the monopole is located has to be excluded from our description because the magnetic field $\vec{B}=-(P/r^2)\vec{e}_r$ is divergent there.}. Let us divide the 3-dimensional space $\mathbb{R}^3/\{0\}$ into two slightly overlapping hemisphere $R^N$ and $R^S$,
\begin{align*}
    R^N&:0\leq\vartheta\leq\frac{\pi}{2}+\frac{\epsilon}{2},\\
    R^S&:\frac{\pi}{2}-\frac{\epsilon}{2}\leq\vartheta\leq\pi,
\end{align*}
with $\epsilon>0$. We can choose the potential \eqref{dirac_pot} to describe the field configuration in upper hemisphere $R^N$ and its gauge transformed version \eqref{dirac_pot_bis} to describe the monopole field in the lower hemisphere $R^S$. Therefore, one consider two definitions of the gauge potential,
\begin{equation}
\label{reg_dirac_pot}
    A^N_\mu dx^\mu=P(\cos\vartheta-1)d\varphi,\quad\quad A^S_\mu dx^\mu=P(\cos\vartheta+1)d\varphi,
\end{equation}
where $A^N_\mu$ (resp. $A^S_\mu$) should be used in the upper hemisphere $R^N$ (resp. lower hemisphere $R^S$). Both of these potentials are regular within their domains of definition, ensuring that the field configuration is now free of any string singularity, as illustrated in the right panel of Fig.~\ref{fig_dirac_mon}. 

In the overlap region $R^N\cap R^S$, both $A^N_\mu$ and $A^S_\mu$ are well-defined and they are related to each other by the gauge transformation \eqref{move_string}. The latter has to be single-valued\footnote{This requirement comes from quantum mechanic considerations that have been skipped here, see for example Ref.~\cite{Shnir2005}.},
\begin{equation}
    U(\varphi+2\pi)+U(\varphi),
\end{equation}
which implies a very important condition for the electric and magnetic charges:
\begin{equation}
\label{charge_quant}
    qP=\frac{n}{2},\quad\text{with}\quad n\in\mathbb{Z}.
\end{equation}
This is the \textit{charge quantization} condition \cite{Dirac1931a}. Dirac suggested that the condition \eqref{charge_quant} could provide an explanation for the quantization of the electric charge. Indeed, the existence of a single magnetic monopole in the Universe is sufficient to justify the quantization electric charges, according to \eqref{charge_quant}.

\subsection{Non-Abelian generalization of the Dirac monopole}
\label{non_ab_intro}

Although being an elegant construction, the Dirac monopole in U(1) electrodynamics still suffers from a singularity in its center rendering its total energy infinite. This is actually not very surprising since the Dirac monopole is a pointlike magnetic charge: all the charge is concentrated in one single point in space. More rigorously, the Dirac potential \eqref{reg_dirac_pot} describing the monopole field is defined everywhere in space except at the origin. In other words, the Dirac theory of monopoles is only a correct description of a magnetic pole at classical scales whereas the monopole "core" should be described in the context of a quantum field theory. However, it turns out that monopole solutions that are regular at the origin and have a finite energy were found later in classical field theories.

The first step was made by Wu and Yang who demonstrated that the Dirac monopole can be embedded into non-Abelian Yang-Mills theories \cite{Wu1969}. This is a natural result since the group U(1) of the electromagnetic theory is a subgroup of SU$(N)$, the gauge group of Yang-Mills theories. Using the non-Abelian gauge transformations, one can then globally remove the Dirac string instead of using two locally regular gauges. However the magnetic field still remains singular at the origin. The expression of the gauge field in the globally regular gauge is sometimes called the Wu-Yang potential \cite{Shnir2005}.

Some times later, 't Hooft \cite{Hooft1974} and Polyakov \cite{Polyakov1974} demonstrated that the singularity at the origin can be removed by adding a Higgs field in the adjoint representation of the gauge group and choosing its potential to be such that the SU$(N)$ symmetry is spontaneously broken down to the U(1) symmetry. The corresponding model is called Yang-Mills-Higgs theory and it is described by the Lagrangian,
\begin{align*}
    \mathcal{L}&=-\frac{1}{2}\text{Tr}(F_{\mu\nu}F^{\mu\nu})-\text{Tr}(D_\mu\Phi D^\mu\Phi)-V(\Phi)\\
\label{ymh}
    &=-\frac{1}{4}F^a_{\mu\nu}F^{a\mu\nu}-\frac{1}{2}D_\mu\phi^a D^\mu\phi^a-V(\Phi).\numberthis{}
\end{align*}
Here $F_{\mu\nu}=F^a_{\mu\nu}T_a$ and $\Phi=\phi^a T_a$ are respectively the non-Abelian field strength tensor and the Higgs field while $T_a$ is a basis of the $\mathfrak{su}(N)$ Lie algebra. Notice that in these expressions, Latin indices refer to components in the $\mathfrak{su}(N)$ Lie algebra and should not be confused with spacetime indices like $\mu$, $\nu$, $\dots$. In the simplest case $N=2$, one can choose the basis to be $T_a=(1/2)\tau^a$, where $\tau^a$ are the Pauli matrices. The field strength tensor is given by
\begin{equation}
    F_{\mu\nu}=\partial_\mu A_\nu-\partial_\nu A_\mu-iq[A_\mu,A_\nu]\quad\text{or}\quad F^a_{\mu\nu}=\partial_\mu A^a_\nu-\partial_\nu A^a_\mu+q\,\varepsilon_{abc}A^b_\mu A^c_\nu,
\end{equation}
where $A_\mu=A^a_\mu T_a$ is the SU(2) gauge field. The covariant derivative is
\begin{equation}
    D_\mu\Phi=\nabla_\mu\Phi-iq[A_\mu,\Phi]\quad\text{or}\quad D_\mu\phi^a=\nabla_\mu\phi^a+q\,\varepsilon_{abc}A^b_\mu\phi^c,
\end{equation}
and finally the potential is,
\begin{equation}
    V(\Phi)=\frac{\lambda}{4}(\phi^a\phi^a-v^2)^2,
\end{equation}
where $\lambda$ is a coupling constant and $v$ is the non-vanishing Higgs vacuum expectation value which allows for a spontaneous breaking of the SU(2) gauge symmetry. The theory is left invariant under the SU(2) gauge transformations,
\begin{align*}
    A_\mu&\rightarrow A'_\mu=UA_\mu U^{-1}+\frac{i}{q}U\partial_\mu U^{-1},\\
    \Phi&\rightarrow \Phi'=U\Phi U^{-1},\quad\text{with}\quad U= e^{iq\,\theta^a(x)T_a}.\numberthis{}
\end{align*}

In the static case, the total energy of a given configuration is,
\begin{equation}
    E=\int{d^3 x\left(\frac{1}{4}F^a_{\mu\nu} F^{a\mu\nu}+\frac{1}{2}D_\mu\phi^a D^\mu\phi^a+\frac{\lambda}{4}(\phi^a\phi^a-v^2)^2\right)}.
\end{equation}
Since every term in this integral is positive definite, we can extract two conditions on the Higgs field for the energy to be minimal,
\begin{equation}
\label{vacuum_condition}
    \phi^a\phi^a=v^2,\quad\quad D_\mu\phi^a=0.
\end{equation}
Thus the fundamental state corresponds to $|\Phi|=v$ and the simplest of these configurations can be expressed as
\begin{equation}
\label{trivial_higgs}
    \phi^a=(0,0,v).
\end{equation}
One can also consider non-trivial fundamental states such as, for example,
\begin{equation}
\label{n_higgs}
    \phi^a=v(\sin\vartheta\cos(n\varphi),\sin\vartheta\sin(n\varphi),\cos\vartheta),
\end{equation}
where $n$ must be an integer\footnote{In topology, the integer $n$ is referred to as the winding number.} to ensure that the Higgs field is single-valued. Then the second condition in Eq.~\eqref{vacuum_condition} is fulfilled if
\begin{align*}
    A^a_\mu T_a dx^\mu=&\frac{1}{q}\left(T_1\sin(n\varphi)-T_2\cos(n\varphi)\right)d\vartheta\\
\label{wu_yang_pot}
    &+\frac{n}{q}\sin\vartheta\left(\cos\vartheta(T_1\cos(n\varphi)+T_2\sin(n\varphi))-T_3\sin\vartheta\right)d\varphi.\numberthis{}
\end{align*}
This is precisely the Wu-Yang potential describing the Dirac monopole embedded in the SU(2) Yang-Mills theory. This gauge field is regular everywhere in space, which can be seen by noting that $A^a_\varphi=0$ for $\vartheta=0,\pi$. One can recover the more familiar expression of the Dirac potential \eqref{dirac_pot} by applying the (singular) gauge transformation,
\begin{equation}
    U(\vartheta,\varphi)= e^{-iT_3 n\varphi} e^{iT_2 \vartheta} e^{iT_3 n\varphi},
\end{equation}
which transforms the gauge field to
\begin{equation}
    A^1_\mu dx^\mu=A^2_\mu dx^\mu=0,\quad A^3_\mu dx^\mu=\frac{n}{q}(\cos\vartheta-1)d\varphi.
\end{equation}
Thus the third isotopic component of non-Abelian gauge field corresponds to the Abelian potential describing the Dirac monopole with a singular string located at $\vartheta=\pi$. By identification with Eq.~\eqref{dirac_pot}, we see that the magnetic charge $P$ in this case is such that,
\begin{equation}
\label{nonab_dirac}
    qP=n,
\end{equation}
which is the non-Abelian version of the Dirac charge quantization \eqref{charge_quant}.

The field configuration \eqref{n_higgs}-\eqref{wu_yang_pot} solves the field equations exactly but, as mentioned above, it still remains singular at the origin and its total energy,
\begin{equation}
    E=\frac{2\pi n^2}{q^2}\int_0^\infty{\frac{dr}{r^2}},
\end{equation}
is divergent. At the same time, the field equations also admit other solutions which only approach \eqref{n_higgs}-\eqref{wu_yang_pot} at spatial infinity and which are perfectly regular at the origin. They correspond to the celebrated 't Hooft-Polyakov monopole. For the lowest allowed value of the magnetic charge, $|n|=1$, the field configuration is spherically symmetric whereas for $|n|>1$ the monopoles are not rotationally invariant in general (in the simplest case, they are only axially symmetric). 

We have only presented here the basic ideas behind the construction of non-Abelian magnetic monopoles. Several important aspects have been omitted, for example, there is no unique definition of the "physical" magnetic field in non-Abelian gauge theories. We thus refer the interested reader to chapter 5 of Ref.~\cite{Shnir2005} for more details. 

It has long been considered that non-Abelian monopoles do not exist in the electroweak theory of Weinberg and Salam. Although its gauge group SU(2)$\times$U(1) is spontaneously broken down to the subgroup U(1), the Higgs field is in the fundamental representation of SU(2) and not in the adjoint representation (see the definition of the electroweak theory in the Sec.~\ref{setting_ews}). Nevertheless in 1996, Cho and Maison finally discovered a spherically symmetric and non-Abelian monopole solution in the electroweak theory \cite{Cho1996}. We shall describe in more details the Cho-Maison monopole and its axially symmetric generalizations in the next chapter.

\chapter[Electroweak black holes with magnetic hair]{Magnetic monopoles in the electroweak theory and their black hole counterparts}
\label{chap_mon}
\subtitle{This chapter is based on \cite{Gervalle2022a,Gervalle2023,GervalleInPrep}.}

\section{Introduction}

The discovery of a globally regular magnetic monopole by 't Hooft \cite{Hooft1974} and Polyakov \cite{Polyakov1974} in the SU(2) Yang-Mills-Higgs theory has triggered both theoretical \cite{Goddard1978,Coleman1983,Konishia,Manton2004,Shnir2005} and experimental \cite{Rajantie2016,Mitsou2019,Mavromatos2020} studies over the last decades (see also Refs.~\cite{Chamseddine1997,Forgacs2004} for particular aspects of monopoles). However the experimental search for magnetic monopoles has always been giving negative results. One possible explanation is that the SU(2) 't~Hooft-Polyakov monopole is not described by the Standard Model. Instead, one must consider extensions of this monopole within Grand Unified Theories. In this case, the mass of the monopole is expected to be of the order of $10^{14}-10^{16}\,\text{GeV}$, thus rendering their detection in colliders impossible \cite{Fairbairn2007}. One can wonder whether there are monopoles within the framework of the Standard Model, without considering an extension of it. The electroweak sector of the Standard Model is characterized by the gauge group SU(2)$\times$U(1), and it is spontaneously broken to the subgroup U(1) via the Higgs mechanism. However, the Higgs field is in the \textit{fundamental representation} of SU(2), which means that it is expressed as $\Phi=(\Phi_1,\Phi_2)^\text{T}$, with $\Phi_1,\Phi_2\in\mathbb{C}$, whereas for the 't Hooft-Polyakov monopole, the Higgs field is in the \textit{adjoint representation} of SU(2), meaning that $\Phi=\phi^a T_a$ with $\phi^a\in\mathbb{R}$ and $a=\{1,2,3\}$ (see Sec.~\ref{non_ab_intro}). The standard topological arguments \cite{Manton2004} for the existence and stability of monopoles only apply when the Higgs field is in the adjoint representation.

Nevertheless, the U(1) Dirac monopole can always be embedded into larger gauge groups, and it follows that it must be a solution of the electroweak theory. The mass of this monopole, however, cannot be predicted by the theory because it has an infinite energy. Another type of electroweak monopoles was discovered in 1977 by Nambu \cite{Nambu1977} who noticed the existence of vortex-type solutions\footnote{In the context of classical gauge theory, vortices are infinitely long "tubular" solutions within which energy and magnetic flux are confined.} in the theory. Nambu used the following form for the Higgs field,
\begin{equation}
\label{nambu_mon}
    \Phi=\phi\begin{pmatrix}
        \sin\frac{\vartheta}{2}\;e^{-i\varphi} \\ -\cos\frac{\vartheta}{2}
    \end{pmatrix},
\end{equation}
where $\phi$ is a real function of the spatial coordinates. This field is not defined on the negative part of the $z$-axis because the first component has no limit for $\vartheta\to\pi$. This can be circumvented by assuming that $\phi$ vanishes at $\vartheta=\pi$. One obtains then a semi-infinite vortex extending along the negative part of the $z$-axis and terminating at a monopole for $z=0$. Analyzing the magnetic flux inside the vortex and that spreading out at infinity through the vortex termination, Nambu arrived at the following condition for the magnetic charge,
\begin{equation}
    qP=n\times\sin^2\theta_\text{W},
\end{equation}
where $\theta_\text{W}$ is the weak mixing angle -- a parameter of the electroweak theory. This resembles the non-Abelian version of the Dirac charge quantization \eqref{nonab_dirac} but with the additional factor of $\sin^2\theta_\text{W}$. However the construction of Nambu cannot be static because the vortex will be pulling the monopole. An equilibrium configuration can be obtained if the vortex has a finite length and terminates some distance away on an antimonopole. Then, the resulting monopole-antimonopole pair configuration will have a finite energy and will be spinning around the center of mass \cite{Urrestilla2002}.

Another possibility to construct monopoles in the electroweak theory was found later, in 1996, by Cho and Maison \cite{Cho1996}. They used the same form \eqref{nambu_mon} as Nambu for the Higgs field, but instead of assuming that $\phi$ vanishes at $\vartheta=\pi$, they avoided the line singularity by using two locally regular gauges, as for the U(1) Dirac monopole (see the Sec.~\ref{removing_string_intro}). One assumes that $\Phi$ in \eqref{nambu_mon} describes the Higgs field in the upper hemisphere while, in the lower hemisphere, one uses its gauge-transformed version $\Phi'=e^{i\varphi}\Phi$ which is regular for $\vartheta\to\pi$. The U(1) gauge transformation $e^{i\varphi}$ which relates the two gauges is regular in the equatorial transition region. This construction provides a globally regular description for a static and spherically symmetric monopole whose magnetic charge corresponds to that of the Dirac monopole \eqref{charge_quant} with $n=\pm 2$. The function $\phi$ in \eqref{nambu_mon} depends only on the radial coordinate, and it is determined by the field equations, together with another unknown function $f(r)$ that is contained in the SU(2) gauge field. We call this field configuration the Cho-Maison (CM) monopole. 

The SU(2) part of the CM monopole is regular at the origin and is similar to that of the 't Hooft-Polyakov monopole. Its U(1) part is divergent at the origin and renders the total energy infinite, as for the Dirac monopole \cite{Cho1996}. Hence the CM monopole can be viewed as a hybrid between a U(1) Dirac monopole and a SU(2) 't Hooft-Polyakov monopole. Several attempts to regularize the monopole energy have been considered, but they require to modify the Lagrangian of the theory \cite{Kimm2015,Pak2015,Blaschke2018,Ellis2021,Hung2021}. In this chapter, we consider a different perspective. Our aim is to investigate also the gravitating counterparts of electroweak monopoles. In this case, it is not necessary to regularize the central singularity because it will be hidden inside an event horizon, just like for a RN black hole. In the spherically symmetric case, electroweak magnetic black holes have been reported in 2021 by Bai and Korwar \cite{Bai2021} and their ADM mass is finite.

In a recent paper, we showed that the CM monopole is stable with respect to arbitrary (small) perturbations \cite{Gervalle2022a}. At the same time, all Dirac monopoles with $|n|>1$ are unstable with respect to perturbations with angular momentum $j=|n|/2-1$. In particular, the Dirac monopole with $|n|=2$ is unstable only in the spherically symmetric sector ($j=0$) while the CM monopole is stable and has the same magnetic charge. This suggests that the CM monopole can be viewed as a stable remnant of the Dirac monopole decay. One may similarly conjecture that stable remnants also exist for monopoles with $|n|>2$, but they cannot be spherically symmetric because the perturbations which grow in time are not.

In this chapter, we confirm this conjecture by constructing generalizations of the CM monopole for higher values of the magnetic charge in the simplest case of axial symmetry. However, we have not yet been able to analyze their stability. It is very plausible that the less energetic configuration -- and thus the most stable -- is a non-Abelian monopole with only discrete symmetries. We also construct the gravitating counterparts of the axially symmetric monopoles, which are static black holes with an axially symmetric and magnetically charged hair. These black holes still contain the U(1) singular part of the magnetic charge within their horizons.

We construct the solutions numerically for various values of the charge, compute their regularized energy, the quadrupole moment, and study their inner structure. This requires solving the underlying system of nonlinear PDEs. Our discretization scheme is based on the finite element method, and the nonlinearities are treated using Newton's method.

The rest of the chapter is organized as follows. In Section~\ref{setting_ews}, we present the field equations of the bosonic sector of the Weinberg-Salam (WS) theory minimally coupled to GR. We also define in this section the physical fields -- the electromagnetic and Z fields. In Section~\ref{axial_ews}, we describe the general form of the axially symmetric fields and their reduction to the spherically symmetric case. This section also presents the procedure for removing the Dirac string singularity, and discuss the gravitational constraint equations along with the gauge condition. Simple analytical solutions which correspond to RN and RN-de Sitter black holes are presented in Section~\ref{anal_sol_ews}. The stability of the RN solution is analyzed in Section~\ref{stab_rn}. The spherically symmetric hairy black holes are described in Section~\ref{spher_hairy_bh}. Here, we provide a more complete description of these solutions as compared to the paper \cite{Bai2021}. After that, we present the axially symmetric monopoles and their black hole counterparts in Section~\ref{axial_sol_ews}. The emphasis is placed on understanding the internal structure of flat space monopoles, and then on the generalization of our results to the black hole case. Finally, our concluding remarks are given in Section~\ref{conclu_ews}. The Appendix~\ref{oscillons} presents an auxiliary result on the existence of neutral, oscillating configurations in the electroweak theory. The remaining three appendices contain technical details such as the asymptotic behavior of the solutions, the local behavior at the origin for flat space monopoles, and the radial coordinate transformation relating the axially symmetric line element to the spherically symmetric one.

\section{Einstein-Weinberg-Salam theory}
\label{setting_ews}

\subsection{Action and field equations}

We consider the bosonic part of the electroweak theory of Weinberg and Salam (WS) minimally coupled to Einstein gravity. The dimensionful action can be represented in the form,
\begin{equation}
\label{action_ews}
    \boldsymbol{S}_\text{EWS}=\frac{1}{\boldsymbol{c}\boldsymbol{g}_0^2}\int\left(\frac{1}{2\kappa}R+\mathcal{L}_\text{WS}\right)\sqrt{-g}\,d^4 x,
\end{equation}
where $R$ is the Ricci scalar associated with the spacetime metric $g_{\mu\nu}$ and
\begin{equation}
    \mathcal{L}_\text{WS}=-\frac{1}{4g^2}W^a_{\mu\nu}W^{a\mu\nu}-\frac{1}{4g'^2}Y_{\mu\nu}Y^{\mu\nu}-(D_\mu\Phi)^\dagger D^\mu\Phi-\frac{\beta}{8}(\Phi^\dagger\Phi-1)^2,
\end{equation}
is WS Lagrangian density. All quantities in the integrand -- the spacetime coordinates $x^\mu$, the metric $g_{\mu\nu}$, the WS fields and the couplings -- are rendered dimensionless by an appropriate rescaling. The Abelian U(1) and non-Abelian SU(2) field strength tensors are respectively
\begin{equation}
\label{field_strenghts_ews}
    Y_{\mu\nu}=\partial_\mu Y_\nu-\partial_\nu Y_\mu,\quad\quad W^a_{\mu\nu}=\partial_\mu W^a_\nu-\partial_\nu W^a_\mu+\epsilon_{abc}W^b_\mu W^c_\nu,
\end{equation}
while the Higgs field $\Phi=(\Phi_1,\Phi_2)^\text{T}$ is in the fundamental representation of SU(2) with the gauge covariant derivative
\begin{equation}
    D_\mu\Phi=\left(\partial_\mu-\frac{i}{2}Y_\mu-\frac{i}{2}\tau^a W^a_\mu\right)\Phi,
\end{equation}
where $\tau^a$ are the Pauli matrices. Notice that we have expanded the SU(2) gauge field $W_\mu=T_a W^a_\mu$ and its field strength tensor $W_{\mu\nu}=T_a W^a_{\mu\nu}$ in the basis $T_a=(1/2)\tau_a$ which satisfies the commutation relation $[T_a,T_b]=i\,\epsilon_{abc}T_c$. The two coupling constants for the gauge fields are $g=\cos\theta_\text{W}$ and $g'=\sin\theta_\text{W}$ where the physical value of the weak mixing angle $\theta_\text{W}$ is such that $g'^2=\sin^2\theta_\text{W}=0.23$.

The dimensionful (boldfaced) parameters in the action \eqref{action_ews} are the speed of light $\boldsymbol{c}$ and $\boldsymbol{g}_0$. The latter is related to the electron charge $\boldsymbol{e}$ via
\begin{equation}
\label{struct_fine}
    \alpha\equiv\frac{\boldsymbol{e}^2}{4\pi\pmb{\hbar}\boldsymbol{c}}=\frac{\pmb{\hbar}\boldsymbol{c}}{4\pi}(gg'\boldsymbol{g}_0)^2\approx\frac{1}{137}\quad\quad\Rightarrow\quad\quad\boldsymbol{e}=\pmb{\hbar}\boldsymbol{c}\boldsymbol{g}_0 e\quad\text{with}\quad e\equiv gg'.
\end{equation}
The dimensionful fields which are commonly used in the literature are $\boldsymbol{Y}_\mu=(\boldsymbol{\Phi}_0/g')Y_\mu$, $\boldsymbol{W}^a_\mu=(\boldsymbol{\Phi}_0/g)W^a_\mu$ and $\boldsymbol{\Phi}=\boldsymbol{\Phi}_0\Phi$ where $\boldsymbol{\Phi}_0=246\,\text{GeV}$ is the Higgs vacuum expectation value. The dimensionful coordinates are $\boldsymbol{x}^\mu=\boldsymbol{l}_0 x^\mu$ where $\boldsymbol{l}_0=1/(\boldsymbol{g}_0\boldsymbol{\Phi}_0)=1.52\times 10^{-16}\,\text{cm}$ is the electroweak length scale. 

The theory is invariant under spacetime diffeomorphisms and SU(2)$\times$U(1) gauge transformations,
\begin{equation}
\label{gauge_trans_ews}
    \Phi\rightarrow U\Phi,\quad\quad\mathcal{W}\rightarrow U\mathcal{W}U^{-1}+iU\partial_\mu U^{-1}dx^\mu,
\end{equation}
with
\begin{equation}
    \mathcal{W}=\frac{1}{2}(Y_\mu+\tau^a W^a_\mu)dx^\mu,\quad\quad U=\text{exp}\left(\frac{i}{2}\theta^0+\frac{i}{2}\tau^a\theta^a\right),
\end{equation}
where $\theta^0$ and $\theta^a$ are functions of the spacetime coordinates $x^\mu$. Varying the action \eqref{action_ews} with respect to the WS fields gives the equations,
\begin{align*}
    \nabla^\mu Y_{\mu\nu}&=g'^2\frac{i}{2}(\Phi^\dagger D_\nu\Phi-(D_\nu\Phi)^\dagger\Phi)\equiv g'^2 J^0_\nu,\\
    \mathcal{D}^\mu W^a_{\mu\nu}&=g^2\frac{i}{2}(\Phi^\dagger\tau^a D_\nu\Phi-(D_\nu\Phi)^\dagger\tau^a \Phi)\equiv g^2 J^a_\nu,\\
\label{eqs_ews}
    D_\mu D^\mu\Phi&-\frac{\beta}{4}(\Phi^\dagger\Phi-1)\Phi=0,\numberthis{}
\end{align*}
where $\nabla_\mu$ is the geometrical covariant derivative with respect to the spacetime metric and $\mathcal{D}_\mu W^a_{\alpha\beta}=\nabla_\mu W^a_{\alpha\beta}+\epsilon_{abc}W^b_\mu W^c_{\alpha\beta}$. Varying the action with respect to the metric yields the Einstein equations,
\begin{equation}
\label{ein_eq_ews}
    E_{\mu\nu}\equiv G_{\mu\nu}-\kappa\,T_{\mu\nu}=0,
\end{equation}
with the energy-momentum tensor,
\begin{equation}
\label{stress_ews}
    T_{\mu\nu}=\frac{1}{g^2}W^a_{\mu\sigma}\tensor{W}{^a_\nu^\sigma}+\frac{1}{g'^2}Y_{\mu\sigma}\tensor{Y}{_\nu^\sigma}+(D_\mu\Phi)^\dagger D_\nu\Phi+(D_\nu\Phi)^\dagger D_\mu\Phi+g_{\mu\nu}\mathcal{L}_\text{WS}.
\end{equation}

The vacuum is defined as the configuration with the flat Minkowski metric $g_{\mu\nu}=\eta_{\mu\nu}$ and with $T_{\mu\nu}=0$. Up to a gauge transformation, the WS fields can be chosen as
\begin{equation}
    W^a_\mu=Y_\mu=0,\quad\quad\Phi=\begin{pmatrix}
        0 \\ 1
    \end{pmatrix}.
\end{equation}
Linearizing the field equations with respect to small fluctuations around the vacuum gives the perturbative mass spectrum of the theory containing the massless graviton and photon and the massive Z, W and Higgs bosons whose dimensionless masses are,
\begin{equation}
\label{boson_masses}
    m_\text{Z}=\frac{1}{\sqrt{2}},\quad\quad m_\text{W}=\frac{g}{\sqrt{2}},\quad\quad m_\text{H}=\sqrt{\frac{\beta}{2}}.
\end{equation}
These masses are expressed in units of the electroweak mass scale $\boldsymbol{m}_0=\pmb{\hbar}/(\boldsymbol{c}\boldsymbol{l}_0)=(\pmb{\hbar}/\boldsymbol{c})\boldsymbol{g}_0\boldsymbol{\Phi}_0$ so that, for example, the dimensionful Z boson mass is $\boldsymbol{m}_\text{Z}=\boldsymbol{m}_0/\sqrt{2}\approx 91.18\,\text{GeV}/\boldsymbol{c}^2$. Using the Higgs mass $\boldsymbol{m}_\text{H}\approx 125\,\text{GeV}/\boldsymbol{c}^2$ yields the value $\beta\approx 1.88$.

The dimensionless gravitational coupling $\kappa$ in the action can be expressed in terms of the Newton constant $\boldsymbol{G}$ or, equivalently, in terms of the Planck mass $\boldsymbol{M}_\text{Pl}$ and Z boson mass $\boldsymbol{m}_\text{Z}$ as follows,
\begin{equation}
\label{kap_val}
    \kappa=\frac{8\pi\boldsymbol{G}\boldsymbol{\Phi}_0^2}{\boldsymbol{c}^4}=\frac{4e^2}{\alpha}\left(\frac{\boldsymbol{m}_\text{Z}}{\boldsymbol{M}_\text{Pl}}\right)^2=5.42\times 10^{-33}.
\end{equation}
This value is very small because the Planck mass is many orders of magnitude larger than the Z boson mass.

Summarizing, the theory contains four parameters which are known from experimental measurements. Their dimensionless values are
\begin{equation}
\label{param_ews}
    g'^2=0.23,\quad g^2=1-g'^2,\quad\beta=1.88,\quad\kappa=5.42\times 10^{-33}.
\end{equation}
The underlying mass and length scales are
\begin{equation}
    \boldsymbol{m}_0=\sqrt{2}\,\boldsymbol{m}_\text{Z}=\frac{\pmb{\hbar}}{\boldsymbol{c}}\boldsymbol{g}_0\boldsymbol{\Phi}_0=128.94\,\text{GeV}/\boldsymbol{c}^2,\quad\boldsymbol{l}_0=\frac{1}{\boldsymbol{g}_0\boldsymbol{\Phi}_0}=1.53\times 10^{-16}\,\text{cm},
\end{equation}
which correspond to the electroweak scales. Another relevant scale in the theory is the Planck scale which is related to the electroweak one by,
\begin{equation}
    \boldsymbol{m}_0=\sqrt{\frac{\alpha\kappa}{2e^2}}\boldsymbol{M}_\text{Pl}=1.05\times 10^{-17}\times\boldsymbol{M}_\text{Pl},\quad\boldsymbol{l}_0=\sqrt{\frac{2e^2}{\alpha\kappa}}\boldsymbol{L}_\text{Pl}=9.46\times 10^{16}\times\boldsymbol{L}_\text{Pl}.
\end{equation}

\subsection{Electromagnetic and Z fields}

In non-Abelian gauge theories, the definition of the electromagnetic tensor is not unique off the Higgs vacuum \cite{Coleman1985b}. We shall adopt the definition of Nambu for the electromagnetic and Z fields \cite{Nambu1977},
\begin{equation}
\label{elec_Z}
    F_{\mu\nu}=\frac{g}{g'}Y_{\mu\nu}-\frac{g'}{g}N^a W^a_{\mu\nu},\quad\quad Z_{\mu\nu}=Y_{\mu\nu}+N^a W^a_{\mu\nu},
\end{equation}
where $N^a=\Phi^\dagger\tau^a\Phi/(\Phi^\dagger\Phi)$. This definition was used by Hindmarsh and James to study the so-called \textit{sphaleron} configuration \cite{Hindmarsh1994}. It should be noted that these 2-forms are not closed in general which means that there is no field potential $A_\mu$ such that $F=dA$. This is related to the fact that Maxwell equations do not hold off the Higgs vacuum. The magnetic field is defined as in the flat space, $B^i=(1/2)\epsilon^{ijk}F_{jk}$.

Using the electromagnetic tensor $F_{\mu\nu}$ and its dual,
\begin{equation}
    \tilde{F}^{\mu\nu}=*F^{\mu\nu}=\frac{1}{2\sqrt{-g}}\epsilon^{\mu\nu\alpha\beta}F_{\alpha\beta},
\end{equation}
one can define the conserved electric and magnetic 4-currents,
\begin{equation}
\label{4_currents}
    J^\mu=\frac{1}{4\pi}\nabla_\nu F^{\mu\nu},\quad\quad\tilde{J}^\mu=\frac{1}{4\pi}\nabla_\nu\tilde{F}^{\mu\nu}.
\end{equation}
By the definition \eqref{elec_Z} of $F_{\mu\nu}$, both $J^\mu$ and $\tilde{J}^\mu$ split into a sum of two separately conserved currents: the U(1) current determined by the contribution of $Y_{\mu\nu}$ and the SU(2) current determined by $W^a_{\mu\nu}$. 

We are interested in purely magnetic\footnote{Purely magnetic configurations are characterized by an identically vanishing electric field, $F^{0i}=0$.} field configurations for which the non-vanishing components of the 4-currents are the electric current $J^k$ and the magnetic charge density $\tilde{J}^0$. The magnetic charge $P$ and its density split into U(1) and SU(2) parts,
\begin{equation}
\label{mag_charge_dens}
    \tilde{J}^0=\frac{1}{4\pi}\vec{\nabla}\cdot\vec{B}\equiv\rho_\text{U(1)}+\rho_\text{SU(2)},
\end{equation}
and
\begin{equation}
\label{split_P}
    P=\int_\Sigma\left(\rho_\text{U(1)}+\rho_\text{SU(2)}\right)\sqrt{-g}\,d^3 x\equiv P_\text{U(1)}+P_\text{SU(2)},
\end{equation}
where $\Sigma$ is a spacelike hypersurface. Here $P_\text{U(1)}$ and $P_\text{SU(2)}$ are separately conserved. Since the $Y$ field is Abelian, one has
\begin{equation}
\label{u1_charge_gen}
    P_\text{U(1)}=\frac{g}{g'}\oint_{\mathbb{S}^2}{dB},
\end{equation}
where the $\mathbb{S}^2$ is a two-sphere at spatial infinity. This integral vanishes unless $Y$ is topologically non-trivial, in which case its value is determined by the topology and does not depend on the radius of the sphere.

\section{Axial symmetry}
\label{axial_ews}

We are interested in static and axially symmetric black hole solutions. For these systems, the spacetime metric admits two Killing vectors, 
\begin{equation}
    \xi=\frac{\partial}{\partial t},\quad\quad\varsigma=\frac{\partial}{\partial\varphi},
\end{equation}
where $t$ is the asymptotically timelike coordinate and $\varphi$ is the azimuthal angle.

Because the Einstein-Weinberg-Salam model contains many fields, we shall describe separately the gravity and the electroweak sectors.

\subsection{Static and axially symmetric gravitational fields}

The metric can be written in terms of the spheroidal\footnote{For $N(r)=1$, the coordinates in the line element \eqref{metric_ews} are said to be \textit{quasi-isotropic} because the polar angles in the hypersurfaces with $t=const.$ are represented without distortion. For \textit{fully} isotropic coordinates, the functions in front of $r^2d\vartheta^2$ and $r^2\sin^2\vartheta d\varphi^2$ would be the same. In the generic case when $N(r)\neq1$, the coordinates are not (quasi-)isotropic.} coordinates $(t,r,\vartheta,\varphi)$ \cite{Delgado2021},
\begin{equation}
\label{metric_ews}
    g_{\mu\nu}dx^\mu dx^\nu=-e^{2U}N(r)\,dt^2+e^{2K}\left(\frac{dr^2}{N(r)}+r^2d\vartheta^2\right)+e^{2 S}\,r^2\sin^2\vartheta\,d\varphi^2,
\end{equation}
where $U$, $K$, $ S$ depend on $r$, $\vartheta$. Except otherwise stated, $N$ is a given function of $r$ only and we shall choose $N(r)=1-r_H/r$, where $r_H$ is the location of the event horizon. This choice is well suited to describe generic black hole solutions with a simple event horizon. 

With the line element \eqref{metric_ews} considered here, the Einstein-Hilbert part of the Lagrangian in Eq.~\eqref{action_ews} can be represented as
\begin{equation}
    \frac{1}{2\kappa}R\sqrt{-g}=L_\text{G}+\partial_r\Gamma_r+\partial_\vartheta\Gamma_\vartheta,
\end{equation}
where,
\begin{align*}
    L_\text{G}=&\frac{1}{\kappa}e^{ S+U}\sin\vartheta\Big[Nr\,\partial_r(K- S)+Nr^2(\partial_r K\,\partial_r S+\partial_r S\,\partial_rU+\partial_rU\,\partial_rK)\\
    &+\frac{r_H}{2}\partial_r(K-U)+\partial_\vartheta K\,\partial_\vartheta S+\partial_\vartheta S\,\partial_\vartheta U+\partial_\vartheta U\,\partial_\vartheta K+\cot\vartheta\,\partial_\vartheta(K- S)\Big],\\
    \Gamma_r=&-\frac{1}{\kappa}e^{ S+U}Nr^2\sin\vartheta\,\partial_r(K+ S+U),\\
\label{red_lag_g}
    \Gamma_\vartheta=&-\frac{1}{\kappa}e^{ S+U}\sin\vartheta\,\partial_\vartheta(K+ S+U).\numberthis
\end{align*}
Here $L_\text{G}$ is the essential part of the Lagrangian that should be kept, while the total derivatives can be integrated by parts and do not affect the field equations.

We shall be considering asymptotically flat black holes with a regular event horizon. Their mass can be determined by the asymptotic behavior of the $g_{00}$ metric coefficient which reads,
\begin{equation}
\label{asymp_mass_ews}
    -g_{00}=1-\frac{2M}{r}+\dots=1-\frac{\boldsymbol{G}\boldsymbol{M}}{\boldsymbol{c}^2\boldsymbol{r}}+\dots,
\end{equation}
where the dots denote the subleading terms. 

Alternatively, the mass also admits the Komar integral representation \eqref{komar_surf}. The latter consists in a surface integral at spatial infinity which can be transformed to a volume integral over a spacelike hypersurface as in Eq.~\eqref{komar_vol}, but for black holes, a boundary term at the horizon surface must be taken into account. The latter can be expressed in terms of the horizon surface gravity, which gives (see for example the section 2.4 in Ref.~\cite{Heusler1996}),
\begin{equation}
\label{vol_mass_ews}
    M=\frac{\kappa_g A_H}{4\pi}+\frac{1}{4\pi}\int_\Sigma{n_\mu\xi_\nu R^{\mu\nu}\sqrt{\gamma}\,d^3 x}=\frac{\kappa_g A_H}{4\pi}-\frac{1}{4\pi}\int_{r>r_H}{\tensor{R}{^0_0}\sqrt{-g}\,d^3x},
\end{equation}
where $\Sigma$ is a spacelike hypersurface in the exterior black hole region, $\gamma$ is the determinant of the induced metric over $\Sigma$, $n=-e^U\sqrt{N}dt$ is the future-directed unit normal vector to $\Sigma$, $A_h$ is the event horizon area and $\kappa_g$ is the horizon surface gravity,
\begin{equation}
    \kappa_g^2=-\left.\frac{1}{2}\nabla_\mu\xi_\nu\,\nabla^\mu\xi^\nu\right|_{r=r_H}.
\end{equation}

The dimensionful mass expressed in physical units is
\begin{equation}
    \frac{\boldsymbol{M}}{\boldsymbol{m}_0}=\frac{\boldsymbol{c}^2\boldsymbol{l}_0}{\boldsymbol{G}\boldsymbol{m}_0}M=\frac{8\pi}{\kappa}\times\frac{M}{\pmb{\hbar}\boldsymbol{c}\boldsymbol{g}_0^2}=\frac{e^2}{4\pi\alpha}\times\frac{8\pi}{\kappa}M\equiv\frac{e^2}{4\pi\alpha}\times\mathcal{M},
\end{equation}
where we have used the relations \eqref{struct_fine} and \eqref{kap_val}. Notice that both $M$ and $\mathcal{M}$ are dimensionless quantities related to the mass that we shall use in this chapter. Using Eq.~\eqref{vol_mass_ews} and taking into account the Einstein equations yields
\begin{equation}
\label{vol_mass_bis_ews}
    \mathcal{M}=\frac{8\pi}{\kappa}M=\frac{2\kappa_g A_H}{\kappa}-\int_{r>r_H}{(2\tensor{T}{^0_0}-T)\sqrt{-g}\,d^3 x},
\end{equation}
where $T$ is the trace of the energy-momentum tensor. We emphasize that this formula is also valid in the flat space limit $\kappa\to 0$. In this case, the first term vanishes because there is no event horizon and the second term can be simplified by using the equation,
\begin{equation}
    \int_{r>0}{(\tensor{T}{^0_0}-T)\sqrt{-g}\,d^3 x}=0,
\end{equation}
which holds for any static solution in flat space field theory \cite{Deser1976,Herdeiro2022}. Substituting $T$ in Eq.~\eqref{vol_mass_bis_ews} and taking $r_H\to 0$ yields
\begin{equation}
\label{flat_space_ener}
    \mathcal{M}_{\kappa\to 0}=-\int_{r>0}{\tensor{T}{^0_0}\sqrt{-g}\,d^3 x},
\end{equation}
which agrees with the standard definition of energy in flat space. 

Using the line element \eqref{metric_ews}, the surface gravity and the horizon area are
\begin{equation}
\label{surf_grav_area}
    \kappa_g=\left.\frac{N'}{2}e^{U-K}\right|_{r=r_H},\quad A_H=2\pi r_H^2\left.\int_0^\pi{e^{K+S}\sin\vartheta\,d\vartheta}\right|_{r=r_H}.
\end{equation}
The surface gravity should be constant at the horizon. This is true if $\left.\partial_\vartheta(U-K)\right|_{r=r_H}=0$ and it turns out that the latter follows from the constraint equations to be described in Sec.~\ref{const_gauge} below. 

\subsection{Static, axially symmetric and purely magnetic electroweak fields}

The axially symmetric and purely magnetic electroweak SU(2) gauge field, U(1) weak hypercharge field and Higgs field can be represented in the form,
\begin{align*}
    W&=T_a W^a_\mu dx^\mu=T_2\left(F_1\,dr+F_2\,d\vartheta\right)+\nu\left(T_3\,F_3-T_1\,F_4\right)d\varphi,\\
\label{ansatz_ews}
    Y&=Y_\mu dx^\mu=\nu\,Y_3\,d\varphi,\quad\quad\Phi=\begin{pmatrix}
        \phi_1 \\ \phi_2
    \end{pmatrix},\numberthis{}
\end{align*}
where $F_1$, $F_2$, $F_3$, $F_4$, $Y_3$, $\phi_1$, $\phi_2$ are 7 real-valued functions of $r$, $\vartheta$ and $\nu$ is a constant parameter. The SU(2) field here corresponds to the purely magnetic ansatz of Rebbi and Rossi \cite{Rebbi1980}. The ansatz \eqref{ansatz_ews} preserves its form under gauge transformations \eqref{gauge_trans_ews} generated by $U=\exp\left\{i\chi(r,\vartheta)T_2\right\}$, which transform the field amplitudes as 
\begin{align*}
    F_1\rightarrow F_1+\partial_r\chi,\quad\quad &F_2\rightarrow F_2+\partial_\vartheta\chi,\quad\quad Y_3\rightarrow Y_3,\\
    F_3\rightarrow F_3\cos\chi-F_4\sin\chi,\quad\quad &F_4\rightarrow F_4\cos\chi+F_3\sin\chi,\\
\label{res_gauge}
    \phi_1\rightarrow\phi_1\cos(\chi/2)+\phi_2\sin(\chi/2),\quad\quad&\phi_2\rightarrow\phi_2\cos(\chi/2)-\phi_1\sin(\chi/2).\numberthis{}
\end{align*}

Inserting \eqref{metric_ews}, \eqref{ansatz_ews} to Eq.~\eqref{stress_ews} yields the energy density of the WS fields
\begin{equation}
\label{ener_dens_ews}
    \mathcal{E}\equiv-\tensor{T}{^0_0}=\frac{e^{-2K}}{2g^2r^2}\mathcal{E}_W+\frac{e^{-2K}}{2g'^2r^2}\mathcal{E}_Y+\mathcal{E}_\Phi=-\mathcal{L}_\text{WS},
\end{equation}
where
\begin{align*}
    \mathcal{E}_W=&e^{-2K}N(\partial_\vartheta F_1-\partial_r F_2)^2\\
    &+e^{-2 S}N\left[(\partial_r F_3+F_1 F_4)^2+(\partial_r F_4-F_1 F_3)^2\right]\frac{\nu^2}{\sin^2\vartheta}\\
    &+e^{-2 S}\left[(\partial_\vartheta F_3+F_2 F_4)^2+(\partial_\vartheta F_4-F_2 F_3)^2\right]\frac{\nu^2}{r^2\sin^2\vartheta},\\
    \mathcal{E}_Y=&e^{-2 S}\left[N(\partial_r Y_3)^2+\frac{1}{r^2}(\partial_\vartheta Y_3)^2\right]\frac{\nu^2}{\sin^2\vartheta},\\
    \mathcal{E}_\Phi=&e^{-2K}N\left[\left(\partial_r\phi_1-\frac{1}{2}F_1\phi_2\right)^2+\left(\partial_r\phi_2+\frac{1}{2}F_1\phi_1\right)^2\right]\\
    &+e^{-2K}\left[\left(\partial_\vartheta\phi_1-\frac{1}{2}F_2\phi_2\right)^2+\left(\partial_\vartheta\phi_2+\frac{1}{2}F_2\phi_1\right)^2\right]\frac{1}{r^2}\\
    &+\e^{-2 S}\left[\left((F_3+Y_3)\phi_1-F_4\phi_2\right)^2+\left((F_3-Y_3)\phi_2+F_4\phi_1\right)^2\right]\frac{\nu^2}{4r^2\sin^2\vartheta}\\
\label{contrib_ews}
    &+\frac{\beta}{8}\left(\phi_1^2+\phi_2^2-1\right)^2.\numberthis{}
\end{align*}
As a consistency check, one can verify that the energy density is invariant under the residual gauge transformation \eqref{res_gauge}.

The zero energy configuration can be expressed as 
\begin{equation}
    F_1=F_2=F_4=\phi_1=0,\quad\phi_2=1,\quad F_3=Y_3=\text{const.}\equiv Y_\infty,
\end{equation}
in which case the solution for the metric corresponds to that of vacuum GR. This configuration keeps its form under gauge transformations $U=\exp\left\{iC\nu\varphi(1+\tau_3)/2\right\}$ with a constant $C$, whose effect is simply a shift $Y_\infty\rightarrow Y_\infty+C$.

The total reduced Lagrangian is the sum of the gravitational part defined in Eq.~\eqref{red_lag_g} and the electroweak part specified in Eq.~\eqref{ener_dens_ews},
\begin{equation}
\label{red_lag_ews}
    L_\text{EWS}=L_\text{G}+L_\text{WS}\quad\quad\text{with}\quad\quad L_\text{WS}=\sqrt{-g}\,\mathcal{L}_\text{WS}.
\end{equation}
The field equations can be obtained by varying $L$ with respect to the 10 unknown functions $U,K, S,F_1,F_2,F_3,F_4,Y_3,\phi_1,\phi_2$.

The fields \eqref{ansatz_ews} can describe two different classes of axially symmetric solutions. Indeed, if we assume the energy density to be invariant under reflections with respect to the equatorial plane, $\vartheta\rightarrow\pi-\vartheta$, then certain field amplitudes must be invariant under the reflections so that they are \textit{even} whereas the others must change sign hence they are \textit{odd}. If we assume further that $\phi_2\to 1$ at spatial infinity, then a direct inspection of Eq.~\eqref{contrib_ews} reveals two different possibilities that we call "monopole case" and "sphaleron case":
\begin{align*}
    \text{\underline{monopole case}}:&\quad\text{odd}\;F_1,\,F_3,\,Y_3,\,\phi_1\;\text{and even}\;F_2,\,F_4,\,\phi_2;\\
\label{parity_sym}
    \text{\underline{sphaleron case}}:&\quad\text{odd}\;F_1,\,F_4,\,\phi_1\;\text{and even}\;F_2,\,F_3,\,Y_3,\,\phi_2.\numberthis{}
\end{align*}
At the same time, all metric functions are always of even parity. The magnetic charge density in the sphaleron case is odd and therefore the total magnetic charge is identically vanishing. In contrast, for monopole solutions, we will see that the total magnetic charge can be non-zero and is proportional to $\nu$. We shall consider in this chapter only the monopole case. We refer the reader to our publication \cite{Gervalle2023} for a discussion of the axially symmetric sphalerons in flat space and their relation to monopoles.

For our purpose, it is convenient to re-express the gauge fields functions as,
\begin{align*}
    F_1=-\frac{H_1(r,\vartheta)}{r\sqrt{N}},\quad F_2=H_2(r,\vartheta),\quad F_3=&\,\cos\vartheta+H_3(r,\vartheta)\sin\vartheta,\\
\label{func_redef}
    F_4=H_4(r,\vartheta)\sin\vartheta,\quad\quad\quad Y_3=&\,\cos\vartheta+y(r,\vartheta)\sin\vartheta.\numberthis{}
\end{align*}
The parity under the reflections $\vartheta\rightarrow\pi-\vartheta$ of the new functions is as follows,
\begin{equation}
\label{parity_hs}
    \text{odd}\;H_1,\,H_3,\,y,\,\phi_1\;\text{and even}\;H_2,\,H_4,\,\phi_2.
\end{equation}
The regularity of the energy density at the polar axis requires that
\begin{equation}
\label{reg_axis}
    H_1=H_3=y=\phi_1=0,\quad H_2=H_4\quad\text{for }\vartheta=0,\pi.
\end{equation}
These conditions also guarantee that the WS fields can be transformed to a regular gauge. Specifically, the $\varphi$-components of $W$ and $Y$ in Eq.~\eqref{ansatz_ews} do not vanish for $\vartheta=0,\pi$. This means that the fields written in this form contain a Dirac string singularity located along the symmetry axis. We shall see below that this line singularity can be gauged away, but only if the parameter $\nu$ in \eqref{ansatz_ews} is an integer.

\subsection{Removing the Dirac string}

The gauge transformation that removes the Dirac string singularity for monopole solutions is generated by
\begin{equation}
\label{reg_gauge_mon}
    U_\pm=e^{-i\nu\varphi T_3}e^{-i\vartheta T_2}e^{\pm i\nu\varphi/2}=e^{\pm i\nu\varphi/2}\begin{pmatrix}
        \cos(\vartheta/2)e^{-i\nu\varphi/2} & -\sin(\vartheta/2)e^{-i\nu\varphi/2} \\
        \sin(\vartheta/2)e^{i\nu\varphi/2} & \cos(\vartheta/2)e^{i\nu\varphi/2}
    \end{pmatrix}.
\end{equation}
It transforms the SU(2) field into
\begin{equation}
\label{reg_W}
    W=T_\varphi\left(-\frac{H_1}{r\sqrt{N}}dr+(H_2-1)d\vartheta\right)+\nu\Big(T_r\,H_3+T_\vartheta(1-H_4)\Big)\sin\vartheta\,d\varphi.
\end{equation}
This is the form of the field that is commonly used in the literature, see \textit{e.g.} Refs.~\cite{Kleihaus1997,Kleihaus1998,Kleihaus1998a}. Here the angle-dependent basis of the $\mathfrak{su}(2)$ Lie algebra is related to the constant one $\{T_1,T_2,T_3\}$ via
\begin{equation}
    T_r=n^a T_a,\quad\quad T_\vartheta=\partial_\vartheta T_r,\quad\quad T_\varphi=\frac{1}{\nu\sin\vartheta}\partial_\varphi T_r,
\end{equation}
where
\begin{equation}
    n^a=\left[\sin\vartheta\cos(\nu\varphi),\sin\vartheta\sin(\nu\varphi),\cos\vartheta\right].
\end{equation}
This basis fulfills the standard commutation relations, for example $[T_r,T_\vartheta]=i\,T_\varphi$. It is clear that the parameter $\nu$ must be an integer since, otherwise, the vector $n^a$ is not single-valued. The conditions \eqref{reg_axis} imply that in the vicinity of the symmetry axis one has $W=(T_1\,dx^1-T_2\,dx^1)(1-H_2)+\dots$ where $x^a=rn^a$ are the Cartesian coordinates and the dots represent terms that vanish for $\vartheta=0,\pi$. This field is perfectly regular and the Dirac string is gone.

The transformation generated by \eqref{reg_gauge_mon} brings $Y$ and $\Phi$ to the form,
\begin{equation}
\label{Y_phi_reg}
    Y_\pm=\nu(\cos\vartheta\pm 1+y\sin\vartheta)d\varphi,\quad\Phi_\pm=e^{\pm i\nu\varphi/2}\begin{pmatrix}
        (\phi_1\cos\frac{\vartheta}{2}-\phi_2\sin\frac{\vartheta}{2})e^{-i\nu\varphi/2}\vspace{0.1cm} \\
        (\phi_1\sin\frac{\vartheta}{2}+\phi_2\cos\frac{\vartheta}{2})e^{+i\nu\varphi/2}
    \end{pmatrix}.
\end{equation}
Here the sign choice "$\pm$" represents two locally regular gauges. On the one hand, $Y_{-}$ and $\Phi_{-}$ are regular for $\vartheta=0$, but $Y_{-}$ shows the Dirac string singularity along the negative $z$-axis at $\vartheta=\pi$ while $\Phi_{-}$ has no limit there. Therefore, this gauge can be used in the upper hemisphere, for $\vartheta\in[0,(\pi+\epsilon)/2]$. On the other hand, $Y_{+}$ and $\Phi_{+}$ are regular for $\vartheta=\pi$ and can be used in the lower hemisphere, for $\vartheta\in[(\pi-\epsilon)/2,\pi]$. As a result, $Y$ and $\Phi$ are completely regular if one uses these two local gauges. In the equatorial region, $(\pi-\epsilon)/2\leq\vartheta\leq(\pi+\epsilon)/2$, the gauge transformation which relates $Y_{-}$, $\Phi_{-}$ to $Y_{+}$, $\Phi_{+}$ is generated by $U=e^{i\nu\varphi}$ and it is single-valued if $\nu\in\mathbb{Z}$.

The U(1) contribution to the magnetic charge in Eq.~\eqref{u1_charge_gen} is defined by the integral
\begin{equation}
    \frac{1}{4\pi}\oint_{\mathbb{S}^2}{dY}=\frac{1}{4\pi}\oint_{\mathbb{S}^1}(Y_{-}-Y_{+})=-\frac{\nu}{2\pi}\oint_{\mathbb{S}^1}{d\varphi}=-\nu,
\end{equation}
where $\mathbb{S}^1$ is a circle around the equatorial region of $\mathbb{S}^2$. The parameter $\nu$ is usually called the \textit{winding number}. The U(1) part of the magnetic charge and the corresponding charge density are then
\begin{equation}
\label{u1_charge}
    P_\text{U(1)}=-\frac{g}{g'}\nu,\quad\quad\rho_\text{U(1)}=P_\text{U(1)}\delta^3({\vec{x}}).
\end{equation}
The U(1) charge is thus pointlike and located at the origin (the latter being hidden inside an event horizon in the gravitating case).

\subsection{Constraint equations and gauge condition}
\label{const_gauge}

Injecting the ansatz \eqref{metric_ews},\eqref{ansatz_ews} to the Eqs.~\eqref{eqs_ews},\eqref{ein_eq_ews} gives the set of PDEs which determines the field amplitudes. These equations must agree with that obtained from the variations of the reduced Lagrangian \eqref{red_lag_ews}. For the Einstein equations, we use the following combinations,
\begin{align}
\label{eqU_ews}
    -\tensor{E}{^0_0}+\tensor{E}{^r_r}+\tensor{E}{^\vartheta_\vartheta}+\tensor{E}{^\varphi_\varphi}&=0,\\
\label{eqK}
    \tensor{E}{^0_0}+\tensor{E}{^r_r}-\tensor{E}{^\vartheta_\vartheta}+\tensor{E}{^\varphi_\varphi}&=0,\\
\label{eqS}
    \tensor{E}{^0_0}+\tensor{E}{^r_r}+\tensor{E}{^\vartheta_\vartheta}-\tensor{E}{^\varphi_\varphi}&=0,
\end{align}
which yields a set of second-order PDEs for, respectively, $U$, $K$ and $S$. At the same time, there are also two non-trivial equations 
\begin{equation}
\label{const_ein}
    \mathcal{C}_1\equiv r\sqrt{-g}\,\tensor{E}{^r_r}=0,\quad\quad\mathcal{C}_2\equiv\frac{\sqrt{-g}}{r}\tensor{E}{^r_\vartheta}=0,
\end{equation}
which still contain second derivatives, but only those of $U$ and $S$. These equations can be resolved with respect to $\partial_r K$ and $\partial_\vartheta K$, yielding
\begin{equation}
\label{eqK_bis}
    \partial_r K=\mathcal{K}_1,\quad\quad\partial_\vartheta K=\mathcal{K}_2,
\end{equation}
where $\mathcal{K}_1$, $\mathcal{K}_2$ depend on $U$, $S$ and their derivatives but not on $K$. In principle, the function $K$ can be obtained by integrating these two equations, which is possible only if $\partial_\vartheta\mathcal{K}_1=\partial_r\mathcal{K}_2$. This condition follows from the Eqs.~\eqref{eqU_ews},\eqref{eqS} while the remaining equation \eqref{eqK} is identically fulfilled by virtue of \eqref{eqK_bis}.

However, in practice, we rather consider the two equations \eqref{const_ein} as constraints instead of solving them directly. Specifically, the two non-trivial Bianchi identities $\nabla_\mu\tensor{E}{^\mu_\nu}=0$ (with $\nu=r,\vartheta$) can be represented as
\begin{equation}
\label{bianchi_ews}
    \partial_\vartheta\mathcal{C}_2=-N\,\partial_r\mathcal{C}_1-\frac{1}{2}N'\mathcal{C}_1+\dots,\quad\partial_\vartheta\mathcal{C}_1=r^2\partial_r\mathcal{C}_2+r\,\mathcal{C}_2+\dots,
\end{equation}
where the dots denote terms linear in the left-hand sides of Eqs.~\eqref{eqU_ews}-\eqref{eqS}. It follows that if the second-order PDEs are imposed and if the constraints \eqref{const_ein} are fulfilled at the symmetry axis, then they are fulfilled everywhere. One has at the symmetry axis,
\begin{equation}
\label{const_axis}
    \vartheta=0,\pi:\;\;\mathcal{C}_1=-r\,e^{S+U}\partial_\vartheta(K-U-2S)+\dots,\quad\mathcal{C}_2=r^2N\,e^{S+U}\partial_r(K-S)+\dots,
\end{equation}
where the dots represent the matter part. These two equations vanish when the appropriate boundary conditions are imposed at the symmetry axis. Therefore, by solving the second-order PDEs \eqref{eqU_ews}-\eqref{eqS} with the boundary conditions to be detailed in Sec.~\ref{bc_ews} below, the two constraints \eqref{const_ein} will be automatically satisfied. Moreover, for $r=r_H$ when $N=0$, the second constraint in Eq.~\eqref{eqK_bis} reduces to,
\begin{equation}
    \partial_\vartheta K\rvert_{r=r_H}=\mathcal{K}_2\rvert_{r=r_H}=\partial_\vartheta U\rvert_{r=r_H}\quad\Rightarrow\quad\partial_\vartheta(U-K)\rvert_{r=r_H}=0,
\end{equation}
hence the difference $U-K$ and the surface gravity $\kappa_g$ in Eq.~\eqref{surf_grav_area} are constant at the horizon.

The field equations admit pure gauge solutions which are associated with the residual gauge invariance \eqref{res_gauge}. Such solutions should be removed by fixing the gauge since, otherwise, the differential operator in the equations is not invertible. We choose a gauge condition commonly employed in the literature \cite{Kleihaus1997,Kleihaus1998,Kleihaus1998a} which consists in setting to zero the covariant divergence of the two-vector $F_1\,dr+F_2\,d\vartheta$ in Eq.~\eqref{ansatz_ews}. This requires that
\begin{equation}
\label{gauge_cond}
    r\sqrt{N}\,\partial_r H_1=\partial_\vartheta H_2\quad\Rightarrow\quad\partial_s H_1=\partial_\vartheta H_2\quad\text{where}\quad s=\int_{r_H}^r{\frac{dr}{r\sqrt{N}}}.
\end{equation}
This gauge condition turns out to yield a good numerical convergence. Its disadvantage, as will be shown in App.~\ref{asymp_mon}, is that it gives rise to a spurious long-range mode in the asymptotic behavior of the solutions at large $r$. This spurious mode can be avoided by using a different gauge condition such as for example the unitary gauge $\phi_1=0$. However the differential operator in the equations is then more complicated and it is therefore preferable to use instead the gauge condition \eqref{gauge_cond}. 

The condition \eqref{gauge_cond} still allows for residual gauge transformations \eqref{res_gauge} with the gauge parameter $\chi$ subject to
\begin{equation}
\label{gauge_eq}
    \big(\partial_{ss}+\partial_{\vartheta\vartheta}\big)\chi=0.
\end{equation}
This equations possesses bounded solutions in the integration domain $s\in[0,\infty)\,$, $\vartheta\in[0,\pi]$ which generate gauge transformations $H_1\to H_1-\partial_s\chi$ and $H_2\to H_2+\partial_\vartheta\chi$. However, as seen in Eq.~\eqref{func_redef}, one must have $H_1=0$ at the horizon since otherwise the gauge amplitude $F_1$ diverges. Therefore, the only allowed solution of Eq.~\eqref{gauge_eq} is $\chi=const.$, but then a non-vanishing value of this constant would change the Higgs field at infinity whereas we assume that $\phi_1=0$ and $\phi_2=1$ (see Eq.~\eqref{bc_fields_ews} below).
 
As a result, the gauge is completely fixed by the condition \eqref{gauge_cond} together with the boundary conditions to be described below in Sec.~\ref{axial_sol_ews}. Using this gauge condition, the field equations become manifestly elliptic: the principal part of the differential operator is diagonal and contains the Laplacian,
\begin{equation}
    \Delta=\frac{1}{\sqrt{-g}}\partial_\mu\left(\sqrt{-g}\,g^{\mu\nu}\partial_\nu\right)=e^{-2K}\left(N\frac{\partial^2}{\partial r^2}+\frac{1}{r^2}\frac{\partial^2}{\partial\vartheta^2}\right)+\dots
\end{equation}

\subsection{Spherical symmetry}

The fields \eqref{metric_ews},\eqref{ansatz_ews} become spherically symmetric when
\begin{equation}
\label{spher_sym_ws_fields}
    H_1=H_3=y=\phi_1=0,\quad H_2=H_4=f(r),\quad\phi_2=\phi(r),
\end{equation}
and for the metric functions,
\begin{equation}
\label{spher_sym_met_fields}
    U=\ln{\sigma(r)},\quad\quad K=S=0.
\end{equation}
Hence the line element reduces to
\begin{equation}
\label{spher_met_ews}
    ds^2=-\sigma^2(r)N(r)dt^2+\frac{dr^2}{N(r)}+r^2\left(d\vartheta^2+\sin^2\vartheta\,d\varphi^2\right).
\end{equation}
Notice that here we consider $N(r)$ as an unknown function subject to the field equations. The precise correspondence between the line elements \eqref{metric_ews} and \eqref{spher_met_ews} is described in App.~\ref{radial_coords}. The WS fields in Eq.~\eqref{ansatz_ews} become
\begin{equation}
\label{spher_ws_fields}
    W=f(r)\left(T_2\,d\vartheta-\nu\,T_1\sin\vartheta\,d\varphi\right)+T_3\,\nu\cos\vartheta\,d\varphi,\quad Y=\nu\cos\vartheta\,d\varphi,\quad\Phi=\begin{pmatrix}
        0 \\ \phi(r)
    \end{pmatrix}.
\end{equation}

The Einstein-Weinberg-Salam equations \eqref{eqs_ews},\eqref{ein_eq_ews} then reduce to a system of ODEs plus two algebraic constraints,
\begin{align*}
    (N\sigma f')'&=\sigma\left(\frac{f^2-1}{r^2}+\frac{g^2}{2}\phi^2\right)f,\\
    (r^2N\sigma\phi')'&=\sigma\left(\frac{\beta r^2}{4}(\phi^2-1)+\frac{1}{2}f^2\right)\phi,\\
    m'&=\frac{\kappa}{2}\left(N\mathcal{U}_0+\mathcal{U}_1+\mathcal{U}_2\right)=-\frac{\kappa}{2}r^2\tensor{T}{^0_0},\\
    \frac{\sigma'}{\sigma}&=\frac{\kappa}{r}\,\mathcal{U}_0,\\
\label{spher_sym_eq}
    (\nu^2-1)f'&=(\nu^2-1)f\phi=0,\numberthis
\end{align*}
where we have introduced the mass function $m$ related to $N$ via $N(r)=1-2m(r)/r$ and
\begin{equation}
\label{Uk_ews}
    \mathcal{U}_0=\frac{1}{g^2}f'^2+r^2\phi'^2,\quad\mathcal{U}_1=\nu^2\frac{(f^2-1)^2}{2g^2r^2}+\frac{1}{2}f^2\phi^2+\frac{\beta r^2}{8}(\phi^2-1)^2,\quad\mathcal{U}_2=\frac{\nu^2}{2g'^2r^2}.
\end{equation}
These equations admit simple analytical solutions that will be presented in the next section. We also emphasize that a generalized ansatz which allows for time dependence is presented in App.~\ref{oscillons}. This ansatz can be used to describe oscillating solutions (\textit{oscillons}).

\section{Analytical solutions}
\label{anal_sol_ews}

In the spherically symmetric case, analytic solutions of Eqs.~\eqref{spher_sym_eq} exist. They describe \textit{Abelian} configurations for which the commutators in the SU(2) field strength vanish\footnote{This amounts to saying that the term $\epsilon_{abc}W^b_\mu W^c_\nu$ in Eq.~\eqref{field_strenghts_ews} vanishes identically for Abelian solutions.}. We shall present these solutions and review their main properties.

\subsection{Reissner-Nordstr{\"o}m solution}

The simplest solution of \eqref{spher_sym_eq} exists for any value of $\nu$,
\begin{equation}
\label{RN_ews}
    f=0,\quad\phi=\sigma=1,\quad N=1-\frac{2M}{r}+\frac{Q^2}{r^2}\equiv N_\text{RN},\quad\text{with}\quad Q^2=\frac{\kappa\nu^2}{2e^2},
\end{equation}
where $M$ is an integration constant. This describes a magnetically charged RN black hole of mass $M$ where the source of the magnetic charge is the Abelian Dirac monopole embedded into the electroweak theory. To avoid naked singularities, one has necessarily $M\geq|Q|$ and the spacetime metric contains two horizons located at $r=r_\pm$ where
\begin{equation}
\label{rpm_ews}
    r_\pm=M\pm\sqrt{M^2-Q^2}.
\end{equation}
The event horizon, located at $r=r_H$, corresponds to the outer horizon of the RN geometry,
\begin{equation}
\label{rh_rn}
    r_H=r_{+}\geq|Q|\equiv r_H^\text{ex},
\end{equation}
where the lower bound $r_H=r_H^\text{ex}$ is the extremal limit in which case one has $N=(1-r_H^\text{ex}/r)^2$.

The WS fields for this configuration are
\begin{equation}
\label{config_dirac}
    Y=\nu\cos\vartheta\,d\varphi,\quad W=T_3\,Y,\quad\Phi=\begin{pmatrix}
        0 \\ 1
    \end{pmatrix}.
\end{equation}
This corresponds to the fields describing a Dirac monopole with the singular string located along the $z$ axis, \textit{i.e.} at $\vartheta=0,\pi$. After the gauge transformation generated by $U_\pm=\exp\left\{\pm i\nu\varphi(1+\tau_3)/2\right\}$ the fields become
\begin{equation}
\label{dirac_in_ews}
    Y_\pm=\nu(\cos\vartheta\pm 1)d\varphi,\quad W_\pm=T_3\,Y_\pm,\quad\Phi=\begin{pmatrix}
        0 \\ 1
    \end{pmatrix}.
\end{equation}
Using $Y_{-}$, $W_{-}$ in the upper hemisphere and $Y_{+}$, $W_{+}$ in the lower hemisphere gives a completely regular description of the field configuration. In particular these two local gauges are related to each other by the transformation
\begin{equation}
    U=\exp(i\nu\varphi(1+\tau_3))=\begin{pmatrix}
        e^{2i\nu\varphi} & 0 \\
        0 & 1
    \end{pmatrix},
\end{equation}
which is well-defined in the equatorial transition region if $\nu$ is an integer or a \textit{half-integer}. Notice that this way of removing the Dirac string singularity works only for \textit{Abelian} configurations. In this very special case, half-integer values of $\nu$ are also allowed. 

Since the Higgs field in Eq.~\eqref{dirac_in_ews} evaluates to its vacuum expectation value, Maxwell electromagnetism should be recovered. In particular, the electromagnetic field $F_{\mu\nu}$ in Eq.~\eqref{elec_Z} admits a potential, $F=dA$ with
\begin{equation}
\label{pot_dirac_ews}
    A=A_\mu dx^\mu=\left(\frac{g}{g'}Y_\mu+\frac{g'}{g}W^3_\mu\right)dx^\mu=\frac{1}{gg'}Y=\frac{\nu}{e}(\cos\vartheta\pm 1)d\varphi,
\end{equation}
from where one can read-off the magnetic charge (see the Sec.~\ref{monop}),
\begin{equation}
    P=-\frac{\nu}{e}.
\end{equation}
Here $e=gg'$ is the dimensionless electron charge defined in Eq.~\eqref{struct_fine}. Since $\nu$ can be an integer or a half-integer, one recover the standard Dirac charge quantization,
\begin{equation}
\label{dirac_quant_ews}
    eP=\frac{n}{2}\quad\text{with}\quad n\in\mathbb{Z}.
\end{equation}

According to Eq.~\eqref{split_P}, the magnetic charge can be split into two parts which correspond to the contributions from $Y_\mu$ and $W^3_\mu$ in \eqref{pot_dirac_ews},
\begin{equation}
\label{mag_charge_dirac}
    P=P_\text{U(1)}+P_\text{SU(2)}\quad\text{with}\quad P_\text{U(1)}=-\frac{g}{g'}\nu,\;\; P_\text{SU(2)}=-\frac{g'}{g}\nu.
\end{equation}
The corresponding charge densities are described by Dirac delta distributions and thus both the U(1) and SU(2) parts of the magnetic charge are pointlike.

In the flat space limit, the event horizon disappears, leaving only the Dirac monopole located at the origin. According to the Eq.~\eqref{flat_space_ener}, the mass of the flat space monopole is
\begin{equation}
\label{ener_dirac}
    \mathcal{M}=\frac{2\pi\nu^2}{g'^2}\int_{0}^\infty{\frac{dr}{r^2}}+\frac{2\pi\nu^2}{g^2}\int_{0}^\infty{\frac{dr}{r^2}}=\left.\frac{2\pi\nu^2}{g'^2 r}+\frac{2\pi\nu^2}{g^2 r}\right|_{r\to 0},
\end{equation}
which is manifestly infinite. 

Summarizing, the magnetically charged RN black hole may be viewed as the gravitating counterpart of the flat space Dirac monopole. The latter has an infinite energy due to its Coulombian singularity at the origin but, in the gravitating case, the presence of an event horizon at $r=r_H>0$ provides a natural cut-off which regularizes the monopole energy.

\subsection{Reissner-Nordstr{\"o}m-de Sitter solution}
\label{sec_rnds}

Another simple solution of Eqs.~\eqref{spher_sym_eq} is described by
\begin{equation}
\label{RNdS}
    f=\sigma=1,\quad\phi=0,\quad N=1-\frac{2M}{r}+\frac{\mathcal{Q}^2}{r^2}-\frac{\Lambda}{3}r^2,\;\;\text{with}\;\;\mathcal{Q}^2=\frac{\kappa\nu^2}{2g'^2},\;\;\Lambda=\frac{\kappa\beta}{8}.
\end{equation}
The Higgs field is in the so-called false vacuum and its potential acts as an effective cosmological constant. This solution corresponds to a magnetic RN black hole but with an asymptotically de Sitter geometry. The SU(2) gauge field expressed in the gauge \eqref{reg_gauge_mon} vanishes, as does the Higgs field, and there remains only the U(1) field,
\begin{equation}
    Y_\pm=\nu(\cos\vartheta\pm 1)d\varphi,\quad\quad W=\Phi=0.
\end{equation}
As in the RN case, the U(1) potential can be expressed in two locally regular gauges that can be related to each other in the equatorial transition region if $2\nu\in\mathbb{Z}$ and which describe together the radial hypercharge field. Because $W=\Phi=0$, only the U(1) sector contributes to the magnetic charge, hence $P=P_\text{U(1)}=-(g/g')\nu$ and the full electroweak gauge symmetry is preserved everywhere in space. Notice that $P$ in the RN-de Sitter case does not fulfill the standard Dirac quantization condition \eqref{dirac_quant_ews} because the system is not in the Higgs vacuum and classical electromagnetism does not apply.

The function $N(r)$ in Eq.~\eqref{RNdS} has three positive roots corresponding to the inner black hole horizon, the outer black hole horizon $r_H$ and the cosmological horizon $r_C$. The value of $r_H$ is bounded from below,
\begin{equation}
\label{rhex_rnds}
    r_H\geq r_H^\text{ex}=\sqrt{\frac{1-\sqrt{1-4\Lambda\mathcal{Q}^2}}{2\Lambda}}=|\mathcal{Q}|\left(1+\frac{\Lambda^2}{2}+\dots\right)=\sqrt{\frac{\kappa}{2}}\frac{|\nu|}{g'}|+\mathcal{O}(|\nu|^3\kappa^{5/2}).
\end{equation}
The lower bound $r_H=r_H^\text{ex}$ corresponds to the extremal limit when the two black hole horizons merge together forming a degenerate horizon so that $N(r)$ factorizes as,
\begin{equation}
\label{extremal_RNdS}
    N(r)=\left(1-\frac{r_H^\text{ex}}{r}\right)^2\left[1-\frac{\Lambda}{3}\left(r^2+2r_H^\text{ex}r+3(r_H^\text{ex})^2\right)\right].
\end{equation}
The positive root of the second factor determines the location of the cosmological horizon,
\begin{equation}
    r_C=\sqrt{\frac{3}{\Lambda}-2(r_H^\text{ex})^2}-r_H^\text{ex}=\sqrt{\frac{24}{\beta\kappa}}+\mathcal{O}(\nu\sqrt{\kappa}).
\end{equation}
The cosmological horizon radius is much larger than the black hole horizon radius $r_H^\text{ex}$ if $|\mathcal{Q}|$ is small but $r_C$ and $r_H^\text{ex}$ approach each other when $|\mathcal{Q}|$ increases. They merge together when $\mathcal{Q}=1/\sqrt{4\Lambda}$ and one has then
\begin{equation}
\label{nu_max}
    |\nu|=\nu_\text{max}\equiv\frac{2g'}{\kappa\sqrt{\beta}}=1.28\times 10^{32},\quad r_C=r_H^\text{ex}=\frac{2}{\sqrt{\beta\kappa}},
\end{equation}
which corresponds to the dimensionful value $\boldsymbol{r}_H^\text{ex}=\boldsymbol{r}_C=3\,\text{cm}$. In this limiting case, $r=r_H^\text{ex}$ corresponds to a triple zero of $N(r)$. Further increasing the magnetic charge would lead to a naked singularity.

The RN-de Sitter solution itself is physically less interesting because it is not asymptotically flat, however, as we shall see below, it provides an approximate description of the near-horizon region of hairy solutions for $|\nu|\gg 1$. For small magnetic charges, \textit{i.e.} when $|\nu|\ll\nu_\text{max}$, the RN-de Sitter solution also describes appropriately the near-horizon region of extremal hairy black holes.

Yet another RN-de Sitter solution can be obtained by setting in Eq.~\eqref{RNdS} $f=0$ and replacing $\mathcal{Q}\rightarrow Q=\mathcal{Q}/g$. However, this solution is not related to the hairy black holes considered below.

\section{Stability of the Reissner-Nordstr{\"o}m black hole}
\label{stab_rn}

Let us consider small perturbations $\delta g_{\mu\nu}$, $\delta W^a_\mu$, $\delta Y_\mu$, $\delta\Phi$ around the RN solution,
\begin{equation}
    g_{\mu\nu}\rightarrow g_{\mu\nu}+\delta g_{\mu\nu},\quad W^a_\mu\rightarrow W^a_\mu+\delta W^a_\mu,\quad Y_\mu\rightarrow Y_\mu+\delta Y_\mu,\quad\Phi\rightarrow\Phi+\delta\Phi,
\end{equation}
where $\{g_{\mu\nu},W^a_\mu,Y_\mu,\Phi\}$ is the background configuration. The metric is thus given by Eq.~\eqref{spher_met_ews} with $\sigma=1$ and $N=N_\text{RN}$ and the WS fields are those in Eq.~\eqref{config_dirac}. Linearizing the field equations with respect to the perturbations and assuming the unitary gauge, $\delta\Phi=(0,\delta\phi)^\text{T}$, one finds that the equations for $\delta W^1_\mu$ and $\delta W^2_\mu$ decouple from the rest and reduce to the complex Proca equation,
\begin{equation}
\label{proca_ews}
    \mathcal{D}^\mu w_{\mu\nu}+ie\,F_{\nu\sigma}w^\sigma=\frac{g^2}{2}w_\nu.
\end{equation}
Here $w_\mu=\delta W^1_\mu+i\,\delta W^2_\mu$ is the complex W-boson field, $w_{\mu\nu}=\mathcal{D}_\mu w_\nu-\mathcal{D}_\nu w_\mu$ is the corresponding field strength and $\mathcal{D}_\mu=\nabla_\mu+ie A_\mu$ where $A_\mu$ is the background electromagnetic field potential \eqref{pot_dirac_ews} with strength $F_{\mu\nu}=\partial_\mu A_\nu-\partial_\nu A_\mu$. 

We consider a generic perturbation one-form, $w_\mu dx^\mu\equiv\text{w}(t,r,\vartheta,\varphi)$, with no rotational symmetry. A rather involved analysis that we do not show explicitly reveals that, after separating the angular variables using spin-weighted spherical harmonics, the 4 equations contained in \eqref{proca_ews} reduce to a single radial equation in the sector with orbital angular momentum $j=|\nu|-1$. The resulting master perturbation equation reads,
\begin{equation}
\label{pert_eq_ews}
    \left[-\frac{d^2}{dr^2_*}+N(r)\left(\frac{g^2}{2}-\frac{|\nu|}{r^2}\right)\right]\psi(r)=\omega^2\psi(r),
\end{equation}
where $\psi$ is the perturbation amplitude, $\omega$ is the frequency entering the harmonic time dependence of the perturbations, and $r_*$ is the tortoise coordinate defined by $dr_*=dr/N$. If this equation admits bounded solutions with $\omega^2<0$ (\textit{a.k.a.} bound states) then perturbations grow in time and the background is unstable.

In the flat space limit where $N(r)=1$, the RN solution reduces to the Dirac monopole in Minkowski space and the equation \eqref{pert_eq_ews} admits infinitely many bound states with $\omega^2<0$. Indeed, if we set $\omega=0$ then the solution close to the origin reads,
\begin{equation}
    \psi(r)=\sqrt{r}\cos\left(\frac{\sqrt{4|\nu|-1}}{2}\ln r\right),
\end{equation}
which oscillates infinitely many times as $r\to 0$. According to the Jacobi criterion, the number of nodes of the solution with $\omega=0$ corresponds to the number of bound states \cite{Amann1995}. It follows that all Dirac monopoles are unstable within the electroweak theory, except those with $\nu=\pm 1/2$ since in this case the value of $j=|\nu|-1$ would be negative. In the Ref.~\cite{Gervalle2022a}, we have shown that the instabilities reside only in the W-sector considered here.

In the gravitating case, the range of the radial coordinate in Eq.~\eqref{pert_eq_ews} has to be restricted to outer black hole region, $r>r_H$, hence the attractive term $-|\nu|/r^2$ cannot be arbitrarily large. Therefore it is expected that the perturbation equation can admit at most only a finite number of bound states. When the horizon radius $r_H$ is large enough, one has
\begin{equation}
    \frac{g^2}{2}-\frac{|\nu|}{r^2}>0\quad\text{for}\quad r>r_H,
\end{equation}
which excludes the existence of bound states. It follows that instabilities can only occur for small black holes.

By varying the horizon radius, one can detect the threshold value of $r_H$ at which a bound state just starts to appear. At this threshold value, the solution is called a \textit{zero mode}: it is a normalizable solution of the perturbation equation with $\omega=0$. The latter property implies that the zero modes are \textit{static} solutions.

To check this, we consider a RN background with $|\nu|=1$. Notice that in this case $j=|\nu|-1=0$ hence the perturbations described are spherically symmetric. According to the Eq.~\eqref{rh_rn}, the horizon radius is bounded from below as
\begin{equation}
    r_H\geq r_H^\text{ex}=\sqrt{\frac{\kappa}{2}}\frac{1}{gg'}=1.23\times 10^{-16}. 
\end{equation}
Starting from a large $r_H$ and decreasing its value, we find that the first zero mode appears for $r_H=r_H^0\approx 0.89$, the corresponding solution $\psi_0(r)$ being \textit{nodeless}. As $r_H$ decreases further, this solution becomes a bound state with $\omega^2<0$, and for $r_H=r_H^1\approx 0.04$ the equation admits a second zero mode $\psi_1(r)$ which has \textit{one node}. Then, when $r_H$ decreases, this solution becomes a bound state with $\omega^2<0$, in addition to the nodeless $\psi_0(r)$, and so on. We find that as $r_H$ descends all the way down to the minimal value $r_H^\text{ex}$, the perturbation equation \eqref{pert_eq_ews} develops in total up to 13 bound states. The values of $r_H$ for which zero modes appear, $r_H^n$, are shown in Table \ref{rhn}.

\begin{table}
    \centering
    \begin{tabular}{|c|c|}
       \hline
       $n$  & $r_H^n$  \\
       \hline
       $0$  &  $0.8983$ \\
       $1$  &  $4.3255\times 10^{-2}$ \\
       $2$  &  $1.1777\times 10^{-3}$ \\
       $3$  &  $3.1323\times 10^{-5}$ \\
       $4$  &  $8.3259\times 10^{-7}$ \\
       \hline
    \end{tabular}
    \begin{tabular}{|c|c|}
       \hline
       $n$  & $r_H^n$  \\
       \hline
       $5$  &  $2.2130\times 10^{-8}$ \\
       $6$  &  $5.8822\times 10^{-10}$ \\
       $7$  &  $1.5634\times 10^{-11}$ \\
       $8$  &  $4.1557\times 10^{-13}$ \\
       $9$  &  $1.1047\times 10^{-14}$ \\
       \hline
    \end{tabular}
    \begin{tabular}{|c|c|}
       \hline
       $n$  & $r_H^n$  \\
       \hline
       $10$  &  $3.3876\times 10^{-16}$ \\
       $11$  &  $1.2766\times 10^{-16}$ \\
       $12$  &  $1.2380\times 10^{-16}$ \\
        & \\
        & \\
       \hline
    \end{tabular}
    \caption{Values of $r_H$ for which the perturbation equation \eqref{pert_eq_ews} admits zero mode with $n$ nodes for $|\nu|=1$.}
    \label{rhn}
\end{table}

\begin{figure}
    \centering
    \includegraphics[width=7.8cm]{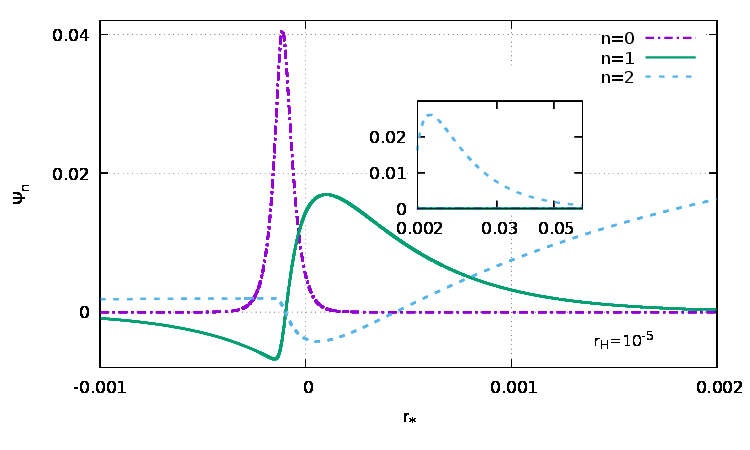}
    \includegraphics[width=7.8cm]{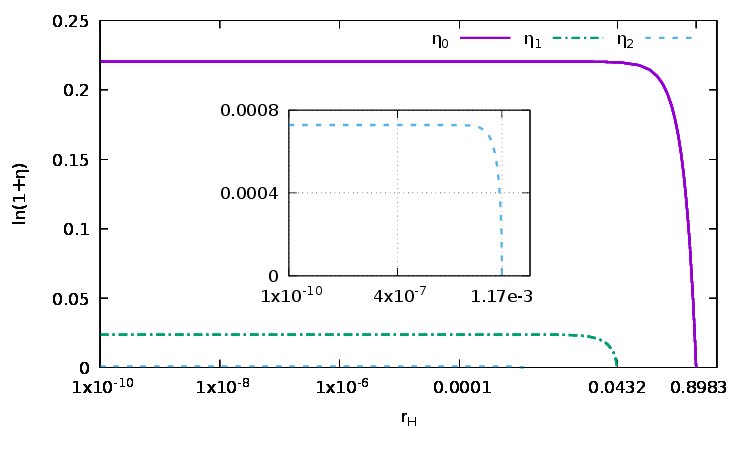}
    \caption[The first three bound state solutions $\psi_n$ of the perturbation equation \eqref{pert_eq_ews} against the tortoise coordinate $r_*$ for $r_H=10^{-5}$ and the quantity $\eta=|\omega|r_H$ against $r_H$ for $n=0,1,2$.]{The first three bound state solutions $\psi_n$ of the perturbation equation \eqref{pert_eq_ews} against the tortoise coordinate $r_*$ for $r_H=10^{-5}$ (left) and the quantity $\eta=|\omega|r_H$ against $r_H$ for $n=0,1,2$ (right).}
    \label{psi_ews}
\end{figure}

Each zero mode becomes a bound state with $\omega^2<0$ when $r_H$ descends below $r_H^n$. For example a black hole with $r_H=3\times 10^{-16}$ admits 11 bound states, a larger one with $r_H=1\times 10^{-15}$ admits only 10 bound states, etc. For $r_H=1\times 10^{-5}$, say, there are four bound states and the corresponding profiles $\psi_n(r)$ of the first three ($n=0,1,2$) are shown in the left panel of Fig.~\ref{psi_ews}. The eigenvalues $\omega^2(r_H)\leq 0$ start from zero for $r_H=r_H^n$ and then decrease as $r_H$ gets smaller, but the product $\eta\equiv|\omega|r_H$ rapidly approaches a constant value, as seen in the right panel of Fig.~\ref{psi_ews}.

As a result, small magnetically charged RN black holes are unstable in the Einstein-Weinberg-Salam theory. At the same time, each new instability settling in at $r_H=r_H^n$ appears first as a static zero mode. The latter can be viewed as a perturbative approximation of a new solution describing a hairy black hole. This new solution bifurcates with the RN black hole for $r_H=r_H^n$ but deviates from it when $r_H<r_H^n$. It is expected to be less energetic than RN and may be stable. Since we are currently discussing the perturbation sector with $j=0$, the new solutions should be spherically symmetric.

\begin{table}
    \centering
    \begin{tabular}{|c|c|c|c|c|c|c|c|c|}
       \hline
       $\nu$  & 1 & 2 & 3 & 5 & 10 & 20 & 50 & 100  \\
       \hline
       $r_H^0$  &  $0.8983$ & $1.4724$ & $1.9362$ & $2.6948$ & $4.1278$ & $6.1928$ & $10.3378$ & $15.0352$  \\
       \hline
       $r_H^1$  &  $0.0432$ & $0.2466$ & $0.5110$ & $1.0597$ & $2.2847$ & $4.2108$ & $8.2401$ & $12.8822$  \\
       \hline
       $r_H^2$  &  $0.0012$ & $0.0251$ & $0.0885$ & $0.3076$ & $1.0714$ & $2.6391$ & $6.3711$ & $10.8785$  \\
       \hline
    \end{tabular}
    \caption{Values $r_H^n(\nu)$ for which the perturbation equation \eqref{pert_eq_ews} admits an $n$-nodes zero mode.}
    \label{rhn_nu}
\end{table}

\begin{figure}
    \centering
    \includegraphics[width=10cm]{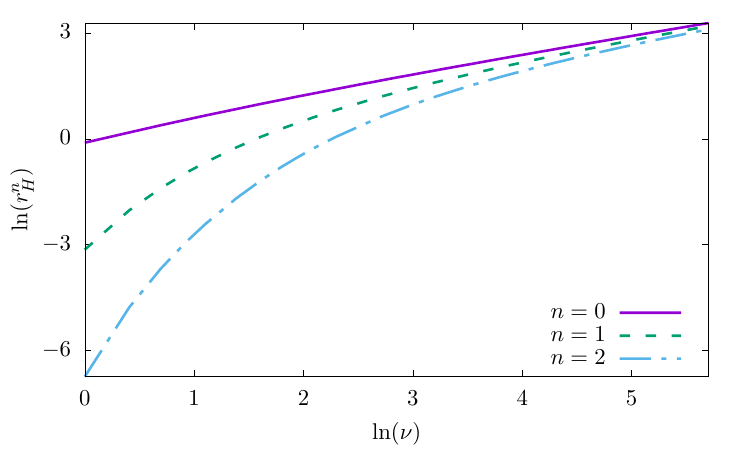}
    \caption[The logarithm of the horizon radius $r_H^n$ of the Reissner-Nordström black hole admitting  an $n$-nodes zero mode against $\ln(\nu)$.]{The logarithm of the horizon radius $r_H^n$ of the RN black hole admitting an $n$-nodes zero mode against $\ln(\nu)$.}
    \label{rh0nu}
\end{figure}

For higher magnetic charges, $|\nu|>1$, the term $-|\nu|/r^2$ in Eq.~\eqref{pert_eq_ews} becomes more attractive and the number of bound states should increase. The radii $r_H^n$ for which an $n$-nodes zero mode appears grow with $\nu$, as seen in the Table~\ref{rhn_nu}. The Fig.~\ref{rh0nu} shows the functions $r_H^n(\nu)$ for $n=0,1,2$, assuming $\nu$ to be a continuous variable, and it seems that $r_H^n\propto\sqrt{|\nu|}$ for large $|\nu|$. It should be emphasized that these zero modes are in the sector with $j=|\nu|-1>0$ hence they are not spherically symmetric. They can be used to approximate new hairy black hole solutions which are also not spherically symmetric.

Specifically, any zero mode solution $\psi(r)$ of the radial equation \eqref{pert_eq_ews} determines the perturbation one-form $w_\mu dx^\mu=\text{w}(r,\vartheta,\varphi)$ whose angular dependence is given by (spin-weighted) spherical harmonics. This one-form depends on the azimuthal quantum number $m$ that can assume $2j+1$ different values, $m=-j,\dots,j$, but the perturbation equation itself does not depends on $m$. Therefore, one zero mode solution $\psi(r)$ corresponds to $2j+1=2|\nu|-1$ different solutions of the Proca equation \eqref{proca_ews}. If $m=0$, then the solution $\textrm{w}(r,\vartheta)$ approximates an axially symmetric hair. We will construct the corresponding hairy black holes at the fully nonlinear level in Sec.~\ref{axial_sol_ews} below. If $m\neq 0$, then the solution $\textrm{w}(r,\vartheta,\varphi)$ approximates a black hole hair with only discrete symmetries. Construction of these solutions at the non-perturbative level remains an open issue.

To recapitulate, the equation \eqref{pert_eq_ews} describing perturbations around a RN black hole admits zero mode solutions for discrete values $r_H^n$ of the horizon radius. These zero modes can be viewed as perturbative approximations of new black hole solutions that exist for $r_H<r_H^n$. The new solutions can be spherically symmetric only for $|\nu|=1$.

\section{Spherically symmetric hairy black holes}
\label{spher_hairy_bh}

The instabilities of the magnetic RN black hole that have been described above indicate the existence of new non-Abelian hairy solutions. In the simplest case, such solutions are spherically symmetric, which is possible only if $\nu=\pm 1$, hence for the magnetic charge number $n=\pm 2$ in the Dirac charge quantization formula \eqref{dirac_quant_ews}. These solutions have already been reported in Ref.~\cite{Bai2021} and they are gravitating counterparts of the flat space monopole of Cho and Maison \cite{Cho1996}. We shall only summarize their essential properties, before discussing in the next sections more general solutions with higher magnetic charges.

\subsection{General properties}

First, one can rewrite the system of ODEs in \eqref{spher_sym_eq} in such a way that the equation for $\sigma$ decouples from the others. The problem then reduces to a system of three coupled ODEs, 
\begin{align*}
    (Nf')'+\frac{\kappa}{r}\,\mathcal{U}_0Nf'&=\left(\frac{f^2-1}{r^2}+\frac{g^2}{2}\phi^2\right)f,\\
    (r^2N\phi')'+\kappa\,r\,\mathcal{U}_0N\phi'&=\left(\frac{\beta r^2}{4}(\phi^2-1)+\frac{1}{2}f^2\right)\phi,\\
\label{spher_sym_red}
    1-(rN)'&=\kappa\left(N\mathcal{U}_0+\mathcal{U}_1+\mathcal{U}_2\right),\numberthis
\end{align*}
where $\mathcal{U}_0$, $\mathcal{U}_1$, $\mathcal{U}_2$ are defined in Eq.~\eqref{Uk_ews}. When these equations are solved, the $\sigma$-equation in \eqref{spher_sym_eq} can be integrated. We shall focus here only on the system of three ODEs \eqref{spher_sym_red} for $N$, $f$ and $\phi$. For a black hole solution one has $N(r_H)=0$, $N'(r_H)\equiv N'_H>0$ (unless for a degenerate horizon for which $N'_H=0$). Taking in Eq.~\eqref{spher_sym_red} the limit $r\to r_H$, one obtains the value $N'_H$ while the derivatives $f''$ and $\phi''$ must be finite at the horizon. Using the latter condition, one can obtain an expression for $f'_H\equiv f'(r_H)$, $\phi'_H\equiv\phi(r_H)$ in terms of $f_H\equiv f(r_H)$, $\phi_H\equiv\phi(r_H)$. One finds that
\begin{align*}
    N'_H&=\frac{1}{r_H}\left(1-\kappa\,\mathcal{U}_{1,H}-\kappa\,\mathcal{U}_{2,H}\right),\\
    f'_H&=\frac{1}{N'_H}\left(\frac{f_H^2-1}{r_H^2}+\frac{g^2}{2}\phi_H^2\right)f_H,\\
\label{bch_ews_spher}
    \phi'_H&=\frac{1}{r_H^2 N'_H}\left(\frac{\beta r_H^2}{4}(\phi_H^2-1)+\frac{1}{2}f_H^2\right)\phi_H,\numberthis
\end{align*}
where $\mathcal{U}_{k,H}$ are the horizon values of $\mathcal{U}_k$. This determines the boundary conditions at the horizon in terms of two free parameters: $f_H$ and $\phi_H$.

At large $r$ the metric should approach Minkowski, $N=1$, and the WS fields should approach the Dirac monopole configuration \eqref{config_dirac} which corresponds to $f=0$ and $\phi=1$. Hence one can set 
\begin{equation}
    N=1+\delta N,\quad f=\delta f,\quad\phi=1+\delta\phi,
\end{equation}
where $\delta N$, $\delta f$ and $\delta\phi$ are small deviations. Injecting this to Eq.~\eqref{spher_sym_red} yields
\begin{equation}
\label{lin_eq_ews}
    (r\delta N)'=\mathcal{N}_N,\quad \delta f''+\left(\frac{1}{r^2}-\frac{g^2}{2}\right)\delta f=\mathcal{N}_f,\quad\delta\phi''+\frac{2}{r}\delta\phi'-\frac{\beta}{2}\delta\phi=\mathcal{N}_\phi,
\end{equation}
where $\mathcal{N}_N$, $\mathcal{N}_f$, $\mathcal{N}_\phi$ are the nonlinear in $\delta N$, $\delta f$, $\delta\phi$ terms. The solution of these equations read
\begin{align*}
    \delta N&=-\frac{2M}{r}+\delta N_\mathcal{N},\\
    \delta f&=C_f\exp(-m_\text{W}r)+\delta f_\mathcal{N},\\
\label{lin_sol_ews}
    \delta\phi&=\frac{C_\phi}{r}\exp(-m_\text{H}r)+\delta\phi_\mathcal{N},\numberthis
\end{align*}
where $M$, $C_f$, $C_\phi$ are integration constants while $\delta N_\mathcal{N}$, $\delta f_\mathcal{N}$, $\delta\phi_\mathcal{N}$ contains sub-leading terms and nonlinear corrections. Neglecting the latter, one can evaluate the Eqs.~\eqref{lin_sol_ews} at some large radius $r=r_\star$ to obtain boundary conditions depending on three free parameters, $M$, $C_f$ and $C_\phi$. Notice that it is also possible to take into account the nonlinear corrections at $r=r_\star$ by converting the Eqs.~\eqref{lin_eq_ews} into integral equations and applying the procedure described in Sec.~\ref{bc2_bigravity}.

\begin{figure}[b!]
    \centering
    \includegraphics[width=7.8cm]{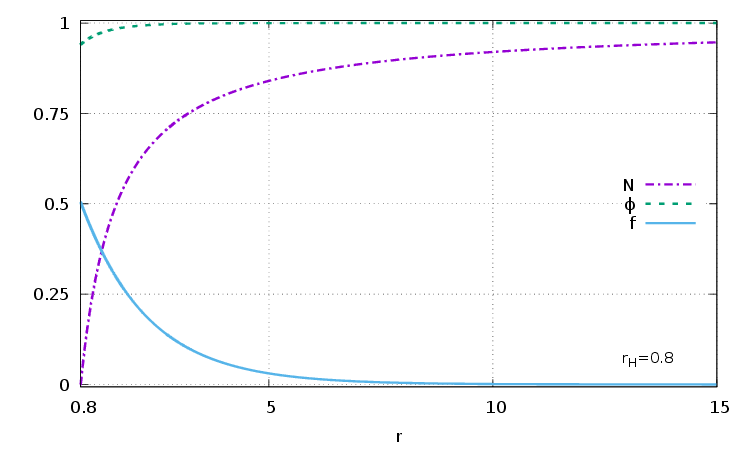}
    \includegraphics[width=7.8cm]{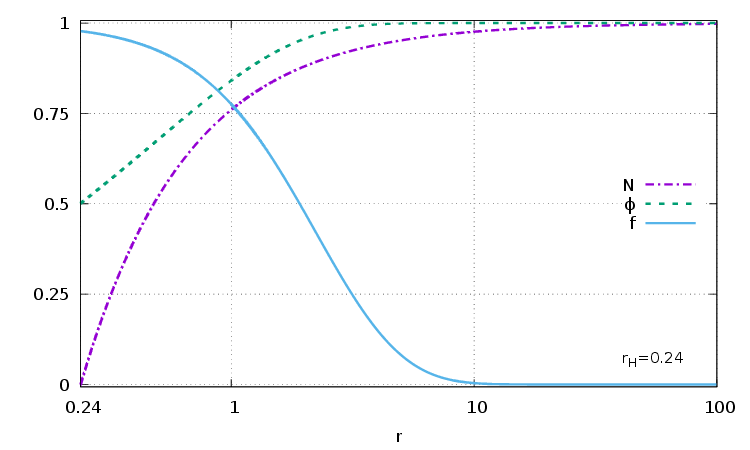}
    \caption[Profiles of the fundamental hairy magnetic black holes with $r_H=0.8$ and $r_H=0.24$.]{Profiles of the fundamental hairy magnetic black holes with $r_H=0.8$ (left) and $r_H=0.24$ (right).}
    \label{hairy_bh_ews}
\end{figure}

The values of the 5 parameters $f_H$, $\phi_H$, $M$, $C_f$, $C_\phi$ entering the boundary conditions can be adjusted by using the \textit{multishooting} method (see the Appendix \ref{num_bvp}). This yields a global solution in the region $r>r_H$. There always exists the trivial RN solution for which $f_H=f(r)=0$ and $\phi_H=\phi(r)=1$. However, choosing a value of $r_H$ slightly below $r_H^0$ in Table \ref{rhn}, for example $r_H=0.89$, the algorithm is able to find another solution for which $f_H$ is small but non-zero while $\phi_H$ is close to unity. This describes a "slightly hairy" black hole. Decreasing then iteratively $r_H$, the horizon value $f_H$ grows while $\phi_H$ decreases, yielding "fully fledged" hairy black holes. Examples of hairy solutions for $r_H=0.8$ and $r_H=0.24$ are shown in the Fig.~\ref{hairy_bh_ews}. We emphasize that the function $N$ is very close to $N_\text{RN}$ because the back-reaction of the WS fields on the spacetime geometry is negligible for $r_H\gg\sqrt{\kappa}\approx 10^{-16}$.

\begin{figure}
    \centering
    \includegraphics[width=7.8cm]{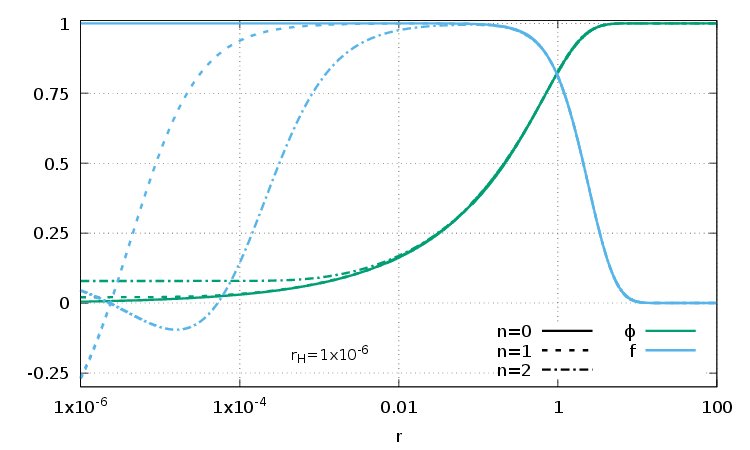}
    \includegraphics[width=7.8cm]{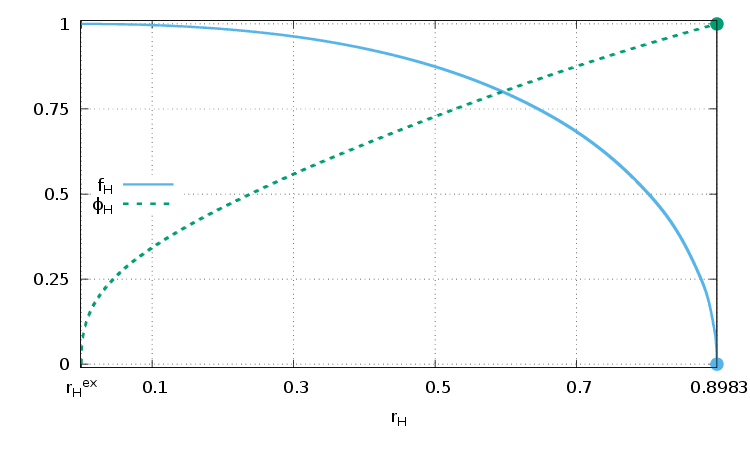}
    \caption[Profile of the fundamental hairy black hole with $r_H=10^{-6}$ compared to that of its first two radial excitations and the horizon values $\phi_H$, $f_H$ for the fundamental hairy solutions.]{Left: profiles of the fundamental hairy black hole with $r_H=10^{-6}$ and of its first two radial excitations. Right: the horizon values $\phi_H$, $f_H$ for the fundamental hairy solutions.}
    \label{hairy_bh_ews_2}
\end{figure}

For smaller values of $r_H$ we also find solutions with oscillating $f$-function. They correspond to the new hairy solutions that appear at the values $r_H^n$ given in Table \ref{rhn} for $n\geq 1$. For example, choosing $r_H=4\times 10^{-2}<r_H^1$, one finds two hairy solutions: the \textit{fundamental} one with a monotone $f$-profile and the \textit{excited} one for which $f(r)$ shows one node. For $r_H<r_H^2$ one finds the fundamental hairy solution and two excitations: the function $f$ exhibit one zero for the first excitation and two zeroes for the second excitation. More generally, the $n$-th excitation would correspond to a solution with $n$ nodes. These excited solutions were not reported in Ref.~\cite{Bai2021}; their profiles for $r_H=10^{-6}$ are compared to that of the fundamental solution in the left panel of Fig.~\ref{hairy_bh_ews_2}. However, the excited solutions are probably unstable and hence less interesting physically. We shall not discuss them anymore.

The horizon radius of the (fundamental) hairy black holes takes values in the interval $r_H\in[r_H^\text{ex},r_H^0]$. The upper bound $r_H^0=0.8983$ corresponds to the bifurcation with the RN solution while the lower bound coincides with the extremal radius \eqref{rhex_rnds} of the RN-de Sitter solution with $|\nu|=1$. One has $r_H^\text{ex}\sim\sqrt{\kappa}\sim 10^{-16}$ which is of the order of the Planck scale. In the right panel of Fig.~\ref{hairy_bh_ews_2}, we present the horizon values $\phi_H$, $f_H$ against $r_H$ for the fundamental hairy black holes. As the horizon radius decreases, the value of $f_H$ approaches unity while $\phi_H$ approaches zero. This is a manifestation of the electroweak symmetry restoration in a strong magnetic field \cite{Ambjoern1990} since for small $r_H$ the U(1) hypercharge field strength $Y_{\mu\nu}$ becomes very large at the horizon. On the other hand, the horizon value of the SU(2) field strength $W^a_{\mu\nu}$ decreases when $f_H$ grows and the SU(2) part of the magnetic charge moves \textit{outside} the horizon. This can be understood as follows.

In the spherically symmetric case, the SU(2) part of magnetic charge density in Eq.~\eqref{mag_charge_dens} is given by
\begin{equation}
\label{su2_charge_dens}
    \rho_\text{SU(2)}=\frac{\nu}{4\pi}\frac{g'}{gr^2\sigma}(f^2)'.
\end{equation}
Hence the SU(2) part of the magnetic charge is now distributed smoothly in space instead of being pointlike as for a RN black hole. The total SU(2) charge outside the horizon is then given by the integral (here $\nu=\pm 1$),
\begin{equation}
    P_\text{SU(2)}^\text{outside}=\int_{r>r_H}{\rho_\text{SU(2)}\sqrt{-g}\,d^3 x}=\nu\,\frac{g'}{g}\int_{r_H}^\infty{(f^2)'dr}=-\nu\,\frac{g'}{g}f_H^2.
\label{P_out_spher}
\end{equation}
This is less than the total SU(2) magnetic charge which can be measured at infinity as in Eq.~\eqref{charge_surf}, $P_\text{SU(2)}=-\nu g'/g$. However the above equation gives only the SU(2) charge distributed in the outer black hole region while the rest of the charge must be inside the event horizon. As a result, the horizon value $f_H^2\in[0,1]$ gives the fraction of the SU(2) magnetic charge distributed outside the horizon. This SU(2) charge distribution is the black hole "hair". In the extremal case, one as $f_H=1$ and thus all the SU(2) charge is in the outer region. As the horizon radius increases, the charge is absorbed by the black hole until $r_H=r_H^0$ in which case the hairy solutions bifurcate with RN. In this limit all the SU(2) charge is contained inside the horizon and the black hole loses its hair.

\subsection{The mass}

To study the mass of spherically symmetric field configurations, it is convenient to use the mass function $m$ related to $N$ via the relation $N(r)=1-2m(r)/r$. The mass function is determined by the differential equation $m'(r)=-(\kappa/2)\,r^2\,\tensor{T}{^0_0}$ and its asymptotic value coincide with the total mass $M$ of the solution as defined in Eq.~\eqref{vol_mass_ews}. We introduce a "rescaled" mass function $\mathcal{M}(r)=(8\pi/\kappa)m(r)$ which fulfills
\begin{equation}
\label{eq_M_scale}
    \mathcal{M}'(r)=-4\pi r^2\tensor{T}{^0_0}=4\pi\left(N\mathcal{U}_0+\mathcal{U}_1+\mathcal{U}_2\right).
\end{equation}
The asymptotic value $\mathcal{M}(\infty)$ coincide with the "rescaled" mass $\mathcal{M}$ defined in Eq.~\eqref{vol_mass_bis_ews}. Integrating the Eq.~\eqref{eq_M_scale} with the boundary condition $N(r_H)=0$ yields
\begin{equation}
    \mathcal{M}(r)=4\pi\left[\frac{r_H}{\kappa}+\frac{1}{2g'^2}\left(\frac{1}{r_H}-\frac{1}{r}\right)\right]+4\pi\int_{r_H}^r{(N\mathcal{U}_0+\mathcal{U}_1)dr}\equiv\mathcal{M}_H(r)+\mathcal{M}_h(r),
\end{equation}
where $M_H(r)$ contains the horizon term and the contribution of the U(1) field while $\mathcal{M}_h(r)$ contains the contribution of the SU(2) and Higgs field, which is the "hair". The total mass $\mathcal{M}$ thus splits into the horizon mass $\mathcal{M}_H$ and the hair mass $\mathcal{M}_h$,
\begin{equation}
    \mathcal{M}\equiv\mathcal{M}(\infty)=\mathcal{M}_H(\infty)+\mathcal{M}_h(\infty)\equiv\mathcal{M}_H+\mathcal{M}_h,
\end{equation}
where
\begin{align*}
    \mathcal{M}_H&=4\pi\left(\frac{r_H}{\kappa}+\frac{1}{2g'^2r_H}\right),\\
\label{M_h_H}
    \mathcal{M}_h&=4\pi\int_{r_H}^\infty{\left[N\left(\frac{1}{g^2}f'^2+r^2\phi'^2\right)+\frac{(f^2-1)^2}{2g^2r^2}+\frac{1}{2}f^2\phi^2+\frac{\beta r^2}{8}(\phi^2-1)^2\right]dr}.\numberthis
\end{align*}
The horizon mass $\mathcal{M}_H$ is the same for RN black holes and for hairy black holes while the hair mass $\mathcal{M}_h$ is not the same. Notice that the denomination "hair mass" is not adequate in the RN case since all the SU(2) charge is inside the horizon and thus RN black holes have no hair.

For the RN solution \eqref{RN_ews} one has $f=0$ and $\phi=1$ so that
\begin{equation}
    \mathcal{M}_h=\frac{4\pi}{2g^2r_H},
\end{equation}
which varies from zero up to a very large value of the order of $1/\sqrt{\kappa}$ when $r_H$ reaches its extremal value. It follows that the total mass of RN black holes is
\begin{equation}
    \mathcal{M}=4\pi\left(\frac{r_H}{\kappa}+\frac{1}{2e^2r_H}\right)\quad\Rightarrow\quad M=\frac{\kappa}{8\pi}\mathcal{M}=\frac{r_H}{2}+\frac{\kappa}{4e^2 r_H},
\end{equation}
the latter being equivalent to the condition $N_\text{RN}(r_H)=0$.

\begin{figure}
    \centering
    \includegraphics[width=10cm]{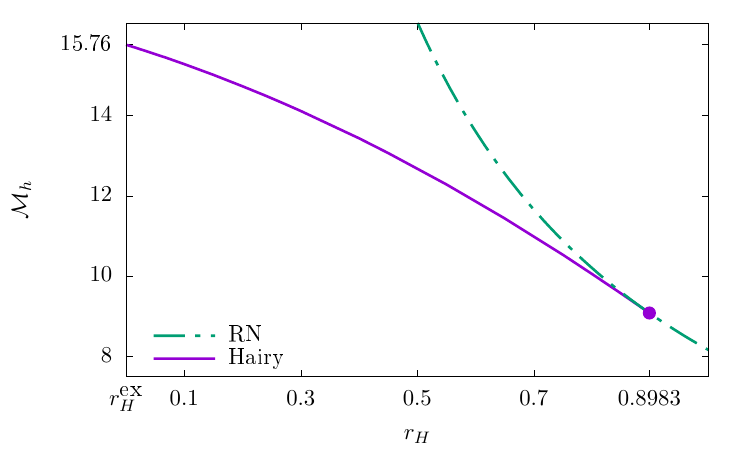}
    \caption{The hair mass $\mathcal{M}_h$ for the hairy black holes and for the RN black holes.}
    \label{fig_M_h_rh}
\end{figure}

For the fundamental hairy black holes, $\mathcal{M}_h$ can only be computed numerically and we show it against $r_H$ on the Fig.~\ref{fig_M_h_rh}. We find that the hair mass varies within the following limits,
\begin{equation}
\label{bound_Mh}
    E_\text{CM}=15.759>\mathcal{M}_h\geq\frac{4\pi}{2g^2r_H^0}=9.083.
\end{equation}
When $r_H$ decreases, $\mathcal{M}_h$ increases from the minimal value 9.083 corresponding to the bifurcation with the RN solution and approaches for small $r_H$ the maximal value $E_\text{CM}=15.759$. As a result, the hairy black holes are less massive than the RN black holes of same size.

\subsection{Hairy black holes versus Cho-Maison monopole}

\begin{figure}
    \centering
    \includegraphics[width=7.8cm]{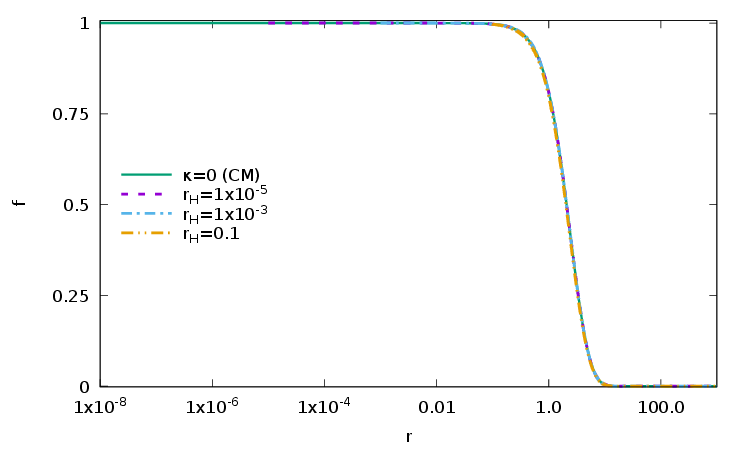}
    \includegraphics[width=7.8cm]{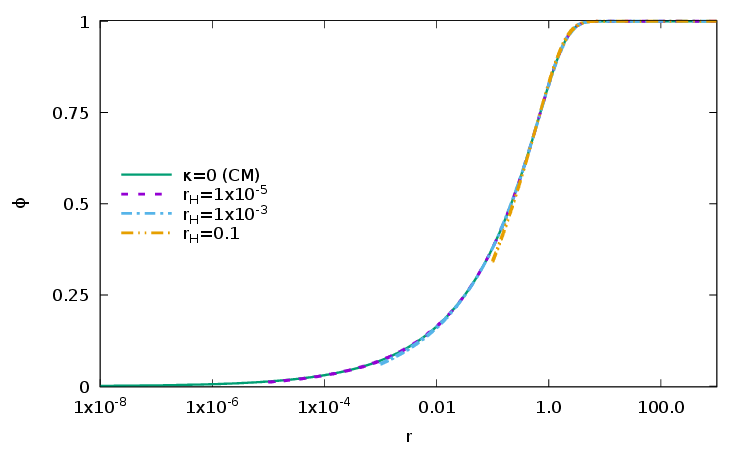}
    \caption[Profiles of $f(r)$ and $\phi(r)$ for the flat space Cho-Maison monopole and for hairy black holes with different values of $r_H$.]{Profiles of $f(r)$ (left) and $\phi(r)$ (right) for the flat space CM monopole ($\kappa=0$) and for hairy black holes with $r_H=\{10^{-1},10^{-3},10^{-5}\}$. For black holes, the profiles are shown for $r>r_H$.}
    \label{cm_bh}
\end{figure}

The value of $E_\text{CM}$ in Eq.~\eqref{bound_Mh} corresponds to the regularized energy of the Cho-Maison monopole \cite{Cho1996}. The latter is a solution in the flat space limit where $\kappa\to 0$ and $N=\sigma=1$, when the equations \eqref{spher_sym_red} reduce to
\begin{equation}
    f''=\left(\frac{f^2-1}{r^2}+\frac{g^2}{2}\phi^2\right)f,\quad\quad(r^2\phi')'=\left(\frac{\beta r^2}{4}(\phi^2-a)+\frac{1}{2}f^2\right)\phi.
\end{equation}
These equations admit a smooth solution which interpolates between the following values for $0\leftarrow r\rightarrow\infty$,
\begin{equation}
\label{asymp_cm}
    1+\mathcal{O}(r^2)\leftarrow f(r)\rightarrow\mathcal{O}(e^{-m_\text{W} r}),\quad\quad\mathcal{O}\left(r^\frac{\sqrt{3}-1}{2}\right)\leftarrow\phi(r)\rightarrow 1+\mathcal{O}(e^{-m_\text{H} r}).
\end{equation}
The corresponding profiles of $f$ and $\phi$ are shown in the Fig.~\ref{cm_bh}. This solution describes a magnetic monopole with the same magnetic charge as for spherically symmetric hairy black holes, that is $P=P_\text{SU(2)}+P_\text{U(1)}=\pm 1/e$. The SU(2) contribution to the magnetic charge is smoothly distributed in space while the U(1) contribution is pointlike and makes the total energy divergent beacause of the term $\nu^2/(2g'^2r^4)$ in the energy density (see the expression of $-\tensor{T}{^0_0}$ in Eq.~\eqref{spher_sym_eq}). Subtracting this divergent part from the energy leaves the finite value $E_\text{CM}=15.759$ (see the Ref.~\cite{Gervalle2023} for more details and also the Sec.~\ref{monop_ws}). This value turns out to be the same as the upper bound in Eq.~\eqref{bound_Mh}. 

The coincidence can be explained by the observation that the profiles $f(r)$ and $\phi(r)$ for the CM monopole are very close to those for hairy black holes in the $r>r_H$ region, as seen in the Fig.~\ref{cm_bh}. When $r_H$ decreases, more and more of the SU(2) charge of the black hole goes outside the horizon and the profiles $f(r)$ and $\phi(r)$ approach those for the CM monopole closer and closer. In the extremal limit, $r_H=r_H^\text{ex}\sim 10^{-16}$, the hairy black hole can be viewed as a CM monopole harboring in its center a tiny RN-de Sitter black hole whose size is microscopically small as compared to the size of the monopole itself; the latter being of order unity. The monopole configuration remains unaffected by the black hole presence, unless at very short distances of the order of $r_H$. However, we emphasize that the black hole presence is important because it renders the energy finite via replacing the divergent U(1) contribution of the flat space monopole by the finite term $\mathcal{M}_H$ defined in Eq.~\eqref{M_h_H}.

The hair mass $\mathcal{M}_h$ of the extremal hairy black hole is given by the integral in Eq.~\eqref{M_h_H} where the lower bound of the integration is $r=r_H^\text{ex}\sim 10^{-16}$. The regularized energy of the CM monopole $E_\text{CM}$ is given by the same integral, but the integration starts at $r=0$. This is why one has $\mathcal{M}_h(r_H^\text{ex})=E_\text{CM}$ up to terms of order $\sqrt{\kappa}$.

\subsection{Extremal limit}

When the horizon radius approaches its lower bound, $r_H\to r_H^\text{ex}$, one has $\phi(r_H)\to 0$ and $f(r_H)\to 0$ so that the near-horizon solution approaches the RN-de Sitter configuration \eqref{RNdS}. In the extremal limit, $r_H$ reaches the minimal value determined by the Eq.~\eqref{rhex_rnds} with $|\nu|=1$,
\begin{equation}
\label{rhex_hairy_spher}
    r_H^\text{ex}=\sqrt{\frac{\kappa}{2}}\frac{1}{g'}+\mathcal{O}(\kappa^{5/2})=1.0855\times 10^{-16},
\end{equation}
and the near-horizon geometry becomes extremal RN-de Sitter. In particular $N(r)$ develops a double zero at the horizon, as seen in Eq.~\eqref{extremal_RNdS}. Therefore one has $N'(r_H)=0$ and the equations \eqref{bch_ews_spher} become meaningless. To describe the near-horizon behavior, one defines
\begin{equation}
    s=\int\frac{dr}{\sqrt{N}},\quad\quad A^2=\sigma^2 N,
\end{equation}
after which the line element \eqref{spher_met_ews} becomes
\begin{equation}
    ds^2=-A(s)^2dt^2+ds^2+r^2(s)\left(d\vartheta^2+\sin^2\vartheta\,d\varphi^2\right).
\end{equation}
Passing in Eq.~\eqref{spher_sym_eq} to the new radial coordinate $s$, the equations admit a solution,
\begin{equation}
\label{near_hor_sol}
    f(s)=1,\quad\quad\phi(s)=0,\quad\quad r(s)=r_H^\text{ex},\quad\quad A(s)=e^{\lambda s},
\end{equation}
where $r_H^\text{ex}$ is determined from
\begin{equation}
    \kappa\left(1+\frac{\beta g'^2}{4}(r_H^\text{ex})^4\right)=2g'^2(r_H^\text{ex})^2\quad\Rightarrow\quad r_H^\text{ex}=\sqrt{\frac{\kappa}{2}}\frac{1}{g'}+\mathcal{O}(\kappa^{5/2}),
\end{equation}
which agrees with Eq.~\eqref{rhex_hairy_spher}, and $\lambda$ is given by
\begin{equation}
    \lambda=\frac{k}{r_H^\text{ex}}\;\;\text{with}\;\; k=\sqrt{1-2(r_H^\text{ex})^2\Lambda}=1+\mathcal{O}(\kappa^2)\quad\Rightarrow\quad\lambda=\frac{1}{r_H^\text{ex}}+\mathcal{O}(\kappa^{3/2}).
\end{equation}
This solution describes the near-horizon region of the extremal hairy black hole, which coincide with the near-horizon limit of the extremal RN-de Sitter black hole. The radial coordinate $s$ is the proper distance and the horizon is located at $s\to-\infty$, but at finite values of $s$ the configuration deviates from its near-horizon limit \eqref{near_hor_sol}. Therefore one can set 
\begin{equation}
    f=1+\delta f(s),\quad\phi=\delta\phi(s),\quad r=r_H^\text{ex}+\delta r(s),
\end{equation}
where the deviations $\delta f$, $\delta\phi$, $\delta r$ are small when $s$ is large and negative. Linearizing the field equations with respect to the deviations, one finds
\begin{equation}
\label{lin_hor_sol}
    \delta f=const.\times e^{\lambda_f s}+\dots,\quad\delta\phi=const.\times e^{\lambda_\phi s}+\dots,\quad\delta r=const.\times e^{\lambda_r s}+\dots,
\end{equation}
where
\begin{align*}
    \lambda_f&=\frac{\sqrt{8+k^2}-k}{2r_H^\text{ex}}=\frac{1}{r_H^\text{ex}}+\mathcal{O}(\kappa^{3/2}),\quad\lambda_r=\lambda=\frac{1}{r_H^\text{ex}}+\mathcal{O}(\kappa^{3/2}),\\
    \lambda_\phi&=\frac{\sqrt{2+k^2-\beta(r_H^\text{ex})^2}-k}{2r_H^\text{ex}}=\frac{\sqrt{3}-1}{2r_H^\text{ex}}+\mathcal{O}(\kappa^{1/2}).\numberthis
\end{align*}
The dots in Eq.~\eqref{lin_hor_sol} denote higher order corrections that start from quadratic terms proportional to $e^{2\lambda_f s}$, $e^{2\lambda_\phi s}$, $e^{2\lambda_r s}$. One might think that these can be neglected, however, the leading terms for $s\to -\infty$ are those with the smallest exponent, and since one has $2\lambda_\phi<\lambda_f$ and $2\lambda_\phi<\lambda_r$, it follows that the leading behavior is given by, 
\begin{equation}
\label{lead_hor_sol}
    \delta f=const.\times e^{2\lambda_\phi s}+\dots,\quad\delta\phi=const.\times e^{\lambda_\phi s}+\dots,\quad\delta r=const.\times e^{2\lambda_\phi s}+\dots
\end{equation}
The last relation implies that $e^{2\lambda_\phi s}=const.\times\delta r+\dots$, therefore
\begin{equation}
\label{lead_hor_sol_bis}
    \delta f=const.\times(r-r_H^\text{ex})+\dots,\quad\delta\phi=const.\times\sqrt{r-r_H^\text{ex}}+\dots,
\end{equation}
where the dots denote higher order terms in $(r-r_H^\text{ex})$. In addition, one has
\begin{equation}
    N=\left(\frac{dr}{ds}\right)^2=\lambda^2(r-r_H^\text{ex})^2+\dots=k^2\left(1-\frac{r_H^\text{ex}}{r}\right)^2+\dots,
\end{equation}
which agrees with the near-horizon limit of the extremal RN-de Sitter solution in \eqref{extremal_RNdS}. 

\begin{figure}
    \centering
    \includegraphics[width=10cm]{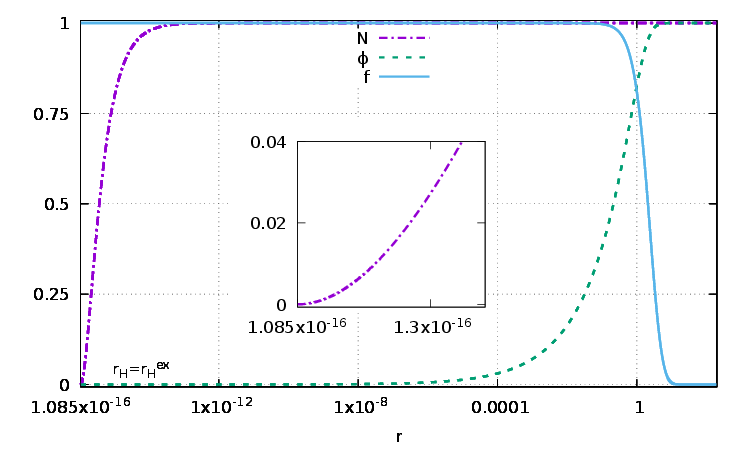}
    \caption{The extremal hairy black hole with $r_H=r_H^\text{ex}=1.0855\times 10^{-16}$.}
    \label{fig_ex_spher}
\end{figure}

As a result, the function $N(r)$ for the asymptotically flat extremal hairy black hole can be represented as
\begin{equation}
\label{N_ex_non_ab}
    N(r)=k^2(r)\left(1-\frac{r_H^\text{ex}}{r}\right)^2\;\;\text{with}\;\; k(r_H)=k=\sqrt{1-2(r_H^\text{ex})^2\Lambda}\;\;\text{and}\;\; k(\infty)=1.
\end{equation}
The solutions is thus determined by four functions: $k(r)=1+\mathcal{O}(\kappa^2)$ and $\sigma(r)=1+\mathcal{O}(\kappa)$, both of which are extremely close to unity, together with $f(r)$ and $\phi(r)$ (shown in Fig.~\ref{fig_ex_spher}), whose profiles closely coincide with those of the flat space CM monopole. For example they show the same short-distance behavior as in Eq.~\eqref{asymp_cm},
\begin{equation}
    r\ll 1\;:\quad\delta f=const.\times r^2+\dots,\quad\delta\phi=const.\times r^\frac{\sqrt{3}-1}{2}+\dots,
\end{equation}
and only for $r\sim r_H^\text{ex}\lll 1$ this changes to \eqref{lead_hor_sol_bis}.

As seen in Fig.~\ref{fig_ex_spher}, the amplitudes $f$, $\phi$ stay close to the horizon values $f\approx 1$, $\phi\approx 0$ for $r< 10^{-4}$, that is in the interval that exceeds the horizon size by 12 orders of magnitude. The spacetime geometry in this region is very close to the extremal RN-de Sitter and the analytical expression \eqref{extremal_RNdS} perfectly approximates the numerical curve $N(r)$ in Fig.~\ref{fig_ex_spher}. In fact, the contribution of the Higgs field to the energy density, $\Lambda=\beta\kappa/8$, is negligible as compared to the U(1) contribution and neglecting in Eq.~\eqref{extremal_RNdS} small terms proportional to $\kappa$ yields $N(r)=(1-r_H^\text{ex})^2$ with $r_H^\text{ex}=\sqrt{\kappa}/(\sqrt{2}g')$. Therefore, with high precision, the near-horizon geometry can be approximated by the extremal RN geometry with magnetic charge $P=P_\text{U(1)}=\pm g/g'$.

The function $N(r)$ quickly relaxes to unity and already for $r\geq 10^2\times r_H^\text{ex}\sim 10^{-14}$ the spacetime geometry is almost flat. It follows that the functions $f(r)$ and $\phi(r)$ show the same behaviour as for the flat space CM solution. In particular, they start to vary only when $r\sim 1$ (see the Fig.~\ref{fig_ex_spher}), that is at a distance which exceeds the horizon size by many orders of magnitude.

\begin{figure}
    \centering
    \includegraphics[width=7.8cm]{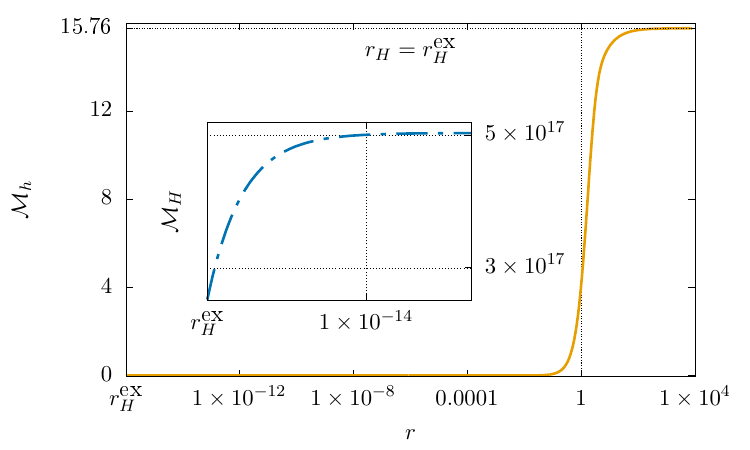}
    \includegraphics[width=7.8cm]{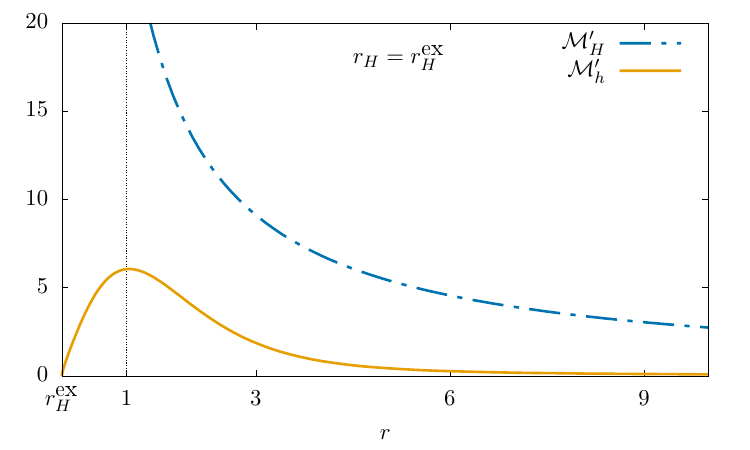}
    \caption[The horizon mass function $\mathcal{M}_H(r)$, the hair mass function $\mathcal{M}_h(r)$, and the corresponding energy densities $M'_H(r)$, $M'_h(r)$ for the extremal hairy black hole.]{The horizon mass function $\mathcal{M}_H(r)$, the hair mass function $\mathcal{M}_h(r)$ (left), and the corresponding energy densities $M'_H(r)$, $M'_h(r)$ (right) for the extremal hairy black hole.}
    \label{fig_M_h}
\end{figure}

As a result, the extremal solution is characterized by two parametrically different scales: the horizon scale which is of order of $10^{-16}$ and the electroweak scale which is of order unity. The separation of scales is exhibited by the mass functions $\mathcal{M}_H(r)$ and $\mathcal{M}_h(r)$ defined in Eq.~\eqref{M_h_H}. On the one hand, the horizon mass function reduces to
\begin{equation}
    \mathcal{M}_H(r)=\frac{4\pi r_H^\text{ex}}{\kappa}\left(2-\frac{r_H^\text{ex}}{r}\right)+\mathcal{O}(\kappa^{5/2}),
\end{equation}
which grows quickly in the horizon vicinity up to the asymptotic value $\mathcal{M}_H=8\pi r_H^\text{ex}/\kappa$ so that it varies essentially only close to the horizon, for $r_H\lll 1$, as seen in the left panel of Fig.~\ref{fig_M_h}. On the other hand, the hair mass function $\mathcal{M}_h(r)$ stays very close to zero for $r<1$ but for $r\sim 1$ the gauge amplitude $f(r)$ starts to vary and approaches zero while the Higgs amplitude $\phi(r)$ approaches unity. This produces a spherical shell of energy containing the non-Abelian hair which forces $\mathcal{M}_h(r)$ to grow and approach the asymptotic value $\mathcal{M}_h=E_\text{CM}=15.759$ (see the left panel of Fig.~\ref{fig_M_h}). Therefore the mass function $\mathcal{M}(r)=\mathcal{M}_H(r)+\mathcal{M}_h(r)$ increases to reach the asymptotic value
\begin{equation}
\label{mass_ex_spher}
    \mathcal{M}=\mathcal{M}_H+\mathcal{M}_h=\frac{8\pi}{\sqrt{2\kappa}g'}+E_\text{CM}=5.033\times 10^{17}+15.759.
\end{equation}
This is the mass of the CM monopole regularized by gravity. The horizon contribution in the above equation diverges in the flat space limit $\kappa\to 0$, but for the physical value of $\kappa$ it is finite, although very large as compared to the hair contribution. However we shall see in the next section that the hair mass becomes more important for higher values of the magnetic charge. 

The distinction between scales is also evident when examining the energy densities associated with the horizon and with the hair. The latter correspond to the derivatives $\mathcal{M}'_H(r)$ and $\mathcal{M}'_h(r)$ which are shown in the right panel of Fig.~\ref{fig_M_h}. The horizon energy density contains U(1) charge contribution, $\mathcal{M}'_H(r)=2\pi/(g'^2r^2)$, and it becomes very large at the horizon. The hair energy density shows a maximum at $r\sim 1$, it approaches zero both at the horizon and at spatial infinity.

\begin{figure}
    \centering
    \includegraphics[width=7.8cm]{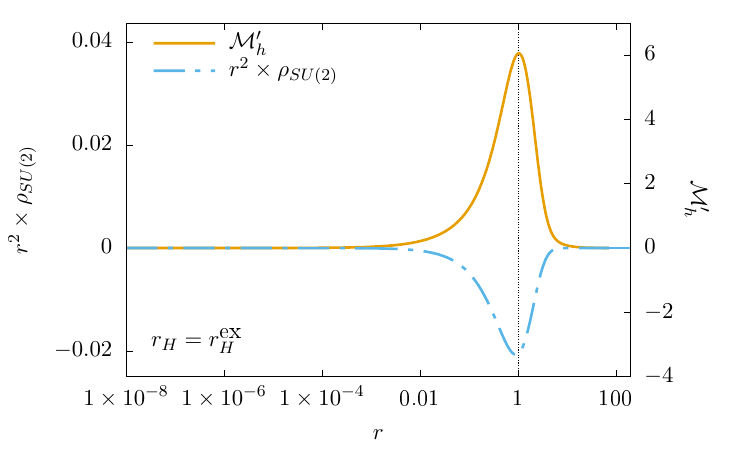}
    \includegraphics[width=7.8cm]{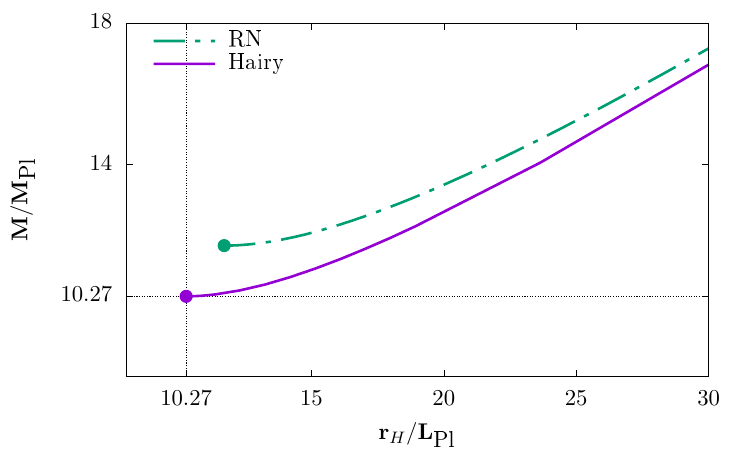}
    \caption[Hair energy density, SU(2) charge density against $r$ for the extremal hairy black hole and the total mass against $r_H$ for the RN and for the hairy black holes close to the extremal limit.]{Left: densities of the hair energy $M'_h(r)$ and of the SU(2) magnetic charge $\rho_\text{SU(2)}(r)$ defined in Eq.~\eqref{su2_charge_dens} (with $\nu$=1) for the extremal hairy solution. Right: the total mass against the horizon radius expressed in Planck unities for the RN and for the hairy black holes close to the extremal limit. The hairy branch is less massive and stays below the RN branch up to $\boldsymbol{r}_H=8.94\times 10^{16}\,\boldsymbol{L}_\text{Pl}$ after which the two branches merge.}
    \label{fig_mag_dens_rh}
\end{figure}

Since $f(r)$ decreases from unit value to zero, the hairy region contains the magnetic charge $P_\text{SU(2)}=\pm g'/g$ whose density is shown in the left panel of Fig.~\ref{fig_mag_dens_rh}. The total magnetic charge is always $P=P_\text{U(1)}+P_\text{SU(2)}=\pm 1/e$, but unlike for non-extremal black holes, the whole SU(2) charge is now distributed in the hairy region outside the black hole. 

To recapitulate, the extremal solution describes the CM monopole with the central pointlike singularity replaced by a tiny black hole. With a high precision, the geometry is the extremal RN which approaches flat space already for $r\lll 1$. This is why the monopole profiles $f(r)$, $\phi(r)$ are insensitive to the black hole presence. The dominant part of the total mass is the U(1) contribution regularized by the cut-off imposed by the event horizon while the hair energy is contained in a spherical shell of radius $r\sim 1$. The U(1) contribution to the magnetic charge is confined inside the black hole whereas the SU(2) part is supported by the hair.

As seen in Eq.~\eqref{mass_ex_spher}, the horizon mass $\mathcal{M}_H$ is largely dominant as compared to the hair mass $\mathcal{M}_h$, hence for the total mass one may use the value $\mathcal{M}\approx\mathcal{M}_H=8\pi/(\sqrt{2\kappa}g')$. The dimensionful mass and horizon size are conveniently expressed in terms of the Planck mass and length,
\begin{equation}
    \boldsymbol{M}=\frac{e^2}{4\pi\alpha}\mathcal{M}\times\boldsymbol{m}_0=\frac{g}{\sqrt{\alpha}}\times\boldsymbol{M}_\text{Pl},\quad\boldsymbol{r}_H^\text{ex}=r_H^\text{ex}\times\boldsymbol{\ell}_0=\frac{g}{\sqrt{\alpha}}\times\boldsymbol{L}_\text{Pl},
\end{equation}
where $g/\sqrt{\alpha}=10.27$. For the extemal RN black hole with the same magnetic charge the mass and radius are larger, $\boldsymbol{M}=\boldsymbol{M}_\text{Pl}/\sqrt{\alpha}=11.71\times\boldsymbol{M}_\text{Pl}$ and $\boldsymbol{r}_H^\text{ex}=\boldsymbol{L}_\text{Pl}/\sqrt{\alpha}$ (see the right panel of Fig.~\ref{fig_mag_dens_rh}) because the whole magnetic charge is contained inside the RN black hole, while the hairy black hole contains inside only the U(1) part of the charge.

As discussed above, the flat space Dirac monopoles and their gravitating RN counterparts of small size are unstable, unless for $\nu=\pm 1/2$. However, the CM monopole, which has $\nu=\pm 1$, is stable \cite{Gervalle2022a}. This suggests that its gravitating counterpart, the extremal hairy black hole, is stable as well. In addition, this extremal solution is stable also quantum-mechanically because its Hawking temperature vanishes which means that it does not radiate.

\section{Axially symmetric solutions}
\label{axial_sol_ews}

The spherically symmetric hairy black holes described above exist only for the magnetic charge $P=\pm 1/e\;\Leftrightarrow\,\nu=\pm 1$. In order to construct their generalizations for higher values of $|P|$, one should relax the assumption of spherical symmetry. The simplest possibility is to consider the axially symmetric fields discussed in Sec.~\ref{axial_ews}. The field equations are then solved numerically in the domain $(r,\vartheta)\in[r_H,\infty)\times[0,\pi/2]$. 

\subsection{Boundary conditions}
\label{bc_ews}

We require the spacetime geometry to be asymptotically flat, symmetric with respect to reflection in the equatorial plane, and globally regular outside the event horizon. In particular, the absence of conical singularity require the condition $K=S$ at the symmetry axis and for this, it is convenient to introduce a new metric function,
\begin{equation}
    h\equiv K-S.
\end{equation}
To describe the boundary conditions at the horizon, $r=r_H$, it is more suitable to introduce a new radial coordinate,
\begin{equation}
\label{def_x_coord}
    x\equiv\sqrt{r^2-r_H^2}\quad\Rightarrow\quad x\in[0,\infty).
\end{equation}
The horizon $r=r_H$ is now located at $x=0$. The derivatives of the metric functions with respect to $r$ should be finite there, which is equivalent to the requirement that the derivatives with respect to $x$ vanish. Summarizing, here are the boundary conditions for $U$, $S$ and $h$:
\begin{equation}
\label{bc_met_ews}
\begin{array}{lrl}
    \text{axis},&\;\;\underline{\vartheta=0}:&\;\;\partial_\vartheta U=\partial_\vartheta S=0,\quad h=0;\\
    \text{equator},&\;\;\underline{\vartheta=\pi/2}:&\;\;\partial_\vartheta U=\partial_\vartheta S=\partial_\vartheta h=0;\\
    \text{horizon},&\;\;\underline{x=0}:&\;\;\partial_x U=\partial_x S=\partial_x h=0;\\
    \text{infinity},&\;\;\underline{x\to\infty}:&\;\;U=S=h=0.
\end{array}
\end{equation}

For the WS fields, one has the conditions \eqref{reg_axis} which guarantee the regularity of the energy density at the symmetry axis and the parity under the reflections $\vartheta\to\pi-\vartheta$ of the field amplitudes is given by the Eq.~\eqref{parity_sym}. At spatial infinity, the field configuration should approach that of the Dirac monopole \eqref{config_dirac}. At the horizon, where $N=0$, the function $H_1$ should vanish since otherwise the radial component of the gauge field \eqref{reg_W} would diverge. The other WS amplitudes can assume any finite values at the (non-degenerate) horizon and their derivatives with respect to $r$ will be finite as well provided that the derivatives with respect to $x$ vanish. Taking all of this into account, the appropriate boundary conditions are,
\begin{equation}
\label{bc_fields_ews}
\begin{array}{rl}
    \underline{\vartheta=0}:&\;H_1=H_3=y=\phi_1=0,\quad\partial_\vartheta H_2=\partial_\vartheta H_4=\partial_\vartheta\phi_2=0;\\
    \underline{\vartheta=\pi/2}:&\;H_1=H_3=y=\phi_1=0,\quad\partial_\vartheta H_2=\partial_\vartheta H_4=\partial_\vartheta\phi_2=0;\\
    \underline{x=0}:&\; H_1=0,\quad\partial_x H_2=\partial_x H_3=\partial_x H_4=\partial_x y=\partial_x \phi_1=\partial_x \phi_2=0;\\
    \underline{x\to\infty}:&\;H_1=H_2=H_3=H_4=y=\phi_1=0,\quad\phi_2=1.
\end{array}
\end{equation}

Moreover, one should have $H_2=H_4$ and $\partial_\vartheta h=0$ at the symmetry axis. It turns out that these relations are automatically fulfilled when the conditions \eqref{bc_met_ews},\eqref{bc_fields_ews} are imposed. One can check that the constraints \eqref{const_axis} are fulfilled at the symmetry axis by virtue of the above boundary conditions.

To summarize, the field equations consist in a set of ten coupled nonlinear PDEs for the functions $(H_1,H_2,H_3,H_4,y,\phi_1,\phi_2,U, S,h)$ which must fulfill the boundary conditions \eqref{bc_met_ews},\eqref{bc_fields_ews}. We solve the equations with the FreeFem finite element solver \cite{MR3043640} together with Newton's method to handle the non-linearities. Details about the numerical methods can be found in the Appendix~\ref{num_pde}. To avoid the use of a cut-off radius, we use a compactified radial coordinate,
\begin{equation}
    \bar{x}\equiv\frac{x}{1+x},
\end{equation}
which maps the semi-infinite interval $x\in[0,\infty)$ to the finite range $\bar{x}\in[0,1]$. 

To assess the numerical accuracy of our solutions, we use the so-called virial identities. The latter are formulated in terms of integrals which are identically vanishing, even though their integrands do not, see for example the Refs~\cite{Gourgoulhon1994,Herdeiro2021b,Herdeiro2022}. These identities are fulfilled for all our solutions with a precision depending on the numbers of discretization points $N_{\bar{x}}$, $N_\vartheta$ along the $\bar{x},\vartheta$ directions. Taking $N_{\bar{x}}=130$ and $N_\vartheta=30$ yields typically errors of the order of $10^{-7}$ or $10^{-8}$.

\subsection{Flat space monopoles}
\label{monop_ws}

We shall begin the study of axially symmetric solutions by considering the flat space monopoles \cite{Gervalle2023}. For this, one should set $\kappa=0$, $N=1$ and $U=S=h=0$. The Einstein equations are then automatically fulfilled and only the WS equations remain. The boundary conditions at the horizon have to be replaced by the following conditions at the origin,
\begin{equation}
\label{cond_ori_ws}
    \underline{x=0}:\quad H_1=H_3=y=\phi_1=\phi_2=0,\quad H_2=H_4=1.
\end{equation}

The equations contain the parameter $\nu$, and for $|\nu|=1$ the solution is known -- it is the spherically symmetric CM monopole for which
\begin{equation}
    H_1=H_3=y=\phi_1=0,\quad H_2=H_4=f(r),\quad \phi_2=\phi(r),
\end{equation}
where the profiles of the functions $f$ and $\phi$ are shown in the Fig.~\ref{cm_bh}. We use this solution as the starting point in an iterative procedure to change the value of $\nu$. Of course, $\nu$ should be an integer to avoid the Dirac string singularity but the equations can be formally solved for any real $\nu$. This allows us to vary the value of $\nu$ by small steps. Our numerical scheme converges well for $|\nu|\neq 1$ and we are able to go as far as $|\nu|=100$, after which the virial identities deteriorate.

\subsubsection{General properties}

\begin{figure}
    \centering
    \includegraphics[width=7.8cm]{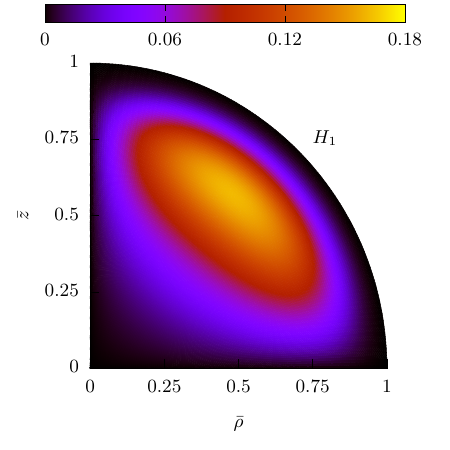}
    \includegraphics[width=7.8cm]{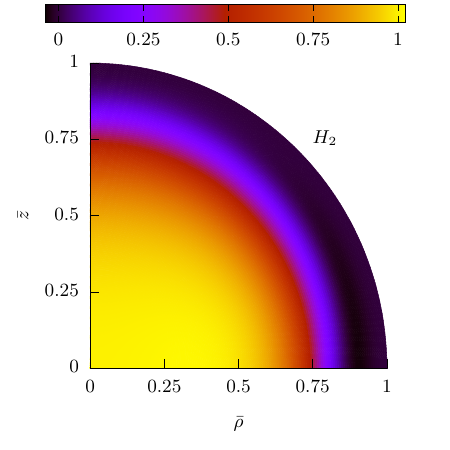}
    \includegraphics[width=7.8cm]{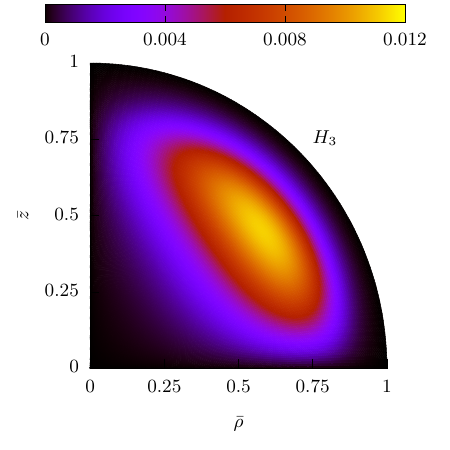}
    \includegraphics[width=7.8cm]{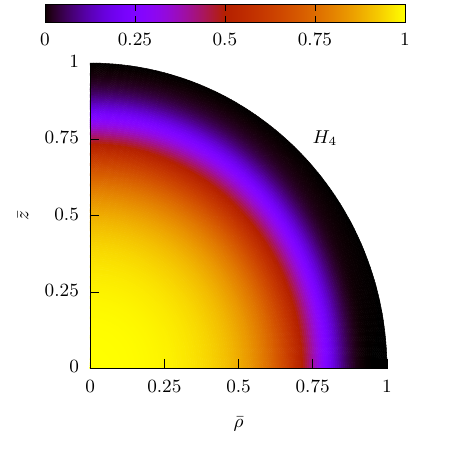}
    \caption{The SU(2) amplitudes for the $\nu=2$ monopole solution against $\bar{\rho}=\bar{x}\sin\vartheta$ and $\bar{z}=\bar{x}\cos\vartheta$.}
    \label{Hks}
\end{figure}

\begin{figure}
    \centering
    \includegraphics[width=7.8cm]{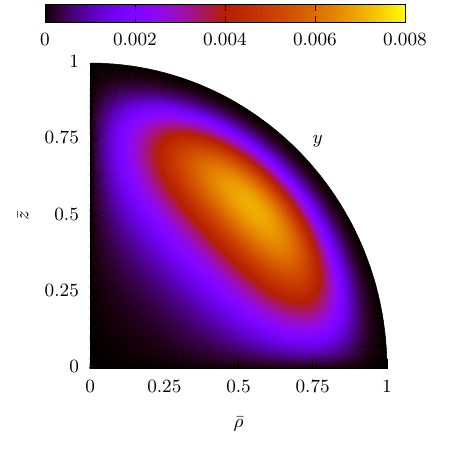}
    \includegraphics[width=7.8cm]{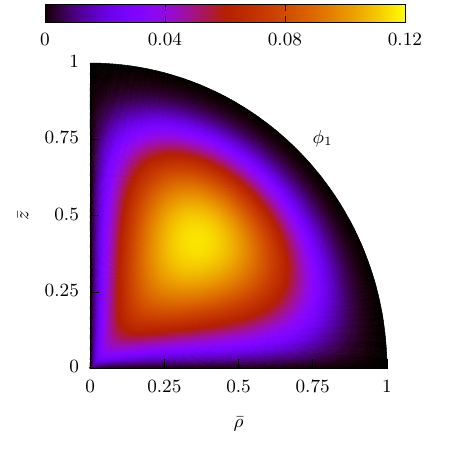}
    \includegraphics[width=7.8cm]{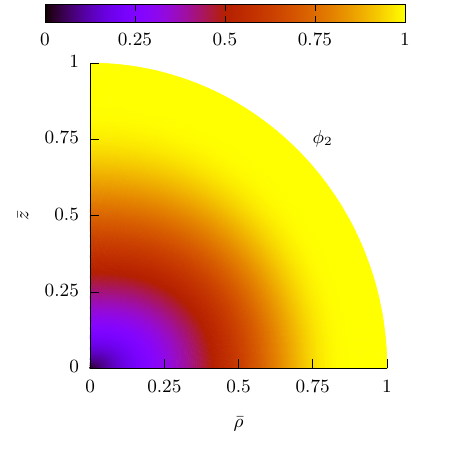}
    \includegraphics[width=7.8cm]{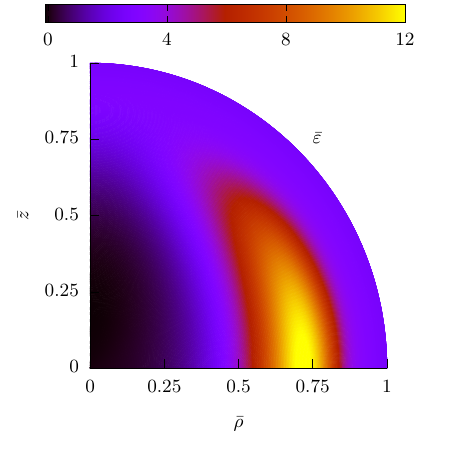}
    \caption{The U(1) and Higgs amplitudes and the regularized energy density for the $\nu=2$ monopole solution.}
    \label{phik_eps}
\end{figure}

The profiles for the $\nu=2$ solution are shown in Fig.~\ref{Hks} and Fig.~\ref{phik_eps}. On the one hand, the functions $H_2$, $H_4$, $\phi_2$ which do not vanish in the spherically symmetric case, $|\nu|=1$, remain essentially the same for the $\nu=2$ solution. In particular they almost do not depend on the polar angle $\vartheta$. The most notable change is that $\phi_2$ now approaches zero at the origin more quickly, as described by Eq.~\eqref{phi_or} below, whereas $H_2$ is no longer positive definite. On the other hand, the functions $H_1$, $H_3$, $y$, $\phi_1$ which vanish for $|\nu|=1$ are now non-zero and show a strong $\vartheta$-dependence. The norm of the Higgs field $|\Phi|$ still vanishes at the origin, according to the boundary conditions \eqref{cond_ori_ws}.

The energy density \eqref{ener_dens_ews} is singular at the origin due to the U(1) contribution $\mathcal{E}_Y$. Since $Y_3=\cos\vartheta+y\sin\vartheta$, one has
\begin{equation}
    \mathcal{E}_Y=\left[(\partial_r Y_3)^2+\frac{1}{r^2}(\partial_\vartheta Y_3)^2\right]\frac{\nu^2}{\sin^2\vartheta}=\frac{\nu^2}{r^2}+\dots,
\end{equation}
where the dots denote terms that are regular at the origin. Injecting this to \eqref{flat_space_ener} yields
\begin{equation}
    \mathcal{M}=\int_{r>0}\mathcal{E}\sqrt{-g}\,d^3x=\frac{2\pi\nu^2}{g'^2}\int_0^\infty{\frac{dr}{r^2}}+E_\text{reg}\equiv E_\text{U(1)}+E_\text{reg}.
\label{def_Ereg}
\end{equation}
Here the first term, $E_\text{U(1)}$ is infinite whereas the second term, $E_\text{reg}$, is finite and contains the finite part of the $\mathcal{E}_Y$ and also contributions of the SU(2) and Higgs fields. In other words, $E_\text{reg}$ is the regularized energy obtained by subtracting the divergent term $E_\text{U(1)}$ to the total energy. For the CM monopole, $\nu^2=1$, the regularized energy reads 
\begin{align*}
    E_\text{reg}=&4\pi\int_0^\infty\Bigg[\frac{1}{g^2}\left(\frac{\nu^2+1}{2}f'^2+\nu^2\frac{(f^2-1)^2}{2r^2}\right)\\
\label{Ereg_CM}
    &+(r\phi')^2+\frac{\nu^2+1}{4}(f\phi)^2+\frac{r^2\beta}{8}(\phi^2-1)^2\Bigg]dr,\numberthis
\end{align*}
and it evaluates to the value $E_\text{reg}=15.759=E_\text{CM}$. Notice that in the above formula $\nu$ is kept arbitrary for the purpose of Sec.~\ref{large_charge_lim_mon}. 

It is convenient to represent the regularized energy density by a function $\bar{\varepsilon}$ as
\begin{equation}
    E_\text{reg}\equiv 2\pi\int_0^\pi{\sin\vartheta\,d\vartheta\int_0^1{\bar{\varepsilon}(\bar{x},\vartheta)\,d\bar{x}}}.
\end{equation}
The profile of $\bar{\varepsilon}$ for the $\nu=2$ solution is shown in bottom right panel of Fig.~\ref{phik_eps} and it shows a strong $\vartheta$-dependence with a marked maximum in the equatorial plane. It is worth noting that the regularized energy density $\bar{\varepsilon}$ is actually not positive definite and can assume negative values in the central region, although the total energy density including the unbounded U(1) contribution is of course always positive. At spatial infinity, it is not difficult to check that the boundary conditions \eqref{bc_fields_ews} imply that
\begin{equation}
    \bar{\varepsilon}(\bar{x},\vartheta)\to\frac{\nu^2}{2g^2r^2}\frac{dr}{d\bar{x}}=\frac{\nu^2}{2g^2}.
\end{equation}
Therefore, the regularized energy density approaches a constant value as $\bar{x}\to 1$.

\begin{figure}
    \centering
    \includegraphics[width=7.8cm]{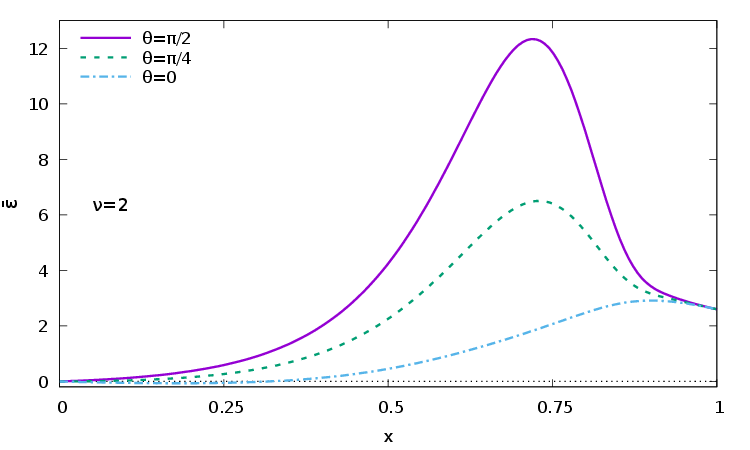}
    \includegraphics[width=7.8cm]{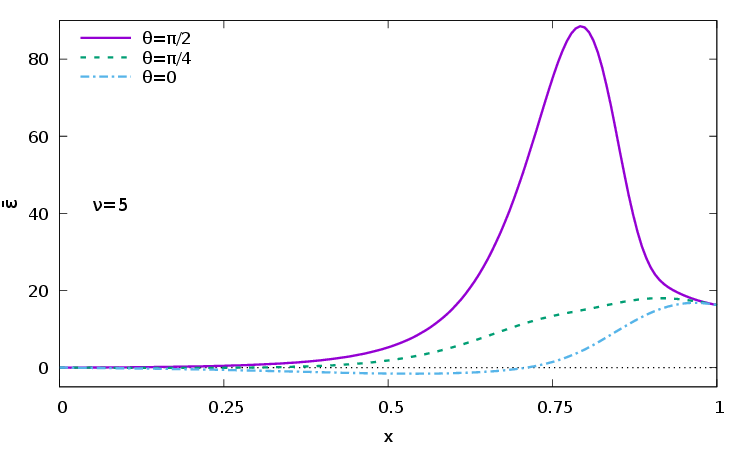}
    \caption[The regularized energy density of the $\nu=2$ and $\nu=5$ monopole solutions for several fixed value of $\vartheta$.]{The regularized energy density $\bar{\varepsilon}(x,\vartheta)$ of the $\nu=2$ (left) and $\nu=5$ (right) monopole solutions for several fixed values of $\vartheta$. In the latter case the maximum is much higher.}
    \label{eps_2d}
\end{figure}

\begin{figure}
    \centering
    \includegraphics[width=12cm]{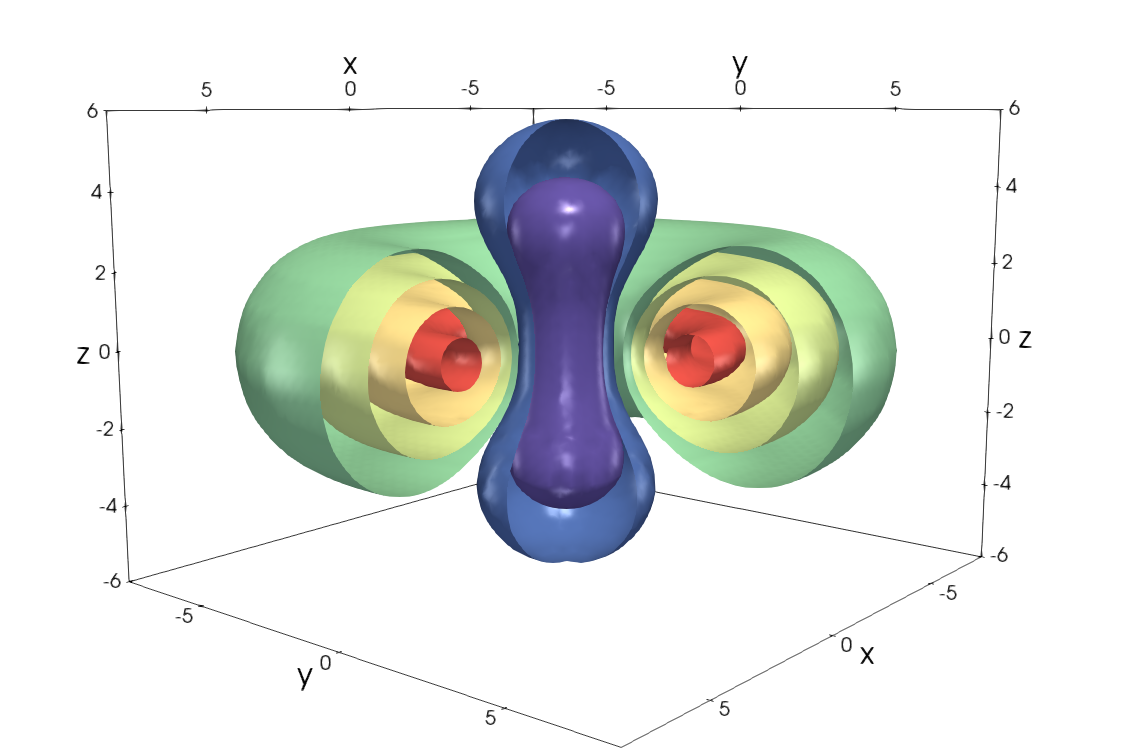}
    \caption[Isosurfaces of the regularized energy density expressed in Cartesian coordinates for the $\nu=5$ monopole solution.]{Isosurfaces of the regularized energy density $\bar{\varepsilon}=\varepsilon_0$ expressed in Cartesian coordinates $(x,y,z)$ for the $\nu=5$ monopole solution and several values of $\varepsilon_0$. For small values of $\varepsilon_0$ the surfaces are deformed ellipsoids and for larger $\varepsilon_0$ they become tori.}
    \label{iso_eps}
\end{figure}

In the Fig.~\ref{eps_2d} we show the regularized energy density $\bar{\varepsilon}$ for fixed values of $\vartheta$. As one can see, $\bar{\varepsilon}$ is an almost monotone function of the radial coordinate along the symmetry axis, $\vartheta=0$, and it shows a marked maximum along the equatorial plane for $\vartheta=\pi/2$. This implies that surfaces with constant energy density $\bar{\varepsilon}=\varepsilon_0$ are similar to ellipsoids if $\varepsilon_0$ is small, but they have a toroidal shape for larger values of $\varepsilon_0$, as seen in the Fig.~\ref{iso_eps}.

\begin{table}[b!]
    \centering
    \begin{tabular}{|c|c|c|c|c|c|c|}
       \hline
       $\nu$  & 1/2 & 1 & 2 & 3 & 4 & 5  \\
       \hline
       $E_\text{reg}$ & 6.94 & 15.76 & 38.12 & 65.76 & 97.92 & 134.13 \\
       \hline
       q & -0.51 & 0 & 3.66 & 10.61 & 20.68 & 33.78 \\
       \hline
    \end{tabular}
    \caption{The regularized energy $E_\text{reg}$ and quadrupole moment $q$ for several monopole solutions.}
    \label{Eq}
\end{table}

Solutions with $|\nu|>2$ have essentially the same structure as the $\nu=2$ solution. The functions $H_2$, $H_4$, $\phi_2$ always depend weakly on the polar angle $\vartheta$ while $H_3$, $H_4$, $y$, $\phi_1$ show more and more pronounced extrema when $\nu$ increases. The Higgs norm vanishes only at the origin and one has close to the origin \cite{Gervalle2023},
\begin{equation}
    \phi_1\sim\phi_2\sim r^\lambda\quad\text{with}\quad\lambda=\frac{\sqrt{1+2\nu}-1}{2}.
\label{phi_or}
\end{equation}
The energy density gets more and more concentrated in the equatorial region and attains higher and higher values there, see for example the Fig.~\ref{eps_2d} where $\bar{\varepsilon}$ is shown for $\nu=2$ and $\nu=5$. The numerical values of the regularized energy $E_\text{reg}$ for several values of the winding number $\nu$ are shown in Table~\ref{Eq}. We include for completeness the case $\nu=1/2$ because it corresponds to the minimal value of the magnetic charge according to the Dirac charge quantization \eqref{dirac_quant_ews}. However for non-Abelian solutions, only integer values of $\nu$ are allowed and thus the solution with $\nu=1/2$ contains the string singularity.

\subsubsection{The interior structure}
\label{interior_struct}

For the CM monopole, which corresponds to $|\nu|=1$ and is spherically symmetric, the interior structure is quite simple. At the origin, one has the pointlike U(1) magnetic charge described by the Eq.~\eqref{u1_charge}. The W bosons condensate in a central region of typical size $\sim 1/m_\text{W}$ which contains all the SU(2) charge. The corresponding charge density $\rho_\text{SU(2)}$ is given by the Eq.~\eqref{su2_charge_dens} (where one has $\sigma(r)=1$ in the flat space theory). The SU(2) part of the magnetic charge is thus distributed smoothly over 2-spheres.

\begin{figure}[t!]
    \centering
    \includegraphics[width=6.5cm]{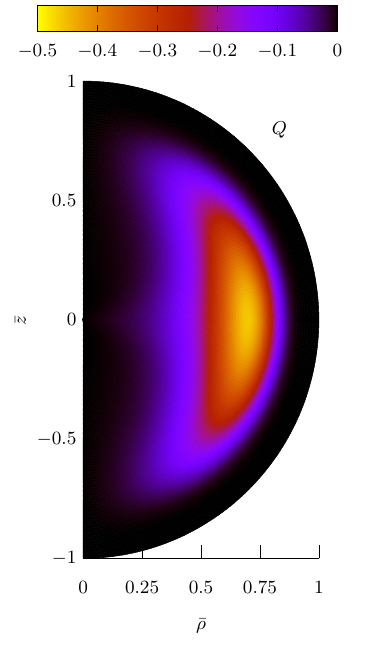}
    \hspace{1cm}
    \includegraphics[width=6.5cm]{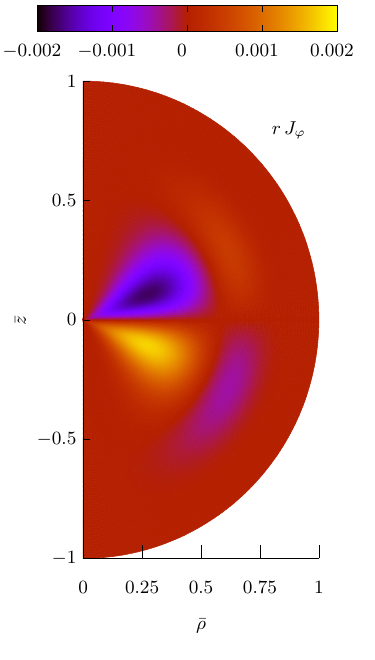}
    \caption[The magnetic charge and electric current densities for the $\nu=2$ monopole solution.]{The magnetic charge (left) and electric current (right) densities for the $\nu=2$ monopole solution.}
    \label{Qj_2d}
\end{figure}

The axially symmetric monopoles show a richer structure. In addition to the smooth SU(2) charge density, we find that the field configuration describes a non-vanishing electric current density. The latter is shown together with the SU(2) magnetic charge density for the $\nu=2$ monopole in the Fig.~\ref{Qj_2d}. The total magnetic charge splits as $P=P_\text{U(1)}+P_\text{SU(2)}$ where the U(1) part is still pointlike and given by the Eq.~\eqref{u1_charge}, while the SU(2) part can be conveniently represented via a function $Q$ as
\begin{equation}
\label{su2_charge}
    P_\text{SU(2)}=\int_{\Sigma}{\rho_\text{SU(2)}\sqrt{-g}\,d^3x}\equiv 2\pi\int_0^\pi{\sin\vartheta\,d\vartheta\int_0^1{Q(\bar{x},\vartheta)\,d\bar{x}}}.
\end{equation}
This part of the charge is smoothly distributed over the space, but its value is the same as for a Dirac monopole, 
\begin{equation}
    P_\text{SU(2)}=-g'^2\nu/e=-\nu g'/g,
\label{su2_charge_val}
\end{equation}
see the Eq.~\eqref{mag_charge_dirac}. This equivalence directly arises from the boundary conditions \eqref{bc_fields_ews} at infinity since the total magnetic charge depends only on the asymptotic field behavior. Additionally, the magnetic charge can be computed numerically by evaluating the integral \eqref{su2_charge} and we check that it yields the correct value \eqref{su2_charge_val}. This is a good consistency check for our procedure.

As one can see in the left panel of Fig.~\ref{Qj_2d}, already for $\nu=2$ the SU(2) charge density is not at all spherical and shows a strong $\vartheta$ dependence with a profound minimum in the equatorial plane. As a result, the SU(2) magnetic charge distribution has a toroidal shape with a maximal value located along a ring in the equatorial plane $z=0$. The solutions with a higher $\nu$ show a similar toroidal distribution of the charge density.

\begin{figure}[t!]
    \centering
    \includegraphics[width=12cm]{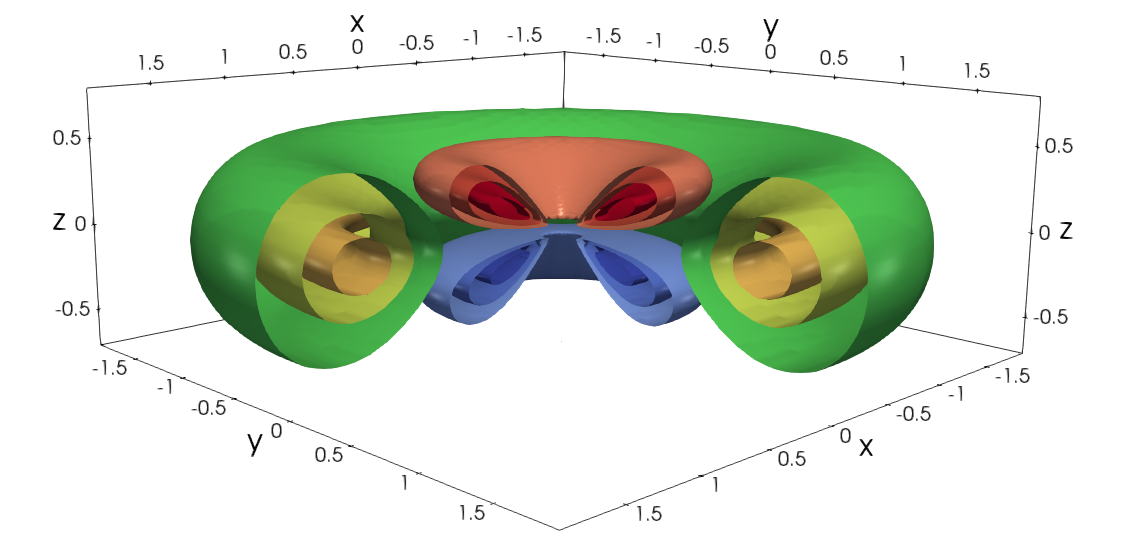}
    \includegraphics[width=12cm]{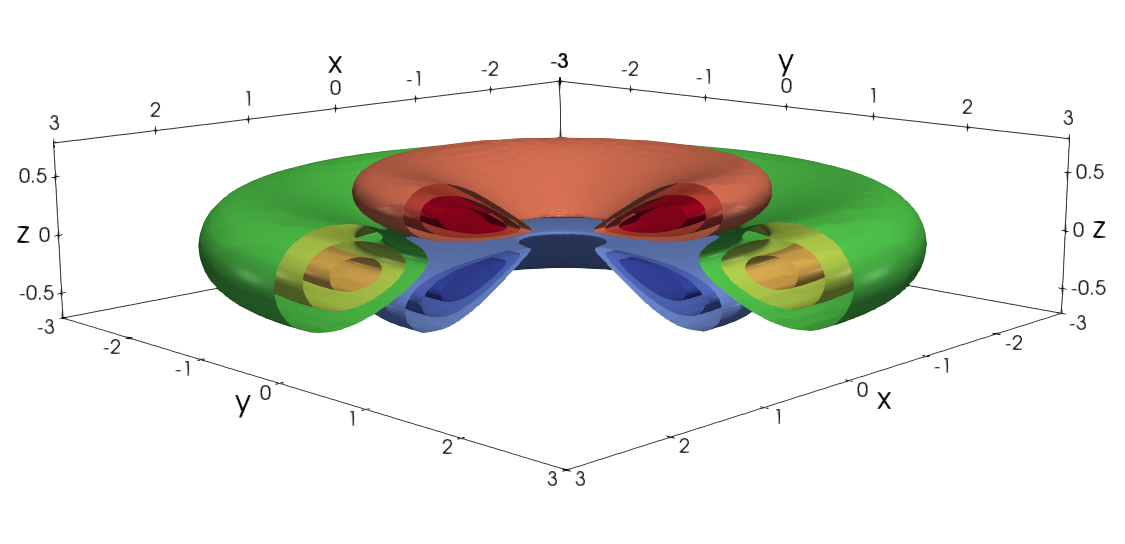}
    \caption[Isosurfaces of the SU(2) magnetic charge density and of the electric current density for the $\nu=2$ and $\nu=4$ monopole solutions.]{Isosurfaces of the SU(2) magnetic charge density (green to orange) and of the electric current density (red and blue) for the $\nu=2$ (upper panel) and $\nu=4$ (lower panel) monopole solutions in Cartesian coordinates of the 3-space. The ring radii in the latter case are twice as large, but their thickness is the same.}
    \label{iso_Qj}
\end{figure}

For the monopoles with $|\nu|>2$, we also find a non-zero azimuthal component $J_\varphi$ of electric current density. The total current through the $\rho-z$ half-plane is zero, but the currents in the $z>0$ and $z<0$ regions do not vanish and compensate each other. This is a direct consequence of the parity of the field amplitudes \eqref{parity_sym} which implies that $J_\varphi$ is an odd function. As a result, the monopoles contain two oppositely directed circular electric currents, which can be viewed as a manifestation of the electroweak superconductivity \cite{Ambjoern1990,Garaud2010a}. One has $J_\varphi\sim 1/r$ close to the origin which does not make singular the total flux but complicates the graphical representation of $J_\varphi$. Therefore, we show in the plots the bounded product $r\times J_\varphi$. As seen in the right panel of Fig.~\ref{Qj_2d}, $J_\varphi$ is antisymmetric with respect to the reflections along the equatorial plane, with a minimum in the upper hemisphere and a maximum in the lower hemisphere. This corresponds to two superconducting azimuthal currents flowing in opposite directions and giving rise to two oppositely directed magnetic moments.

In the Fig.~\ref{iso_Qj} we show level surfaces for the SU(2) charge density $Q$  defined in Eq.~\eqref{su2_charge} and for the current density $rJ_\varphi$ for the $\nu=2$ and $\nu=4$ monopole solutions. The thick toroidal region centered in the equatorial plane contains the non-Abelian magnetic charge. Although solutions with $|\nu|>1$ can be viewed as superpositions of $\nu$ Cho-Maison monopoles, these monopoles cannot be distinguished from each other and merge together into a toroidal condensate. The two other systems of tori above and below the equatorial plane correspond to two oppositely directed distributions of azimuthal electric current -- superconducting rings. As one can see in the Fig.~\ref{iso_Qj}, the whole picture is qualitatively the same for $\nu=2$ and $\nu=4$, and the same picture is found for other (even or odd) values of $\nu$.

All of this suggests the following description of the inner structure of the monopole solutions. The SU(2) part of their magnetic charge is distributed inside a torus centered in the equatorial plane (the U(1) part is always located at the origin and is pointlike). This magnetic ring produces a magnetic field which is mostly anti-parallel to the $z$-axis for $z<0$ (assuming that $\nu>0$, the charge of the ring is negative) and mostly parallel to $z$-axis for $z<0$. This magnetic field forces the electrically charged W bosons constituting the condensate inside the monopole to Larmor orbit in one direction in the upper hemisphere and in the opposite direction in the lower hemisphere\footnote{Since we are describing a purely magnetic configuration, the condensates in each hemisphere consist of both $W^{+}$ and $W^{-}$ bosons, thereby avoiding the presence of an electric dipole moment. The $W^{-}$ bosons orbit in the opposite direction to the $W^{+}$ bosons.}. This constitutes the two circular superconducting electric currents. The latter produce two oppositely directed magnetic dipole moments repelling each other but attracted to the magnetic ring. Each dipole creates a magnetic field directed oppositely to that of the magnetic ring (Lenz's law), hence pushing the individual CM monopoles (or rather their SU(2) charge) toward the equatorial plane. This field overcomes the mutual repulsion of the individual monopoles and squeezes them into a toroidal condensate.

Of course, this electromagnetic description of the inner structure is not completely accurate since standard electromagnetism applies only in the Higgs vacuum, whereas the Higgs field is not in its vacuum state inside the monopole.

\subsubsection{Quadrupole moment}

The electromagnetic description shows that the total magnetic dipole moment of the monopoles is zero (see the App.~\ref{asymp_mon}). Indeed, the oppositely directed electric currents produce two dipole moments which compensate each other while the magnetic charge density is everywhere sign definite. However the magnetic quadrupole moment does not vanish. The latter is described by a traceless tensor $q_{ik}$ receiving contributions from the magnetic charge and electric current \cite{Raab2004},
\begin{equation}
\label{quadru_moment}
    q_{ik}=\int_\Sigma(3x_i x_k-r^2\delta_{ik})\rho_\text{SU(2)}\,d^3 x+\int_\Sigma\left[x_i(\vec{r}\times\vec{J})_k+x_k(\vec{r}\times\vec{J})_k\right]d^3 x,
\end{equation}
where $x_k=(x,y,z)$ are the Cartesian coordinates and $\vec{J}$ is the spatial part of the electric 4-current. Owing to the axial symmetry, this tensor has the structure $q_{ik}=\text{diag}[-q/2,-q/2,q]$, where the only independent component is
\begin{equation}
\label{quadru_moment_2}
    q=q_{zz}=\int_\Sigma\left[3z^2-r^2\right]\rho_\text{SU(2)}\,d^3x+\int_\Sigma 2zJ_\varphi\,d^3x.
\end{equation}
The first integral here gives the dominant contribution and for the \textit{oblate} systems shown in Fig.~\ref{iso_Qj}, one has $q>0$ since $\rho_\text{SU(2)}$ is negative when $\nu$ is positive. We can get the value of $q$ for our solutions as follows. The quadrupole moment \eqref{quadru_moment} determines the non-spherically symmetric part of the asymptotic behavior of the magnetic field \cite{Raab2004},
\begin{equation}
    \delta B_idx^i=\frac{1}{2r^7}\left[5x_i x_j x_k-r^2(x_i\delta_{jk}+x_j\delta_{ik}+x_k\delta_{ij})\right]q_{jk}dx^i.
\end{equation}
Notice that the spherically symmetric part of the magnetic field is that of the Dirac monopole associated with the potential \eqref{pot_dirac_ews}. In the axially symmetric case, passing to spherical coordinates, this reduces to
\begin{equation}
    \delta B_i\,dx^i=\frac{3q}{4r^4}\left[(3\cos^2\vartheta-1)dr+r\sin(2\vartheta)d\vartheta\right].
\label{non_spher_b}
\end{equation}
On the other hand, as shown in Eq.~\eqref{asymp_q}, the asymptotic form of the electromagnetic vector potential is
\begin{equation}
    \delta A_\mu dx^\mu=\frac{\nu}{gg'}y_\gamma\sin\vartheta\,d\varphi=\frac{\nu}{gg'}\frac{C_\gamma}{r^2}\sin^2\vartheta\cos\vartheta\,d\varphi,
\end{equation}
where the value of the coefficient $C_\gamma$ is determined by the numerics. Computing then the magnetic field $\delta\vec{B}=\vec{\nabla}\times\delta\vec{A}$ yields exactly the same expression as in Eq.~\eqref{non_spher_b} with
\begin{equation}
\label{quadrupole_mom}
    q=\frac{4\nu}{3gg'}C_\gamma.
\end{equation}
We can therefore read-off the quadrupole moment from the asymptotic behavior of our solutions and we show its values for the lowest $\nu$ in Table~\ref{Eq}. As one can see, the value of $q$ increases with $\nu$ which corresponds to the fact that the oblateness of the solutions increases with growing magnetic charge. It is also worth noting that $q$ becomes negative for $|\nu|<1$, and we checked that solutions becomes \textit{prolate} in this case: the magnetic charge density isosurfaces become stretched along the $z$-axis.

\subsubsection{The limit of large magnetic charge}
\label{large_charge_lim_mon}

Both the regularized energy $E_\text{reg}$ and the quadrupole moment $q$ increase with $\nu$. We propose in this subsection to derive analytical estimates in the large magnetic charge limit.

First, it is known that the Higgs field should approach zero when the magnetic field becomes very strong. The electroweak gauge symmetry is then fully restored \cite{Ambjoern1989,Ambjoern1990a}. This can be seen at the level of classical solutions by studying their inner structure \cite{Ambjoern1990,Garaud2010a}. In our case, when the magnetic charge $P\propto\nu$ increases, the magnetic field gets stronger, hence the Higgs field in the central region of the monopole is expected to approach zero. This expectation is confirmed already at the perturbative level since close to the origin one has
\begin{equation}
    \phi_1\sim r^\lambda\left[\left(\sin\frac{\vartheta}{2}\right)^{\nu+1}+\left(\cos\frac{\vartheta}{2}\right)^{\nu+1}\right],\quad\phi_2\sim\partial_\vartheta\phi_1,
\end{equation}
with $\lambda=(\sqrt{1+2\nu}-1)/2$, hence the Higgs field gets smaller in the central region when $\nu$ increases.

\begin{figure}
    \centering
    \includegraphics[width=7.8cm]{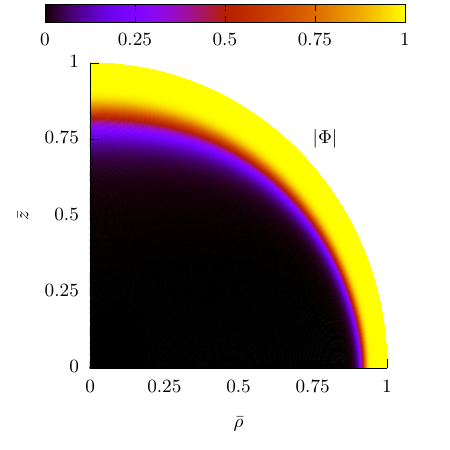}
    \includegraphics[width=7.8cm]{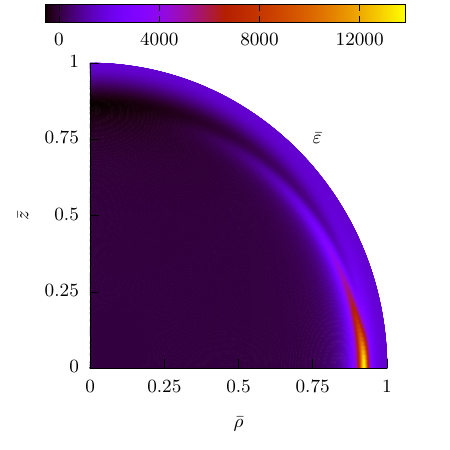}
    \caption[The norm of the Higgs field $|\Phi|$ and the regularized energy density $\bar{\varepsilon}$ for the monopole solution with $\nu=50$.]{The norm of the Higgs field $|\Phi|$ (left) and the regularized energy density $\bar{\varepsilon}$ (right) for the monopole solution with $\nu=50$.}
    \label{phi_eps_2d}
\end{figure}

The numerical analysis confirms the expectation at the non-perturbative level and shows that for large $\nu$ the monopoles develop in the central region a spheroidal bubble where the norm of the Higgs $|\Phi|=\sqrt{\phi_1^2+\phi_2^2}$ is very close to zero. Hence the system is in the so-called Higgs false vacuum there. This can be seen in the left panel of Fig.~\ref{phi_eps_2d}. The SU(2) gauge field also vanishes inside the bubble since $H_1$, $H_3$ are very close to zero while $H_2$, $H_4$ are very close to unity which implies that $W^a_\mu=0$, see Eq.~\eqref{reg_gauge_mon}. The $y$ function is very close to zero too. As a result, inside the bubble there remains only the U(1) hypercharge field,
\begin{equation}
    \text{\underline{inside}:}\quad\quad Y_\mu dx^\mu=\nu(\cos\vartheta\pm 1)d\varphi,\quad W^a_\mu=0,\quad\Phi=0.
\end{equation}
In view of the Eq.~\eqref{elec_Z}, this describes the electromagnetic field $F_{\mu\nu}=(g'/g)Y_{\mu\nu}$ associated with the pointlike U(1) charge $P_\text{U(1)}=-\nu g/g'$ and the Z field is $Z_{\mu\nu}=Y_{\mu\nu}$. Since the gauge symmetry is restored, the Z field is massless inside the bubble.

Outside the bubble, the Higgs field approach its vacuum expectation value, $|\Phi|=1$, generating non-zero masses for the fields, and being now massive, the latter tend to zero at large distances exponentially fast except for the photon that remains massless and long-ranged. The field configuration then approaches that in Eq.~\eqref{dirac_in_ews},
\begin{equation}
    \text{\underline{outside}:}\quad\quad Y_\mu dx^\mu=\nu(\cos\vartheta\pm 1)d\varphi,\quad T_a W^a_\mu=T_3\,Y_\mu dx^\mu,\quad\Phi=\begin{pmatrix}
        0 \\ 1
    \end{pmatrix}.
\end{equation}
This corresponds to the Dirac monopole with charge $P_\text{U(1)}+P_\text{SU(2)}=-\nu/e$.

\begin{figure}
    \centering
    \includegraphics[width=16cm]{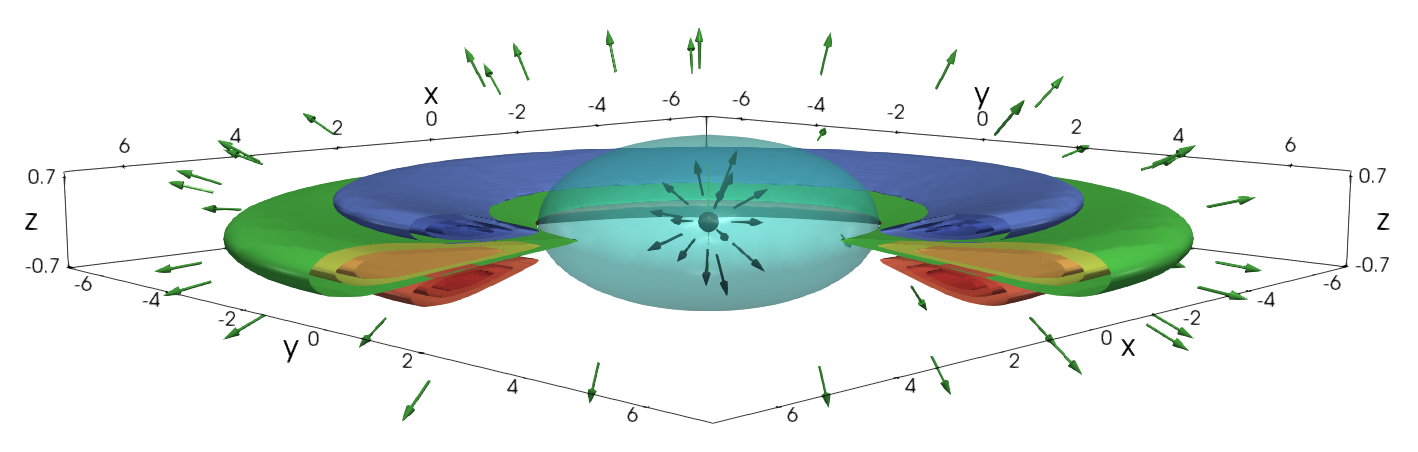}
    \caption[Internal srtructure of the non-Abelian magnetic monopole with $\nu=-20$.]{Internal structure of the non-Abelian magnetic monopole with $\nu=-20$. The central region is occupied by a spheroidal bubble (cyan) containing the U(1) hypercharge field generated by the pointlike magnetic charge $P_\text{U(1)}=-\nu g/g'$ in the center. This field is strong enough to suppress all other fields. Outside the bubble, the nonlinear fields interact and produce a condensate forming a ring of non-Abelian magnetic charge $P_\text{SU(2)}=-\nu g'/g$ squeezed between two superconducting rings of oppositely directed electric currents. Still farther away, the nonlinear fields die away and there remains only the magnetic field of a Dirac monopole with total charge $P_\text{U(1)}+P_\text{SU(2)}=-\nu/e$.}
    \label{iso_Qj_large_nu}
\end{figure}

Therefore, the Higgs field interpolates between $|\Phi|=0$ and $|\Phi|=1$ in a transition region -- the "bubble crust". This region contains a W-condensate in the form of rings close to the equatorial plane, as shown in Fig.~\ref{iso_Qj_large_nu} for $\nu=-20$. The condensate consists of a magnetically charged ring squeezed between two rings of oppositely directed electric currents. Comparing with the similar picture in Fig.~\ref{iso_Qj}, one can see that the rings become large and strongly "squashed" for large $|\nu|$, while their thickness in the $z$-direction visibly does not change. The total non-Abelian magnetic charge contained in the crust is $P_\text{SU(2)}=-\nu g'/g$.

\begin{figure}
    \centering
    \includegraphics[width=7.8cm]{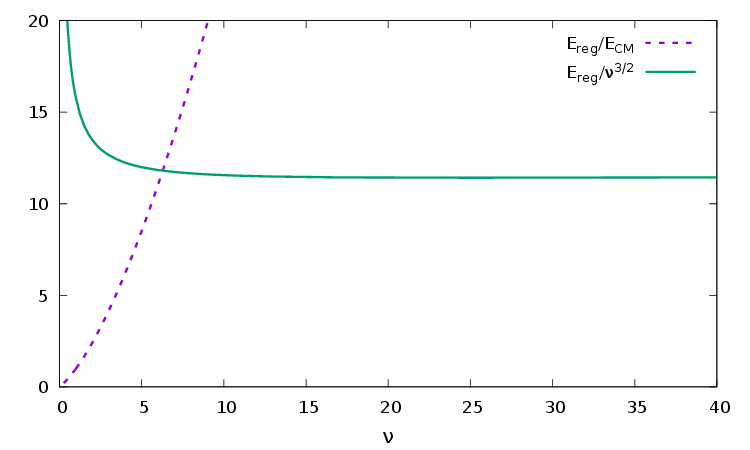}
    \includegraphics[width=7.8cm]{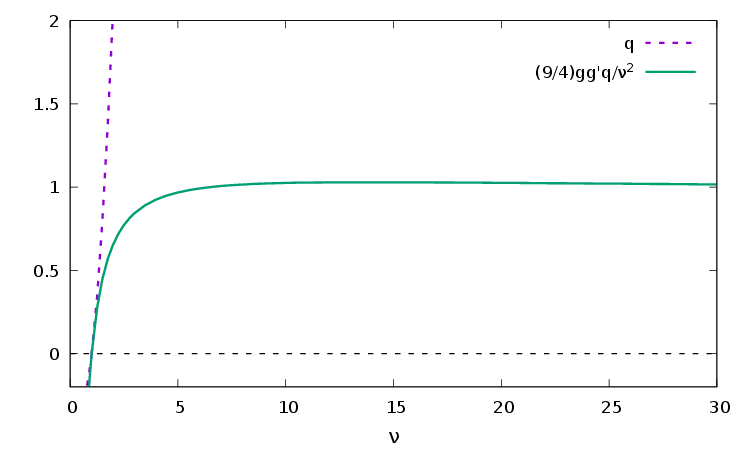}
    \caption[The regularized energy and the quadrupole moment against the winding number $\nu$.]{Left: the regularized energy $E_\text{reg}(\nu)$ in units of the CM monopole energy $E_\text{CM}=E_\text{reg}(1)$ and $E_\text{reg}(\nu)$ divided by $\nu^{3/2}$. Right: the quadrupole moment $q$ and also $q$ divided by $4\nu^2/(9gg')$ against the winding number $\nu$.}
    \label{plot_Eq}
\end{figure}

As seen in the left panel of Fig.~\ref{phi_eps_2d} where the region of Higgs false vacuum corresponds to the black region, the bubble is not exactly spherical. However we can obtain reasonable estimates by approximating the fields by their spherically symmetric expressions \eqref{spher_sym_ws_fields} with the functions $f$, $\phi$ given by
\begin{equation}
    f(r)=1\;\;\text{if}\;\;r<R,\quad f(r)=0\;\;\text{if}\;\;r>R,\quad \phi(r)=1-f(r),
\end{equation}
where $R$ is the bubble radius. Injecting this to Eq.~\eqref{Ereg_CM} where $\nu$ is kept arbitrary yields the energy,
\begin{equation}
    E_\text{reg}=\frac{\beta}{8}\frac{4\pi R^3}{3}+4\pi\frac{\nu^2}{2g^2R}.
\end{equation}
Here the first term is the contribution of the constant Higgs energy density inside the bubble, and the second one is the non-Abelian magnetic energy outside the bubble. Minimizing with respect to $R$ yields the following estimates for the bubble size and energy,
\begin{equation}
    R=\left(\frac{4}{\beta g^2}\right)^{1/4}\sqrt{|\nu|}\approx 1.29\sqrt{|\nu|},\quad\quad E_\text{reg}=\frac{8\pi}{3}\left(\frac{\beta}{4g^2}\right)^{1/4}|\nu|^{3/2}\approx 7.4|\nu|^{3/2}.
\label{estim_RE}
\end{equation}
We can identify the bubble size and hence the position of the bubble crust with the radius of the magnetic ring in the Fig.~\ref{iso_Qj_large_nu}. The values of the bubble size that can be obtained from the numerical solutions are in good agreement with $R$ in Eq.~\eqref{estim_RE}. Moreover, as seen in the left panel of Fig.~\ref{plot_Eq}, the numerically obtained ratio $E_\text{reg}/\nu^{3/2}$ converges indeed toward a constant value at large $\nu$. This value, 11.4, is larger than 7.4 as suggested by the Eq.~\eqref{estim_RE}. This is because the above analytical estimates take into account only the energy inside and outside the bubble without considering the energy in the crust. More accurate estimates could be obtained by introducing a finite transition region where $f(r)$ and $\phi(r)$ interpolate between their inside and outside values.

Our numerics suggests that for large $\nu$ the constant $C_\gamma$ in the asymptotic formula \eqref{asymp_q} approaches the value $\nu/3$. Hence the quadrupole moment defined by Eq.~\eqref{quadrupole_mom} is
\begin{equation}
    q=\frac{4}{9gg'}\nu^2.
\end{equation}
This can be seen in the right panel of Fig.~\ref{plot_Eq}. This can be represented as
\begin{equation}
    q=-\frac{4\nu}{9g'^2}P_\text{SU(2)}\approx -1.16\times P_\text{SU(2)}R^2,
\end{equation}
with $R$ given in Eq.~\eqref{estim_RE}. Thus one has $q\approx P_\text{SU(2)}R^2$ which is the quadrupole moment of a homogeneously charged torus of radius $R$ and charge $P_\text{SU(2)}$. This shows again that the above estimate for the bubble size $R$ is sensible, because the dominant contribution in the quadrupole moment formula \eqref{quadru_moment_2} is the first integral containing the magnetic charge density while the second integral containing the contribution of the electric current is negligible for large $|\nu|$. Specifically the currents $I_\pm$ in each hemisphere can be computed as
\begin{equation}
    I_{+}=-I_{-}=\int_0^\infty{dz\int_0^\infty{(\vec{J}\cdot\vec{n}_\varphi)\rho\,d\rho}},
\end{equation}
where $\vec{n}_\varphi$ is the unit vector in the azimuthal direction. We find that these currents approach the constant value $I_\pm=\mp 0.095$ for large $|\nu|$. Since the radius $R$ of the superconducting rings is proportional to $\sqrt{\nu}$, the dipole moments produced by each rings scale as $\pi R^2I_\pm\propto\nu$. These dipole moments are separated in space and their fields do not exactly compensate each other but instead produce a quadrupole moment. However the separation between the two superconducting rings is almost independent on $\nu$ and therefore their quadrupole moment grows slower than $\nu^2$ and is sub-dominant as compared to that produced by the magnetic ring.

One can finally wonder why the condensate which constitutes the magnetic ring and the two superconducting rings are strongly squashed for large $\nu$. As seen in the Eq.~\eqref{W_modes_decomp}, the angular dependence of the W-modes in the far field zone is given in terms of the associated Legendre polynomials $P^\nu_j(\cos\vartheta)$ and $P^{\nu\pm 1}_j(\cos\vartheta)$. For large $\nu$ one has $\nu\pm 1\approx\nu$ and since the leading contribution corresponds to the value $j=|\nu|$, the angular dependence of the W-modes is given by
\begin{equation}
    P^\nu_{|\nu|}(\cos\vartheta)\propto(\sin\vartheta)^{|\nu|}.
\end{equation}
These modes are strongly localized around $\vartheta=\pi/2$, which agrees with the rings in the equatorial region shown in Fig.~\ref{iso_Qj_large_nu}. On the other hand, the angular dependence of the Z, Higgs and electromagnetic modes is different. It follows that the electric currents and the SU(2) magnetic charge must be supported mainly by a condensate of W bosons.

It is also worth reminding that the Dirac monopole embedded in the electroweak theory is unstable with respect to perturbations with angular momentum $j=|\nu|-1$ and the instability resides in the W-sector (see the discussion in Sec.~\ref{stab_rn}). The Dirac monopole can be viewed as a superposition of two pointlike charges, $P_\text{U(1)}$ and $P_\text{SU(2)}$. It seems plausible that the instability growth affects the SU(2) field configuration by radiating away all its central part, and what remains condenses to the rings in the equatorial region. The total magnetic charge does not change but its SU(2) part gets distributed over the volume of the central torus. Of course, it remains to verify that the non-Abelian monopoles with $|\nu|>1$ are indeed stable, in which case they may be viewed as remnants of the decay of Dirac monopoles. For the CM monopole ($|\nu|=1$), numerical evidences regarding its stability are provided in our publication \cite{Gervalle2022a}. One should also remember that the instabilities are not necessarily axially symmetric for $|\nu|>1$. Therefore the non-Abelian monopoles that have been considered here are only a special case of more general solutions which have no symmetry at all, or perhaps show only discrete symmetries as for the spherical harmonics. The stable remnant of the Dirac monopole decay may be one of these monopoles rather than our axially symmetric configurations.

\subsection{Axially symmetric hairy black holes}

Let us move on to the black hole case. The set of PDEs to be solved contains the parameter $\nu$, and for $|\nu|=1$ the solutions are already known. These are the spherically symmetric black holes expressed by the fields \eqref{spher_met_ews},\eqref{spher_ws_fields} in terms of 4 functions $N$, $\sigma$, $f$, $\phi$ of the radial coordinate $r$. The event horizon is located at $r=r_H$ where $N(r_H)=0$. These solutions can also be expressed in the axially symmetric form \eqref{metric_ews},\eqref{reg_W},\eqref{Y_phi_reg} in terms of 10 functions $U$, $K$, $S,\dots$ depending only on the radial coordinate. The line element \eqref{metric_ews} also contains $N=1-r_H/r$ which is a \textit{given} function of $r$ (\textit{i.e.} not subject to the field equations). This function is introduced for convenience, and can always be gauged away by passing to the new radial coordinate $\tilde{r}$ such that $d\tilde{r}/\tilde{r}=dr/(r\sqrt{N})$. The horizon is located at $r=r_H$ where the 10 functions assume values $U_H$, $K_H$, $S_H$, etc.

Naively, the relation between the two descriptions is provided by the Eqs.~\eqref{spher_sym_ws_fields},\eqref{spher_sym_met_fields} which provides the method to rapidly descend from the axially symmetric to the spherically symmetric case. However, this does not determine the precise correspondence between the two descriptions since the radial coordinate $r$ is not the same in the line elements \eqref{metric_ews} and \eqref{spher_met_ews}. The relation between the two coordinate systems is given in App.~\ref{radial_coords}.

Summarizing here the results presented in the App.~\ref{radial_coords}, the spherically symmetric hairy solutions can be described using the line element \eqref{spher_met_ews} as was done in the Sec.~\ref{spher_hairy_bh}. They are labeled by the Schwarzschild radius of their event horizon which takes value in the interval, $r_H\in[r_H^\text{ex},r_H^0]$. The same solutions can be described using instead the line element \eqref{metric_ews}. The parameter $r_H$ in this case does not correspond to the Schwarzschild radius anymore. It takes value in the interval, $r_H\in[0,r_H^0-r_{-}^0]$, where $r_{-}^0=\nu^2\kappa/(2e^2r_H^0)$ is the inner horizon radius of the RN geometry. Hence, the bifurcation of hairy solutions with RN does not occur for $r_H=r_H^0$; instead, it happens for $r_H=r_H^0-r_{-}^0$. In the extremal limit, $r_H$ now approaches zero and the horizon values $U_H,K_H,S_H$ become infinite. Consequently, one has to use a different function $N(r)$ in the ansatz \eqref{metric_ews}. 

The same conclusion applies to axially symmetric hairy black holes for which $|\nu|>1$. The parameter $r_H$ and the function $N(r)$ in the line element \eqref{metric_ews} assume the following values,
\begin{align*}
    \text{non-extremal case:}&\quad 0<r_H\leq r_H^0-r_{-}^0,\quad\; N(r)=1-\frac{r_H}{r},\\
    \text{extremal case:}&\quad r_H=r_H^\text{ex},\quad\quad\quad\quad\quad\; N(r)=k^2(r)\left(1-\frac{r_H}{r}\right)^2,\numberthis
\end{align*}
where $r_H^0$ is given in the Table~\ref{rhn_nu} for a few values of $|\nu|$, $r_H^\text{ex}$ is the extremal horizon radius for RN-de Sitter black holes which is defined in Eq.~\eqref{rhex_rnds}, and
\begin{equation}
\label{k_ext}
    k^2(r)=1-\frac{\Lambda}{3}\left(r^2+2r_H r+3r_H^2\right)\times\frac{1+r_H^4}{1+r^4}.
\end{equation}
The spherically symmetric solutions with $|\nu|=1$ can be used as the starting point in an iterative procedure to change the value of $\nu$. Our numerical scheme converges well for $|\nu|\neq 1$ and we are able to go as far as $|\nu|=100$, after which the virial identities deteriorate.

\subsubsection{General properties}

The simplest non-spherical solutions are obtained when $\nu=\pm 2$. To illustrate their profiles, we set $\kappa=10^{-3}$ instead of choosing the physical value $\kappa\sim 10^{-33}$ since otherwise it is difficult to see the deviation from spherical symmetry. The parameter $r_H$ ranges from zero up to the maximal value $r_H^0-r_{-}^0\approx 1.4657$ corresponding to the bifurcation with the RN solution.

\begin{figure}[b!]
    \centering
    \includegraphics[height=6.4cm]{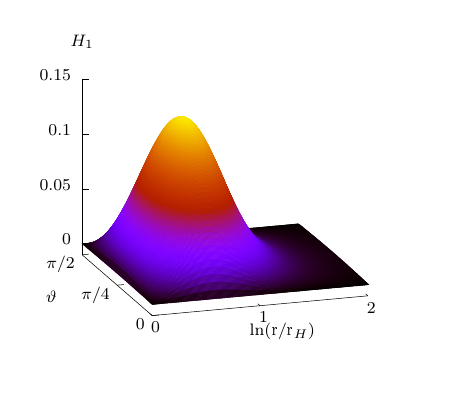}
    \includegraphics[height=6.4cm]{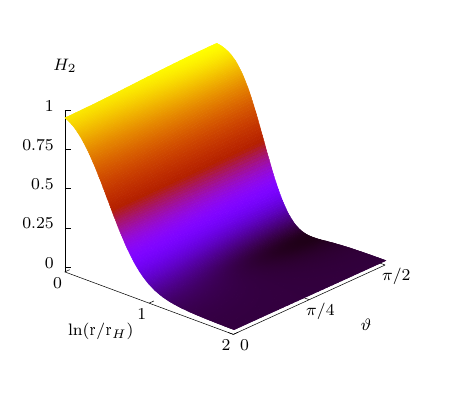}\\
    \vspace{-1cm}
    \includegraphics[height=6.4cm]{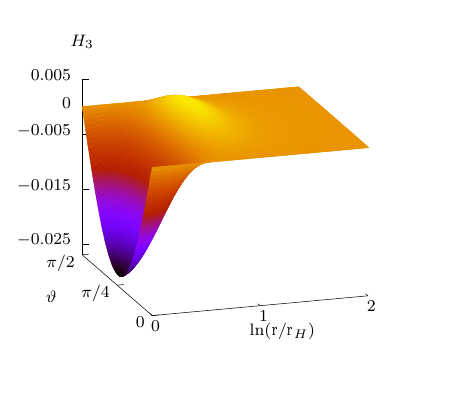}
    \includegraphics[height=6.4cm]{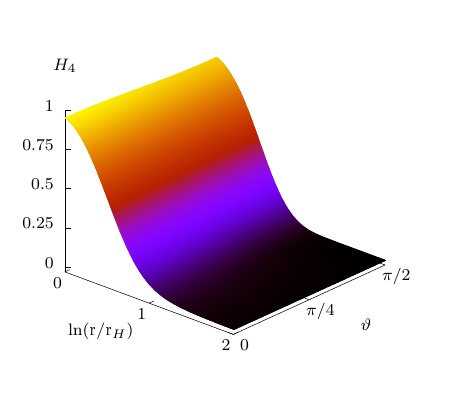}
    \caption{The SU(2) amplitudes $H_1,H_2,H_3,H_4$ for the hairy black hole solution with $\kappa=10^{-3}$, $r_H=0.7$, $\nu=2$ against $\ln(r/r_H)$ and $\vartheta$.}
    \label{fig_bh_07_1}
\end{figure}

\begin{figure}
    \centering
    \includegraphics[height=6.4cm]{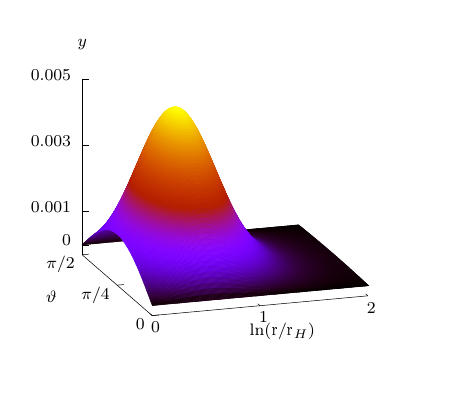}
    \includegraphics[height=6.4cm]{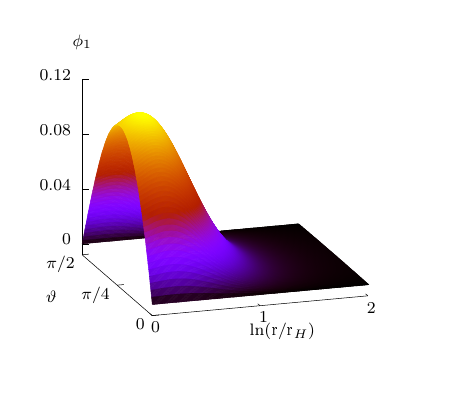}\\
    \vspace{-1cm}
    \includegraphics[height=6.4cm]{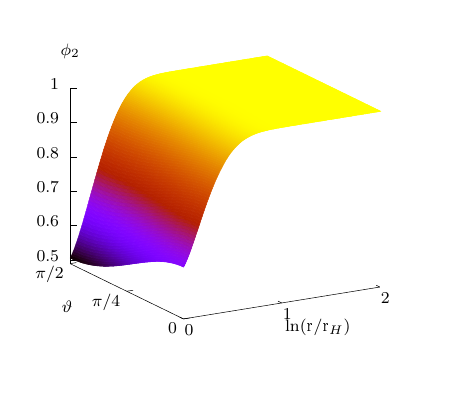}
    \includegraphics[height=6.4cm]{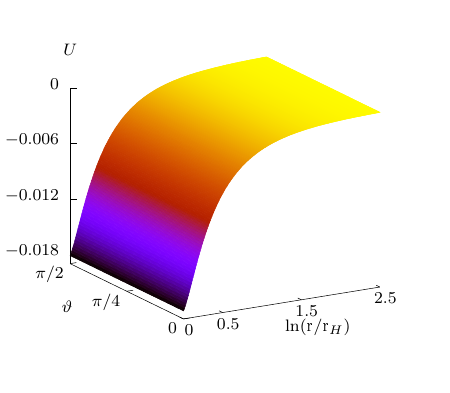}
    \caption{The U(1) amplitude $y$, the Higgs amplitudes $\phi_1$ and $\phi_2$, and the metric function $U$ for the hairy black hole solution with $\kappa=10^{-3}$, $r_H=0.7$, $\nu=2$.}
    \label{fig_bh_07_2}
\end{figure}

\begin{figure}
    \centering
    \includegraphics[height=6.4cm]{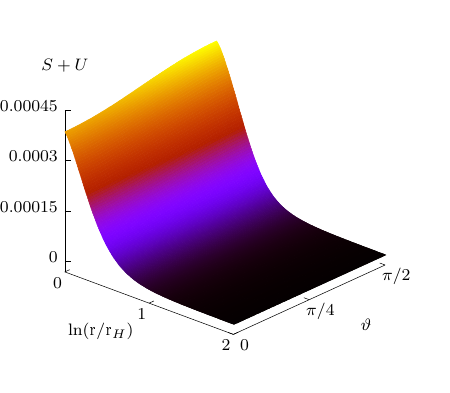}
    \includegraphics[height=6.4cm]{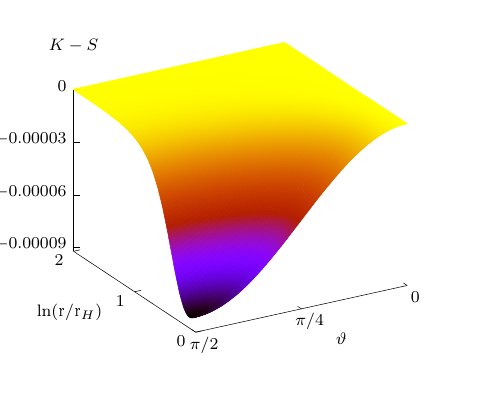}
    \caption{The metric functions $S+U$ and $K-S$ for the hairy black hole solution with $\kappa=10^{-3}$, $r_H=0.7$, $\nu=2$.}
    \label{fig_bh_07_3}
\end{figure}

If $r_H$ is close to the upper bound, $r_H^0-r_{-}^0$, then the black hole is only "slightly hairy" and the spacetime geometry is essentially RN with magnetic charge $P=\pm 2/e$. We therefore choose an intermediate value, $r_H=0.7$, and show in Figs.~\ref{fig_bh_07_1}-\ref{fig_bh_07_3} the profiles for the solution with $\nu=2$ against $\ln(r/r_H)$ and $\vartheta$.

The functions $H_2,H_4$ and $\phi_2$, which do not vanish in the spherically symmetric case when $|\nu|=1$, exhibit only a weak dependence on $\vartheta$ for the $\nu=2$ solution. Their $\vartheta$-dependence is more visible at the horizon, as seen in the profile of $\phi_2$ in the Fig.~\ref{fig_bh_07_2}. Although the profiles of $H_2$ and $H_4$ closely resemble each other, the maximum value of $H_2$ occurs in the equatorial plane, $\vartheta=\pi/2$, whereas the maximum of $H_4$ is located at the symmetry axis, $\vartheta=0$. On the other hand, the functions $H_1,H_3,y$ and $\phi_1$, which are zero for $|\nu|=1$, no longer vanish when $|\nu|>1$, and they show a strong $\vartheta$-dependence. They exhibit a distinct maximum in between the symmetry axis and the equatorial plane. At the horizon, their dependence on $\vartheta$ remains pronounced, except for $H_1$ which vanishes at $r=r_H$ -- see the boundary conditions \eqref{bc_fields_ews}.

The metric functions $U,K$ and $S$ almost do not depend on the polar angle $\vartheta$. In the case of the RN solution, one has $K=S=-U$, as shown in the Eq.~\eqref{RN_axisym}. Therefore, to emphasize the deviation from the RN geometry, we present in the Fig.~\ref{fig_bh_07_3} the combinations $S+U$ and $K-S$. These are non-zero only for the non-Abelian hairy black holes. As one can see, $S+U$ exhibits a slight maximum in the equatorial plane for $r=r_H$, whereas $K-S$ shows a pronounced minimum there. The latter illustrates the deviation from spherical symmetry since $K=S$ holds for a spherically symmetric spacetime. This deviation is always small, typically of the order of $10^{-5}$. For $\vartheta=0$, one has $K=S$ and also $\partial_\vartheta U=\partial_\vartheta K=\partial_\vartheta S=0$. In view of Eq.~\eqref{const_axis}, this guarantees that the two gravitational constraints $\mathcal{C}_1$ and $\mathcal{C}_2$ vanish at the symmetry axis, while \eqref{bianchi_ews} then implies that they should vanish everywhere. This is indeed confirmed by our numerics with the same precision as the virial identities. 

\begin{figure}
    \centering
    \includegraphics[width=7.8cm]{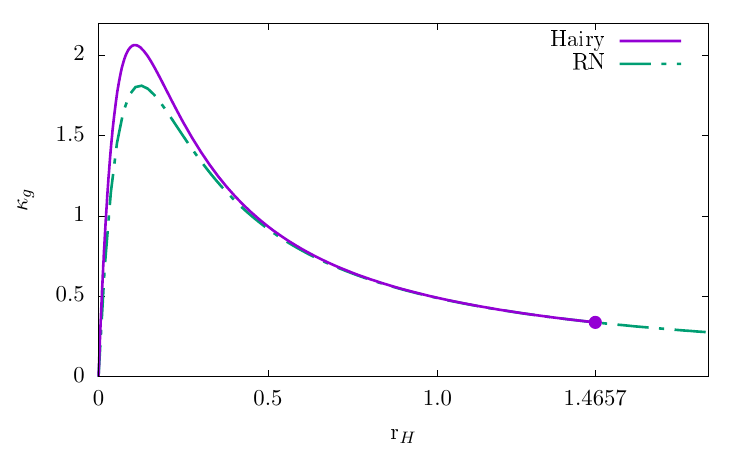}
    \includegraphics[width=7.8cm]{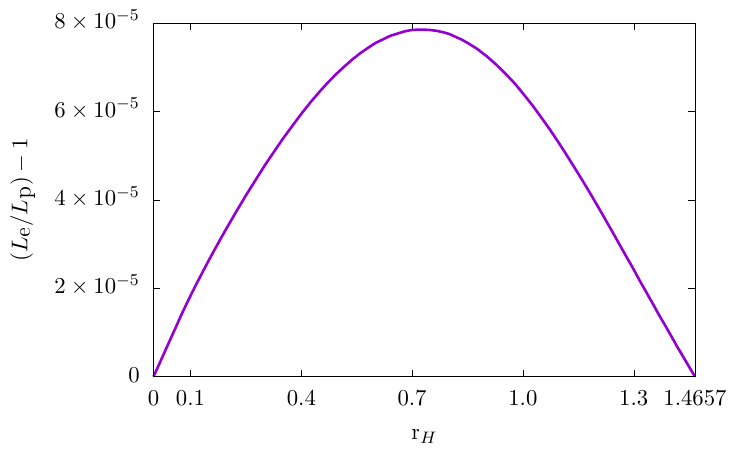}\\
    \vspace{0.2cm}
    \includegraphics[width=7.8cm]{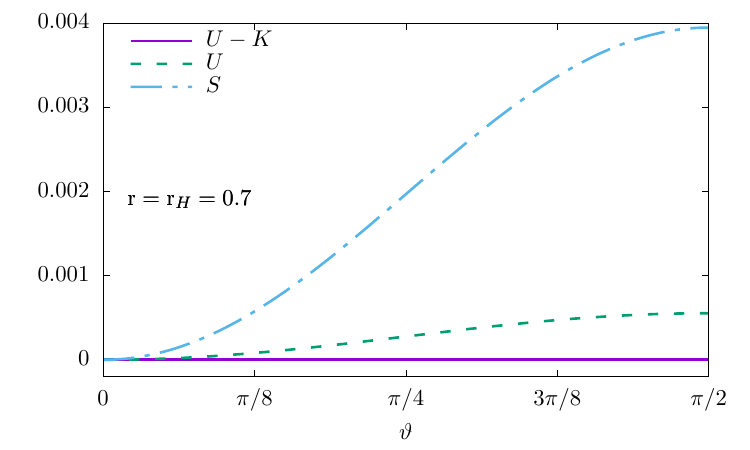}
    \caption[The surface gravity $\kappa_g$, the ratio of the horizon circumferences $L_\text{e}/L_\text{p}$ as functions of $r_H$ and the angular dependence of the metric functions $U-K,U,S$ for the hairy black hole solution with $r_H=0.7$.]{Top: the surface gravity $\kappa_g$ (left) and the ratio of circumferences $L_\text{e}/L_\text{p}$ (right) as functions of $r_H$ for the hairy black holes with $\kappa=10^{-3}$ and $\nu=2$. Bottom: the function $f(\vartheta)=\left(X(r_H,\vartheta)-X(r_H,0)\right)/X(r_H,\vartheta)$ with $X=\{U-K,U,S\}$ for the hairy black hole solution with $r_H=0.7$.}
    \label{amph_hor}
\end{figure}

We also check that the combination $U-K$ which determines the surface gravity,
\begin{equation}
\label{surf_grav_bis}
    \kappa_g=\left.\frac{N'}{2}e^{U-K}\right|_{r=r_H}=\left.\frac{1}{2r_H}e^{U-K}\right|_{r=r_H},
\end{equation}
is constant at the horizon. To illustrate this, we compare in the bottom panel of Fig.~\ref{amph_hor} the angular dependence of $U-K$ at $r=r_H$ with that $U$ and $S$ for the solution with $r_H=0.7$. As one can see, $U-K$ remains constant -- confirming once again that the constraints are fulfilled -- while the dependence on $\vartheta$ of $U(r_H)$ and $S(r_H)$ is clearly visible. We do not show the angular dependence of $K(r_H)$ as it closely resembles that of $S(r_H)$, see for example the difference $K-S$ shown in the Fig.~\ref{fig_bh_07_3}. One has $U-K\rvert_{r=r_H}=-0.0366$ which yields the value $\kappa_g=0.6885$.
\medskip

In the top left panel of Fig.~\ref{amph_hor}, we compare the surface gravity as a function of $r_H$ for the RN and for the hairy black holes. In both cases, when the horizon size decreases, the surface gravity first grows, then attains a maximal value and starts decreasing, approaching zero in the extremal limit. In this limit, one has $r_H\to 0$ so that the factor $1/r_H$ in Eq.~\eqref{surf_grav_bis} diverges but it is damped by the exponent $e^{U-K}$ approaching zero because the horizon value of $U-K$ becomes large and negative. The surface gravity for hairy black holes is always greater than or equal to that for RN black holes. When $r_H\to r_H^0-r_{-}^0$, the hairy solutions bifurcate with RN and the surface gravity evaluates to $\kappa_g=0.3376$.

In our system of coordinates, the horizon resides at a surface of constant radial coordinate, $r=r_H$. However, for the hairy black hole with $\nu=2$, the WS fields are angle-dependent at the horizon, see the Figs.~\ref{fig_bh_07_1} and \ref{fig_bh_07_2}. This suggests that the horizon is deformed. The deformation is revealed when measuring the horizon circumference along the equator, $L_\text{e}$, and its circumference along the poles, $L_\text{p}$,
\begin{equation}
    L_\text{e}=\left.2\pi r_H\,e^{S}\right|_{r=r_H,\vartheta=\pi/2},\quad\quad L_\text{p}=\left.2r_H\int_0^\pi{e^K\,d\vartheta}\right|_{r=r_H}.
\end{equation}
The deviation of the ratio $L_\text{e}/L_\text{p}$ from unity is shown against $r_H$ in the top right panel of Fig.~\ref{amph_hor}. This ratio is often referred to as the horizon \textit{sphericity} \cite{Delgado2018,Delgado2019}. For a spherical horizon, one has $L_\text{e}=L_\text{p}$, but for the hairy black hole solutions we find that
\begin{equation}
    0\;\leq\;\frac{L_\text{e}}{L_\text{p}}-1\;\lesssim\; 8\times 10^{-5}\quad\quad\Rightarrow\quad\quad L_\text{e}\geq L_\text{p}.
\end{equation}
Hence the horizon is an oblate ellipsoid but the deviation from spherical symmetry is small. The ratio $L_\text{e}/L_\text{p}$ approaches unity as $r_H$ reaches its upper bound, $r_H\to 1.4657$, and in the extremal limit, $r_H\to 0$. This can be easily understood since in the former case the hairy solutions bifurcate with RN, which is a spherically symmetric field configuration; whereas in the later case, the near-horizon geometry is that of a RN-de Sitter black hole, which is again spherically symmetric. For the intermediate solution with $r_H=0.7$ shown in Figs.~\ref{fig_bh_07_1}-\ref{fig_bh_07_3}, the deviation from spherical symmetry is almost maximal, $(L_\text{e}/L_\text{p})-1\approx  8\times 10^{-5}$.

\begin{figure}
    \centering
    \includegraphics[width=7.8cm]{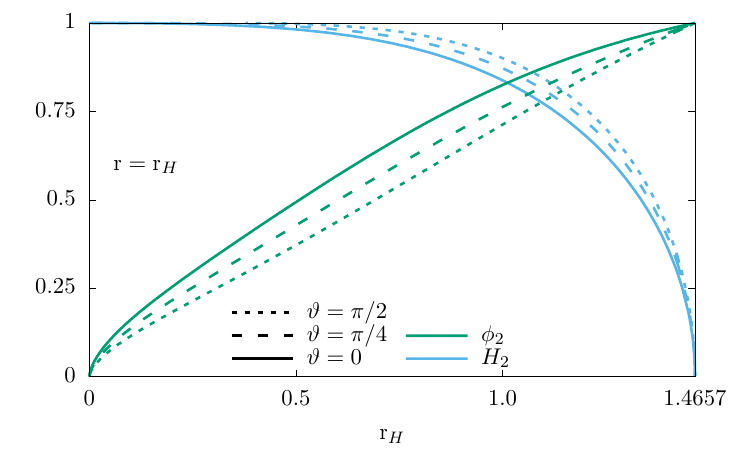}
    \includegraphics[width=7.8cm]{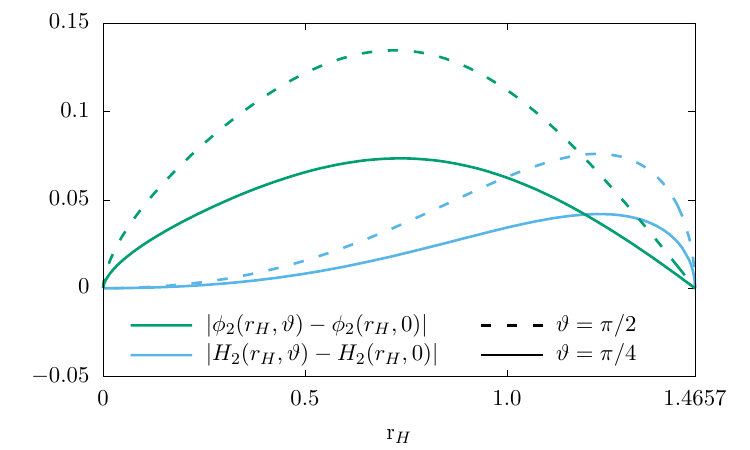}\\
    \vspace{0.2cm}
    \includegraphics[width=7.8cm]{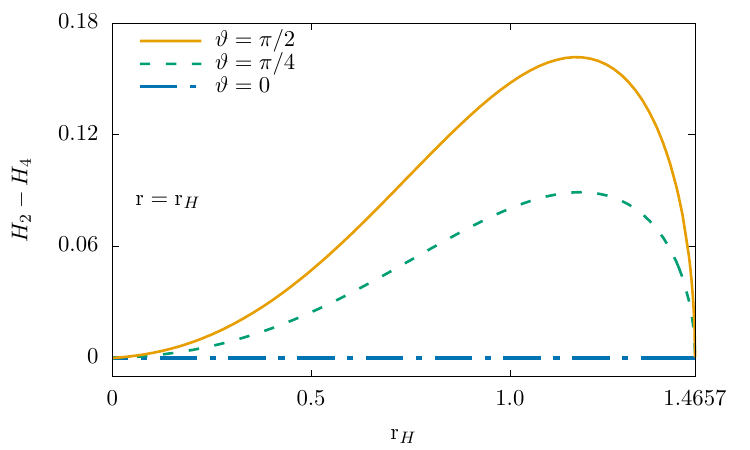}
    \includegraphics[width=7.8cm]{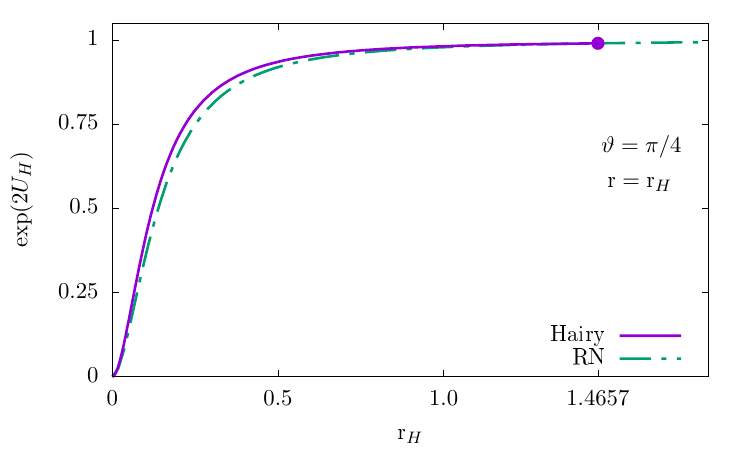}
    \caption[The horizon values of $\phi_2$ and $H_2$, the differences $|\phi_2(r_H,\vartheta)-\phi_2(r_H,0)|$, $|\phi_2(r_H,\vartheta)-\phi_2(r_H,0)|$, $H_2(r_H,\vartheta)-H_4(r_H,\vartheta)$ and the horizon value of $e^{2U}$ as functions of $r_H$.]{The horizon values of $\phi_2$ and $H_2$ (top left), the differences $|\phi_2(r_H,\vartheta)-\phi_2(r_H,0)|$, $|\phi_2(r_H,\vartheta)-\phi_2(r_H,0)|$ (top right), $H_2(r_H,\vartheta)-H_4(r_H,\vartheta)$ (bottom left) and the horizon value of $e^{2U}$ (bottom right) as functions of $r_H$.}
    \label{amph_hor_bis}
\end{figure}

In the whole range $r_H\in[0,1.4657]$, the functions $L_{e}(r_H)$ and $L_{p}(r_H)$ vary monotonically. For $r_H=0$ and $r_H=1.4657$, the horizon is spherical, hence the circumferences $L_\text{e}$, $L_\text{p}$ coincide and become directly related to the usual Schwarzschild horizon radius. In particular, we check that
\begin{equation}
    \left.\frac{L_\text{e}}{2\pi}\right|_{r_H=0}=r_H^\text{ex}\approx 0.0932,\quad\quad\left.\frac{L_\text{e}}{2\pi}\right|_{r_H=1.4657}=r_H^0\approx 1.4734,
\end{equation}
where $r_H^\text{ex}$ is the Schwarzschild horizon radius of the extremal RN-de Sitter black hole with $\nu=2$, while $r_H^0$ is the Schwarzschild radius of the RN black hole that bifurcates with the hairy solutions. This show again that the results obtained with our code are consistent. 

Additional information about the horizon values of the fields is presented in the Fig.~\ref{amph_hor_bis}. In the top left panel, the horizon values of $H_2$ and $\phi_2$ are shown for $\vartheta=\{0,\pi/2,\pi/4\}$. Remembering that these are functions which are non-vanishing in the spherically symmetric case ($|\nu|=1$), this panel is analogous to the right panel of Fig.~\eqref{hairy_bh_ews_2}. Specifically, $H_2(r_H,\vartheta)$ and $\phi_2(r_H,\vartheta)$ both lie in the range $[0,1]$. As the horizon size decreases, the value of $H_2$ approaches unity, while $\phi_2$ approaches zero. The key difference is that now, these horizon values depend on $\vartheta$ for intermediate values of $r_H$. The angular dependence of $H_2$ and $\phi_2$ at the horizon is highlighted in the top right panel of the Fig.~\ref{amph_hor_bis}. We compare their values at $\vartheta=0$ with those at $\vartheta=\{\pi/4,\pi/2\}$. Notably, the $\vartheta$-dependence of $\phi_2(r_H)$ is more pronounced when $r_H\approx 0.7$ -- coinciding with the maximal deviation from spherical symmetry of the horizon geometry. On the other hand, the maximal angular variations of $H_2(r_H)$ occur at $r_H\approx 1.2$.

In the spherically symmetric limit, one has also $H_4=H_2$. Therefore the difference $H_2-H_4$ also provides information about the deviation from spherical symmetry. The horizon value of this difference is shown in the bottom left panel of Fig.~\ref{amph_hor_bis} for $\vartheta=\{0,\pi/2,\pi/4\}$. One can see that $H_2-H_4$ vanishes both for $r_H=0$ and $r_H=1.4657$ -- in these limits the horizon is spherically symmetric. The difference between $H_4(r_H)$ and $H_2(r_H)$ becomes maximal at $r_H\approx 1.2$. This figure also demonstrates that the regularity condition $H_2=H_4$ for $\vartheta=0$ holds at the horizon. Of course we also check that it is fulfilled for $r>r_H$.

Finally, the bottom right panel of Fig.~\ref{amph_hor_bis} presents the horizon value of $e^{2U}$ at $\vartheta=\pi/4$ for RN and for hairy black holes. We do not show this horizon value for $\vartheta\neq\pi/4$ because $U$ is almost angle-independent. The two curves are very close to each other and the hairy branch bifurcates with RN at $r_H=1.4657$. In the extremal limit, $r_H\to 0$, one has $\exp[2U(r_H,\vartheta)]\to 0$ and therefore $U(r_H,\vartheta)\to -\infty$. 

\begin{figure}[b!]
    \centering
    \includegraphics[height=6.4cm]{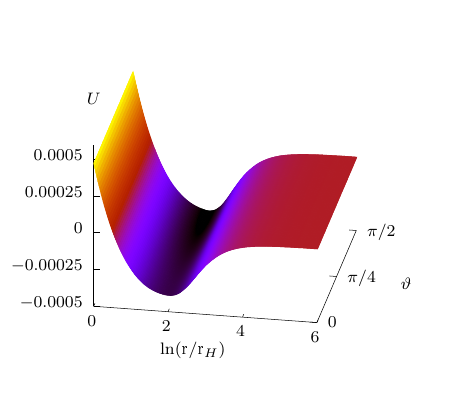}
    \includegraphics[height=6.4cm]{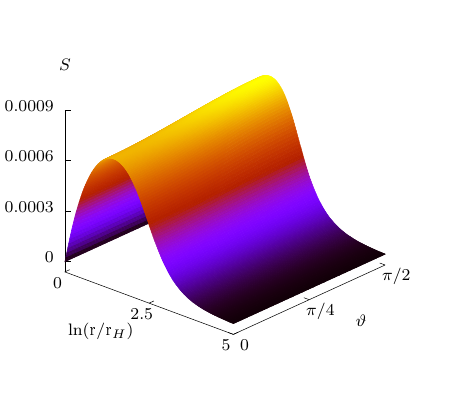}\\
    \vspace{-0.5cm}
    \includegraphics[height=6.4cm]{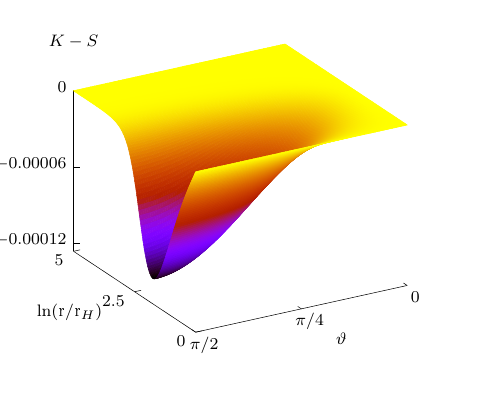}
    \includegraphics[height=6.4cm]{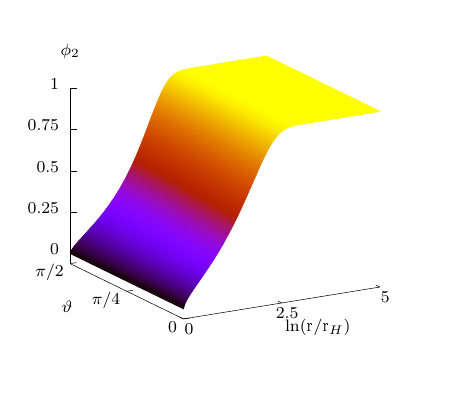}
    \caption{The metric functions $U$, $S$, $K-S$ and the Higgs amplitude $\phi_2$ for the extremal hairy black hole solution with $\kappa=10^{-3}$ and $\nu=2$.}
    \label{fig_bh_ex_nu2}
\end{figure}

It follows that the extremal solution cannot be obtained by using the metric parameterization \eqref{metric_ews} with the function $N(r)=1-r_H/r$. Instead, one should set $N(r)=k^2(r)(1-r_H/r)^2$, where $k(r)$ is given in Eq.~\eqref{k_ext} and $r_H=r_H^\text{ex}$ is determined by Eq.~\eqref{rhex_rnds}. One has then $N'(r_H)=0$ so that the surface gravity \eqref{surf_grav_bis} identically vanishes and no large values of the metric functions are needed. The profiles of $U$, $S$, $K-S$ and $\phi_2$ for the extremal solution with $\nu=2$ and $\kappa=10^{-3}$ are shown in the Fig.~\ref{fig_bh_ex_nu2}. While the metric function $U$ still shows little variation with respect to the polar angle, its $r$-dependence is no longer monotonic. The horizon becomes spherical so that $K-S$ now vanishes at $r=r_H$ and, furthermore, $S(r_H)=0$, indicating that $r_H$ coincides with the Schwarzschild horizon radius, $r_H=L_\text{e}/(2\pi)=L_\text{p}/(2\pi)=r_H^\text{ex}$. All the WS amplitudes are now constant at the horizon: $H_1,H_3,y,\phi_1,\phi_2$ all vanish, while $H_2,H_4$ equal to unity. Hence, close to the horizon, the spacetime geometry is extremal RN-de Sitter with the WS fields given by,
\begin{equation}
    Y=\nu\cos\vartheta\,d\varphi,\quad\quad W=\Phi=0.
\end{equation}
This is again a manifestation of the electroweak symmetry restoration in strong magnetic fields \cite{Ambjoern1990}.

\subsubsection{Non-abelian magnetic charge distribution}

\begin{figure}[b!]
    \centering
    \includegraphics[width=10cm]{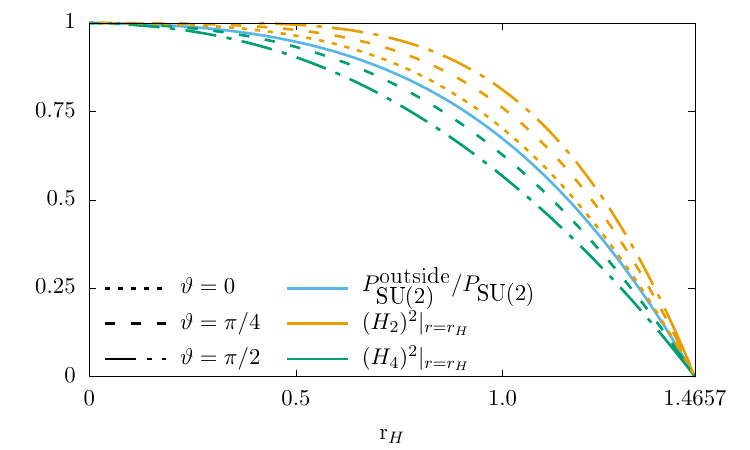}
    \caption[The ratio between the SU(2) charge in the outer black hole region, $P_\text{SU(2)}^\text{outside}$, and the total SU(2) charge, $P_\text{SU(2)}=-\nu g'/g$, for the hairy black holes with $\kappa=10^{-3}$ and $\nu=2$.]{The ratio between the SU(2) charge in the outer black hole region, $P_\text{SU(2)}^\text{outside}$, and the total SU(2) charge, $P_\text{SU(2)}=-\nu g'/g$, for the hairy black holes with $\kappa=10^{-3}$ and $\nu=2$. This ratio is compared to the horizon values of $(H_2)^2$ and $(H_4)^2$ at $\vartheta=\{0,\pi/4,\pi/2\}$.}
    \label{fig_p}
\end{figure}

The SU(2) part of the magnetic charge moves in the outer black hole region as the horizon size decreases. In the spherically symmetric case, the fraction of the SU(2) charge distributed in the outer region was given by the horizon value of a single function, according to the Eq.~\eqref{P_out_spher}. For axially symmetric hairy black holes, the situation is more complicated. The integral giving the SU(2) magnetic charge distributed outside the horizon,
\begin{equation}
    P_\text{SU(2)}^\text{outside}=\int_{r>r_H}{\rho_\text{SU(2)}\sqrt{-g}\,d^3x},
\end{equation}
cannot be evaluated analytically. Therefore, we compute this integral numerically and show in the Fig.~\ref{fig_p} the ratio $P_\text{SU(2)}^\text{outside}/P_\text{SU(2)}$, where $P_\text{SU(2)}=-\nu\,g'/g$ is the total SU(2) charge. At $r_H=1.4657$, when the hairy solutions bifurcates with RN, the SU(2) charge is confined inside the horizon so that $P_\text{SU(2)}^\text{outside}=0$. In the extremal limit, $r_H\to 0$, one has $P_\text{SU(2)}^\text{outside}=P_\text{SU(2)}$, hence the whole SU(2) charge is distributed in the exterior region. 

In the spherically symmetric limit ($|\nu|\to 1$), the functions $H_2$ and $H_4$ reduce to one single radial function $f(r)$, as shown in Eq.~\eqref{spher_sym_ws_fields}. The horizon value of $f^2$ coincide with the ratio $P_\text{SU(2)}^\text{outside}/P_\text{SU(2)}$. To understand what happens in the axially symmetric case, we also show in the Fig.~\ref{fig_p} the horizon values of $(H_2)^2$ and $(H_4)^2$ for $\vartheta=\{0,\pi/4,\pi/2\}$. It is instructive to see that the curve corresponding to the fraction of the SU(2) charge in the outer region lies between the curves for $(H_2)^2$ and $(H_4)^2$. For $\nu=1$, the same figure would show that all curves coincide with  each other. 

\begin{figure}
    \centering
    \includegraphics[width=5.5cm]{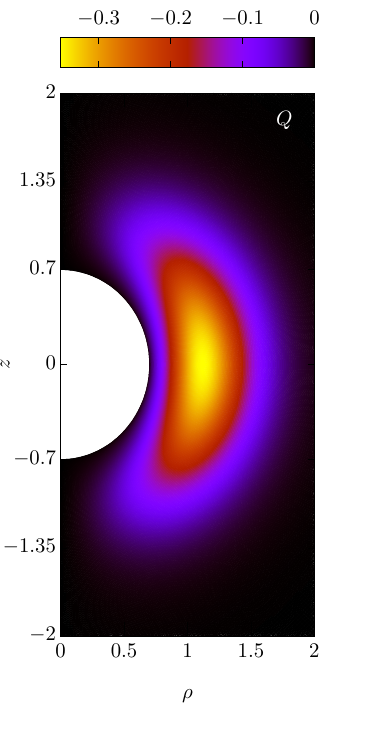}
    \hspace{1.5cm}
    \includegraphics[width=5.5cm]{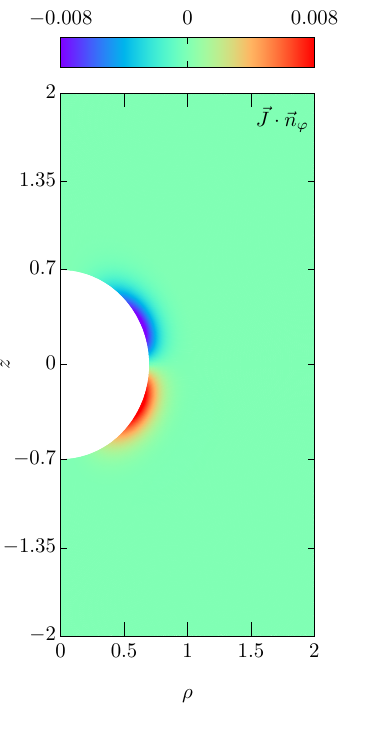}
    \vspace{-0.5cm}
    \caption[The magnetic charge and electric current densities against the cylindrical coordinates $\rho=r\sin\vartheta$, $z=r\cos\vartheta$, for the hairy black hole solution with $\kappa=10^{-3}$, $r_H=0.7$, $\nu=2$.]{The magnetic charge (left) and electric current (right) densities against the cylindrical coordinates $\rho=r\sin\vartheta$, $z=r\cos\vartheta$, for the hairy black hole solution with $\kappa=10^{-3}$, $r_H=0.7$, $\nu=2$.}
    \label{Qj_bh}
\end{figure}

For the flat space monopoles with $|\nu|>1$, the SU(2) charge distribution has a toroidal shape with a maximum located along a ring in the equatorial plane. In the black hole case, we can represent the magnetic charge distribution by the same function $Q$ as in the flat space theory, see the Eq.~\eqref{su2_charge}. The profile of $Q$ for our reference black hole solution with $\nu=2$ is shown in the left panel of Fig.~\ref{Qj_bh}. We plot this profile using non-compact cylindrical coordinates $(\rho,z)=(r\sin\vartheta,r\cos\vartheta)$ rather than their compactified counterparts, $(\bar{\rho},\bar{z})$, to provide a clearer view of the horizon, located at $r=0.7$. Remember, however, that the horizon is not spherical. The SU(2) magnetic charge is still confined within a torus centered in the equatorial plane; but this torus now contains only a fraction of the total charge. Notice that the magnetic charge density $Q$ vanishes at $r=r_H$. 

In the right panel of the Fig.~\ref{Qj_bh}, we present the electric current density for the reference solution. As for flat space monopoles, the only non-vanishing component of the electric current density is $J_\varphi$ and therefore, we plot the scalar product $\vec{J}\cdot\vec{n}_\varphi$ where $\vec{n}_\varphi$ is the unit vector in the azimuthal direction. The electric current density exhibits a minimum in the upper hemisphere and a maximum in the lower hemisphere, both very close to the horizon. They correspond to the superconducting rings of oppositely directed electric currents discussed in Sec.~\ref{interior_struct}. The current density clearly does not vanish at $r=r_H$. 

The magnetic torus containing a fraction of the SU(2) charge and the rings of oppositely directed electric currents constitute the black hole hair. It is made of a condensate of W bosons. When decreasing the horizon size, we observe that the magnetic ring remains fixed at $\rho\approx 1$, while the electric currents rings remain stuck to the horizon and decrease in size. This is consistent with the picture of the Fig.~\ref{iso_Qj} in flat space which shows that the circular electric currents are very close to the origin while the magnetic ring has a radius of order unity. 

As a result, the main feature of the magnetic hairy black holes as compared to their flat space counterparts is the presence of the horizon which regularizes the monopole energy. The horizon hides the singular U(1) magnetic charge as well as a fraction of the SU(2) magnetic charge from all observers in the exterior black hole region. The WS field configuration remains very similar to that of the monopoles, especially in the extremal limit when the horizon size shrinks, causing the entire SU(2) charge to be distributed outside the horizon.

\subsubsection{The mass}

In the spherically symmetric case, we decomposed the rescaled mass $\mathcal{M}$ of hairy black holes in terms of the "hair mass" $\mathcal{M}_h$ and the "horizon mass" $\mathcal{M}_H$ which are defined in Eq.~\eqref{M_h_H}. However, these cannot be defined in the general case when $|\nu|>1$ since their definitions rely on the ODE for the mass function $m(r)$ in \eqref{spher_sym_eq}. Instead, one should use the general definition \eqref{vol_mass_bis_ews} of the mass which can be represented as,
\begin{equation}
    \mathcal{M}=\frac{2\kappa_g A_H}{\kappa}+\mathcal{M}_\text{bulk}\quad\quad\text{with}\quad\quad\mathcal{M}_\text{bulk}=-\int_{r>r_H}{(2\tensor{T}{^0_0}-T)\sqrt{-g}\,d^3 x}.
\end{equation}
We find it convenient to decompose $\mathcal{M}_\text{bulk}$ into two parts,
\begin{equation}
    \mathcal{M}_\text{bulk}=\mathcal{M}_\text{U(1)}+\mathcal{M}_\text{SU(2)},
\end{equation}
where $\mathcal{M}_\text{U(1)}$ contains the contribution of the Abelian field $Y_\mu$ in the stress-energy tensor,
\begin{equation}
    \accentset{\text{U(1)}}{T}_{\mu\nu}=\frac{1}{g'^2}Y_{\mu\sigma}\tensor{Y}{_\nu^\sigma}-\frac{1}{4g'^2}g_{\mu\nu}Y_{\alpha\beta}Y^{\alpha\beta},
\end{equation}
while $\mathcal{M}_\text{SU(2)}$ accounts for the contributions of the non-Abelian field $W^a_\mu$ and the Higgs field $\Phi$. For the sake of completeness, we also consider the regularized energy of the WS fields that was introduced for the flat space monopoles,
\begin{equation}
    E_\text{reg}=\int_{r>r_H}{\left(\mathcal{E}-\frac{\nu^2}{2g'^2}\frac{e^{-2(K+S)}}{r^4}\right)\sqrt{-g}\,d^3x}.
\end{equation}
This integral coincide with the definition \eqref{def_Ereg} of $E_\text{reg}$ in the flat space limit.

\begin{figure}
    \centering
    \includegraphics[width=7.8cm]{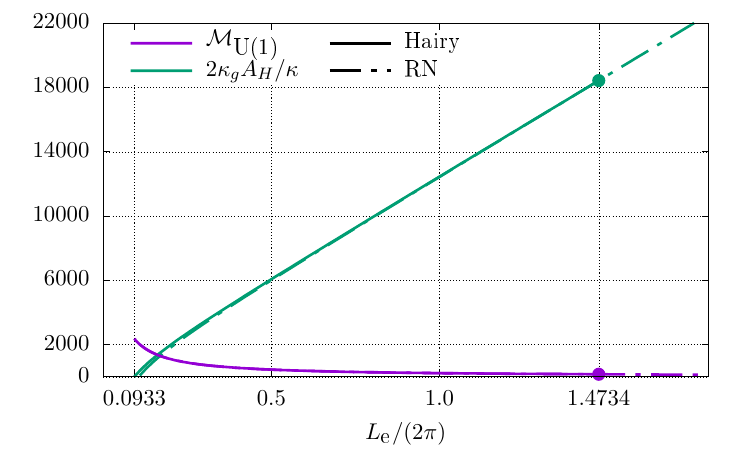}
    \includegraphics[width=7.8cm]{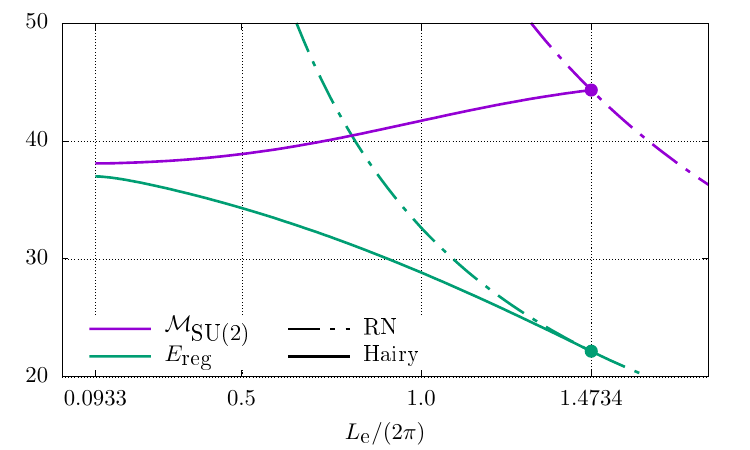}\\
    \vspace{0.2cm}
    \includegraphics[width=7.8cm]{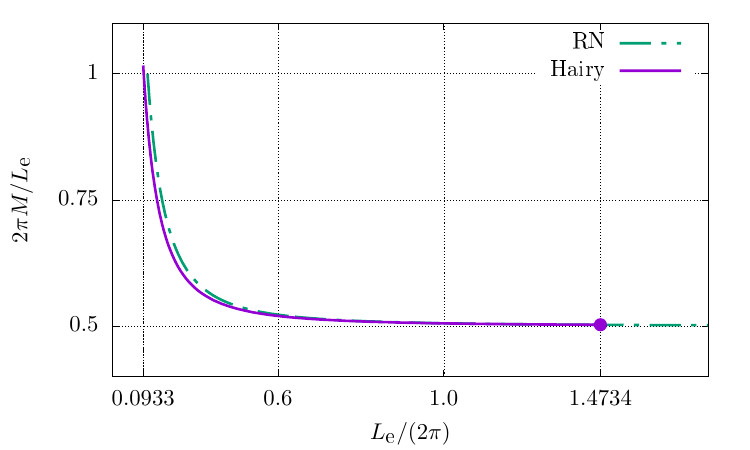}
    \caption[The mass of the Abelian field $\mathcal{M}_\text{U(1)}$, the horizon contribution to the mass, the mass of the non-Abelian fields $\mathcal{M}_\text{SU(2)}$, the regularized energy $E_\text{reg}$ and the total mass $M$ divided by $L_\textrm{e}/(2\pi)$ as functions of the equatorial horizon radius, $L_\textrm{e}/(2\pi)$, for RN and hairy black holes.]{The mass of the Abelian field $\mathcal{M}_\text{U(1)}$, the horizon contribution to the mass (top left), the mass of the non-Abelian fields $\mathcal{M}_\text{SU(2)}$, the regularized energy $E_\text{reg}$ (top right) and the total mass $M$ divided by $L_\textrm{e}/(2\pi)$ (bottom) as functions of the equatorial horizon radius, $L_\textrm{e}/(2\pi)$, for RN and hairy black holes with $\kappa=10^{-3}$ and $\nu=2$.}
    \label{masses_bh}
\end{figure}

For the RN solution, one has
\begin{equation}
\label{masses_rn}
    \frac{2\kappa_g A_H}{\kappa}=\frac{4\pi}{\kappa}(r_{+}-r_{-}),\quad\mathcal{M}_\text{U(1)}=\frac{4\pi\nu^2}{g'^2r_{+}},\quad\mathcal{M}_\text{SU(2)}=\frac{4\pi\nu^2}{g^2r_{+}}=2E_\text{reg},
\end{equation}
where $r_{+}=L_\text{e}/(2\pi)$ is the outer horizon radius of the RN geometry, and $r_{-}=\nu^2\kappa/(2e^2r_{+})$ is the inner horizon radius. For the hairy solutions with $\nu=2$, $\kappa=10^{-3}$, these quantities are presented as functions of the equatorial radius $L_\text{e}/(2\pi)$ in the top panels of Fig.~\ref{masses_bh}. They interpolate within the following limits,
\begin{align*}
    2.342\times 10^{3}\geq&\;\mathcal{M}_\text{U(1)}\geq\frac{4\pi\nu^2}{g'^2r_H^0}=148.3,\quad\quad0\leq\;\frac{2\kappa_g A_H}{\kappa}\leq\frac{4\pi}{\kappa}(r_H^0-r_{-}^0)=18.42\times 10^{3},\\
    38.08\leq&\;\mathcal{M}_\text{SU(2)}\leq\frac{4\pi\nu^2}{g^2r_H^0}=44.30,\quad\quad36.98\geq\;E_\text{reg}\geq\frac{2\pi\nu^2}{g^2r_H^0}=22.15.\numberthis
\end{align*}
The curves representing $\mathcal{M}_\text{U(1)}$ and $2\kappa_g A_H/\kappa$ for the hairy solutions almost coincide with those for RN black holes. Close to the bifurcation, the horizon contribution to the total mass is
\begin{equation}
    \frac{2\kappa_g A_H}{\kappa}=\frac{4\pi r_{+}}{\kappa}-\frac{4\pi r_{-}}{\kappa}=\frac{2L_\text{e}}{\kappa}-\frac{4\pi^2\nu^2}{e^2L_\text{e}}\approx\frac{2L_\text{e}}{\kappa}\quad\text{if}\quad\kappa\ll 1,
\end{equation}
which is the value for a Schwarzschild black hole. The distinction between the RN and the hairy branches becomes evident when looking at the non-Abelian mass $\mathcal{M}_\text{SU(2)}$, or the regularized energy $E_\text{reg}$. For RN black holes, these quantities grow up to large values of order of $1/\sqrt{\kappa}$ when the equatorial horizon radius $L_\text{e}/(2\pi)$ descends to its lower bound. For the hairy black holes, the non-Abelian mass \textit{decreases} with the equatorial radius down to the minimal value $\mathcal{M}_\text{SU(2)}\approx 38.08$, as seen in the top right panel of Fig.~\ref{masses_bh}. Therefore, just as in the spherically symmetric case, the hairy solutions are less energetic than the RN black holes of same size. The same conclusion holds if one is considering $E_\text{reg}$ instead of $\mathcal{M}_\text{SU(2)}$.

The total mass $M$ can be represented as,
\begin{equation}
\label{M_decompo}
    M=\frac{\kappa}{8\pi}\mathcal{M}=\frac{\kappa_g A_H}{4\pi}+\frac{\kappa}{8\pi}\left(\mathcal{M}_\text{U(1)}+\mathcal{M}_\text{SU(2)}\right).
\end{equation}
At the bifurcation with RN, the Abelian and non-Abelian masses are negligible in front of the horizon term, one has $M=L_\text{e}/(4\pi)+\mathcal{O}(\kappa)$ and hence $2\pi M/L_\text{e}$ approaches $1/2$, as seen in the bottom panel of Fig.~\ref{masses_bh}. In the extremal limit, the horizon term vanishes and the dominant contribution to the mass is from the Abelian field. The figure shows that the ratio $2\pi M/L_\text{e}$ approaches a value close to unity. Hence, one has for the extremal hairy black hole with $\nu=2$ and $\kappa=10^{-3}$,
\begin{equation}
\label{approx_mass_any_nu}
    M\approx\frac{L_\text{e}}{2\pi}=r_H^\text{ex}=\sqrt{\frac{\kappa}{2}}\frac{|\nu|}{g'}+\frac{\beta\,\kappa^{5/2}|\nu|^3}{32\sqrt{2}\,g'^3}+\mathcal{O}(|\nu|^5\kappa^{9/2}),
\end{equation}
where we have used the expression of $r_H^\text{ex}$ given in Eq.~\eqref{rhex_rnds}, and expanded up to the cubic order in $|\nu|$. We emphasize that this cubic term is negligible for the values of the parameters considered here. For the extremal RN black hole, the mass and radius are larger, given by $M=L_\text{e}/(2\pi)=\sqrt{\kappa}|\nu|/(\sqrt{2}gg')$, where the difference with the hairy solution is essentially the factor $1/g\approx 1.14$. 

\begin{figure}[b!]
    \centering
    \includegraphics[width=7.8cm]{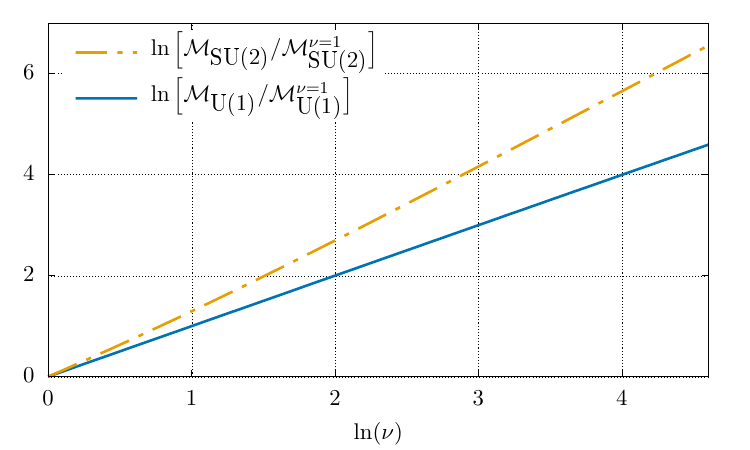}
    \includegraphics[width=7.8cm]{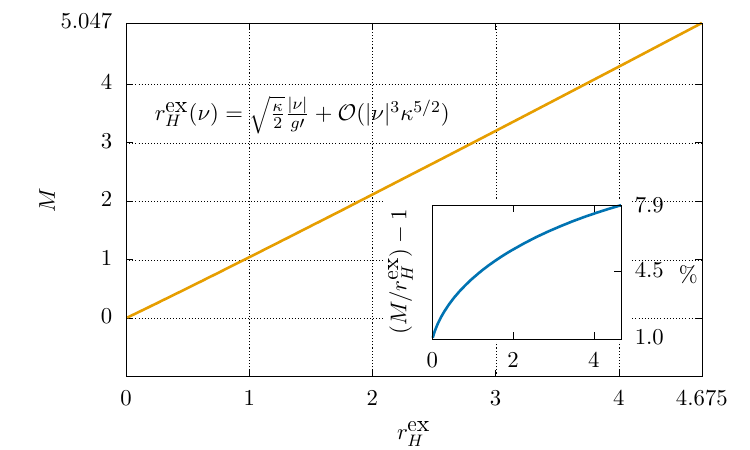}
    \caption[The logarithm of $\mathcal{M}_{SU(2)}/\mathcal{M}_{SU(2)}^{\nu=1}$ and of $\mathcal{M}_{U(1)}/\mathcal{M}_{U(1)}^{\nu=1}$ against $\ln(\nu)$ for the extremal hairy black hole and the total mass $M$ against the extremal horizon radius $r_H^\text{ex}$.]{Left: the logarithm of $\mathcal{M}_{SU(2)}/\mathcal{M}_{SU(2)}^{\nu=1}$ and of $\mathcal{M}_{U(1)}/\mathcal{M}_{U(1)}^{\nu=1}$ against $\ln(\nu)$ for the extremal hairy black hole solutions with $\kappa=10^{-3}$. Right: the total mass $M$ against the extremal horizon radius $r_H^\text{ex}$; the insertion shows the relative difference between $M$ and $r_H^\text{ex}$ in percentage.}
    \label{masses_bh_nu}
\end{figure}

Heretofore, we have provided a description of axially symmetric hairy black holes with $\nu=2$. We shall now consider only extremal solutions and increase the value of $\nu$ to see whether the mass hierarchy $\mathcal{M}_\text{U(1)}\gg\mathcal{M}_\text{SU(2)}$ still holds for $\nu>2$. For this, we consider values of $\nu$ ranging in the interval $[0,100]$. The solutions with $\nu<0$ are essentially identical to those with positive $\nu$, up to an overall change of sign in the magnetic charge and electric current densities. In the left panel of Fig.~\ref{masses_bh_nu}, we present the logarithm of the Abelian (resp. non-Abelian) mass $\mathcal{M}_\text{U(1)}$ (resp. $\mathcal{M}_\text{SU(2)}$) divided by its value for $\nu=1$. The figure indicates that the non-Abelian mass grows faster than the Abelian mass when $\nu$ increases. The slope of the curve for $\mathcal{M}_\text{U(1)}$ is almost equal to unity in the whole range of $\nu$ and therefore $\mathcal{M}_\text{U(1)}\propto|\nu|$. This agrees with our previous observation of the Abelian mass being approximately the same in the hairy case and in the RN case. Indeed, according to Eqs.~\eqref{masses_rn} and \eqref{approx_mass_any_nu}, one has in the extremal limit,
\begin{equation}
\label{approx_ab_mass}
    \mathcal{M}_\text{U(1)}=\frac{4\pi\nu^2}{g'^2r_H^\text{ex}}=\frac{4\pi\sqrt{2}}{\sqrt{\kappa}g'}|\nu|-\frac{\pi}{4}\frac{\beta\,\kappa^{3/2}}{\sqrt{2}g'^3}|\nu|^3+\mathcal{O}(|\nu|^5\kappa^{7/2}),
\end{equation}
where the cubic term is negligible compared to the linear term, even when $\nu=100$. Hence the linear approximation is valid.

For the non-Abelian mass, we do not have an analytical estimate. Therefore, we can only rely on the numerical results and find that the slope of the curve for $\mathcal{M}_\text{SU(2)}$ in the left panel of Fig.~\ref{masses_bh_nu} is approximately 1.5 for large $\nu$. Hence, it appears that $\mathcal{M}_\text{SU(2)}\propto\nu^{3/2}$, exhibiting the same behavior as the regularized energy of flat space monopoles, see Eq.~\eqref{estim_RE}. However, for the range of $\nu$ considered here, the Abelian mass is still larger than the non-Abelian mass. In the right panel of the Fig.~\ref{masses_bh_nu}, we present the total mass $M$ as function of the extremal horizon radius $r_H^\text{ex}$. The orange curve illustrates that the approximation $M\approx r_H^\text{ex}$ is no longer valid when $\nu$ increases. This suggests a growing significance of the non-Abelian mass in front of the Abelian mass. Indeed, using the estimate \eqref{approx_ab_mass} for $\mathcal{M}_\text{U(1)}$, the total mass $M$ for extremal hairy black holes can be represented as,
\begin{align*}
\label{estim_total_M}
	M=&\sqrt{\frac{\kappa}{2}}\frac{|\nu|}{g'}+\frac{\kappa}{8\pi}\mathcal{M}_\text{SU(2)}+\mathcal{O}(|\nu|^3\kappa^{5/2})=r_H^\text{ex}+\frac{\kappa}{8\pi}\mathcal{M}_\text{SU(2)}+\mathcal{O}(|\nu|^3\kappa^{5/2}),\numberthis
\end{align*}
and therefore the essential difference between $M$ and $r_H^\text{ex}$ comes from $\mathcal{M}_\text{SU(2)}$. The insertion in the right panel of the Fig.~\ref{masses_bh_nu} presents the relative difference $(M/r_H^\text{ex})-1$. It reaches approximately 8\% for $r_H^\text{ex}=\left.r_H^\text{ex}\right|_{\nu=100}\approx 4.675$. All this suggests that the non-Abelian mass will eventually become comparable to the Abelian mass for a value of $\nu>100$.

Summarizing, for the range $|\nu|\in[0,100]$ considered here, one always has $\mathcal{M}_\text{U(1)}>\mathcal{M}_\text{SU(2)}$ but the non-Abelian contribution to the total mass becomes more and more important so that the approximation $M\approx r_H^\text{ex}$ is no longer valid when $|\nu|\gg 1$. What happens for very large values of $|\nu|$ remains unclear, but we expect the solutions to approach a limiting configuration as $|\nu|\to\nu_\text{max}=2g'/(\kappa\sqrt{\beta})$, see Eq.~\eqref{nu_max}. We leave this analysis for a future project \cite{GervalleInPrep}.

\subsubsection{Physical value of $\kappa$ and limit of large magnetic charge}

So far, we have been discussing axially symmetric hairy black holes with the nonphysical value $\kappa=10^{-3}$. This is actually sufficient to understand many properties of the hairy black holes with the physical value $\kappa=5.42\times 10^{-33}$. For example, the event horizon is spherical at the bifurcation with RN and in the extremal limit, which correspond respectively to the maximum and minimum horizon sizes,
\begin{equation}
    \left.\frac{L_\text{e}}{2\pi}\right|_\text{max}=r_H^0\propto\sqrt{|\nu|},\quad\quad\left.\frac{L_\text{e}}{2\pi}\right|_\text{min}=r_H^\text{ex}=\sqrt{\frac{\kappa}{2}}\frac{|\nu|}{g'}+\mathcal{O}(|\nu|^3\kappa^{5/2}).
\label{min_max_Le}
\end{equation}
The values of $r_H^0(\nu)$ for the physical value of $\kappa$ are given in the Table~\ref{rhn_nu} or in the Fig.~\ref{rh0nu}. The deviation from spherical symmetry of the event horizon geometry is maximal at an intermediate value of its size, when the black hole hair is close to the horizon. For $|\nu|=2$, this maximal deviation is very small, already for $\kappa=10^{-3}$. Hence, small magnetic charge hairy black holes with the physical value of $\kappa$ have an almost spherical horizon -- the back-reaction of the WS field configuration on the spacetime geometry being smaller when $\kappa$ decreases. Of course, the maximal deviation from a spherically symmetric horizon is expected to grow when $|\nu|$ increases. 

The physically most interesting hairy magnetic black holes are those which are close to extremality, since the horizon size shrinks when the black hole evaporates. According to Maldacena \cite{Maldacena2021}, the Hawking radiation is even enhanced for magnetically charged black holes, which accelerates their evaporation. Our analysis actually highlighted that the extremal hairy black holes are very similar to their flat space counterparts which have been extensively studied in the Sec.~\ref{monop_ws}. The main difference is the presence of the event horizon which is of great importance as it regularizes the monopole energy. The horizon contains the U(1) part of the magnetic charge and it is surrounded by a toroidal SU(2) magnetic charge distribution and two rings of oppositely directed electric currents. We expect the picture of Fig.~\ref{iso_Qj_large_nu} describing the internal structure of monopoles to accurately describes as well the extremal hairy black holes as long as the winding number $|\nu|$ is not too large. The central U(1) singularity in this figure just has to be replaced by a small horizon. 

Eq.~\eqref{min_max_Le} actually reveals that extremal hairy magnetic black holes cannot exist for arbitrarily large values of $|\nu|$. Indeed, the minimal value of horizon size is proportional to $|\nu|$ whereas its maximal value is $r_H^0\propto\sqrt{|\nu|}$. Hence the extremal horizon radius for hairy black holes grows faster than the horizon size corresponding to the bifurcation with RN. Moreover, the Eq.~\eqref{estim_RE} shows that the region of Higgs false vacuum for flat space monopoles has a radius $R\propto\sqrt{|\nu|}$. Assuming this estimate to hold in the black hole case, the bubble of Higgs false vacuum grows with $|\nu|$ slower than the extremal horizon radius. However, the latter cannot be larger than $R$ since the WS fields are supposed to be in the Higgs false vacuum at the extremal horizon. This again shows that more complicated effects are expected to occur when $|\nu|$ becomes very large. 

To estimate at which value of $|\nu|$ our description of extremal hairy black holes breaks down, we extrapolate our numerical results on the values $r_H^0(\nu)$ (Fig.~\ref{rh0nu}) and find that
\begin{equation}
    r_H^0(\nu) < r_H^\text{ex}(\nu)\quad\quad\text{for}\quad\quad|\nu|\gtrsim 10^{32}.
\end{equation}
This is of the same order of magnitude as the maximal value of $\nu$ mentioned in \eqref{nu_max}, which prevents the occurrence a naked singularity for RN-de Sitter black holes. It also agrees with the estimates made by Maldacena \cite{Maldacena2021}. Understanding the behavior of the fields near this limit is a current challenge \cite{GervalleInPrep}, as it requires considering the Einstein-Weinberg-Salam equations assuming very large values for the winding number $\nu$, and very small values for the gravitational coupling $\kappa$.

As a result, we are confident in our understanding of extremal hairy black holes if their magnetic charge $|P|=|\nu|/e$ satisfies $|\nu|\ll 10^{32}$. In this case, the backreaction of the non-Abelian hair on the spacetime geometry is negligible and the dominant contribution to the total mass is from the Abelian field. The dimensionful mass and horizon size are, in Planck units,
\begin{equation}
\label{approx_mass_gen}
    \boldsymbol{M}=\frac{2e^2}{\kappa\alpha}M\times\boldsymbol{m}_0\approx\frac{|\nu|g}{\sqrt{\alpha}}\times\boldsymbol{M}_\text{Pl},\quad\frac{\boldsymbol{L}_\text{e}}{2\pi}=r_H^\text{ex}\times\boldsymbol{\ell}_0\approx\frac{|\nu|g}{\sqrt{\alpha}}\times\boldsymbol{L}_\text{Pl}.
\end{equation}
Hence one has $\boldsymbol{M}/\boldsymbol{M}_\text{Pl}\approx\boldsymbol{L}_e/(2\pi\boldsymbol{L}_\text{Pl})$, as for extremal RN black holes. The latter are larger in mass and size than their hairy counterparts by a factor of $1/g\approx 1.14$. The approximation \eqref{approx_mass_gen} holds as long as Abelian mass $\mathcal{M}_\text{U(1)}$ in \eqref{M_decompo} is dominant in front of the non-Abelian mass $\mathcal{M}_\text{SU(2)}$. For large $|\nu|$, we find that $\mathcal{M}_\text{SU(2)}$ scales as $|\nu|^{3/2}$, the non-Abelian mass therefore becomes more and more significant, and it could ultimately overcome the Abelian mass $\mathcal{M}_\text{U(1)}$, which can be estimated using the Eq.~\eqref{approx_ab_mass}.  

\section{Summary and concluding remarks}
\label{conclu_ews}

We have constructed static black holes with a non-Abelian magnetic hair in the framework of the bosonic part of the Weinberg-Salam electroweak theory minimally coupled to GR. In the simplest case, these black holes are spherically symmetric and carry a total magnetic charge of $P=\pm 1/e$. They were previously reported in Ref.~\cite{Bai2021} and are gravitating counterparts of the electroweak monopole of Cho and Maison \cite{Cho1996}. The new solutions we have constructed are axially symmetric and exist for any value of $P$, but they are free of the Dirac string singularity only if their magnetic charge is an integer multiple of $1/e$, specifically, $P=P_\text{U(1)}+P_\text{SU(2)}=-\nu/e$ with $\nu\in\mathbb{Z}$. Far away from the event horizon, the solutions become purely electromagnetic and the field configuration approaches that of a RN black hole with magnetic charge $P$. At some finite distance from the horizon, the solutions support a non-Abelian hair made of a condensate of W bosons, which carries a fraction of the total SU(2) magnetic charge. The remaining part of the SU(2) charge, along with the U(1) contribution to the magnetic charge, is confined inside the event horizon.

For $|\nu|=1$, the black hole hair is contained within a spherical shell whose radius is $\boldsymbol{R}\sim 10^{-16}\,\text{cm}$, which is of the order of the electroweak length scale. The horizon has a size ranging from the minimal value $\boldsymbol{r}_H^\text{ex}\sim 10\,\boldsymbol{L}_\text{Pl}\sim 10^{-32}\,\text{cm}$ in the extremal limit, up to the maximal value $\boldsymbol{r}_H^0\sim 10^{-16}\,\text{cm}$. Therefore, in this upper limit, the black hole hair is absorbed inside the horizon and the solution becomes the Abelian RN configuration. Of course, it is essential to emphasize that the physical relevance of spherically symmetric hairy black holes should be considered in context, as quantum effects, which are entirely disregarded in this study, are expected to play a significant role at these scales. For $|\nu|>1$, the black hole hair consists of a toroidal distribution of SU(2) magnetic charge plus two superconducting rings of oppositely directed electric currents. If $|\nu|\ll 10^{32}$, the maximal value of the horizon size scales as $\sqrt{|\nu|}$ and it corresponds again to the bifurcation of hairy solutions with the RN branch. The minimal value of the horizon size is $\boldsymbol{r}_H^\text{ex}\sim 10|\nu|\boldsymbol\times\boldsymbol{L}_\text{Pl}$, it corresponds to extremal hairy black holes. What happens for extremely large magnetic charges remains unclear, but we expect the hairy solutions to approach a limiting configuration with $|P|\sim 10^{32}/e$ and with an event horizon radius of the order of 3 cm.

The flat space Dirac monopoles and their gravitating RN counterparts of small sizes are unstable for $|P|\geq 1/e$. On the contrary, the CM monopole, which has $P=\pm 1/e$, is stable~\cite{Gervalle2022a}. Therefore, the spherically symmetric hairy black holes, which generalize the CM monopole in the presence of gravity and are always less energetic than RN, are expected to be stable as well. Now, let us consider a RN black hole of large horizon size with $P=\pm 1/e$. Its horizon radius shrinks as it evaporates through the Hawking radiation process until $\boldsymbol{r}_H=\boldsymbol{r}_H^0$. At this moment, the instability is triggered, and the black hole begins to radiate part of its energy. The perturbations that grow in time during this process are spherically symmetric and hence it is conceivable that the remnant of the RN black hole decay is the hairy solution with the same magnetic charge, since it is itself spherically symmetric and presumably stable. The Hawking radiation then continues to reduce the size of the hairy black hole, until it reaches the extremal state for which $\boldsymbol{r}_H=\boldsymbol{r}_H^\text{ex}$. At this juncture, the black hole can no longer undergo further evaporation as its temperature is vanishing.

A similar scenario is expected to apply for black hole with larger magnetic charges, and it leads to the (extremal) axially symmetric hairy solutions considered in this chapter. However, one should keep in mind that the instability of small RN black holes is characterized by spherical harmonics with angular momentum $j=|\nu|-1$. The azimuthal quantum number $m$ can assume $2j+1$ different values, $m=-j,\dots,j$, and only for $m=0$ the perturbations are axially symmetric. One may therefore conjecture that this instability leads to the formation of non-Abelian hairy black holes with no spatial symmetry at all, or perhaps only discrete symmetries. The possible existence of such black holes was recently advocated by Maldacena \cite{Maldacena2021}. While we have focused on the special case of axial symmetry, the numerical construction of more general solutions with no continuous isometries remains an open challenge, as it requires solving the underlying full 3D elliptic problem. 

The physically most interesting solutions are the extremal ones as they are the end state of the Hawking evaporation process. The extremal hairy black holes are very similar to the flat space monopoles. The major difference is that the central U(1) Coulombian singularity of monopoles is now hidden inside the event horizon so that everything is regular from the outside. In particular the ADM mass of hairy black holes is finite, and in the extremal limit, it is of the order of $|\nu|g/\sqrt{\alpha}\times\boldsymbol{M}_\text{Pl}\approx 10|\nu|\times\boldsymbol{M}_\text{Pl}$. Close to the horizon, the hypercharge U(1) field is strong enough to suppress all other fields and restore the full electroweak gauge symmetry within a spheroidal bubble of size $\boldsymbol{R}\sim\sqrt{|\nu|}\times 10^{-16}\,\text{cm}$. Inside this bubble of Higgs false vacuum, the spacetime geometry is that of a RN-de Sitter black hole. Outside the bubble, the non-linearly interacting fields produce the SU(2) magnetically charged ring sandwiched between the loops of oppositely directed electric currents. The magnetic ring gives the leading contribution to the electromagnetic quadrupole moment $\boldsymbol{q}\propto|P_\text{SU(2)}|\boldsymbol{R}^2$.

It is likely that our solutions can be generalized to describe pairs of magnetic black holes with opposite magnetic charges, as was the case for the 't Hooft-Polyakov monopoles \cite{Kleihaus2004,Kleihaus2000}.

Finally, it remains to provide a scenario for the formation of these electroweak hairy black holes. Production of \textit{magnetically charged} black holes from primordial fluctuations in the early Universe is unlikely \cite{Bousso1996}. A more plausible scenario would involve the separate production of, first, monopoles and then, neutral black holes. Since monopoles in the electroweak theory are always singular, these primordial monopoles must be considered within a GUT framework \cite{Zeldovich1978,Preskill1979}. Subsequently, primordial black holes could swallow the monopoles, resulting in magnetically charged black holes (see Ref.~\cite{Zhang2023} for a recent discussion). This mechanism was proposed by Maldacena \cite{Maldacena2021}, and it draws inspiration from the work in Refs.~\cite{Stojkovic2005,Bai2020a}. Once they are produced, the Hawking evaporation of the magnetic black holes would certainly yield to the formation of the non-Abelian hair and they may have reached extremality by now. For a general discussion about methods to detect them, we refer to the work of Bai \textit{et al.}~\cite{Bai2020} (see also the Refs.~\cite{Bai2021a,Ghosh2021a,Estes2023}).

Since extremal hairy magnetic black holes are thermodynamically stable, it is very appealing to promote them as dark matter candidate. The astrophysical constraints on the abundance of magnetically charged black holes are qualitatively similar to those for monopoles. The first constraint is called the \textit{Parker bound}, and it arises from the observation of large magnetic fields within galaxies \cite{Parker1970,Turner1982}. These magnetic fields would accelerate magnetically charged objects, causing them to extract energy from the field. If the abundance of monopoles is too high, it could lead to the disappearance of the magnetic field. A similar constraint can be derived from the recent observation of primordial magnetic fields in the intergalactic void \cite{Kobayashi2022}. These constraints on the abundance of monopoles greatly limit their relevance as dark matter candidates. Although considering magnetically charged black holes instead of monopoles changes quantitative aspects of the Parker bound \cite{Stojkovic2005}, a recent preprint suggests that this would not be sufficient to resolve the monopole problem \cite{Zhang2023a}. Additional constraints arise from the potential catalysis of proton decay by monopoles \cite{Rubakov1982,Callan1984} and the implications of this phenomenon for astronomical bodies. For a discussion of this effect in the context of magnetic black holes, see for example the Ref.~\cite{Diamond2022}.

As a result, considering electroweak hairy black holes as dark matter candidates seems to be very speculative for the moment. Nonetheless, these objects remain of particular interest as they provide insights into very high energy phenomena, with a restoration of the full electroweak symmetry in their near-horizon region.

\begin{subappendices}

\section{Ansatz with time dependence}
\label{oscillons}

In this Appendix, we present a generalized ansatz for spherically symmetric solutions which allows for time dependence. For this, the metric can be chosen in the form
\begin{equation}
\label{spher_met_time}
    g_{\mu\nu}dx^\mu dx^\nu=-\sigma^2N dt^2+\frac{dr^2}{N}+R^2\left(d\vartheta^2+\sin^2\vartheta\,d\varphi^2\right),
\end{equation}
where $\sigma$, $N$ and $R$ are functions of $t$, $r$. It is invariant under the action of the SO(3) spatial rotations.

The spherically symmetric gauge fields should be invariant under the combined action of spatial rotations and gauge transformations \cite{Forgacs1980}. It follows that the most general SO(3)-invariant SU(2) gauge field can be represented in the form
\begin{align*}
    W=T_a W^a_\mu dx^\mu=T_3(a_0\,dt+a_1\,dr)&+(w_2\,T_1+w_1\,T_2)d\vartheta \\
\label{spher_ansatz_w}
    &+\nu(w_2\,T_2-w_1\,T_1)\sin\vartheta\,d\varphi+T_3\,\nu\cos\vartheta\,d\varphi,\numberthis{}
\end{align*}
where $a_0$, $a_1$, $w_1$, $w_2$ are functions of $t$, $r$ and $\nu$ is a constant parameter. This field generalizes the well-known ansatz of Witten \cite{Witten1977}. Written in this gauge, the field is singular at the $z$-axis, but the singularity can be removed if $\nu\in\mathbb{Z}$ in the general case, or if $2\nu\in\mathbb{Z}$ in the special case when $w_1=w_2=0$ \cite{Gervalle2022a}.

The spherically symmetric U(1) gauge field is
\begin{equation}
\label{spher_ansatz_y}
    Y_\mu dx^\mu=b_0\,dt+b_1\,dr+\nu\,\cos\vartheta\,d\varphi,
\end{equation}
where $b_0$ and $b_1$ depend on $t$, $r$. This field is also singular at the $z$-axis unless $2\nu\in\mathbb{Z}$. Finally, the spherically symmetric Higgs field is
\begin{equation}
\label{spher_ansatz_higgs}
    \Phi=\phi\,e^{i\xi/2}\begin{pmatrix}
        0 \\ 1
    \end{pmatrix},
\end{equation}
where $\phi$ and $\xi$ depend on $t$, $r$.

The ansatz for the WS fields \eqref{spher_ansatz_w}-\eqref{spher_ansatz_higgs} preserves its form under gauge transformations \eqref{gauge_trans_ews} generated by
\begin{equation}
    U=\exp\left(\frac{i}{2}\lambda+\frac{i}{2}\gamma\,\tau_3\right),
\end{equation}
where $\lambda$, $\gamma$ are arbitrary functions of $t$, $r$. Its effect on the 8 field amplitudes is
\begin{align*} 
    w_1&\to w_1\cos\gamma-w_2\sin\gamma,\quad w_2\to w_1\sin\gamma+w_2\cos\gamma,\quad a_0\to a_0+\dot{\gamma},\\
    a_1&\to a_1+\gamma',\quad b_0\to b_0+\dot{\lambda},\quad b_1\to b_1+\lambda',\quad\phi\to\phi,\quad\xi\to\xi+\lambda-\gamma.\numberthis
\end{align*}
It is then convenient to set $w_1=f\cos\alpha$ and $w_2=f\sin\alpha$ so that $f\to f$, $\alpha\to\alpha+\gamma$. It follows that the combinations
\begin{equation}
    \Omega_0\equiv a_0-\dot{\alpha},\quad\Omega_1\equiv a_1-\alpha',\quad\Theta_0\equiv a_0-b_0+\dot{\xi},\quad\Theta_1\equiv a_1-b_1+\xi',
\end{equation}
and also $f$, $\phi$ are gauge-independent. Injecting the ansatz \eqref{spher_met_time}-\eqref{spher_ansatz_higgs} to the field equations \eqref{eqs_ews}, \eqref{ein_eq_ews}, the angular dependence separates and the gauge-dependent amplitudes $\alpha$, $\xi$ drop out from the equations.

The field equations consist of second-order PDEs depending on $t$, $r$, but also first-order differential constraints and an algebraic constraint. The latter is
\begin{equation}
\label{alg_const}
    (\nu^2-1)f\phi=0,
\end{equation}
hence either $\nu^2=1$, $f=0$ or $\phi=0$. Then the equations for $f$ and $\phi$ read
\begin{align*}
    \frac{\partial}{\partial r}\big(N\sigma f'\big)-\frac{\partial}{\partial t}\left(\frac{\dot{f}}{N\sigma}\right)&=\sigma\left(\frac{f^2-1}{R^2}+N\Omega_1^2-\frac{\Omega_0^2}{N\sigma^2}+\frac{g^2}{2}\phi^2\right)f,\\
\label{gen_f_phi_eq}
    \frac{\partial}{\partial r}\big(R^2 N\sigma\phi'\big)-\frac{\partial}{\partial t}\left(\frac{R^2\dot{\phi}}{N\sigma}\right)&=\frac{1}{4}\sigma R^2\left(\beta(\phi^2-1)+N\Theta_1^2-\frac{\Theta_0^2}{N\sigma^2}+\frac{2}{R^2}f^2\right)\phi,\numberthis
\end{align*}
while the equations for $\Omega_0$, $\Omega_1$, $\Theta_0$, $\Theta_1$ are 
\begin{align*}
    \frac{\partial}{\partial r}\left(\frac{R^2}{\sigma}\left(\Omega_0'-\dot{\Omega}_1\right)\right)&=\frac{2}{N\sigma}f^2\,\Omega_0+\frac{g^2}{2N\sigma}R^2\phi^2\,\Theta_0,\\
    \frac{\partial}{\partial t}\left(\frac{R^2}{\sigma}\left(\Omega_0'-\dot{\Omega}_1\right)\right)&=2N\sigma f^2\,\Omega_1+\frac{g^2}{2}N\sigma R^2\phi^2\,\Theta_1,\\
    \frac{\partial}{\partial r}\left(\frac{R^2}{\sigma}\left(\Theta_0'-\dot{\Theta}_1\right)\right)&=\frac{2}{N\sigma}f^2\,\Omega_0+\frac{1}{2N\sigma}R^2\phi^2\,\Theta_0,\\
\label{elec_eq}
    \frac{\partial}{\partial t}\left(\frac{R^2}{\sigma}\left(\Theta_0'-\dot{\Theta}_1\right)\right)&=2N\sigma f^2\,\Omega_1+\frac{1}{2}N\sigma R^2\phi^2\,\Theta_1.\numberthis
\end{align*}
In addition there are two first-order differential constraints,
\begin{equation}
    \frac{\partial}{\partial r}\big(N\sigma f^2\,\Omega_1\big)=\frac{\partial}{\partial t}\left(\frac{f^2}{N\sigma}\Omega_0\right),\quad\frac{\partial}{\partial r}\big(N\sigma R^2\phi^2\Theta_1\big)=\frac{\partial}{\partial t}\left(\frac{R^2\phi^2}{N\sigma}\Theta_0\right),
\end{equation}
but these are not independent and follow from \eqref{elec_eq} since the $t$-derivatives of the first and third equations in \eqref{elec_eq} must coincide with the $r$-derivatives of the second and fourth equations. The Einstein equations are rather involved and we shall not show them explicitly. They consist of three second-order PDEs for $N$, $\sigma$, $R$, plus one differential constraint. 

In the static and purely magnetic case, one can set 
\begin{equation}
    \Omega_0=\Omega_1=\Theta_0=\Theta_1=0,\quad R=r,
\end{equation}
and the remaining amplitudes should depend only on $r$. The equations \eqref{gen_f_phi_eq}, \eqref{elec_eq} and the Einstein equations then reduce to the system of ODEs \eqref{spher_sym_eq}.

As discussed in the main text, the field configuration always contains the Coulombian  U(1) singularity at $r=0$. The only way to avoid it is to set $\nu=0$, choosing then $f=0$ to satisfy the algebraic constraint \eqref{alg_const}, the Eqs.\eqref{gen_f_phi_eq}, \eqref{elec_eq} reduce to
\begin{align*}
    \frac{\partial}{\partial r}\big(R^2 N\sigma\phi'\big)-\frac{\partial}{\partial t}\left(\frac{R^2\dot{\phi}}{N\sigma}\right)&=\frac{1}{4}\sigma R^2\left(\beta(\phi^2-1)+N\Theta_1^2-\frac{\Theta_0^2}{N\sigma^2}\right)\phi,\\
    \frac{1}{R^2}\frac{\partial}{\partial r}\left(\frac{R^2}{\sigma}\left(\Omega_0'-\dot{\Omega}_1\right)\right)&=\frac{g^2}{2N\sigma}\phi^2\,\Theta_0,\quad\frac{1}{R^2}\frac{\partial}{\partial t}\left(\frac{R^2}{\sigma}\left(\Omega_0'-\dot{\Omega}_1\right)\right)=\frac{g^2}{2}N\sigma\phi^2\,\Theta_1,\\
\label{gen_oscillon}
    \frac{1}{R^2}\frac{\partial}{\partial r}\left(\frac{R^2}{\sigma}\left(\Theta_0'-\dot{\Theta}_1\right)\right)&=\frac{1}{2N\sigma}\phi^2\,\Theta_0,\quad\frac{1}{R^2}\frac{\partial}{\partial t}\left(\frac{R^2}{\sigma}\left(\Theta_0'-\dot{\Theta}_1\right)\right)=\frac{1}{2}N\sigma\phi^2\,\Theta_1,\numberthis
\end{align*}
while the Einstein equations remain complicated. One can show that non-trivial static solutions of these equations have infinite energy. However, there are time-dependant solutions with a finite energy. In the flat space limit, solutions of this type were previously studied assuming a different ansatz for the Higgs field, in which case the fields are not spherically symmetric unless for $\theta_\text{W}=0$ \cite{Ratra1988,Farhi2005,Graham2007,Graham2007a}. On the other hand, the Eq.~\eqref{gen_oscillon} describes spherically symmetric systems for any value of the Weinberg angle.

Let us set the gauge amplitudes to zero, $\Omega_0=\Omega_1=\Theta_0=\Theta_1=0$, and consider the flat space limit, $\kappa=0$, $N=\sigma=1$, $R=r$. Rescaling the spacetime coordinates via $t\to(2/\sqrt{\beta})\,t$ and $r\to(2/\sqrt{\beta})\,r$, the remaining Higgs equation reduces to
\begin{equation}
    \frac{1}{r^2}\left(r^2\phi'\right)'-\ddot{\phi}=(\phi^2-1)\phi.
\end{equation}
This equation has been extensively studied in the literature, and it is known to describe \textit{oscillons} -- oscillating quasi-periodic field configurations with a finite energy \cite{Copeland1995,Honda2002,Fodor2006}. Oscillons are well-localized in space during a certain period of time but finally they decay into pure radiation. However their lifetime, that is the period when they remain localized, can be as large as the age of the universe. We therefore discover that the electroweak theory admits spherically symmetric oscillons, which, to the best of our knowledge, has never been observed before. 

Although being out of the scope of the present thesis, the study of the system of equations \eqref{gen_oscillon} could reveal the existence of more general oscillons. This has never been considered before, even in the flat space theory.

\section{Far field region}
\label{asymp_mon}

In this Appendix, we analyze the asymptotic behavior of monopole solutions at spatial infinity. This shows in particular that monopoles have a magnetic quadrupole moment, but no dipole moment. We will also see that the gauge condition \eqref{gauge_cond} used in our numerical calculations gives rise to a spurious long-range mode of pure gauge origin. 

We consider the flat space theory: $\kappa=0$, $N=1$, and $U=S=K=0$. At large distances, the WS fields approach that of the Dirac monopole, we can therefore set 
\begin{align*}
    H_1&=\delta H_1,\;\; H_2=\delta H_2,\;\; H_3=\delta H_3,\;\; H_4=\delta H_4,\\
    \phi_1&=\delta\phi_1,\;\;\phi_2=1+\delta\phi_2,\;\;y&=\delta y,\numberthis
\end{align*}
where $\delta H_1,\dots,\delta\phi_2$ are small deviations. It will be convenient to use the original field equations where the gauge is not fixed. Then, the linearized equations admit the gauge symmetry,
\begin{align*}
    \delta H1&\to\delta H_1-r\partial_r\chi,\quad\delta H_2\to\delta H_2+\partial_\vartheta\chi,\quad\delta H_4\to\delta H_4+\chi\cot\vartheta,\\
\label{lin_gauge}
    \delta \phi_1&\to\delta\phi_1+\chi/2,\quad\delta H_3\to\delta H_3,\quad\delta y\to\delta y,\quad\delta\phi_2\to\delta\phi_2,\numberthis
\end{align*}
which is obtained by linearizing the transformations \eqref{res_gauge} and assuming the gauge parameter $\chi$ to be small.

The linearized equation for $\delta\phi_2$ decouples from the others and reads,
\begin{equation}
    \left(\frac{\partial^2}{\partial r^2}+\frac{2}{r}\frac{\partial}{\partial r}+\frac{1}{r^2}\frac{\partial^2}{\partial\vartheta^2}+\frac{\cot\vartheta}{r^2}\frac{\partial}{\partial\vartheta}-\frac{\beta}{2}\right)\delta\phi_2=0.
\end{equation}
The angular dependence separates if we set
\begin{equation}
    \delta\phi_2=\frac{R_H(r)}{r}P_j(\cos\vartheta),
\end{equation}
where $P_j(\cos\vartheta)$ are the Legendre polynomials. It remains then only a radial equation for $R_H$,
\begin{equation}
    \left(\frac{d^2}{dr^2}-\frac{j(j+1)}{r^2}-\frac{\beta}{2}\right)R_H=0.
\end{equation}
The orbital quantum number $j$ can take any positive value, $j=0,1,2,\dots$, hence the general solution is a superposition of modes with different $j$, but the $j=0$ mode decays slower than the others at large $r$. Therefore, the leading behavior of $\delta\phi_2$ is described by the Yukawa potential,
\begin{equation}
    \delta\phi_2=\frac{C_H}{r}e^{-m_\text{H}r},
\end{equation}
where $C_\text{H}$ is an integration constant and $m_\text{H}$ is the Higgs boson mass defined in Eq.~\eqref{boson_masses}.

The equations for $\delta H_3$ and $\delta y$ comprise a closed system and setting $\delta y=y_\gamma+g'^2 y_Z$, $\delta H_3=y_\gamma-g^2 y_Z$, the system splits into two independent equations,
\begin{equation}
    \hat{\mathcal{D}}_1 y_\gamma=0,\quad\quad\left(\hat{\mathcal{D}}_1-\frac{1}{2}\right)y_Z=0,
\end{equation}
where the differential operator is defined by
\begin{equation}
    \hat{\mathcal{D}}_m=\frac{\partial^2}{\partial r^2}+\frac{1}{r^2}\left(\frac{\partial^2}{\partial\vartheta^2}+\cot\vartheta\frac{\partial}{\partial\vartheta}-\frac{m^2}{\sin^2\vartheta}\right).
\end{equation}
The eigenfunctions of the angular part of this operator are the associated Legendre polynomials $P_j^m(\cos\vartheta)$ and the corresponding eigenvalues are $-j(j+1)$ with $j=|m|,|m|+1,\dots$. Hence we can set 
\begin{equation}
    y_\gamma=R_\gamma(r)P_j^1(\cos\vartheta),\quad\quad y_Z=R_Z(r)P_j^1(\cos\vartheta),
\end{equation}
where the two radial functions are determined by the equations,
\begin{equation}
    \left(\frac{d^2}{dr^2}-\frac{j(j+1)}{r^2}\right)R_\gamma=0,\quad\quad\left(\frac{d^2}{dr^2}-\frac{j(j+1)}{r^2}-\frac{1}{2}\right)R_Z=0.
\end{equation}
This describes the massless photon and massive Z modes. One should have $y=H_3=0$ for both $\vartheta=0$ and $\vartheta=\pi/2$, hence the minimal value $j=1$ is not allowed since $P_1^1=-\sin\vartheta$ does not vanish for $\vartheta=\pi/2$. Therefore the first non-vanishing mode is the one with $j=2$ which defines the leading behavior,
\begin{equation}
\label{asymp_q}
    y_\gamma=\frac{C_\gamma}{r^2}\sin\vartheta\cos\vartheta,\quad\quad y_Z=C_Z e^{-m_\text{Z}r}\sin\vartheta\cos\vartheta+\dots,
\end{equation}
and this corresponds to the magnetic quadrupole moment. 

The four amplitudes $\delta H_1$, $\delta H_2$, $\delta H_4$, $\delta\phi_1$ fulfill a system of four equations admitting the gauge symmetry \eqref{lin_gauge}. The analysis of this sector can be simplified by imposing the unitary gauge, $\delta\phi_1=0$. Then the angular dependence separates by setting
\begin{align*}
    \delta H_1&=\nu\frac{f_1(r)}{r}P_j^\nu(\cos\vartheta),\\
    \delta H_2&=\nu f_3(r)P_j^{\nu-1}(\cos\vartheta)+\nu f_2(r)P_j^{\nu+1}(\cos\vartheta),\\
\label{W_modes_decomp}
    \delta H_4&=f_3(r)P_j^{\nu-1}(\cos\vartheta)-f_2(r)P_j^{\nu+1}(\cos\vartheta),\numberthis
\end{align*}
and using the recurrence relations
\begin{equation}
    (\partial_\vartheta\mp m\cot\vartheta)P_j^m(\cos\vartheta)=\lambda_\pm P_j^{m\pm 1}(\cos\vartheta),
\end{equation}
where $\lambda_{+}=1$ and $\lambda_{-}=m(m-1)-j(j+1)$. The equations for $f_1(r)$ and $f_2(r)$ become
\begin{align*}
    \left(\frac{d^2}{dr^2}+\frac{\nu^2-j(j+1)}{r^2}-\frac{g^2}{2}\right)f_1&=0,\\
\label{eq_w_modes}
    \left(\frac{d^2}{dr^2}+\frac{\nu^2-j(j+1)}{r^2}-\frac{g^2}{2}\right)f_2&=\frac{f_1}{r^3},\numberthis
\end{align*}
and the remaining equations reduce to the constraint
\begin{equation}
    f_3=f1'+(j-\nu)(j+1+\nu)f_2,
\end{equation}
which determines the amplitude $f_3$. Denoting $C^{(1)}_W$ and $C^{(2)}_W$ two integration constants, one obtains from the Eq.~\eqref{eq_w_modes},
\begin{equation}
\label{W_modes}
    f_1=C^{(1)}_W e^{-m_\text{W}r}+\dots,\quad f_2=C^{(2)}_W e^{-m_\text{W}r}+\dots.
\end{equation}
This solution describes massive W boson modes.

Summarizing, all field amplitudes approach their asymptotic values exponentially fast, except $\delta H_3$ and $\delta y$ which decay as $1/r^2$. This agrees with properties of the perturbative spectrum of the theory. However, this behavior is manifest only in the unitary gauge, while we rather use the gauge condition \eqref{gauge_cond} in our numerical computations. This does not change the asymptotic solutions for $\delta\phi_2$, $\delta H_3$, $\delta y$ since these amplitudes are gauge invariant, see the Eq.~\eqref{lin_gauge}. For the gauge-dependent amplitudes $\delta H_1$, $\delta H_2$, $\delta H_4$, $\delta\phi_1$, the situation changes and one finds that 
\begin{align*}
    \delta H_1&=\frac{A}{r^2}\sin(2\vartheta)+\dots,\quad\delta H_2=\frac{A}{r^2}\cos(2\vartheta)+\dots,\\
    \delta H_4&=\frac{A}{r^2}\cos^2\vartheta+\dots,\quad\delta\phi_1=\frac{A}{4r^2}\sin(2\vartheta)+\dots.
\end{align*}
Here $A$ is an integration constant and the dots denote subleading terms containing the exponentially small massive modes \eqref{W_modes}. As a result, our numerical solutions exhibit this second long-range tail in addition to the electromagnetic one. One can remove this spurious long-range mode by applying the gauge transformation \eqref{lin_gauge} generated by the gauge transformation
\begin{equation}
    \chi=-2\delta\phi_1=-\frac{A}{2r^2}\sin(2\vartheta)+\dots.
\end{equation}
One may wonder why not to have used the unitary gauge, $\phi_1=0$, from the beginning in our numerics. As shown below in Appendix.~\ref{local_origin}, it turns out that this gauge is singular at the origin, whereas the gauge \eqref{gauge_cond} is globally regular. Moreover the principal part of the differential operator in the nonlinear field equations is not diagonal in the unitary gauge.

The asymptotic analysis of the full gravitating theory is a topic for future publication~\cite{GervalleInPrep}. In the gravitating case, the electromagnetic sector couples to the gravity sector, which contains a gravitational quadrupole moment.

\section{Local solution at origin for flat space monopoles}
\label{local_origin}

For black hole solutions, the behavior of amplitudes at the horizon can be obtained by expanding the field equations in power series with the radial coordinate $x$ defined in Eq.~\eqref{def_x_coord}. In this Appendix, we analyze the behavior at the origin for flat space monopoles. The complete analysis turns out to be rather involved, and we shall consider only the behavior of the Higgs field which will lead to important conclusions.

Close to the origin, the monopole fields approach
\begin{equation}
\label{false_vac}
    H_1=H_3=y=\phi_1=\phi_2=0,\quad\quad H_2=H_4=1,
\end{equation}
which is often dubbed Higgs false vacuum. This in an exact solution of the equations for any $r$, $\vartheta$, but the monopole fields approach it only for $r\to 0$. Therefore we set
\begin{align*}
    H_1&=\delta H_1,\quad H_2=1+\delta H_2,\quad H_3=\delta H_2,\quad H_4=1+\delta H_4,\\
    y&=\delta y,\quad\phi_1=\delta\phi_1,\quad\phi_2=\delta\phi_2,\numberthis
\end{align*}
where $\delta H_1,\dots,\delta\phi_2$ are small deviations. Injecting this to the field equations and linearizing with respect to the deviations, we find that the equations for $\delta\phi_1$ and $\delta\phi_2$ decouple from the rest. One can neglect in these two equations terms proportional to the Higgs coupling $\beta$ since they are small as compared to the other terms if $r$ is small. After this, the equations become homogeneous in $r$ so that we can set 
\begin{equation}
    \delta\phi_1=r^\lambda S_1(\vartheta),\quad\quad\delta\phi_2=r^\lambda S_2(\vartheta).
\end{equation}
The variables separate and the equations reduce to
\begin{align*}
    0&=\left(\lambda(\lambda+1)+\frac{d^2}{d\vartheta^2}+\cot\vartheta\frac{d}{d\vartheta}-\frac{\nu^2}{\sin^2\vartheta}+\frac{3\nu^2-1}{4}\right)S_1-\left(\frac{d}{d\vartheta}+\frac{1-\nu^2}{2}\cot\vartheta\right)S_2,\\
\label{eigenval_orig}
    0&=\left(\lambda(\lambda+1)+\frac{d^2}{d\vartheta^2}+\cot\vartheta\frac{d}{d\vartheta}-\frac{\nu^2+1}{4}\right)S_2+\left(\frac{d}{d\vartheta}+\frac{1+\nu^2}{2}\cot\vartheta\right)S_1.\numberthis
\end{align*}
This constitutes an eigenvalue problem to determine $\lambda$.

In the spherically symmetric case, one has $|\nu|=1$, $S_1=0$, $S_2=const.$, and the equations reduce to
\begin{equation}
    \lambda(\lambda-1)-\frac{1}{2}=0\quad\Rightarrow\quad\lambda=\frac{\sqrt{3}-1}{2},
\end{equation}
which reproduces the small $r$ behavior of the CM monopole \cite{Cho1996}. If $|\nu|\neq 1$ then the solution is obtained by choosing (assuming that $\nu>0$),
\begin{equation}
    \lambda=\frac{\sqrt{1+2\nu}-1}{2},\quad\quad S_1(\vartheta)=-\frac{2}{\nu+1}\frac{d}{d\vartheta}S_2(\vartheta).
\end{equation}
Injecting this to \eqref{eigenval_orig}, the two equations reduce to 
\begin{equation}
    \left(\frac{d^2}{d\vartheta^2}-\nu\cot\vartheta\frac{d}{d\vartheta}+\frac{1-\nu^2}{4}\right)S_2=0,
\end{equation}
whose solution is
\begin{equation}
\label{s1s2}
    S_2(\vartheta)=\left(\sin\frac{\vartheta}{2}\right)^{\nu+1}+\left(\cos\frac{\vartheta}{2}\right)^{\nu+1}.
\end{equation}
Since the derivative $dS_2/d\vartheta$ vanishes for both $\vartheta=0$ and $\vartheta=\pi/2$, the deviations $\delta\phi_1$ and $\delta\phi_2$ satisfy the correct boundary conditions at the symmetry axis and in the equatorial plane.

This result has an interesting consequence. The gauge transformation \eqref{res_gauge} changes the Higgs amplitude as
\begin{align*}
    \delta\phi_1&\to\delta\tilde{\phi}_1=\delta\phi_1\cos\frac{\chi}{2}+\delta\phi_2\sin\frac{\chi}{2},\\
    \delta\phi_2&\to\delta\tilde{\phi}_2=\delta\phi_2\cos\frac{\chi}{2}-\delta\phi_1\sin\frac{\chi}{2},\numberthis
\end{align*}
and if we require the new gauge to be unitary, $\delta\tilde{\phi}_1=0$, this implies that
\begin{equation}
\label{param_unitary}
    \tan\frac{\chi}{2}=-\frac{\delta\phi_1}{\delta\phi_2}=-\frac{S_1(\vartheta)}{S_2(\vartheta)}.
\end{equation}
This determines the $r\to 0$ limit of the gauge parameter $\chi$ which put the solution to the unitary gauge. Notice that although $\delta\phi_1$ and $\delta\phi_2$ are small close to the origin, their ratio and hence $\chi$ are not small.

\begin{figure}
    \centering
    \includegraphics[width=7.7cm]{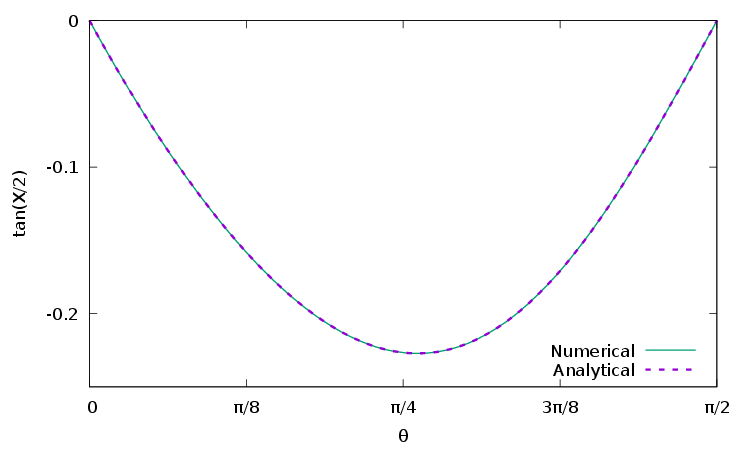}
    \includegraphics[width=8.1cm]{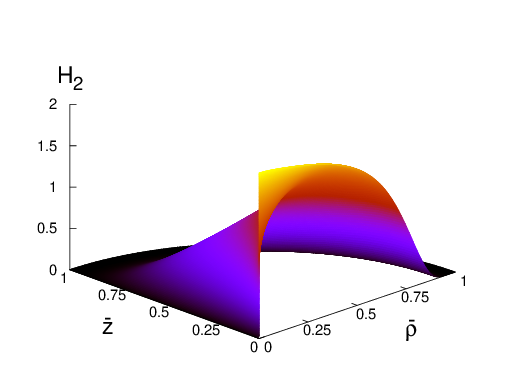}
    \caption[The gauge parameter that transforms the solutions to the unitary gauge in the limit $r\to 0$ and the function $H_2$ for $\nu=2$ monopole solution in the unitary gauge.] {Left: plots of $\tan(\chi/2)$ obtained analytically and numerically for $\nu=2$ in the limit $r\to 0$. Right: the amplitude $H_2$ of the $\nu=2$ monopole solution in the unitary gauge.}
    \label{fig_unit_gauge}
\end{figure}

We can use this to check the quality of our numerical solutions obtained in the gauge \eqref{gauge_cond}. In order to transform a given solution to the unitary gauge, one should perform the gauge transformation \eqref{res_gauge} with the parameter,
\begin{equation}
\label{param_unitary_2}
    \tan\frac{\chi}{2}=-\frac{\phi_1}{\phi_2},
\end{equation}
where $\phi_1$ and $\phi_2$ are obtained numerically. This gauge parameter must agree for small $r$ with the one in Eq.~\eqref{param_unitary} for the procedure to be consistent, and this is indeed the case. In the left panel of Fig.~\ref{fig_unit_gauge}, we plot $\tan(\chi/2)$ given by the analytical formula \eqref{s1s2}, \eqref{param_unitary} and also $\tan(\chi/2)$ numerically obtained from the Eq.~\eqref{param_unitary_2} in the limit $r\to 0$. As one can see, the two curves exactly coincide with each other so that our procedure is indeed consistent.

At the same time, the gauge transformation associated with the parameter $\chi$ in \eqref{param_unitary} changes the false vacuum configuration \eqref{false_vac} to
\begin{align*}
    H_1&=0,\quad\quad H_2=1+\frac{d\chi}{d\vartheta},\quad\quad y=\phi_1=\phi_2=0,\\
    H_3&=(\cos\chi-1)\cot\vartheta-\sin\chi,\quad H_4=\cos\chi+\sin\chi\cot\vartheta,\numberthis
\end{align*}
which is the $r\to 0$ limit of the solution expressed in the unitary gauge. As one can see, this limit is $\vartheta$-dependent since $\chi$ in \eqref{param_unitary} depends on $\vartheta$. However, $r=0$ is a single point in flat space where nothing should depend on $\vartheta$. Therefore the solution is not single-valued at the origin in the unitary gauge. To illustrate this, we show in the right panel of Fig.~\ref{fig_unit_gauge} the function $H_2$ for the $\nu=2$ solution, the same as in Fig.~\ref{Hks} but transformed to the unitary gauge. One can see that $H_2$ does not have a definite limit at the origin, $\bar{\rho}=\bar{z}=0$ but assumes there all values from the interval $[0,2]$. For example, approaching the origin by keeping $\vartheta=0$ constant gives the value $H_2=0$ whereas for $\vartheta=\pi/2$ the $r\to 0$ is $H_2=2$. Therefore the unitary gauge is singular at small $r$, although it is well adapted to describe the large $r$ region. 

\section{Radial coordinate transformation}
\label{radial_coords}

The spherically symmetric solutions can be expressed using either the line element \eqref{spher_met_ews}, or the more general metric \eqref{metric_ews}. In this Appendix, we provide the precise correspondence between the two descriptions. Let us denote by $\tilde{r}$ the radial coordinate in the spherically symmetric spacetime metric \eqref{spher_met_ews},
\begin{equation}
\label{tilded_spher_met}
    ds^2=-\sigma^2(\tilde{r})\tilde{N}(\tilde{r})dt^2+\frac{d\tilde{r}^2}{\tilde{N}(\tilde{r})}+\tilde{r}^2\left(d\vartheta^2+\sin^2\vartheta\,d\varphi^2\right).
\end{equation}
Notice that we have also introduced a tilde ($\sim$) notation for the function $\tilde{N}(\tilde{r})$ to prevent any confusion with $N(r)=1-r_H/r$ in the axially symmetric line element \eqref{metric_ews}. 

A direct comparison of Eq.~\eqref{tilded_spher_met} with Eq.~\eqref{metric_ews} reveals that $\tilde{r}=r\,e^{S(r)}=r\,e^{K(r)}$. Hence, for spherically symmetric solutions, one has $S(r)=K(r)$ and $\tilde{r}=\tilde{r}_H$ corresponds to the Schwarzschild radius of the event horizon. However, $r_H$ does not have a direct physical interpretation since $\tilde{r}_H=r_H\,e^{S_H}=r_H\,e^{K_H}$ with $K_H=K(r_H)$ and $S_H=S(r_H)$ generally not equal to zero. A further inspection of the line elements \eqref{tilded_spher_met} and \eqref{metric_ews} reveals that the coordinates $r$ and $\tilde{r}$ are related by
\begin{equation}
\label{dr_drt}
    \frac{dr}{r\sqrt{N(r)}}=\frac{d\tilde{r}}{\tilde{r}\sqrt{\tilde{N}(\tilde{r})}},
\end{equation}
integrating which yields $r=r(\tilde{r})$. For the non-Abelian hairy black holes presented in Sec.~\ref{spher_hairy_bh}, the function $\tilde{N}$ is only known numerically. The 10 functions in the axial ansatz are then given by
\begin{align*}
    e^{2U}&=\frac{\tilde{N}(\tilde{r})}{N(r)}\sigma^2(\tilde{r}),\quad e^K=e^S=\frac{\tilde{r}}{r},\\
\label{rel_spher_axial}
    H_2&=H_4=f(\tilde{r}),\quad H_1=H_3=y=\phi_1=0,\quad\phi_2=\phi(\tilde{r}),\numberthis
\end{align*}
where $\phi,f$ and $\sigma$ are only known numerically for the non-Abelian hairy black holes.

It is instructive to consider first the RN solution \eqref{RN_ews} for which 
\begin{equation}
    \tilde{N}(\tilde{r})=\left(1-\frac{\tilde{r}_{-}}{\tilde{r}}\right)\left(1-\frac{\tilde{r}_{+}}{\tilde{r}}\right),\quad \sigma(\tilde{r})=\phi(\tilde{r})=1,\quad f(\tilde{r})=0.
\end{equation}
Here, $\tilde{r}_{-}$ and $\tilde{r}_{+}=\tilde{r}_H$ correspond respectively to the inner and outer horizons radii, as given in Eq.~\eqref{rpm_ews}. Integrating \eqref{dr_drt} then yields,
\begin{equation}
    \sqrt{r}+\sqrt{r-r_H}=\sqrt{\tilde{r}-\tilde{r}_{+}}+\sqrt{\tilde{r}-\tilde{r}_{-}}\quad\Rightarrow\quad r=\tilde{r}-\tilde{r}_{-},\quad r_H=\tilde{r}_{+}-\tilde{r}_{-}.
\end{equation}
Notice that $r_H=\tilde{r}_{+}-\tilde{r}_{-}$ vanishes in the extremal limit when the two horizons merge. Then, it follows from \eqref{rel_spher_axial} that,
\begin{equation}
\label{RN_axisym}
    e^{U(\tilde{r})}=1-\frac{\tilde{r}_{-}}{\tilde{r}},\quad\quad e^{K(\tilde{r})}=e^{S(\tilde{r})}=\frac{\tilde{r}}{\tilde{r}-\tilde{r}_{-}}=e^{-U(\tilde{r})}.
\end{equation}
Therefore, $S=K=-U$ and the horizon values $U_H$, $S_H$, $K_H$ are given by
\begin{equation}
    e^{U_H}=\frac{\tilde{r}_{+}-\tilde{r}_{-}}{\tilde{r}_{+}}=e^{-K_H}=e^{-S_H}.
\end{equation}
In the extremal limit, one has $M\to|Q|$ hence $r_H=\tilde{r}_{+}-\tilde{r}_{-}\to 0$ while $\tilde{r}_H\to|Q|$ so that $U_H\to -\infty$ and $S_H,K_H\to +\infty$. As a result, the horizon values $U_H$, $S_H$, $K_H$ do not correspond to the naive formula \eqref{spher_sym_met_fields}. 

Finally, $\tilde{r}_{-}$ can be expressed in terms of $r_H$ using the fact that $\tilde{r}_\pm=M\pm\sqrt{M^2-Q^2}$,
\begin{equation}
	\tilde{r}_{+}\tilde{r}_{-}=\tilde{r}_{-}^2+r_H\tilde{r}_{-}=Q^2\equiv\frac{\kappa\nu^2}{2e^2}\quad\Rightarrow\quad \tilde{r}_{-}=\frac{1}{2}\left(\sqrt{r_H^2+4Q^2}-r_H\right),
\end{equation}
and this complete the expression of the RN solution in terms of the axially symmetric variables,
\begin{equation}
\label{rn_axial}
	e^{U}=e^{-K}=e^{-S}=\frac{r}{r+\tilde{r}_{-}},\;\; H_1=H_2=H_3=H_4=y=\phi_1=0,\;\;\phi_2=1.
\end{equation}

\begin{figure}[b!]
    \centering
    \includegraphics[height=6.4cm]{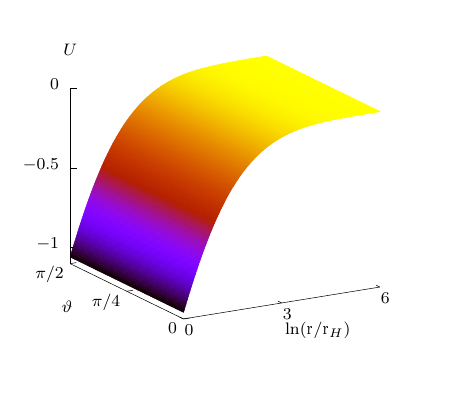}
    \includegraphics[height=6.4cm]{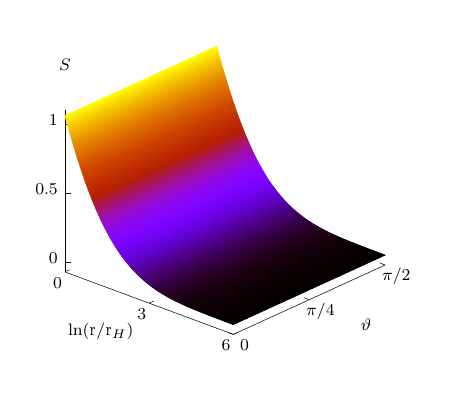}
    \caption[The metric functions $U$ and $S$ for the spherically symmetric hairy black hole with $\nu=1$, $\kappa=10^{-3}$ and $r_H=0.02$ against $\ln(r/r_H)$ and $\vartheta$.]{The metric functions $U$ (left) and $S$ (right) for the spherically symmetric hairy black hole with $\nu=1$, $\kappa=10^{-3}$ and $r_H=0.02$ against $\ln(r/r_H)$ and $\vartheta$.}
    \label{fig_U_S_nu_1}
\end{figure}

To integrate \eqref{dr_drt} for the non-Abelian solutions, it is convenient to represent $\tilde{N}(\tilde{r})$ as
\begin{equation}
    \tilde{N}(\tilde{r})=\frac{1}{\alpha^2(\tilde{r})}\left(1-\frac{\tilde{r}_H}{\tilde{r}}\right),
\end{equation}
where $\alpha(\tilde{r}_H)\equiv\alpha_H$ and $\alpha(\infty)=1$. Then the integration yields,
\begin{equation}
    X(r,r_H)=\left(X(\tilde{r},\tilde{r}_H)\right)^{\alpha(\tilde{r})}\times\mathcal{F}(\tilde{r})\;\;\text{with}\;\;\mathcal{F}(\tilde{r})=\exp\left(\int_{\tilde{r}}^\infty{\alpha'(\tilde{r})\ln\left(X(\tilde{r},\tilde{r}_H)\right)d\tilde{r}}\right),
\end{equation}
with $X(x,x_H)\equiv(\sqrt{x}+\sqrt{x-x_H})/2$. This implies that at infinity when $r,\tilde{r}\to\infty$ one has $r/\tilde{r}\to 1$ and $U,S,K\to 0$. At the horizon when $r\to r_H$ and $\tilde{r}\to\tilde{r}_H$ one has
\begin{equation}
    \sqrt{r_H}=2\left(\frac{\sqrt{\tilde{r}_H}}{2}\right)^{\alpha_H}\times\mathcal{F}(\tilde{r}_H),
\end{equation}
which determines the relation between the horizon radius expressed in terms of $r$ and $\tilde{r}$ radial coordinates. This also shows that the value of $e^{S_H}=\tilde{r}_H/r_H$ is not equal to unity. Both $N(r)$ and $\tilde{N}(\tilde{r})$ vanish at the horizon. Using the chain rule and l'Hopital's rule, one finds
\begin{equation}
\label{non_ab_Uh}
    \lim_{\tilde{r}\to\tilde{r}_H}\frac{d\tilde{N}(\tilde{r})}{d\tilde{r}}=\frac{1}{\alpha^2_H\tilde{r}_H},\quad\lim_{\tilde{r}\to\tilde{r}_H}\frac{dN(r(\tilde{r}))}{d\tilde{r}}=\frac{\alpha_H^2}{\tilde{r}_H}\quad\Rightarrow\quad e^{U_H}=\frac{\sigma_H}{\alpha_H^2},
\end{equation}
where $\sigma_H\equiv\sigma(\tilde{r}_H)$. Therefore, $e^{U_H}$ is not equal to $\sigma_H$, as suggested by the naive formula \eqref{spher_sym_met_fields}. It includes the additional factor $1/\alpha_H^2$ which is related to the surface gravity. Computing the later using Eq.~\eqref{surf_grav_area} in the spherically symmetric and axially symmetric formalisms yields,
\begin{equation}
    \kappa_g=\frac{1}{2r_H}e^{U_H-K_H}=\frac{\sigma_H}{2\alpha_H^2\tilde{r}_H}=\frac{1}{2\tilde{r}_H}e^{U_H},
\end{equation}
where in the last equality we have used the Eq.~\eqref{non_ab_Uh}. In the extremal limit $\tilde{r}_H\to\tilde{r}_H^\text{ex}>0$, when the horizon becomes degenerate, one has $\kappa_g\to 0$ but at the same time $\sigma_H$ approaches a non-zero value. It follows that $\alpha_H\to\infty$ and
\begin{equation}
\label{hor_val_ex}
    \tilde{r}_H\to\tilde{r}_H^\text{ex},\quad r_H\to 0,\quad e^{U_H}\to 0,\quad e^{K_H}=e^{S_H}=\frac{\tilde{r}_H}{r_H}\to\infty.
\end{equation}

Having computed $U,S,\dots$ from the Eqs.~\eqref{dr_drt},\eqref{rel_spher_axial}, the same values are obtained by directly solving the axially symmetric problem for $\nu=1$. The Fig.~\ref{fig_U_S_nu_1} shows the profiles of $U$ and $S$ for $\kappa=10^{-3}$ and $r_H=0.02$. For this solution, one has $\tilde{r}_H=r_H\,e^{S_H}=0.0577>\tilde{r}_H^\text{ex}=0.0466$ while $r_H<\tilde{r}_H^\text{ex}$. When $r_H$ decreases further, the profiles remain essentially the same but $U_H$ becomes more and more negative whereas $S_H$ grows, as suggested by the Eq.~\eqref{hor_val_ex}.

Since $U_H,S_H,K_H$ become unbounded when approaching the extremal limit, the latter should be considered separately. The horizon is degenerate and the function $\tilde{N}(\tilde{r})$ has the structure shown in Eq.~\eqref{N_ex_non_ab}. The equations \eqref{dr_drt},\eqref{rel_spher_axial} then imply that the functions $U,K,S$ will be finite at the horizon and approach zero at infinity only if $N(r)$ has the same structure as $\tilde{N}(\tilde{r})$,
\begin{equation}
\label{N_Nt_ext}
    \tilde{N}(\tilde{r})=\tilde{k}^2(\tilde{r})\left(1-\frac{\tilde{r}_H}{\tilde{r}}\right)^2,\quad N(r)=k^2(r)\left(1-\frac{r_H}{r}\right)^2,
\end{equation}
where, according to the Eq.~\eqref{N_ex_non_ab}, $k(r_H)=\tilde{k}(\tilde{r}_H)\equiv k=\sqrt{1-2(\tilde{r}_H^\text{ex})^2\Lambda}$ and $k(\infty)=\tilde{k}(\infty)=1$, whereas,
\begin{equation}
    r_H=\tilde{r}_H=\tilde{r}_H^\text{ex}.
\end{equation}
Injecting this to \eqref{dr_drt} and integrating by parts yields,
\begin{equation}
\label{r_rt_ex}
    \frac{\ln(\tilde{r}-r_H)}{\tilde{k}(\tilde{r})}-\frac{\ln(r-r_H)}{k(r)}=\int_{\tilde{r}}^\infty{\ln(\tilde{r}-r_H)\frac{\tilde{k}'(\tilde{r})}{\tilde{k}^2(\tilde{r})}d\tilde{r}}-\int_r^\infty{\ln(r-r_H)\frac{k'(r)}{k^2(r)}dr}.
\end{equation}
The right-hand side of this equation approaches zero when $r,\tilde{r}\to\infty$ so that $r/\tilde{r}\to 1$ and \eqref{rel_spher_axial} implies that $U,K,S\to 0$. At the horizon, $r,\tilde{r}\to r_H$, the right-hand side of \eqref{r_rt_ex} approaches a finite value that can be denoted by $C$, hence close to the horizon one has
\begin{equation}
    \frac{\tilde{r}-r_H}{r-r_H}=\sqrt{\frac{\tilde{N}(\tilde{r})}{N(r)}}=e^{kC}.
\end{equation}
Finally, the Eq.~\eqref{rel_spher_axial} implies that,
\begin{equation}
\label{hor_val_reg}
    e^{U_H}=\sigma_H e^{kC},\quad e^{K_H}=e^{S_H}=1,
\end{equation}
so that now everything is finite at the horizon. Notice that the above formulas become very simple for the extremal RN solution. In this case, one has $k(r)=\tilde{k}(\tilde{r})=1$, $r=\tilde{r}$ and $U=K=S=0$.

\begin{figure}
    \centering
    \includegraphics[height=6.4cm]{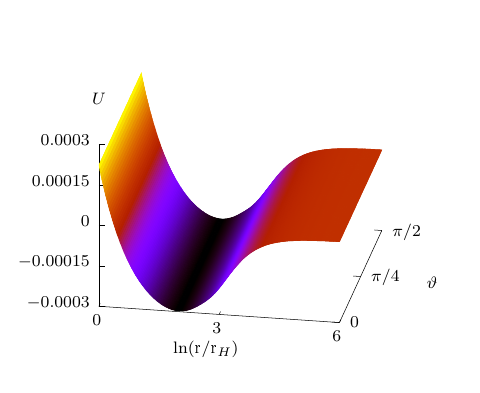}
    \includegraphics[height=6.4cm]{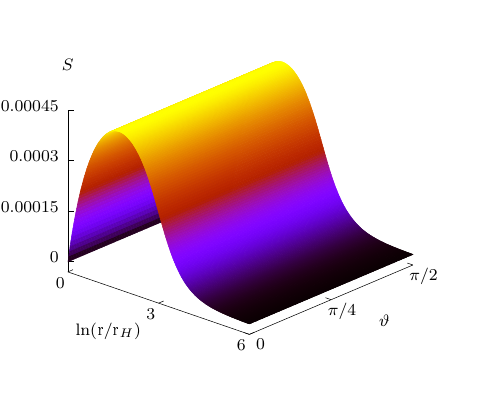}
    \caption[The metric functions $U$ and $S$ for the extremal spherically symmetric hairy black hole with $\nu=1$ and $\kappa=10^{-3}$, in which case $r_H=r_H^\text{ex}=0.046$.]{The metric functions $U$ (left) and $S$ (right) for the extremal spherically symmetric hairy black hole with $\nu=1$ and $\kappa=10^{-3}$, in which case $r_H=r_H^\text{ex}=0.046$.}
    \label{fig_U_S_nu_1_ex}
\end{figure}

To summarize, for extremal hairy solutions, close to the horizon, one should have $N=\tilde{N}$, and the spacetime geometry should be that of an extremal RN-de Sitter black hole. According to Eq.~\eqref{extremal_RNdS}, the function $N$ is therefore given by,
\begin{equation}
    N(r)=\left(1-\frac{r_H}{r}\right)^2\left[1-\frac{\Lambda}{3}\left(r^2+2r_H r+3r_H^2\right)\right],
\end{equation}
with $r_H=r_H^\text{ex}$ defined in Eq.~\eqref{rhex_rnds}. For the numerical computations, we choose $k^2(r)$ in Eq.~\eqref{N_Nt_ext} to be the same as the expression inside the square brackets above, but with an additional cut-off factor,
\begin{equation}
    k^2(r)=1-\frac{\Lambda}{3}\left(r^2+2r_H r+3r_H^2\right)\times\frac{1+r_H^4}{1+r^4}.
\end{equation}
Hence $k^2(r_H)=k^2$ and $k^2(\infty)=1$. Solving the axially symmetric equations for the extremal hairy black hole with $\nu=1$ and $\kappa=10^{-3}$ then yields the profiles of $U$ and $S$ shown in the Fig.~\ref{fig_U_S_nu_1_ex}. They look quite different from those for the non-extremal solution in the Fig.~\ref{fig_U_S_nu_1} but they agree with Eq.~\eqref{hor_val_reg}. In particular, $S=\ln(\tilde{r}/r)$ now vanishes both at the horizon and at infinity.

\end{subappendices}

\chapter{Chains of rotating boson stars}
\label{chap_bs}
\subtitle{This chapter is based on \cite{Gervalle2022}.}

\section{Introduction}

So far we have focused on \textit{black hole} solutions in GR and in the massive bigravity theory. Another class of solutions to explore is the so-called gravitating \textit{solitons}. These are nontrivial solutions without an event horizon in a theory of gravitation that is minimally coupled to some matter fields. Generically, hairy black holes and solitons are closely related to each other: when the horizon radius of a hairy black hole shrinks to zero, one usually obtains a soliton configuration (see for example the Sec.~\ref{zero_size_bh} in the context of the massive bigravity theory). In the previous chapter, we have considered electroweak hairy black holes. In this case, the horizon size could not be arbitrarily small. Instead, one approaches the extremal hairy solution at a finite value of the horizon size. This is related to the absence of regular monopoles in the flat space electroweak theory. In essence, the presence of an event horizon is necessary to avoid a naked U(1) singularity. In this chapter, we shall consider the well-known class of \textit{nontopological} solitons in GR.

Nontopological solitons were first introduced in flat space field theory by Rosen \cite{Rosen1968} and later by Friedberg \textit{et al.} \cite{Friedberg1976} in the 70s. In this context, they consist in nontrivial stationary solutions of a nonlinear theory which have a finite energy and are spatially localized. Unlike topological solitons such as (regular) monopoles, their existence is guaranteed by the conservation of a Noether charge rather than a topological charge\footnote{In the case of monopoles, the topological charge is the winding number.}. A simple example of these solitons are $Q$-\textit{balls}, introduced by Coleman \cite{Coleman1985} in 1985. $Q$-balls are nontopological solitons in a complex scalar field theory with a self-interacting potential. The theory proposed by Coleman exhibits a global phase invariance which yields to a conserved charge $Q$, according to Noether's theorem. Upon quantization, the theory describes interacting massive bosons. In light of this, $Q$-balls can be seen as an agglomerate of massive bosons in flat space tied together by their self-interactions and the charge $Q$ is related to the number of particles \cite{Friedberg1976,Coleman1985}. For a given charge $Q$, the minimization of the energy favors a configuration in which particles are glued together in a soliton, rather than a configuration with individual free particles. This ensures the existence of $Q$-balls \cite{Coleman1985}.

Spherically symmetric $Q$-balls are described by a complex scalar field of the form $\Phi=\phi(r)\e^{i\omega t}$ where $\phi$ is a real function of the radial coordinate, $\omega$ is a constant frequency and $t$ is the time coordinate. The harmonic time dependence is a crucial ingredient which, together with a well-chosen potential, allows the existence of solitons \cite{Derrick1964}. Because of the spherical symmetry, surfaces with constant energy density are spheres. Solutions exist in a finite frequency range $\omega_\text{min}<\omega<\omega_\text{max}$: the upper bound is related to the mass of the scalar field whereas the lower bound depends on the shape of the potential \cite{Friedberg1976,Coleman1985,Lee1992a}. The mass $M$ of a $Q$-ball and its charge $Q$ reach their minimal values at a critical frequency $\omega_\text{cr}\in(\omega_\text{min},\omega_\text{max})$ and they both diverge at the boundaries of the frequency domain. For a fixed frequency $\omega$, there are infinitely many spherically symmetric $Q$-balls labeled by the number $n$ of nodes of their $\phi$ profile \cite{Volkov2002a}. Solutions with $n\geq 1$ are called radially excited $Q$-balls and are unstable, whereas solutions with $n=0$ are referred to as fundamental $Q$-balls and are stable in the range $\omega_\text{min}<\omega<\omega_\text{cr}$ \cite{Friedberg1976}. In general, finding $Q$-ball solutions requires solving a nonlinear differential equation numerically; however, it has been found recently that very accurate analytical approximations of stable $Q$-balls can be obtained \cite{Heeck2021}.

Rotating generalizations of $Q$-balls exist \cite{Volkov2002a,Kleihaus2005}. The rotation is achieved by adding a extra phase factor $\e^{im\varphi}$ to the scalar field, where $\varphi$ is the azimuthal angle and $m$ is the integer rotational number. The rotating solutions are classified according to the behavior of their $\phi$ profile under reflections with respect to the equatorial plane. The energy density isosurfaces consist in one or more tori. The solutions can thus be interpreted as one single rotating $Q$-ball or a chain of several rotating $Q$-balls. These solutions carry an angular momentum $J$ directly related to the charge by $J=mQ$. Their angular momentum is thus quantized at the classical level and as a consequence, for a fixed $Q$, there are no slowly rotating $Q$-balls (with arbitrarily small $J$). Apart from a different $\phi$ profile and a nonzero angular momentum, spinning $Q$-balls share the same properties as the spherically symmetric ones regarding the frequency dependence of their mass or charge.

\textit{Boson stars} (BSs) arise when $Q$-balls are coupled to gravity \cite{Lee1992a,Friedberg1987,Mielke1998}. BSs are globally regular solitonic configurations, namely, they have neither horizons nor singularities. Because the coupling to gravity provides an attractive interaction, BSs exist also when the scalar field potential contains only a mass term \cite{Kaup1968,Ruffini1969}. The corresponding solutions are sometimes called \textit{mini}-boson stars because their maximum mass is very small, except for tiny values of the boson mass \cite{Jetzer1992}. BSs can be rotating and their angular momentum obeys the quantization relation $J=mQ$ \cite{Kleihaus2005,Schunck1996,Kleihaus2008}. Although the spacetime geometry of BSs does not have horizons, ergoregions may be appear for rotating BSs \cite{Kleihaus2008,Cardoso2008}. It should be also mentioned that self-gravitating solitons exist also if the scalar field is replaced by a vector or a spinor field \cite{Herdeiro2017,Herdeiro2019,Herdeiro2020}.

The study of BSs was motivated by different elements. First, the discovery of the Higgs boson \cite{Aad2012,Chatrchyan2012} has shown that fundamental scalar fields exist in nature. In light of this, BSs can be seen as a toy model for solitonic configurations of more realistic fields. Second, BSs are known to be good black holes mimickers \cite{Guzman2009,Herdeiro2021c}, hence they can have applications in astrophysics. Third, in cosmology, scalar fields are the fundamental ingredient in many models such as, for example, in primordial inflation models \cite{PatrickPeter2013} or in quintessence models of dark energy \cite{Tsujikawa2013}. We refer the reader to the Refs.~\cite{Lee1992a,Jetzer1992,Schunck2003,Shnir2022,Liebling2023} for reviews about BSs.

In this chapter, we consider BSs which have flat space counterparts ($Q$-balls): the scalar field potential contains self-interactions carried by terms quartic and sextic in $|\Phi|$. The coupling with gravity has a crucial influence on the frequency dependence of the mass and the charge. Solutions exist below a maximal frequency $\omega_\text{max}$ for which BSs emerge from the flat vacuum. In this limit, the mass $M$ and the charge $Q$ vanish (rather than diverge in the $Q$-ball case). For fundamental BSs, the $(\omega,M)$ and $(\omega,Q)$ curves present a spiraling behavior so that the solutions are no longer uniquely determined by their scalar field frequency. The configurations approach a limiting solution at the center of the spirals with finite mass and charge \cite{Friedberg1987}. For excited BSs, the spiraling pattern occurs for nonrotating configurations \cite{Collodel2017}. For a single excited and rotating BS, the spiraling behavior is replaced by a loop patern: BS sequences start and end at $\omega_\text{max}$ where they approach the vacuum configuration. 

In this chapter, we present the construction of chains of rotating BSs and analyze their properties. In the nonrotating case, chains with two constituents were first considered by Yoshida and Eriguchi \cite{Yoshida1997} and then generalized to larger numbers of constituents by Herdeiro \textit{et al.} \cite{Herdeiro2021}. Their solutions consist in static, axisymmetric equilibrium configurations interpreted as several BSs located along the symmetry. The different constituents in the chains are tied together by the gravitational attraction and are kept apart from each other by the scalar repulsion \cite{Palenzuela2007}. Adding rotation to these solutions is a very natural generalization and we numerically construct the rotating chains \cite{Gervalle2022}. 

The rest of the chapter is organized as follows. In Section~\ref{model_bs} we describe the model, recalling the action and the general field equations. We also present the axisymmetric ansatz for the fields, the conserved quantities and the boundary conditions. In Section~\ref{num_bs}, we describe our numerical procedure which is based on the finite element solver \textit{FreeFem} \cite{MR3043640}. In Section~\ref{sol_bs}, we reproduce sequences of already known solutions--rotating BSs with even/odd parity \cite{Kleihaus2005,Kleihaus2008} and chains of nonrotating BSs \cite{Herdeiro2021}--to test our numerical solver. We also present our chains of rotating BSs, analyze their properties and their flat space limit. Section \ref{conclusion_bs} gives our conclusions and perspectives. Finally we give in the Appendix \ref{exp_eq_bs} the coupled set of elliptic PDEs to be solved.

\section{The model}

\label{model_bs}

\subsection{Action and conserved quantities}

We consider the theory of a complex scalar field $\Phi$ minimally coupled to GR. The dimensionless\footnote{In the dimensionfull action, the potential $U(|\Phi|^2)$ contains an overall factor that has been set to unity by a rescaling.} action is given by
\begin{equation}
\label{action_bs}
    S_\text{BS}=\int{d^4 x\sqrt{-g}\left[\frac{R}{4\alpha^2}-\frac{1}{2}g^{\mu\nu}\left(\partial_\mu\Phi^\ast\partial_\nu\Phi+\partial_\nu\Phi^\ast\partial_\mu\Phi\right)-U(|\Phi|^2)\right]},
\end{equation}
where $R$ is the Ricci scalar associated with the spacetime metric $g_{\mu\nu}$, $\alpha$ is the gravitational coupling and $U$ is the scalar field potential. Here we consider a potential with self-interactions,
\begin{equation}
\label{pot_bs}
    U(|\Phi|^2)=|\Phi|^6-\lambda|\Phi|^4+u_0^2|\Phi|^2,
\end{equation}
where $u_0$ is the mass of the scalar field (\textit{i.e.}, the boson mass) and $\lambda>0$ is a parameter determining the self-interactions.

By taking the limit $\alpha\to 0$ and setting the background metric to be Minkowski, gravity decouples and the theory reduces to that of flat space $Q$-balls. In this sense, BS solutions in the theory \eqref{action_bs} with the sextic potential \eqref{pot_bs} can be seen as gravitating $Q$-balls. We choose the following values for the parameters in the potential,
\begin{equation}
    u_0^2=1.1,\quad\quad\lambda=2,
\end{equation}
which are the values commonly used in the literature (see for example Refs.~\cite{Volkov2002a,Kleihaus2005,Kleihaus2008}). With these values, the potential has a global minimum at $|\Phi|=0$ and a local minimum at some finite value of $|\Phi|$. Hence the potential \eqref{pot_bs} does not allow for a spontaneous symmetry breaking.

The field equations are obtained by varying the action \eqref{action_bs}. Variations with respect to the scalar field yield the nonlinear Klein-Gordon equation,
\begin{equation}
\label{kg_bs}
    \left(\nabla_\mu\nabla^\mu-\frac{dU}{d|\Phi|^2}\right)\Phi=0.
\end{equation}
Variations with respect to the spacetime metric give the Einstein equation,
\begin{equation}
\label{ein_bs}
    E_{\mu\nu}\equiv R_{\mu\nu}-\frac{1}{2}g_{\mu\nu}R-2\alpha^2 T_{\mu\nu}=0,
\end{equation}
where $T_{\mu\nu}$ is the stress-energy tensor of the scalar field,
\begin{equation}
    T_{\mu\nu}=\partial_\mu\Phi^\ast\partial_\nu\Phi+\partial_\nu\Phi^\ast\partial_\mu\Phi-g_{\mu\nu}\left[\frac{1}{2}g^{\alpha\beta}\left(\partial_\alpha\Phi^\ast\partial_\beta\Phi+\partial_\beta\Phi^\ast\partial_\alpha\Phi\right)+U(|\Phi|^2)\right].
\end{equation}

We are interested in solutions which are axisymmetric and stationary. These symmetries are associated with two Killing vector fields which can be expressed in adapted coordinates as,
\begin{equation}
\label{killing_bs}
    \xi=\partial_t,\quad\quad\chi=\partial_\varphi,
\end{equation}
where $t$, $\varphi$ are respectively the asymptotic time and the azimuthal angle. We also assume that spacetime is asymptotically flat so that the mass $M$ and the angular momentum $J$ of the solutions can be computed by the Komar integrals (see Sec.~\ref{global_quant}),
\begin{align}
\label{mass_komar}
    M=&\frac{1}{\alpha^2}\int_\Sigma{d^3 x\sqrt{\gamma}\,n_\mu\xi_\nu R^{\mu\nu}}=2\int_\Sigma{d^3 x\sqrt{\gamma}\,n_\mu\xi_\nu\left(T_{\mu\nu}-\frac{1}{2}g^{\mu\nu}T\right)},\\
\label{j_komar}
    J=&-\frac{1}{2\alpha^2}\int_\Sigma{d^3 x\sqrt{\gamma}\,n_\mu\chi_\nu R^{\mu\nu}}=-\int_\Sigma{d^3 x\sqrt{\gamma}\,n_\mu\chi_\nu\left(T_{\mu\nu}-\frac{1}{2}g^{\mu\nu}T\right)},
\end{align}
where $\Sigma$ is a spacelike hypersurface, $\gamma$ is the determinant of the induced metric on $\Sigma$ and $n$ is the normal vector to $\Sigma$ such that $n_\mu n^\mu=-1$.

The theory \eqref{action_bs} is invariant under global phase transformations $\Phi\to\e^{i\theta}\Phi$, with a constant $\beta$. This global U(1) symmetry is associated with a conserved 4-current,
\begin{equation}
    j_\mu=-i(\Phi\partial_\mu\Phi^\ast-\Phi^\ast\partial_\mu\Phi),\quad\quad\nabla_\mu j^\mu=0.
\end{equation}
The integration of this 4-current over a spacelike hypersurface gives the conserved Noether charge,
\begin{equation}
\label{charge_bs}
    Q=\int_\Sigma{d^3 x\sqrt{\gamma}\,n_\mu j^\mu}.
\end{equation}

\subsection{Ansatz and boundary conditions}

In a system of coordinates such that the Killing vectors are given by Eq.~\eqref{killing_bs}, the stationary and axisymmetric spacetime metric is independent of $t$, $\varphi$. Then, the line element can be put in the Lewis-Papapetrou form \cite{Kleihaus1998,Stephani2003},
\begin{equation}
\label{metric_bs}
    ds^2=-f\,dt^2+\frac{\ell}{f}\left[h(dr^2+r^2d\vartheta^2)+r^2\sin^2\vartheta\left(d\varphi-\frac{w}{r}dt\right)^2\right],
\end{equation}
where $f$, $\ell$, $h$ and $w$ are four functions of the quasi-isotropic coordinates $(r,\vartheta)$. The symmetry axis of spacetime is the set of points such that $\chi_\mu\chi^\mu=(\ell/f)r^2\sin^2\vartheta$ is vanishing. It corresponds to the $z$ axis in cylindrical coordinates or, equivalently, $\vartheta=0,\pi$. The Minkowski metric is recovered when $f=\ell=h=1$ and $w=0$. Notice that the line element \eqref{metric_ews}, which was used in the previous chapter to describe the electroweak black holes, reduces to the one above after gauging away the function $N(r)$ and when $w=0$ (this is the condition for spacetime to be static). One has then the identifications, $f=e^{2U}$, $\ell=e^{2(S+U)}$ and $h=e^{2(K-S)}$.

The axisymmetric ansatz for the scalar field $\Phi$ is \cite{Kleihaus2005,Kleihaus2008},
\begin{equation}
\label{phi_bs}
    \Phi=\phi(r,\vartheta)\e^{i\omega t+im\varphi},
\end{equation}
where $\phi$ is a real function, $\omega$ is the constant frequency parameter and $m$ is the constant rotational number. The latter has to be integer to ensure that the scalar field is single-valued. We will assume without loss of generality that $m$ is positive. Although the field \eqref{phi_bs} depends explicitly on $t$ and $\varphi$, a direct computation reveals these dependencies disappear in the stress-energy tensor and in the field equations. 

With the line element \eqref{metric_bs}, the normal vector in Eqs.~\eqref{mass_komar},\eqref{j_komar} and \eqref{charge_bs} is $n=\sqrt{f}\,dt$ and $\sqrt{\gamma}=\sqrt{-g}/\sqrt{f}$. Then the mass and the charge are
\begin{align}
    Q=&4\pi\int_0^\infty{dr\int_0^\pi{d\vartheta\,r^2\sin\vartheta\frac{\ell^{3/2}h}{f^2}\left(\omega+\frac{mw}{r}\right)\phi^2}},\\
\label{mass_vol_bs}
    M=&4\pi\int_0^\infty{dr\int_0^\pi{d\vartheta\,r^2\sin\vartheta\frac{\ell^{3/2}h}{f^2}\left[f\,U(\phi^2)-2\omega\left(\omega+\frac{mw}{r}\right)\phi^2\right]}},
\end{align}
while for the angular momentum, one finds the quantization relation
\begin{equation}
\label{j_quant}
    J=mQ,
\end{equation}
which was first derived in Ref.~\cite{Schunck1996}.

Alternatively, $M$ and $J$ can be read off from the asymptotic expansions of the functions $f$ and $w$ \cite{Kleihaus2001}
\begin{equation}
\label{asymp_MJ}
    M=\frac{2\pi}{\alpha^2}\lim_{r\to\infty} r^2\partial_r f,\quad\quad J=\frac{2\pi}{\alpha^2}\lim_{r\to\infty} r^2 w.
\end{equation}
This provides a way to check the numerical accuracy of the solutions since the computation of the mass and angular momentum from the volume integrals in Eqs.~\eqref{mass_vol_bs}-\eqref{j_quant} should agree with the values measured at infinity in Eq.~\eqref{asymp_MJ}.

Injecting the fields \eqref{metric_bs}-\eqref{phi_bs} into the Einstein-Klein-Gordon equations \eqref{kg_bs}-\eqref{ein_bs} yields a coupled set of five PDEs for the unknown functions $\phi$, $f$, $\ell$, $h$, $w$ whose explicit expressions are given in the Appendix \ref{exp_eq_bs}. The set of PDEs is elliptic and is treated as a boundary value problem with appropriate boundary conditions.

The theory \eqref{action_bs} depends only on the norm of the scalar field $|\Phi|$. As a consequence, the solutions can be classified according to the behavior of the function $\phi$ under reflections with respect to the equatorial place $\vartheta=\pi/2$,
\begin{align*}
    \mathcal{P}=1\quad\text{(even parity)}\;:&\;\phi(r,\pi-\vartheta)=\phi(r,\vartheta),\\
\label{par_phi}
    \mathcal{P}=-1\quad\text{(odd parity)}\;:&\;\phi(r,\pi-\vartheta)=-\phi(r,\vartheta).\numberthis{}
\end{align*}
At the same time, the geometry is left invariant under these reflections
\begin{align*}
    f(r,\pi-\vartheta)&=f(r,\vartheta),\quad\ell(r,\pi-\vartheta)=\ell(r,\vartheta),\\
\label{par_met}
    h(r,\pi-\vartheta)&=h(r,\vartheta),\quad w(r,\pi-\vartheta)=w(r,\vartheta).\numberthis{}
\end{align*}
As a result, we can reduce the integration domain to $(r,\vartheta)\in[0,\infty)\times[0,\pi/2]$.

As we are assuming asymptotic flatness, the metric should approach Minkowki when $r\to\infty$ while the scalar field goes to its vacuum configuration,
\begin{equation}
    f\rvert_{r\to\infty}=\ell\rvert_{r\to\infty}=h\rvert_{r\to\infty}=1,\quad w\rvert_{r\to\infty}=\phi\rvert_{r\to\infty}=0.
\end{equation}
The regularity of the solutions at the origin requires that
\begin{equation}
    \partial_r f\rvert_{r=0}=\partial_r\ell\rvert_{r=0}=0,\quad h\rvert_{r=0}=1,\quad w\rvert_{r=0}=0,
\end{equation}
while for the scalar field
\begin{align*}
    \partial_r\phi\rvert_{r=0}&=0\quad\text{if}\;\;m=0\quad\text{and}\quad\mathcal{P}=1,\\
    \phi\rvert_{r=0}&=0\quad\text{otherwise}.\numberthis{}
\end{align*}
Reflection symmetry with respect to the equatorial plane \eqref{par_phi}-\eqref{par_met} requires that
\begin{equation}
    \partial_\vartheta f\rvert_{\vartheta=\pi/2}=\partial_\vartheta\ell\rvert_{\vartheta=\pi/2}=\partial_\vartheta h\rvert_{\vartheta=\pi/2}=\partial_\vartheta w\rvert_{\vartheta=\pi/2}=0,
\end{equation}
while the conditions for the scalar field depend on the parity,
\begin{align*}
    \partial_\vartheta\phi\rvert_{\vartheta=\pi/2}&=0\quad\text{if}\;\;\mathcal{P}=1,\\
\label{bc_phi_par}
    \phi\rvert_{\vartheta=\pi/2}&=0\quad\text{if}\;\;\mathcal{P}=-1.\numberthis{}
\end{align*}
Finally, axial symmetry and regularity imply the following conditions on the symmetry axis,
\begin{equation}
    \partial_\vartheta f\rvert_{\vartheta=0}=\partial_\vartheta\ell\rvert_{\vartheta=0}=\partial_\vartheta w\rvert_{\vartheta=0}=0,\quad h\rvert_{\vartheta=0}=1,
\end{equation}
and for the scalar field
\begin{align*}
    \partial_\vartheta\phi\rvert_{\vartheta=0}&=0\quad\text{if}\;\;m=0,\\
    \phi\rvert_{\vartheta=0}&=0\quad\text{if}\;\;m\geq 1.\numberthis{}
\end{align*}
In addition, the absence of a conical singularity also requires that $\partial_\vartheta h\rvert_{\vartheta=0}=0$. This constraint is not imposed in practice but we checked that it holds once a solution is obtained, up to numerical accuracy.

Let us finally mention that the scalar field function $\phi$ exponentially vanishes at infinity only if,
\begin{equation}
\label{om_max}
    \omega\leq u_0\equiv\omega_\text{max}.
\end{equation}
This provides the upper bound for the field frequency.

\section{Numerical approach}

\label{num_bs}

To construct the chains of rotating BS, we solve numerically the set of five coupled nonlinear PDEs for the functions $(\phi,f,\ell,h,w)$ with the boundary conditions defined in the previous section. For this, we use the finite element solver FreeFem \cite{MR3043640} together with the Newton's method to deal with nonlinear equations. Details about the numerical method employed here can be found in the Appendix~\ref{num_pde}. It should be emphasized that the numerical schemes commonly used in the literature are based on the finite difference method \cite{Schonauer1989,Schauder1992} or on spectral methods \cite{Grandclement2010,Grandclement2014}. To our knowledge, there are very few examples of using the finite element method in GR.

The finite element approach requires equations to be in a weak form. As a first step, Eqs.~\eqref{eqphi}-\eqref{eqw} are multiplied by an overall factor such that the second derivative terms are
\begin{equation}
    \frac{f}{\ell h}\frac{\partial^2 X}{\partial r^2}+\frac{f}{r^2\ell h}\frac{\partial^2X}{\partial\vartheta^2}=[\dots],
\end{equation}
where $X$ denotes the functions $\phi$, $f$, $\ell$, $h$ and $w$. This is the structure of the Laplace-Beltrami operator $\Delta\phi\equiv\frac{1}{\sqrt{-g}}\partial_\mu(\sqrt{-g}\,g^{\mu\nu}\partial_\nu\phi)$ for the metric \eqref{metric_bs}. Then we introduce a compactified radial coordinate 
\begin{equation}
    x\equiv\frac{r}{1+r},
\end{equation}
which maps the semi-infinite interval $r\in[0,\infty)$ to the finite range $x\in[0,1]$. Finally, the PDEs are multiplied by test functions, integrated over the domain $(x,\vartheta)\in[0,1]\times[0,\pi/2]$, and the second derivates are integrated by parts as described in the Appendix \ref{weak_app}.

\begin{figure}
    \centering
    \includegraphics[width=7.8cm]{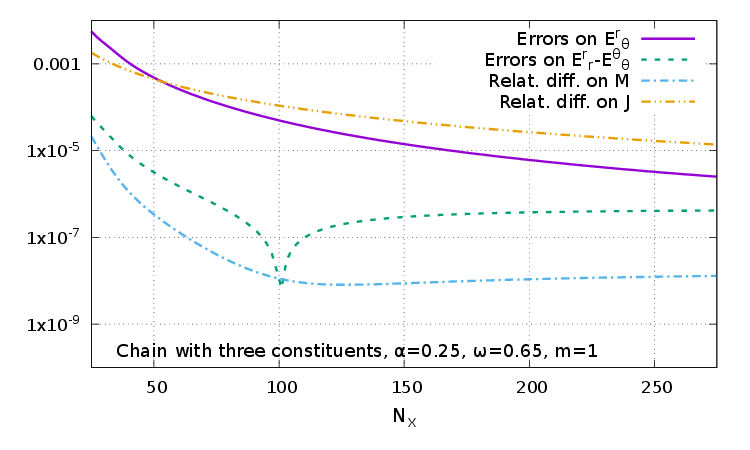}
    \includegraphics[width=7.8cm]{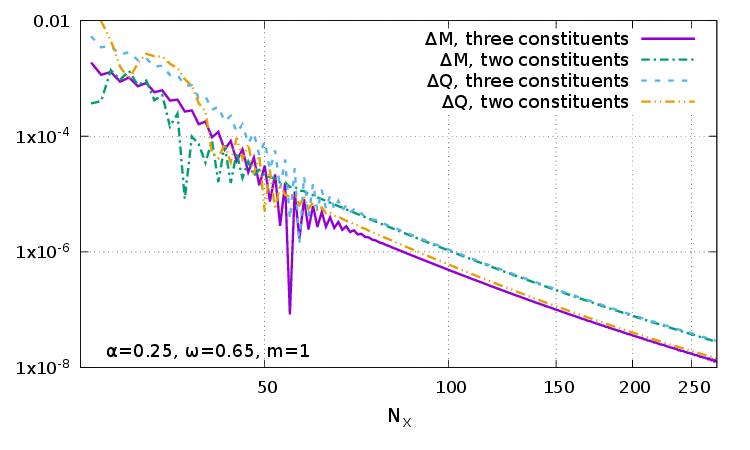}
    \caption[Error indicators against the the number of triangles $N_x$ in the $x$-direction and the convergence of global quantities values when increasing the number of triangles.]{Left: Various error indicators against the number of triangles $N_x$ in the $x$-direction for a typical chain of three BSs. Right: Differences $\Delta M\equiv M(N_x+1)-M(N_x)$ and $\Delta Q\equiv Q(N_x+1)-Q(N_x)$ for typical chains with two and three constituents against $N_x$. The peaks of the curves occur because of the logarithmic scale when the plotted quantities are coincidentally close to zero.}
    \label{err_bs}
\end{figure}

The Einstein equations \eqref{ein_bs} contains two additional constraints, $\tensor{E}{^r_\vartheta}=0$ and $\tensor{E}{^r_r}-\tensor{E}{^\vartheta_\vartheta}=0$, which are not solved in practice. If the numerical procedure is consistent, they should be satisfied by the numerical solutions. Therefore we integrate them over a spacelike hypersurface to obtain an estimation of numerical errors. In addition, we also evaluate the relative difference on the mass and angular momentum computed from \eqref{mass_vol_bs}-\eqref{j_quant} and from \eqref{asymp_MJ}. We have noticed that increasing the number of triangles $N_\vartheta$ in the $\vartheta$-direction does not change significantly the errors. Therefore, we set $N_\vartheta=25$ and present the dependence of the different error indicators on the number of triangles $N_x$ in the $x$-direction in the left panel of Fig.~\ref{err_bs}. In the following, we fix $N_x=200$ so that our errors are typically of the order of $10^{-5}$. The computation time for obtaining one solution on a personal computer with a parallelized code is about thirty seconds.

The right panel of Fig.~\ref{err_bs} shows a convergence test of our code. We evaluate the mass and the charge for increasing values of $N_x$ and compute the differences $\Delta M\equiv M(N_x+1)-M(N_x)$ and $\Delta Q\equiv Q(N_x+1)-Q(N_x)$. The latter are shown against $N_x$ with a logarithmic scale for both axis. After an oscillating phase when the number of triangles is too small, all curves becomes straight lines with a slope of $-4$. Hence the convergence is of fourth order in the number of triangles used to construct the mesh. 

Our code contains 3 input parameters: the gravitational coupling $\alpha$, the scalar field's frequency $\omega$ and the rotational number $m$. The parity is imposed via the appropriate boundary condition at $\vartheta=\pi/2$ as in Eq.~\eqref{bc_phi_par}. The number of individual constituents of the chain is fixed by a suitable initial guess of the $\phi$-function. We compute first a solution for a small value of $\alpha$, choosing Minkowski as the initial guess for the metric, and then increase $\alpha$ iteratively. Generally, it is more easy to start with $\omega$ close to $\omega_\text{max}$; the full sequences of solutions are then obtained by varying $\omega$ by small steps.

\section{The solutions}

\label{sol_bs}

We have constructed numerically solutions corresponding to chains of rotating BSs. The chains with one and two constituents have already been considered in the literature, they correspond respectively to the even and odd parity configurations of Ref.~\cite{Kleihaus2008}. The non-rotating chains with $m=0$ have also been considered in Ref.~\cite{Herdeiro2021}. We have reproduced these solutions to test our code before moving on to solutions never studied before.

What we call the number of constituents in the chains corresponds to the number of extrema of the scalar field function $\phi$ or, equivalently, the number of maxima of the energy density $\tensor{T}{^0_0}$. The solutions are classified according to the parity of the scalar field amplitude: chains with an even (odd) number of constituents have an odd (even) $\phi$ profile.

\begin{figure}
    \centering
    \includegraphics[scale=1]{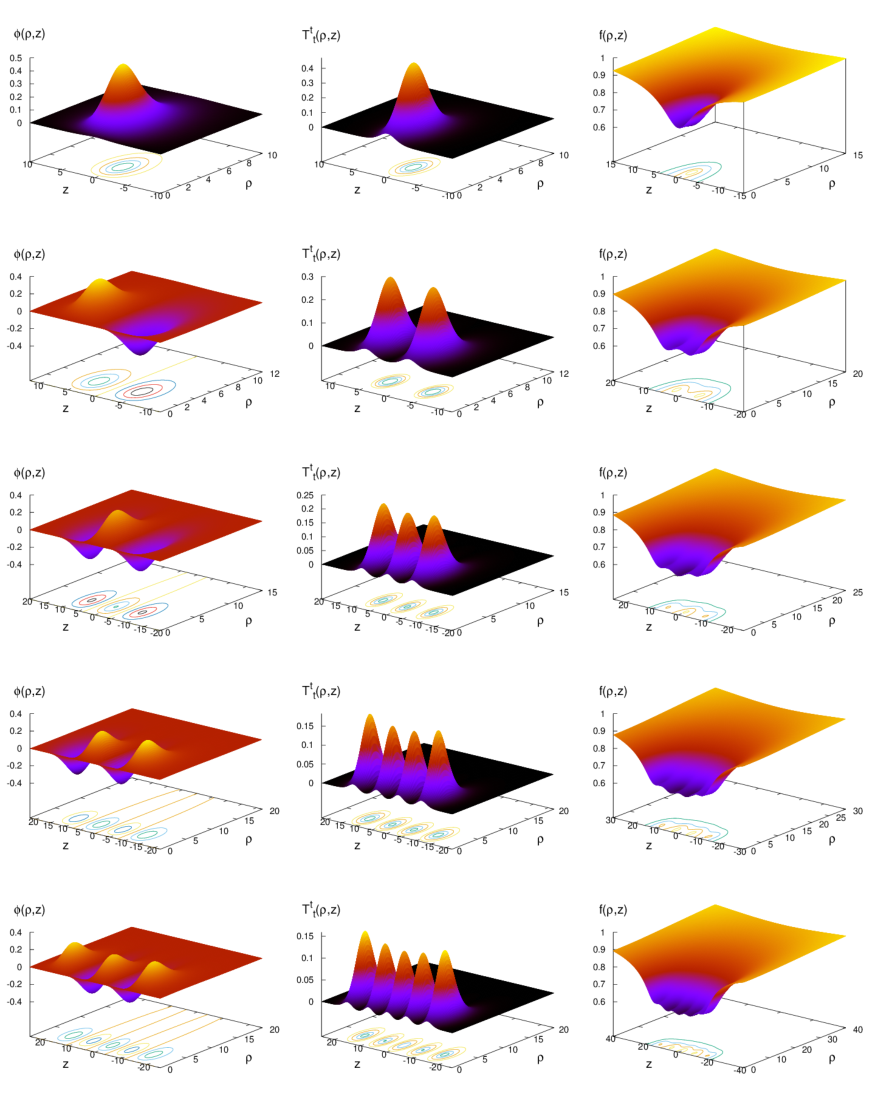}
    \caption[Chains of rotating boson stars with one to five constituents on the fundamental branch of solutions for $\alpha=0.25$, $\omega=0.9$ and $m=1$.]{Chains of rotating BSs with one to five constituents (from top to bottom) on the fundamental branch of solutions for $\alpha=0.25$, $\omega=0.9$ and $m=1$. The plots represent the scalar field amplitude $\phi$ (left), the energy density $\tensor{T}{^0_0}$ (middle) and the metric function $f$ (right) against the cylindrical coordinates $(\rho,z)$.}
    \label{plot_fundam}
\end{figure}

To illustrate this, we show in Fig.~\ref{plot_fundam} typical examples of solutions where the profiles of $\phi$, $\tensor{T}{^0_0}$ and $f$ are plotted against the quasi-isotropic cylindrical coordinates $\rho=r\sin\vartheta$ and $z=r\cos\vartheta$. A single BS is shown in the first row, the $\phi$ function and the energy density show a single peak and have a parity $\mathcal{P}=1$. A pair of BSs is shown in the second row, the scalar field amplitude is now antisymmetric: its parity is $\mathcal{P}=-1$. The number of extrema is two (one peak and one trough) while the energy density presents two symmetric maxima. We also show the lapse squared function $f$ (right column); its profile exhibits as many minima as the number of constituents in the chain and they are located where the maxima of the energy density are. The profiles for chains with a larger number of constituents present similar features: the $\phi$ amplitude shows alternating peaks troughs which are related to symmetric peaks for the energy density and symmetric troughs for the lapse. The shape of the surfaces with constant energy density gives the spatial structure of the solutions. Single BSs have a typical torus shape just like rotating $Q$-balls (see \textit{e.g.} the Ref.~\cite{Radu2008}), pairs have a double tori shape, triplets correspond to triple tori and so on. All these solutions are rotating generalizations of the static chains presented in Ref.~\cite{Herdeiro2021}.

\subsection{Single boson stars and pairs}

\begin{figure}
    \centering
    \includegraphics[width=7.8cm]{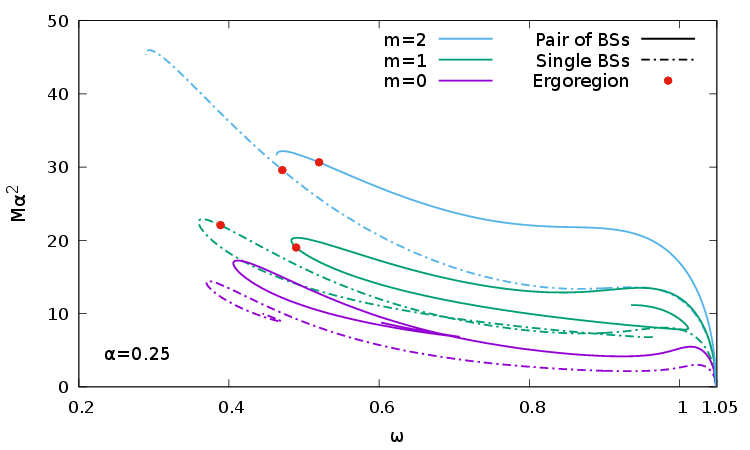}
    \includegraphics[width=7.8cm]{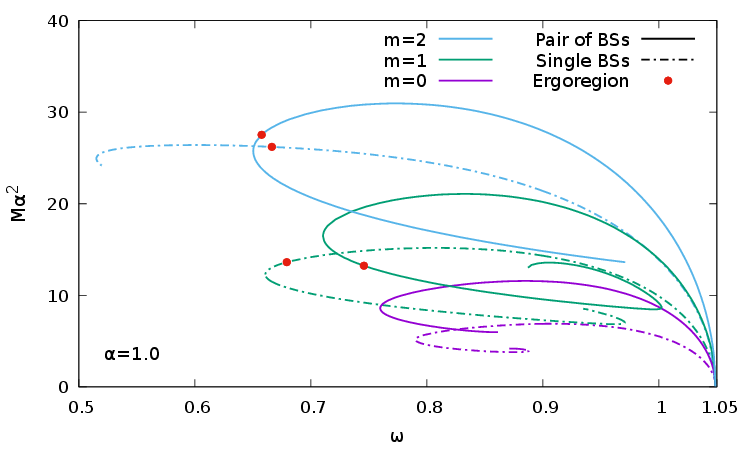}
    \includegraphics[width=7.8cm]{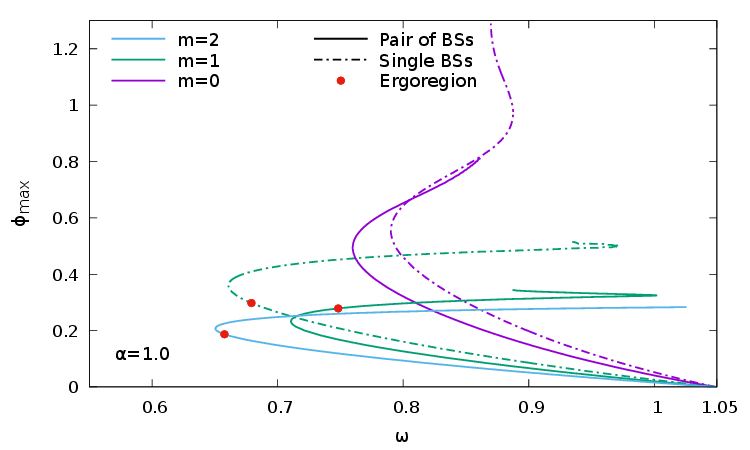}
    \includegraphics[width=7.8cm]{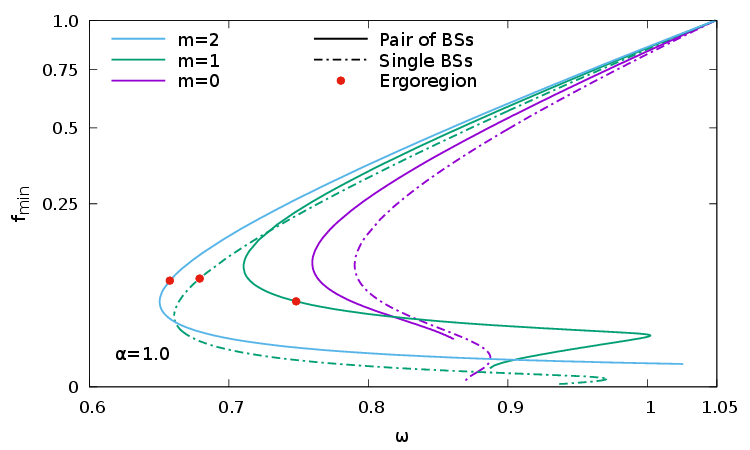}
    \caption[The mass $M$, the maximal value of the scalar field amplitude $\phi_\text{max}$ and the minimal value of the metric function $f_\text{min}$ againt the frequency $\omega$ for singles and pairs of boson stars.]{Scaled mass $M$ (upper panels), maximal value of the scalar field amplitude $\phi_\text{max}$ (lower left panel), and minimal value of the metric function $f_\text{min}$ (lower right panel) against the frequency $\omega$ for single BSs (dash-dotted curves) and pairs (solid curves). Different values of the rotational number $m$ and gravitational coupling $\alpha$ are presented. The red dots indicate the onset of ergoregions. Note the quadratic scale for $f_\text{min}$.}
    \label{single_pairs}
\end{figure}

We will now recall the main results on single and pairs of BSs \cite{Kleihaus2005,Kleihaus2008}. The solutions emerge from the flat vacuum at the maximal value of the scalar field frequency $\omega_\text{max}=u_0$. The mass $M$ and the charge $Q$ of BSs vanish in this upper limit (rather than diverge in the $Q$-balls case). Decreasing $\omega$ spans the first (or fundamental) branch of solutions; it ends at a frequency $\omega_\text{min}$ whose value depends on the gravitational coupling $\alpha$ and the rotational number $m$. As seen on the upper panels of Fig.~\ref{single_pairs}, the mass $M$ remains finite at $\omega_\text{min}$ and the curves present their first backbending. The sequences of solutions can be extended by moving toward larger frequencies. Second branches start after the first backbendings, third branches after the second backbendings, etc.

\begin{figure}
    \centering
    \includegraphics[scale=1]{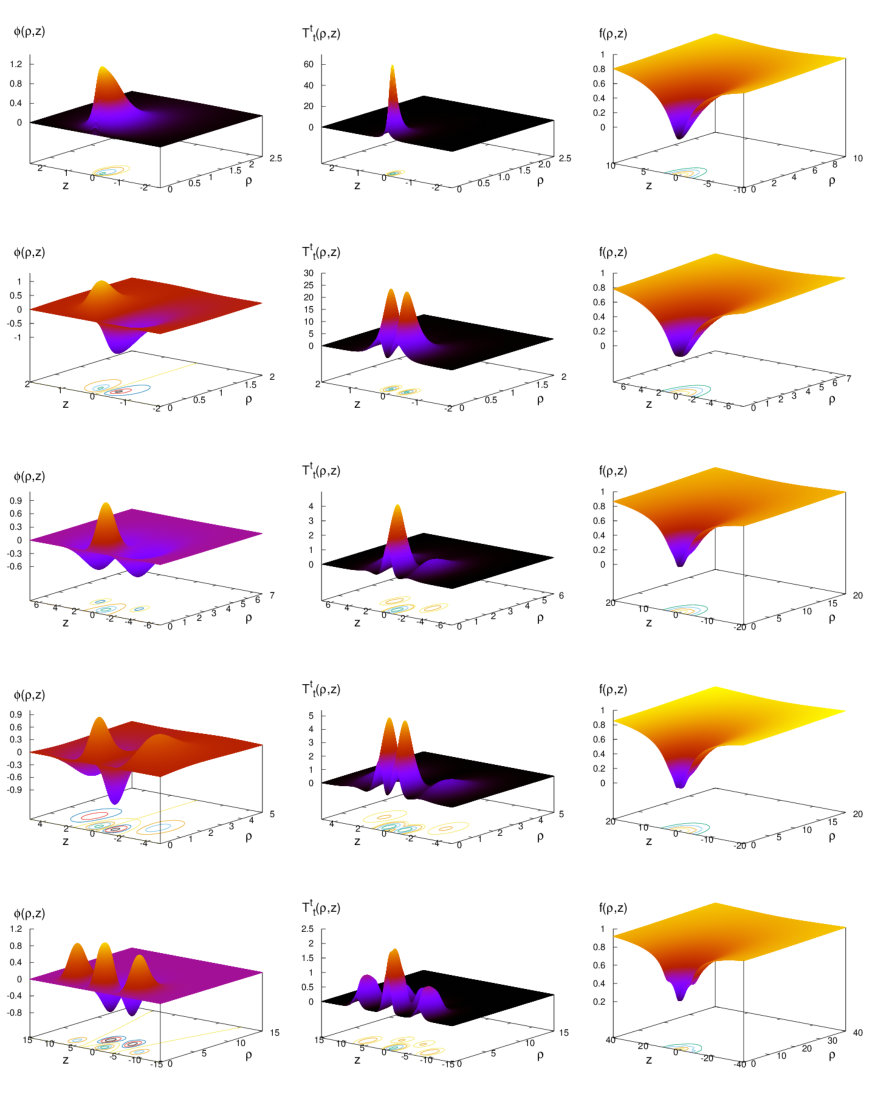}
    \caption[Chains of rotating boson stars with one to five constituents on the second branch of solutions with $m=1$ and various values of $\alpha$ and $\omega$.]{Chains of rotating BSs with one to five constituents on the second branch of solutions with $m=1$, $\alpha=\{0.25,0.25,0.25,0.25,0.15\}$ and $\omega=\{0.5,0.7,0.7,0.7,0.7\}$ (from top to bottom). The plots show the scalar field function $\phi$ (left), the energy density $\tensor{T}{^0_0}$ (middle) and the metric function $f$ (right) against the cylindrical coordinates $(\rho,z)$.}
    \label{plot_2nd}
\end{figure}

The curves for single and pairs of BSs exhibit an inspiraling behavior, approaching a limiting solution at the center of the spirals. In practice, the complete determination of spirals is extremely difficult because the profiles of the solutions become more and more sharp \cite{Kleihaus2005}. Therefore to avoid very time-consuming numerical computations, we only present the first few branches in the figures. The $(\omega,Q)$ diagrams are now shown here but they show similar inspiraling pattern, see the Refs.~\cite{Kleihaus2005,Kleihaus2008}. Examples of profiles for solutions on the second branch can be seen in Fig.~\ref{plot_2nd}. For the chains with one or two constituents considered in this subsection, the main features remains qualitatively the same. The extrema of the scalar field function and the energy density are sharper and closer to the $z$ axis as compared to solutions on the fundamental branch. Hence the $\phi$ function presents in this region very high second derivatives, rendering the numerical computations more challenging.

We also show in Fig.~\ref{single_pairs} the frequency dependence of the maximal value of the scalar field amplitude $\phi_\text{max}$ (lower left panel) and the minimal value of the metric function $f_\text{min}$ (lower right panel). The curves present damped oscillations instead of an inspiraling behavior. On the one hand, the maximal value of the scalar field function goes from zero at $\omega_\text{max}$ when the solutions emerge and then grows as we move to the different branches. On the other hand, the minimal value of the lapse function begins from the vacuum value $\sqrt{f_\text{min}}=1$, and then appraoches zero after (presumably) infinitely many oscillations.

From the comparison between the curves for $\alpha=0.25$ (upper left panel) and $\alpha=1$ (upper right panel), one can see that a larger gravitational coupling increases the minimal value of the frequency $\omega_\text{min}$ and thus reduces the domain of existence of solutions. The rotational number $m$ has a more complicated influence on the value of $\omega_\text{min}$. For rotating ($m\geq 1$) single BSs and pairs of BSs, increasing $m$ decreases the value of $\omega_\text{min}$, but when we pass from nonrotating ($m=0$) to rotating solutions ($m\geq 1$), whether the minimal frequency increases or decreases depends on the value of gravitational coupling $\alpha$. For example, $\omega_\text{min}$ increases for BS pairs with $\alpha=0.25$ if we compare the $m=0$ to the $m=1$ sequence (upper left panel of Fig.~\ref{single_pairs}). In contrast for $\alpha=1$ (upper right panel), $\omega_\text{min}$ decreases when $m$ goes from $0$ to $1$.

\begin{figure}
    \centering
    \includegraphics[width=7.8cm]{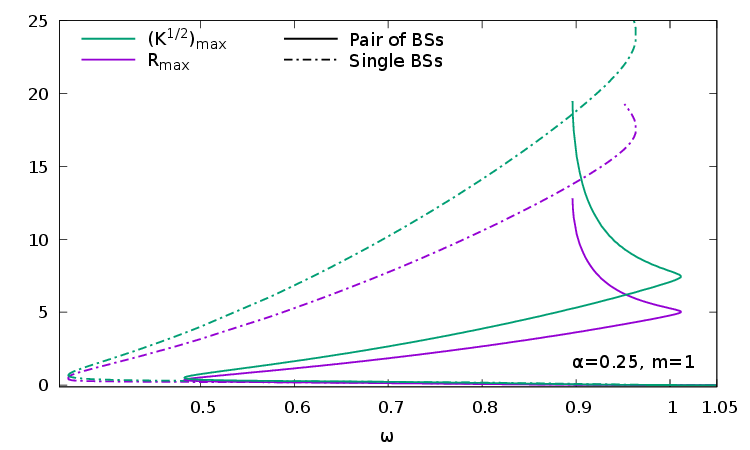}
    \includegraphics[width=7.8cm]{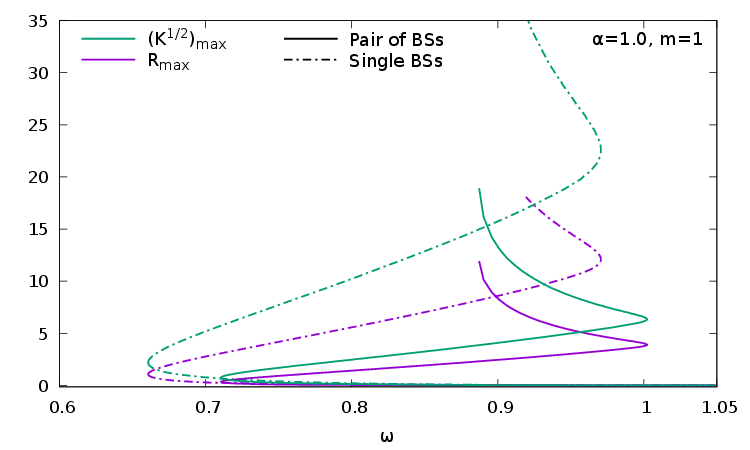}
    \caption[Maximal values of the Ricci scalar and of the Kretschmann scalar against $\omega$ for single boson stars and pairs with $m=1$.]{Maximal values of the Ricci scalar $R$ and of the Kretschmann scalar $K$ against the frequency $\omega$ for single BSs (dash-dotted curves) and pairs (solid curves) with $m=1$ and two different values of the gravitational coupling $\alpha$.}
    \label{RK_sing}
\end{figure}

Another remarkable property of rotating BSs is that they can have an ergoregion. The stationary limit surface (or ergosurface) is defined as the set of points such that
\begin{equation}
    g_{tt}=-f+\frac{\ell}{f}w^2\sin^2\vartheta=0,
\end{equation}
and the ergoregion resides inside this hypersurface, where $g_{tt}>0$ (see Sec.~\ref{intro_kerr}). The onset of ergoregions is indicated by red dots on the curves. Generically, an ergoregion appears on the first or second branch and then the solutions further down the spiral all possess one. However the presence of ergoregions for objects without horizon is associated with a superradiant instability \cite{Cardoso2008} and a light ring instability \cite{Ghosh2021}. Therefore the physical relevance of BSs that have an ergoregion is limited. 

Finally, we have also computed two curvature invariants: the Ricci scalar $R$ and the Kretschmann scalar $K=R_{\alpha\beta\mu\nu}R^{\alpha\beta\mu\nu}$. The Fig.~\ref{RK_sing} show their maximal values as functions of $\omega$ for single BSs and pairs. This provides new information about the limiting configuration at the center of the spirals. Indeed the increase of $R_\text{max}$ and $K_\text{max}$ as one moves towards the different branches becomes larger and larger. Therefore, if the spirals are infinite, it strongly suggests that the curvature invariants diverge and become infinite for the limiting solutions. As a result, the latter are certainly singular and thus unlikely to be numerically obtained.

\subsection{Chains with odd numbers of constituents}

\begin{figure}[t!]
    \centering
    \includegraphics[width=7.8cm]{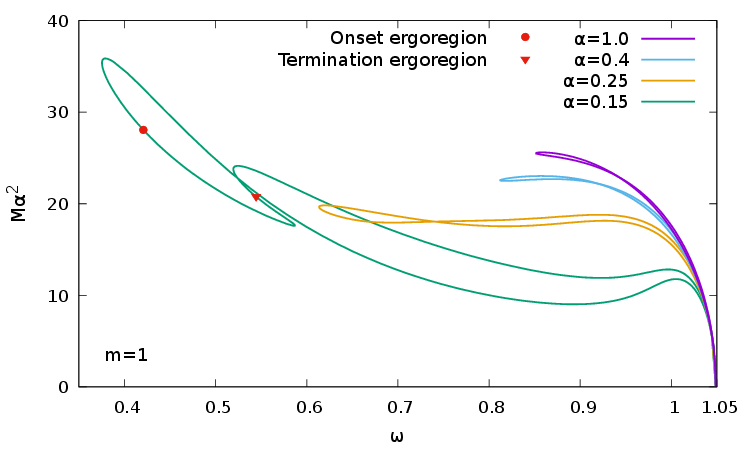}
    \includegraphics[width=7.8cm]{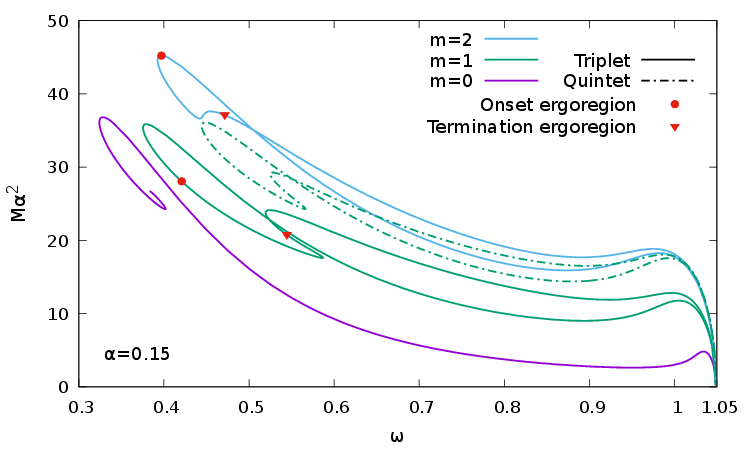}
    \includegraphics[width=7.8cm]{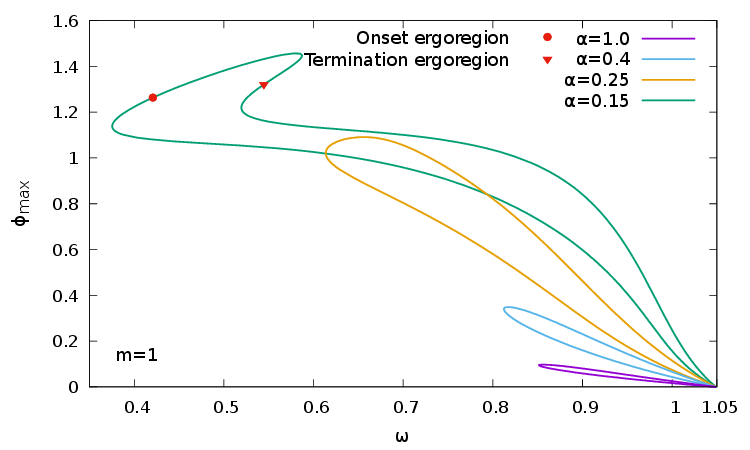}
    \includegraphics[width=7.8cm]{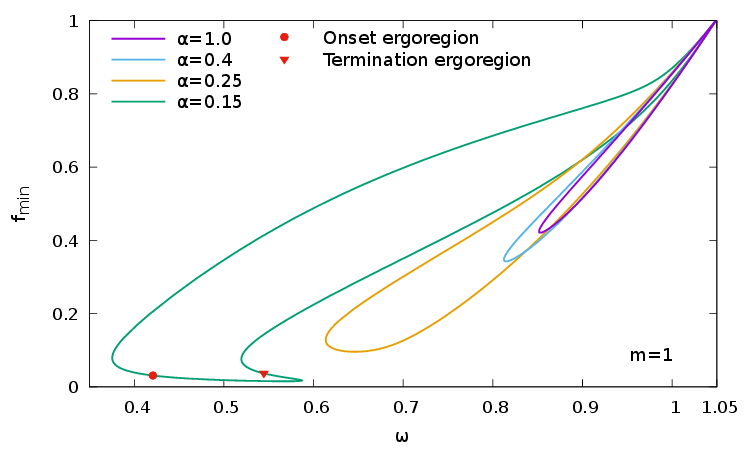}
    \includegraphics[width=7.8cm]{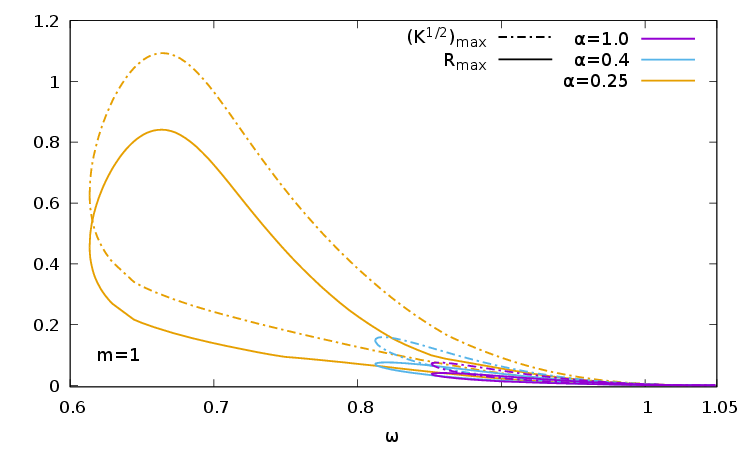}
    \caption[The mass $M$, the maximal value of the scalar field amplitude $\phi_\text{max}$, the minimal value of the metric function $f_\text{min}$ and the maximal values of curvature invariants against $\omega$ for boson star triplets with $m=1$.]{Triplets of BSs: the $\omega$ dependence of the scaled mass $M$ for different gravitational coupling $\alpha$ and rotational number $m$ (upper panels), maximal value of the scalar field amplitude $\phi_\text{max}$ (middle left panel), minimal value of the metric function $f_\text{min}$ (middle right panel) and maximal values of the curvature invariants $R$, $K$ (lower panels) for different values of $\alpha$ and $m=1$. The onset and termination of ergoregions are indicated respectively by red dots and triangles. In the upper right panel, the curve for BS quintets with $m=1$ is also shown for comparison.}
    \label{plot_odd}
\end{figure}

Let us now consider chains of BSs with a higher odd number of constituents. These solutions are characterized by an even parity ($\mathcal{P}=1$) of their $\phi$ profile: one BS is centered in the equatorial plane and the other ones are located symmetrically in the upper and lower semispaces as one can see in Fig.~\ref{plot_fundam} where a triplet and a quintet are shown. It turns out that these chains exhibit a different frequency dependence of their mass (or charge) as compared to single BSs or pairs. The $(\omega,M)$ diagrams for triplets are shown in Fig.~\ref{plot_odd}. For all rotating solutions ($m>0$) the curves no longer present the inspiraling behavior but instead form nontrivial loops with only two branches. A first branch of solution starts at the maximal value of scalar field frequency $\omega_\text{max}$, extends until the backbending is reached, and a second branch leads all the way back to the vacuum configuration.

The terminology \textit{first} and \textit{second} branches can thus seem ambiguous in this case but one can actually distinguish solutions belonging to one or the other. Indeed, for solutions on the fundamental branch when $\omega$ is close to its maximal value (high-frequency regime), the peaks and troughs of the $\phi$ function (left column of Fig.~\ref{plot_fundam}) are similar in shape and located along a line parallel to the $z$ axis. On the contrary, for solutions on the second branch, the central BS dominates in amplitude and the different constituents no longer form a line. For example in the third row of Fig.~\ref{plot_2nd}, the two satellites are located at a larger $\rho$ coordinate than the central BS. Moreover, the central trough of the $f$ profile overlaps with that of the satellites for solutions on the second branch. Regarding the $(\omega,M)$ diagrams, the fundamental branch is always the one with a lower mass for $\omega$ close to $\omega_\text{max}$. This suggests that the fundamental branch in the high-frequency regime is more stable than the second branch. However for lower values of $\omega$, this mass hierarchy is inverted. 

In the upper left panel of Fig.~\ref{plot_odd}, we show the $(\omega,M)$ diagrams for different values of the gravitational coupling $\alpha$. The two branches always intersect at some value of the scalar field frequency but the curves in the low-frequency regime can present more complicated patterns when $\alpha$ is small. For example the curves for $\alpha\gtrsim 0.25$ are very similar in shape but for $\alpha=0.15$, the second branch have two successive backbendings before returning to the vacuum configuration. We also see that ergoregions do not necessarily occur: for BS triplets with $m=1$, they only appear below a critical value $\alpha_\text{cr}\lesssim 0.25$. Since the second branch terminates at the flat vacuum configuration, if there is an onset of ergoregion (red dots on the figure), a termination (red triangles) necessarily occurs. Finally, as for single BSs, the domain of existence of solutions increases when $\alpha$ decreases.

The influence of the rotational number $m$ on odd chains is presented in the upper right panel of Fig.~\ref{plot_odd} where sequences of solutions have been constructed with a fixed value of $\alpha$ and $m=0,1,2$. A larger rotational number seems to increase the minimal value of the scalar field frequency and thus reduces the domain of existence of the solutions. We also note that rotating odd chains have a different branch structure than the static ones. Indeed, the nonrotating ($m=0$) triplets with $\alpha=0.15$ exhibit the same inspiraling pattern as single BSs and pairs. In fact, the authors in Ref.~\cite{Herdeiro2021} have noticed that static triplets can have a loop structure but only for large gravitational coupling whereas rotating triplets seem to never show the spiral pattern, even when $\alpha$ is small. Moreover, it is shown in Ref.~\cite{Herdeiro2021} that when nonrotating odd chains present the loop structure (large $\alpha$), the second branch overlaps with the fundamental branch of a radially excited and spherically symmetric single BS sequence.

\begin{figure}[b!]
    \centering
    \includegraphics[width=11cm]{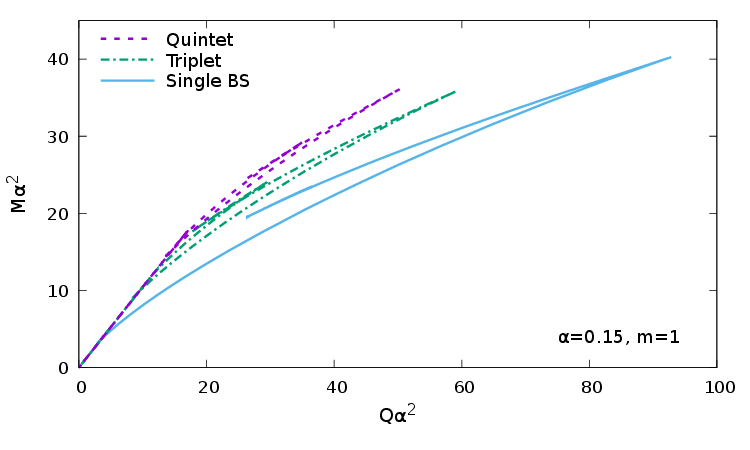}
    \caption[The mass $M$ against the charge $Q$ for chains of boson stars with odd numbers of constituents with $m=1$ and $\alpha=0.15$.]{Scaled mass $M$ against the scaled charge $Q$ for chains of BSs with odd numbers of constituents with $m=1$ and $\alpha=0.15$.}
    \label{plot_M_Q_odd}
\end{figure}

We expected a similar scenario for the rotating odd chains because they present a loop structure. However it turns out that our solutions coincide exactly with excited rotating single BSs previously constructed in Ref.~\cite{Collodel2017}. On the one hand, we have found no other solutions different from the radially excited BSs of Ref.~\cite{Collodel2017}. On the other hand, the chain structure of the $\phi$ profile (see Fig.~\ref{plot_fundam}) is in total agreement with their nonrotating counterparts -- the static chains of Ref.~\cite{Herdeiro2021}. We conclude that rotating chains with an odd number of constituents exist, and correspond to radial excitations of single BSs. To support this result, we present in Fig.~\ref{plot_M_Q_odd} the mass of single BSs, triplets and quintets against their charge $Q$. For a given charge, the less energetic solution always lies on the sequence of single BSs. Even in the region where the different curves overlap, the single BS sequence is still below the two others. Therefore if triplets and higher odd chains are unstable, they could possibly radiate their energy keeping their charge fixed, and decay into a single BS. Of course to rigorously confirm this scenario, fully time-dependent simulations would be required, which is out of the scope of this work.

\medskip
For the sake of completeness, we also present in the middle and lower panels of Fig.~\ref{plot_odd} the maximal value of the scalar field function $\phi_\text{max}$, the minimal value of the metric function $f_\text{min}$ and the maximal values of the curvature invariants $R_\text{max}$, $K_\text{max}$ against the frequency $\omega$ for BSs triplets. These curves also form loops but contrary to the $(\omega,M)$ diagrams they do not intersect. Therefore the point of intersection on a $(\omega,M)$ curve corresponds to two distinct solutions. It worth noting that whereas single BSs and pairs with fixed rotational number $m$ could be uniquely parametrized by $\phi_\text{max}$, $f_\text{min}$, $R_\text{max}$ or $K_\text{max}$, this is not possible for higher odd chains because of the loop structure. This peculiarity was already mentioned in Ref.~\cite{Collodel2017} but with a different parameter choice.

We have checked that the loop pattern occurs for the BS quintets (as seen in the upper right panel of Fig.~\ref{plot_odd}) and expect this to be generic for all rotating chains with a higher odd number of constituents.

\subsection{Chains with even numbers of constituents}

\begin{figure}[b!]
    \centering
    \includegraphics[width=7.8cm]{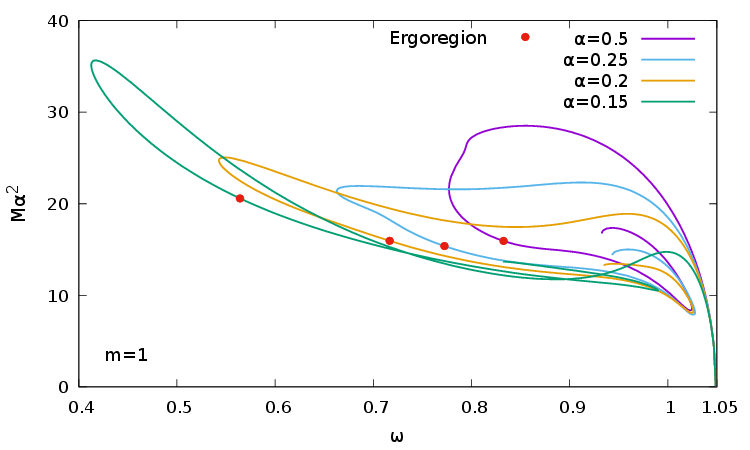}
    \includegraphics[width=7.8cm]{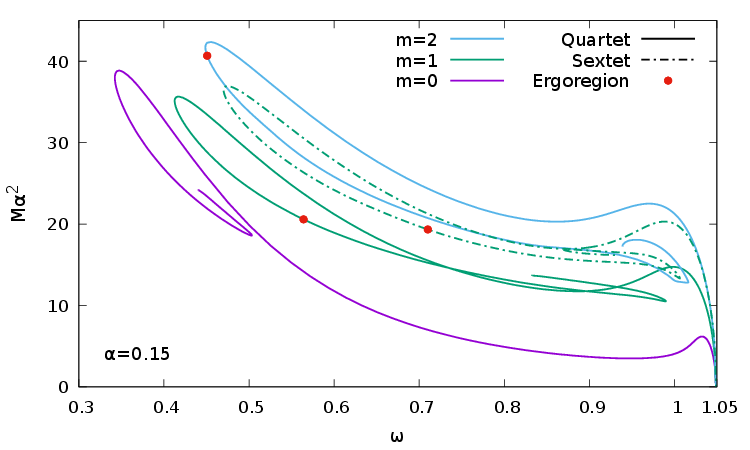}
    \includegraphics[width=7.8cm]{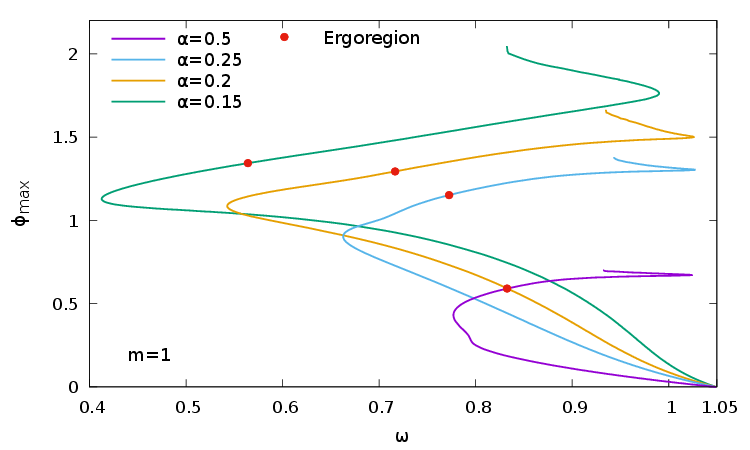}
    \includegraphics[width=7.8cm]{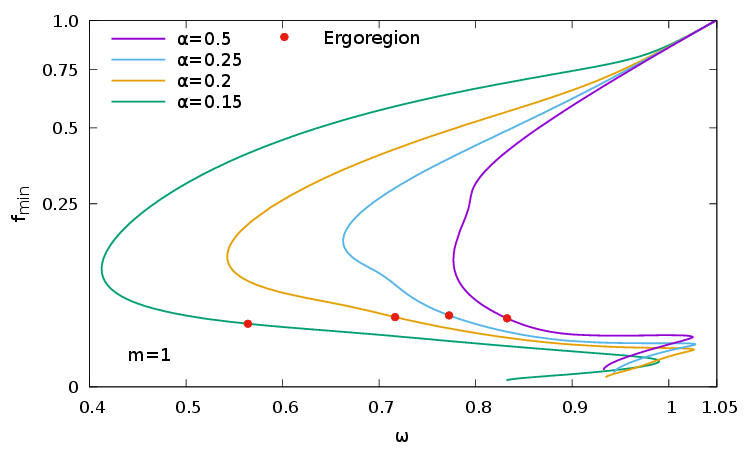}
    \includegraphics[width=7.8cm]{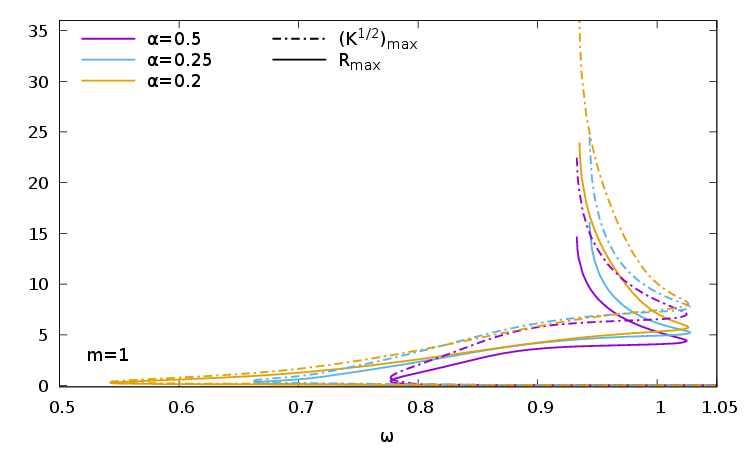}
    \caption[The mass $M$, the maximal value of the scalar field amplitude $\phi_\text{max}$, the minimal value of the metric function $f_\text{min}$ and the maximal values of curvature invariants against $\omega$ for boson star quartets with $m=1$.]{Quartets of BSs: the $\omega$ dependence of the scaled mass $M$ for different gravitational coupling $\alpha$ and rotational number $m$ (upper panels), maximal value of the scalar field amplitude $\phi_\text{max}$ (middle left panel), minimal value of the metric function $f_\text{min}$ (middle right panel) and maximal values of the curvature invariants $R$, $K$ (lower panels) for different values of $\alpha$ and $m=1$. The onset of ergoregions are indicated by red dots. In the upper right panel, the curve for BS sextets with $m=1$ is also shown for comparison.}
    \label{plot_even}
\end{figure}

We now turn to rotating chains with a higher even number of constituents. These solutions have a odd parity $\phi$ profile ($\mathcal{P}=-1$) just like BS pairs. Although these are very natural generalizations of the configurations considered in Refs.~\cite{Kleihaus2008,Herdeiro2021}, this is the first time such solutions have been explicitly constructed. It turns out that the $(\omega,M)$ diagrams for even chains are all similar to those for BS pairs. As one can see on the upper panels of Fig.~\ref{plot_even}, the curves present the inspiraling behavior with (presumably) infinitely many branches. The upper left panel show the $(\omega,M)$ diagrams for $m=1$ and different values for the gravitational coupling. Again, larger values for $\alpha$ yield a smaller domain of existence of the solutions. The sequences present an ergoregion onset (red dots) and all solutions located further down the spiral possess one. We expect the spirals to occur for all even chains, although so far we have checked it only for quartets and sextets.

The higher even chains also share some properties with the odd chains. For example the upper right panel of Fig.~\ref{plot_even} shows the $(\omega,M)$ diagrams for quartets with $\alpha=0.15$ and $m=0,1,2$. One can see that increasing the rotational number seems to decrease the domain of existence just like for the triplets of Fig.~\ref{plot_odd}. Another similarity with the odd chains is the profile of the configurations as we move on the second branch. The solutions of the fundamental branch have all constituents aligned and parallel to the $z$ axis (fourth row of Fig.~\ref{plot_fundam}) while on the second branch, the central BS pair dominates in amplitude and the other constituents are located at a larger $\rho$ coordinate than the two central ones (fourth row of Fig.~\ref{plot_2nd}). However the situation between odd and even chains becomes different when we pursue the sequence. For odd chains, the amplitude of all extrema decreases as the solutions dissolve to the flat vacuum: this can be seen on the middle left panel of Fig.~\ref{plot_odd} where the value of $\phi_\text{max}$ for triplets is shown. In contrast for even chains, only the amplitude of the outer constituents decreases while the central pair continues to grow in amplitude as we move to the different higher branches, see the middle left panel of Fig.~\ref{plot_even}.

Unfortunately, we cannot conclude on the finiteness of $\phi_\text{max}$ for the limiting configuration at the center of spirals since the $\phi$ function becomes extremely sharp for higher branches. Regarding the minimal value of the metric function $f_\text{min}$, it seems to approach zero as we move towards more involved spirals, as seen in the middle right panel of Fig.~\ref{plot_even}. 

Even though it is not clear whether the scalar field function remains finite or not as we approach the center of spirals, the limiting solutions are certainly singular. Indeed, we show in the lower panel of Fig.~\ref{plot_even} the maximal values of curvature invariants $R$ and $K$ as functions of $\omega$. Their values increase as we move towards the different branches, and the increase itself becomes larger and larger. This suggests that the curvature invariants diverge for the limiting solution. 

\begin{figure}
    \centering
    \includegraphics[width=11cm]{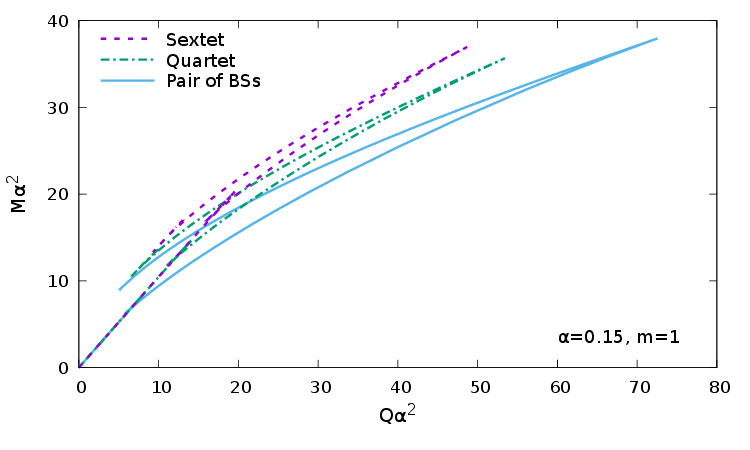}
    \caption[The mass $M$ against the charge $Q$ for chains of boson stars with even numbers of constituents with $m=1$ and $\alpha=0.15$.]{Scaled mass $M$ against the scaled charge $Q$ for chains of BSs with even numbers of constituents with $m=1$ and $\alpha=0.15$.}
    \label{plot_M_Q_even}
\end{figure}

Finally, as for the chains with odd numbers of constituents, we conjecture that the higher even chains coincide with excitations of BS pairs. To support this conjecture, we present in Fig.~\ref {plot_M_Q_even} the masses of pairs of BSs, quartets and sextets as functions of the charge $Q$. For all the solutions with an even number of constituents greater than or equal to four (quartets and sextets in the figure) there is a pair configuration with the same charge and lower mass. As for the odd chains in Fig.~\ref{plot_M_Q_odd}, the sequence of BS pairs is still below the two others even in the region where the curves almost coincide. As a result, chains with a higher even number of constituents can decay into a pair of BSs keeping their charge fixed. It is therefore plausible that they are unstable and correspond to excitations of BS pairs.

\section{Flat space limit}

The model of BSs we are considering here have a flat space limit: $Q$-balls. This limit is reached as the gravitational coupling $\alpha$ approaches zero. However, constructing full sequences of rotating solutions with small value of $\alpha$ turns out to be numerically challenging \cite{Kleihaus2005,Kleihaus2008}; for the chains configurations, only a part of the fundamental branch can be easily constructed. We have thus fixed the value of the boson frequency $\omega$ by choosing a solution on the fundamental branch, and then varied only $\alpha$. Our results are expected to apply for every BS chains belonging to the fundamental branches.

\begin{figure}
    \centering
    \includegraphics[width=7.8cm]{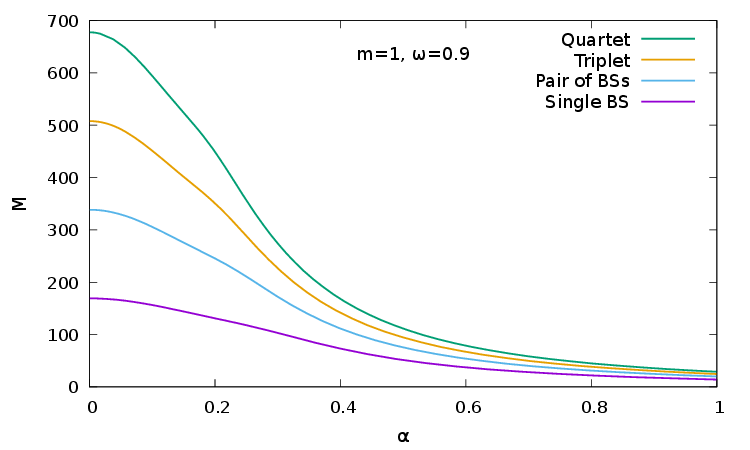}
    \includegraphics[width=7.8cm]{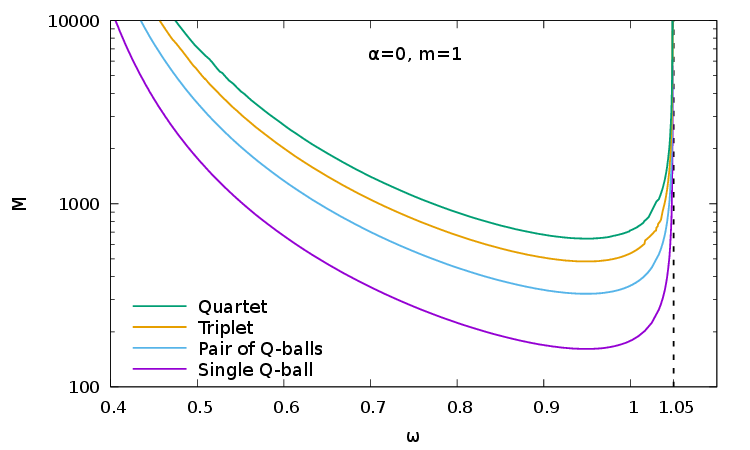}
    \caption[The mass $M$ of boson stars chains with one to four constituents against the gravitational coupling $\alpha$ for $\omega=0.9$, $m=1$ and the mass $M$ of $Q$-chains with one to four constituents with $m=1$.]{Left: mass $M$ of BS chains with one to four constituents on the fundamental branch against the gravitational coupling $\alpha$ for $\omega=0.9$ and $m=1$. Right: mass $M$ against the frequency $\omega$ for the chains of $Q$-balls with one to four constituents with $m=1$.}
    \label{M_flat_fundam}
\end{figure}

We present the mass $M$ of the chains up to four constituents against the gravitational coupling $\alpha$ in the left panel of Fig.~\ref{M_flat_fundam}. One can see that the flat space limit is approach smoothly when $\alpha\to 0$. For single BSs and pairs, the flat space solutions are the rotating $Q$-balls with even and odd parity presented in Refs.~\cite{Volkov2002a,Kleihaus2005,Kleihaus2008}. More interestingly, we find that triplets and quartets also have $Q$-ball counterparts in the absence of gravity; the limiting solutions thus corresponds to chains of $Q$-balls (also referred to as $Q$-chains). To our knowledge, these configurations have never been reported before in the literature although chains of nonrotating $Q$-balls coupled to a U(1) gauge field (\textit{a.k.a.} gauged $Q$-balls) have been constructed recently \cite{Loiko2021}.

\begin{figure}
    \centering
    \includegraphics[width=7.8cm]{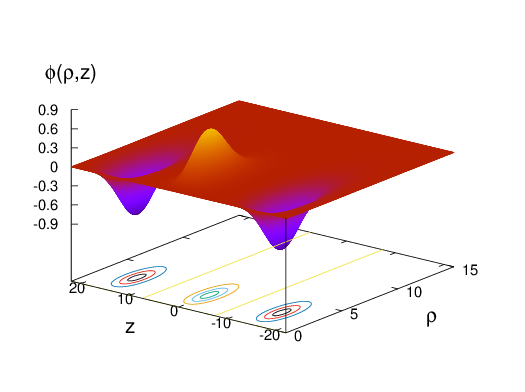}
    \includegraphics[width=7.8cm]{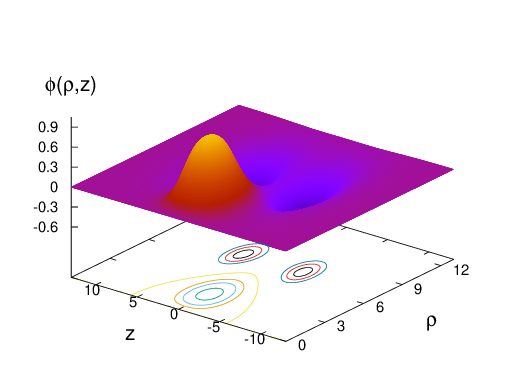}
    \caption[Profiles of $Q$-ball triplets of the first and second family for $\omega=0.9$ and $m=1$.]{$Q$-ball triplets of the first family (left) and second family (right) with $\omega=0.9$ and $m=1$. The scalar field function $\phi$ is plotted against cylindrical coordinates $(\rho,z)$.}
    \label{profile_Qballs}
\end{figure}

\vspace{1cm}
As seen in the right panel of Fig.~\ref{profile_Qballs}, the $\phi$ profile for the $Q$-chains is very similar to that of BSs. We have checked that our solutions are numcerically stable by increasing the resolution of the mesh used for the discretization. Increasing the number of grid points for a given solution does not affect the $\phi$ profile and in particular, the distance between neighboring constituents does not change.

The sequences of $Q$-cahins up to four constituents are presented in the right panel of Fig.~\ref{M_flat_fundam} where the mass $M$ is plotted against $\omega$. The frequency dependence of the charge $Q$ is qualitatively similar. We recover the main properties that are known for single $Q$-balls or pairs. The mass and the charge both diverge at the upper bound of the frequency domain, $\omega=\omega_\text{max}$, where $\omega_\text{max}$ still corresponds to the boson mass $u_0$ as in Eq.~\eqref{om_max}. The domain is also bounded from below at a finite value of the frequency, but now the lower bound depends only on the parameters entering the potential \cite{Volkov2002a,Kleihaus2005,Kleihaus2008},
\begin{equation}
    \omega_\text{min}^\text{$Q$-ball}=\sqrt{u_0^2+\frac{\lambda^2}{4}}.
\end{equation}
Mass and charge diverge as well in this limit. It is worth mentioning that contrary to the BS case, the different constituents in the $Q$-chains remain aligned throughout the sequence.

A remarkable observation is that the mass of a $Q$-chain with $n$ constituents coincides with $n$ times the mass of a single $Q$-ball. This can be understood in the following way. In the absence of gravity, the different constituents interact with each other via the scalar interaction which is short ranged because the scalar field in massive. Therefore, if the neighboring $Q$-balls in chains are far enough apart, then they almost do not interact. As a consequence, the energy of the whole $Q$-chain should be very close to the sum of the energies of the different constituents taken indivudally.

It remains to clarify what happens to the higher branches of BS chains in the flat space limit. For single BSs, this analysis has already been carried out by the authors in Ref.~\cite{Kleihaus2005}. They found that the spiral is shifted to the low frequencies so that the minimal frequency for the BSs becomes smaller than the one for $Q$-balls as $\alpha$ approaches zero. Therefore, for small values of $\alpha$, the solutions belonging to higher branches in the spiral have their frequency in the range
\begin{equation}
    \omega\in\left[\omega_\text{min}(\alpha\to 0),\omega_\text{min}^\text{$Q$-ball}\right],
\end{equation}
and they do not admit a flat space limit with finite charge or mass.

The chains of rotating BSs with even numbers of constituents also exhibit the spiraling frequency dependence of their mass (or change). Although we have not been able to construct full sequences for very small values of the gravitational coupling, we conjecture that the higher branches of BS chains with even numbers of constituents do not have a regular flat space limit just like the single BSs.

\begin{figure}
    \centering
    \includegraphics[width=7.8cm]{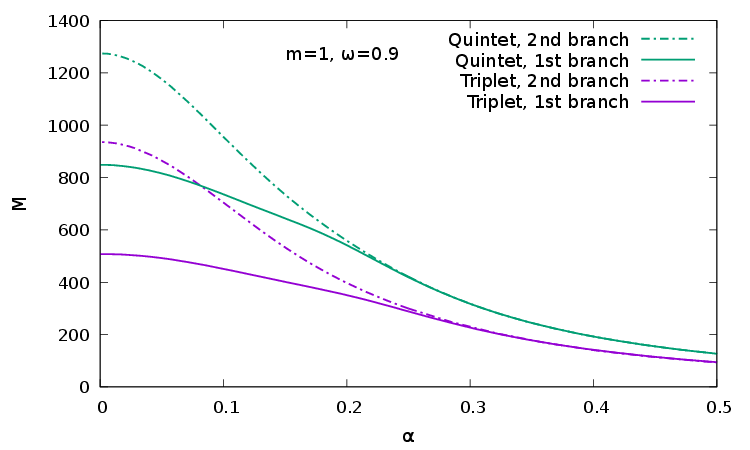}
    \includegraphics[width=7.8cm]{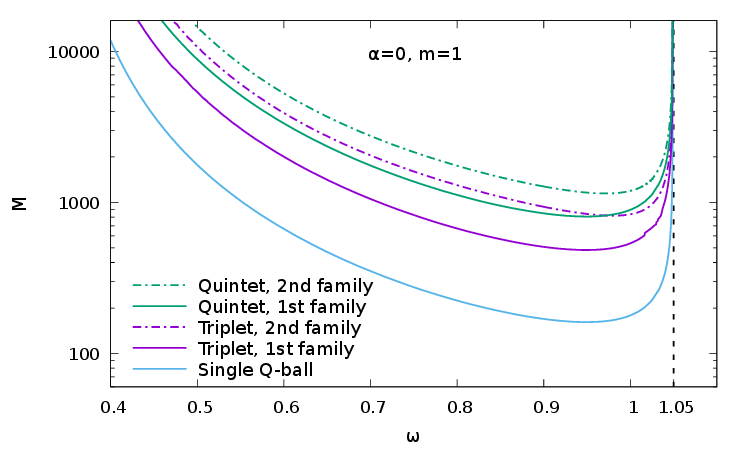}
    \caption[The mass $M$ of boson stars triplets and quintets on the first and second branches against the gravitational coupling $\alpha$ for $\omega=0.9$, $m=1$ and the mass $M$ of the corresponding $Q$-chains against $\omega$.]{Left: mass $M$ of BS triplets and quintets on the first and second branches against the gravitational coupling $\alpha$ for $\omega=0.9$ and $m=1$. Right: mass $M$ against the frequency $\omega$ for the two families of $Q$-balls triplets and quintets with $m=1$. For comparison, the curve for single $Q$-balls is also shown.}
    \label{M_flat_2nd}
\end{figure}

The situation is different for the chains of BSs with an odd number of constituents. These configurations have their second branch starting at $\omega_\text{min}$ and ending at $\omega_\text{max}$ where the solutions converge to the flat vacuum. Choosing a solution on the second branch, we find that a regular limiting solution is approached when $\alpha\to 0$. The $\alpha$ dependence of the mass for BS triplets and quintets on the second branch are presented in the left panel of Fig.~\ref{M_flat_2nd}. We also compare to the curves for solutions on the fundamental branch. It turns out that the two flat space solutions are different and thus new families of $Q$-balls exist.

We construct the sequences of these new families of solutions for triplets and quintets and show them in right panel of Fig.~\ref{M_flat_2nd}. The second families are always more energetic than the first ones, for any value of the scalar field frequency. In the right panel of Fig.~\ref{profile_Qballs}, we present the $\phi$ profile for a typical triplet of the second family: the central constituent is centered in the equatorial place and is surrounded by the other companions but the different constituents are not aligned. The two satellites are close to the central $Q$-ball and thus the scalar interaction between them is no longer negligible. As a result, the energy of $Q$-chains of the second family is higher than the sum of the energies of the individual constituents.

We conclude that the set of rotating solutions in the $Q$-ball theory is richer than expected. There is still an open question whether the $Q$-chains are stable or not. If the BS chains are unstable as we conjecture, it seems very unlikely that their flat space counterparts would be stable in the absence of the attractive gravitational interaction.

Our result are expected to generalize for higher number of constituents and for all rotational numbers $m\geq 1$. However the flat space limit of the nonrotating ($m=0$) chains of BSs found by Herdeiro \textit{et al.} \cite{Herdeiro2021} remains an open issue. A proof of non-existence have been reported recently for nonrotating $Q$-dipole \cite{Cunha2022}. Their argument holds for all solutions with an odd $\phi$ profile (even numbers of constituents) but not for even $\phi$ profile. Our numerical approach indicates that chains of nonrotating $Q$-balls are unlikely to exist for any parity of the scalar field function. Indeeed, it is possible to obtain numerical solutions for $\alpha\to 0$, but the configurations are numerically unstable: increasing the grid resolution leads to an increase of distances between the neighboring constituents. Moreover, it seems unlikely that an equilibrium field configuration exists since there is no dipole-dipole interaction in the absence of rotation. The only possible interaction between static $Q$-balls is either a scalar attraction (if they are in phase), or a scalar repulsion (if they are in opposite phase) \cite{Battye2000,Bowcock2009}.

\section{Conclusion and perspectives}

We have addressed the rotating generalization of the static chains of BSs reported in Ref.~\cite{Herdeiro2021}. They are regular configurations of a self-interacting complex scalar field minimally coupled to GR. The scalar field has a harmonic time and azimuthal dependencies associated with the boson frequency $\omega$ and the rotational number $m$. Similar chains were also considered in Refs.~\cite{Herdeiro2021a,Sun2023} but for a scalar field potential without self-interactions. In our model, this limit is reached when $\alpha\to\infty$. A single rotating BS has a toroidal shape \cite{Kleihaus2005} and the chains we have constructed here consist of a stack of multiple tori.

The scalar field amplitude changes sign between two neighboring constituents in a chain. This can be viewed as two BSs with opposite phase and it yields a repulsive scalar interaction \cite{Battye2000,Bowcock2009,Cunha2022}. This repulsion between neighboring constituents must be balanced by an attractive interaction for an equilibrium configuration to exist. In the present case, there are actually two attractive interactions: gravitation and the dipole-dipole interaction which is due to the non-zero angular momentum.

The rotating chains of BSs have been constructed by using the FreeFem finite element solver. To test the robustness of our code, we have also reproduced known solutions: rotating single BSs, pairs \cite{Kleihaus2005,Kleihaus2008} and nonrotating chains \cite{Herdeiro2021}. The convergence of our code when computing global quantities is of fourth-order in the number of points used for the discretization. 

We have constructed chains of rotating BSs with up to six constituents but they are likely to exist for an arbitrarily large number of constituents. The frequency range for which solutions exist is finite, just like for the single BSs or the pairs. The upper bound is completely determined by the mass of scalar excitations around the vacuum (namely, the coefficient in front of the quadratic term in the potential) whereas the lower bound depends in the gravitational coupling $\alpha$ and on the rotational number $m$. We have constructed sequences of solutions with fixed values $\alpha$, $m$ and focused our interest on the frequency dependence of quantities such as the mass, the maximal value of the scalar field amplitude or the maximal values of curvature invariants. Different properties emerge depending on the parity of the scalar field function. We have also discussed a qualitative argument regarding the stability of the chains and analyzed the flat space limit when $\alpha\to 0$.

On the one hand, the chains with an \textit{odd} number of constituents do not present the spiraling frequency dependence of their mass that is typical for single BSs. Instead, the $(\omega,M)$ diagrams form non-trivial loops which start and terminate at the flat vacuum. As a consequence, these solutions cannot be uniquely parametrized by a single parameter. It turns out that the BS triplets were previously obtained in Ref.~\cite{Collodel2017} in which the solutions are referred to as radially excited BSs. It should be emphasized that the starting point of this work was the static chains of Ref.~\cite{Herdeiro2021}. It was only after constructing the BS triplets that we discovered that they correspond to the solutions of Ref.~\cite{Collodel2017}. Hence all the chains with an odd number constituents may correspond to excitations of single BSs. To illustrate this, we have shown that triplets and quintets are more energetic than single BSs with the same Noether charge. The latter is related to the number of particles in the field configuration and must be conserved over time. Therefore, triplets, quintets and other chains with higher odd numbers of constituents are likely to radiate their energy and decay into a single BS with the same number of particles or, equivalently, with the same charge. The odd chains admit a flat space limit when the gravitational coupling approaches zero. In this limit, the solutions reduce to two different families of $Q$-ball chains ($Q$-chains) which have never been reported in the literature.

On the other hand, the chains with an \textit{even} number of constituents exhibit a spiraling frequency dependence of their mass. At the same time, the scalar field amplitude and the curvature invariants grow as we move along the different branches of the spiral. In particular the growth of the curvature invariants indicates that the limiting configurations at the spiral centers certainly have a singular spacetime geometry. We have also found $Q$-chains with even numbers of constituents which correspond to the flat space limit of even BS chains. Although we have checked these properties only for pairs, quartets and sextets, we expect our results to be generic for all chains with higher even numbers of constituents. Finally, for a given charge, the quartets and the sextets are more energetic than BS pairs so they are likely to be unstable. Hence we conjecture that all the chains with higher even numbers of constituents correspond to excitations of BS pairs. 

A dynamical study of the chains of BSs is still lacking. This would confirm their possible decay into single BSs or pairs. On the top of that, dynamical simulations could help to find potential scenarios for the formation of the chains. The time evolution of such a nonlinear physical system is certainly very complex and cannot be inferred from the present work. Time evolution and stability analysis of BSs with a self-interacting potential have already been carried out in the literature \cite{Liebling2023,Becerril2007,Kleihaus2012,Siemonsen2021}, but only for nonrotating or rotating \textit{single} BSs. For BSs without self-interactions, dynamical simulations of binaries can be found, for example in Ref.~\cite{Palenzuela2007}. Nevertheless, we found strong evidences that chains with more than two constituents are certainly unstable. In any case, the onset of ergoregions in the sequences of BS chains indicates the presence of an instability for the configurations which possess one.

Another perspective would be to include rotating black holes. For a single rotating BS, it is known that the configuration can support a black hole at the center \cite{Herdeiro2014}, which is not possible in the non-rotating case. The combination of a black hole with a pair of rotating BSs has also been considered in Ref.~\cite{Kunz2019}. Therefore, it is natural to consider a generalization of these configurations for rotating chains with more than two constituents. One could place a black hole either at the center of the chain, or include a black hole at the center of each of the constituents in the chain.
\vfill

\begin{subappendices}

\section{System of partial differential equations}
\label{exp_eq_bs}

The set of coupled elliptic PDEs we obtain after injecting the ansatz \eqref{metric_bs}, \eqref{phi_bs} to the field equations is the following
\begin{align*}
    r^2\phi_{,rr}+\phi_{,\theta\theta}+2r\,\phi_{,r}+\text{cot}\,\theta\,\phi_{,\theta}=&-\frac{h}{f^2}\left(\ell(r\,\omega+m\,w)^2-r^2\ell\,f\,U'(\phi^2)-f^2m^2\text{csc}^2\theta\right)\phi\\
    \label{eqphi}
    &-\frac{1}{2\ell}\left(r^2\,\ell_{,r}\phi_{,r}+\ell_{,\theta}\phi_{,\theta}\right),\numberthis{}\\
    r^2 f_{,rr}+f_{,\theta\theta}+2r\,f_{,r}+\text{cot}\,\theta\,f_{,\theta}=&\frac{\ell}{f}\big(8\alpha^2 h\left(r\,\omega+m\,w\right)^2\phi^2-4r^2\alpha^2f\,h\,U(\phi^2)\\
    &+\sin^2\theta(w-r\,w_{,r})^2+\sin^2\theta\,w_{,\theta}^2\big)\\
    \label{eqf}
    &+\frac{1}{f}\left(r^2 f_{,r}^2+f_{,\theta}^2\right)-\frac{1}{2\ell}\left(r^2\ell_{,r}f_{,r}+\ell_{,\theta}f_{,\theta}\right),\numberthis{}\\
    r^2 \ell_{,rr}+\ell_{,\theta\theta}+3r\,\ell_{,r}+2\,\text{cot}\,\theta\,\ell_{,\theta}=&\frac{8\alpha^2\ell^2h}{f^2}\left((r\,\omega+m\,w)^2\phi^2-r^2f\,U(\phi^2)\right)\\
    &-8\alpha^2 m^2\text{csc}^2\theta\,\ell\,h\,\phi^2+\frac{1}{2\ell}\left(r^2\ell_{,r}^2+\ell_{,\theta}^2\right),\numberthis{}\\
     r^2 h_{,rr}+h_{,\theta\theta}+r\,h_{,r}=&\frac{4\alpha^2 h^2}{f^2}\left(3m^2\text{csc}^2\theta\,f^2-\ell(r\,\omega+m\,w)^2\right)\phi^2\\
    &+\frac{4\alpha^2 h^2\ell}{f}r^2 U(\phi^2)-4\alpha^2 h\left(r^2\phi_{,r}^2+\phi_{,\theta}^2\right)\\
    &-\frac{h}{2f^2}\left(r^2 f_{,r}^2+f_{,\theta}^2\right)+\frac{3h\,\ell}{2f^2}\sin^2\theta\left((w-r\,w_{,r})^2+w_{,\theta}^2\right)\\
    \label{eqh}
    &+\frac{h}{2\ell^2}\left(r^2 \ell_{,r}^2+\ell_{,\theta}^2\right)+\frac{1}{h}\left(r^2 h_{,r}^2+h_{,\theta}^2\right)\\
    &+\frac{2h}{\ell}\left(r\,\ell_{,r}+\text{cot}\,\theta\,\ell_{,\theta}\right),\numberthis{}\\
    r^2 w_{,rr}+w_{,\theta\theta}+2r\,w_{,r}+3\,\text{cot}\,\theta\,w_{,\theta}=&8\alpha^2\text{csc}^2\theta\,m\,h(r\,\omega+m\,w)\phi^2+2w-\frac{2}{f}\big(r\,f_{,r}(w-r\,w_{,r})\\
    \label{eqw}
    &-f_{,\theta}w_{,\theta}\big)+\frac{3}{2\ell}\left(r\,\ell_{,r}(w-r\,w_{,r})-\ell_{,\theta}w_{,\theta}\right),\numberthis{}
\end{align*}
where we have introduced the compact notation $\phi_{,\mu}\equiv\partial_\mu\phi$. The Eq.~\eqref{eqphi} corresponds to the Klein-Gordon equation, \eqref{kg_bs}, whereas the Eqs.~\eqref{eqf}-\eqref{eqw} correspond respectively to the following combinations of the Einstein equation \eqref{ein_bs},
\begin{align*}
    \tensor{E}{^r_r}+\tensor{E}{^\vartheta_\vartheta}+\tensor{E}{^\varphi_\varphi}-\tensor{E}{^t_t}-\frac{2}{r}w\,\tensor{E}{^t_\varphi}&=0,\\
    \tensor{E}{^r_r}+\tensor{E}{^\vartheta_\vartheta}-\tensor{E}{^\varphi_\varphi}+\tensor{E}{^t_t}+\frac{2}{r}w\,\tensor{E}{^t_\varphi}&=0,\\
    \tensor{E}{^r_r}+\tensor{E}{^\vartheta_\vartheta}-\tensor{E}{^\varphi_\varphi}-\tensor{E}{^t_t}&=0,\\
    \tensor{E}{^t_\varphi}&=0.\numberthis{}
\end{align*}

If we fix the metric to be Minkowski $f=\ell=h=1$, $w=0$, and set to zero the gravitational coupling $\alpha$, only one PDE remains, Eq.~\eqref{eqphi}, and it describes rotating $Q$-balls.

\end{subappendices}

\label{conclusion_bs}

     \renewcommand{\chaptermark}[1]{\MakeUppercase{\markboth{}{#1}}}

\addcontentsline{toc}{chapter}{Conclusion}\chaptermark{Conclusion}
\chapter*{Conclusion}

Dans cette thèse, nous avons considéré et analysé deux exemples de trous noirs statiques avec cheveux. Tout d'abord, nous avons étudié des trous noirs chevelus à symétrie sphérique dans la théorie de la bigravité massive. Ces derniers avaient été découverts pour la première fois par Brito, Cardoso et Pani \cite{Brito2013a} en 2013, mais leur existence avait été remise en question par un groupe Suédois \cite{Torsello2017} quelques années plus tard. Nous confirmons les résultats de Brito et al., et avons élucidé les raisons pour lesquelles Torsello, Kocic et Mörtsell n'ont pas été en mesure de reproduire ces solutions. Nous renvoyons à la \hyperref[conclusion_bigravity]{conclusion du premier chapitre} pour une discussion à ce sujet. Nous avons aussi mis en évidence que ces trous noirs chevelus peuvent être stables, ce qui permet de résoudre le problème d'instabilité de la solution de Schwarzschild en bigravité \cite{Babichev2013}. Afin d'obtenir des masses physiquement acceptables pour ces trous noirs, il est nécessaire de choisir une valeur extrêmement faible pour la constante de couplage de l'une des deux métriques. Dans cette limite, l'équation qui détermine la métrique physique, celle qui est choisie pour décrire la géométrie de l'espace-temps, est extrêmement proche de l'équation d'Einstein dans le vide. Ainsi, la géométrie des trous noirs chevelus que nous considérons comme physiquement pertinents est très proche de la géométrie de Schwarzschild. Les cheveux de ces trous noirs résident dans la seconde métrique, qui n'est pas directement mesurable. Cela soulève la question de leur détection par le biais d'observations astronomiques.

Bien que les solutions statiques (ou stationnaires) de la bigravité s'éloignent peu de la relativité générale, des effets considérablement différents, tant qualitativement que quantitativement, sont attendus lors de processus dynamiques. En effet, la dynamique de la relativité générale est caractérisée par la propagation de deux degrés de liberté associés au graviton sans masse, tandis que la bigravité massive propage 7 degrés de liberté -- 5 pour le graviton massif et 2 pour le graviton sans masse. À titre d'exemple, les travaux de Max, Platscher et Smirnov \cite{Max2017} mettent en évidence un phénomène de modulation des signaux d'ondes gravitationnelles en bigravité qui est absent en relativité générale. Dans une pré-publication récente \cite{Cardoso2023}, Cardoso et ses collaborateurs établissent également la présence d'un mode dipolaire nouveau dans les signaux décrits par la bigravité. Il convient cependant de noter qu'aucune simulation de fusion de trous noirs n'a encore été réalisée dans le cadre de cette théorie alternative de la gravitation. De telles simulations pourraient potentiellement révéler la signature des cheveux lors de la fusion de deux trous noirs chevelus. 

La solution de Kerr qui est la plus pertinente pour décrire des trous noirs réalistes existe également en bigravité \cite{Babichev2014b}. Il serait intéressant de vérifier si des versions chevelues de ce trou noir existent.

Nous avons ensuite étudié des trous noirs chevelus portant une charge magnétique dans le cadre du secteur bosonique de la théorie électrofaible couplé à la relativité générale (modèle Einstein-Weinberg-Salam). Une version à symétrie sphérique de ces trous noirs avait été établie par Bai et Korwar \cite{Bai2021} en 2021. Leurs solutions correspondent à la plus petite charge magnétique admissible pour une configuration non-Abélienne du champ. Nous avons construit des généralisations de ces trous noirs pour des charges magnétiques plus élevées dans le cas à symétrie axiale. Cependant, il est attendu que des solutions encore plus générales, sans symétrie continue, existent. De telles solutions avaient été construites au niveau perturbatif dans une théorie similaire par Ridgway et Weinberg \cite{Ridgway1995} en 1995. De plus, notre analyse des perturbations de la solution de Reissner-Nordström pour de grandes charges magnétiques donne également des indications en ce sens. En effet, nous avons identifié des instabilités dont la dépendance angulaire est caractérisée par des harmoniques sphériques $Y_{j,m}(\vartheta,\varphi)$ avec $j>0$. Lorsque $m=0$, ces perturbations présentent une symétrie axiale, et leur développement au niveau non-linéaire conduit à la formation de cheveux axisymétriques. Néanmoins, la croissance de modes plus généraux avec $m\neq 0$ ne peut être exclue, ce qui pourrait mener à des trous noirs chevelus présentant seulement des symétries discrètes. Pour obtenir ces solutions de manière explicite, il est nécessaire de résoudre numériquement un problème elliptique tridimensionnel. En principe, cette tâche est réalisable avec la librairie FreeFem, à condition de disposer de suffisamment de puissance de calcul. 

Si plusieurs trous noirs chevelus existent avec la même charge magnétique mais différentes symétries, il restera à déterminer lequel est stable. Une bonne indication peut être l'argument énergétique : la configuration du champ la plus stable est, en général, celle qui a la plus basse énergie. Une autre possibilité serait de laisser évoluer dynamiquement un trou noir (instable) de Reissner-Nordström pour observer quel est l'état stable final. Cette perspective est toutefois difficile à réaliser en pratique dans le cas gravitationnel. Puisque les propriétés de stabilité des trous noirs électrofaibles sont similaires à celles de leurs analogues en espace-temps plat, un bon point de départ serait de réaliser cette évolution pour les monopôles, sans tenir compte de la gravitation. Des évolutions en théorie classique des champs sans gravité ont déjà été réalisées avec FreeFem, voir par exemple les travaux de Garaud, Radu et Volkov sur les vortons \cite{Garaud2013}. 

Enfin, il serait également intéressant d'étudier l'existence d'états liés constitués d'une paire de trous noirs électrofaibles de charges opposées. De telles configurations existent au niveau statique dans le modèle Einstein-Yang-Mills-Higgs \cite{Kleihaus2000}, mais elles sont instables. Aucune étude avec dépendance temporelle n'a cependant été réalisée. Dans le cadre du modèle Einstein-Weinberg-Salam, cela se traduirait par des systèmes liés de deux trous noirs magnétiques en orbite l'un autour de l'autre. Il est fort probable que l'émission d'ondes gravitationnelles empêche un tel système d'exister éternellement, mais sa durée de vie pourrait potentiellement être importante. Si les trous noirs constituant ce système sont extrémaux (ou quasi-extrémaux), leur petite taille, de l'ordre du centimètre, pourrait en faire de potentiels candidats pour la matière noire. En effet, contrairement aux trous noirs magnétiques isolés, ces binaires de trous noirs ont une charge magnétique totale nulle, et n'ont donc pas de champ magnétique monopolaire. Cependant, leur moment dipolaire non nul pourrait laisser une signature observable qui reste à déterminer.

Pour conclure sur une note plus générale, cette thèse a été l'occasion de développer un algorithme robuste basé sur la méthode des éléments finis pour résoudre des problèmes gravitationnels elliptiques. Nous avons pu tester notre algorithme pour des configurations avec ou sans horizon des évènements (comme illustré dans les chapitres \ref{chap_mon} et \ref{chap_bs}). {\`A} notre connaissance, les algorithmes employés pour des problèmes elliptiques dans la communauté de la relativité numérique sont basées soit sur les différences finies \cite{Schonauer1989,Schauder1992}, soit sur les méthodes spectrales \cite{Grandclement2010}. La librairie FreeFem \cite{MR3043640} sur laquelle est basée notre algorithme offre une alternative intéressante. Cette librairie est en constante évolution, avec de nouvelles fonctionnalités régulièrement présentées lors d'un workshop annuel\footnote{Voir \url{https://freefem.org}.}. La prise en main de FreeFem est très accessible à quiconque a déjà codé en C++, et les codes réalisés pour un problème spécifique peuvent être facilement adaptés à d'autres théories classiques des champs. La parallélisation des algorithmes est également simple à mettre en place puisque la plupart des fonctions natives proposent cette fonctionnalité. Il est à noter que nous avons utilisé FreeFem exclusivement pour des problèmes à deux dimensions effectives. Il sera intéressant de tester ses capacités pour résoudre des problèmes gravitationnels à trois dimensions.

    \renewcommand{\chaptermark}[1]{\markboth{\MakeUppercase{\chaptername\ \thechapter. {#1}}}{}}
	
	\appendix
	
\chapter[Numerical methods]{Numerical methods for solving differential equations}
\label{app_num}

In classical field theory, the equations of motion generically consist in a system of nonlinear Partial Differential Equations (PDEs). These equations have to be equipped with appropriate conditions at the boundaries of the integration domain in order to be solved. The boundary conditions follow from physical assumptions such as requiring the energy to be finite, regularity, asymptotic flatness, or symmetry considerations. If we are interested in stationary and spherically symmetric field configurations, the PDEs can be reduced to a system of Ordinary Differential Equations (ODEs). 

In general, analytical solutions to such nonlinear differential equations (ODEs or PDEs) \textit{do not} exist. Numerical analysis is then required to obtain approximate solutions at the nonlinear level. In this appendix, we present the numerical techniques that have been used throughout this thesis to solve ordinary and partial differential equations. We refer the reader to Refs.~\cite{Stoer2002,Press2007,Cook2007,Hutton2003,Reddy2005} for further reading about numerical analysis.

\section{Ordinary Differential Equations}

An ODE is defined on a domain $D\subseteq\mathbb{R}$ and can be written as
\begin{equation}
\label{generalode}
    y'(r)=f(r,y),\quad r\in D,
\end{equation}
where $y$ is the unknown function depending on the variable $r$ and $f$ is an arbitrary function which can depend non-linearly on $r$ and $y$. The unknown function can be scalar-valued, $y:\mathbb{R}\rightarrow\mathbb{R}$, or vector-valued, $y:\mathbb{R}\rightarrow\mathbb{R}^n$ with $n\in\mathbb{N}$. In the latter case, Eq.~\eqref{generalode} then describes a system of coupled ODEs where the unknown functions are the components of $y=\left(y^1,y^2,\dots,y^n\right)$. 

Although the differential equation in the form \eqref{generalode} is of first order, it can also describes an arbitrary order equation since any ODE of order $n$ can be rewritten as a system of $n$ first order ODEs. To illustrate this, consider the following second order equation for the unknown function $u:\mathbb{R}\rightarrow\mathbb{R}$,
\begin{equation}
\label{ex2order}
    u''+\alpha\,u'+\beta\,u=0,\quad \alpha,\beta\in\mathbb{R}.
\end{equation}
One can define a new function $v(r)\equiv u'(r)$. Then Eq.~\eqref{ex2order} can be rewritten as
\begin{align*}
    u'&=v,\\
    v'&=-\alpha\,v -\beta\,u.\numberthis{}
\end{align*}
This is a system of two first order equations which can be cast in the form \eqref{generalode} by the identification $y\equiv(u,v)$, $f\equiv(v,-\alpha\,v -\beta\,u)$.

In what follows we will describe how the general problem formulated by Eq.~\eqref{generalode} can be solved using numerical analysis. We emphasize that all the ODEs encountered in this thesis belong to this class of problems although they may not be written explicitly in the form \eqref{generalode} in the main text. 

The differential equation alone does not have a \textit{unique} solution. It has to be equipped with conditions to be fulfilled by the unknown function $y$ at the boundaries of the integration domain $D$. Generically, there must be as many conditions as there are first order equations in the system (in other words, the appropriate number of conditions is the dimension $n$ of the vector-valued function $y$). Depending on the location at which the conditions are imposed, the ODE is said to be an \textit{initial value problem} or a \textit{boundary value problem}. We shall treat these two types of problems separately as the latter are substantially more difficult to solve.

\subsection{Initial value problem}
\label{num_ivp}

In the context of ODEs, an initial value problem (also known as a Cauchy problem) is a differential equation for which the unknown functions have to satisfy some conditions at one of the two boundaries of the integration domain. To be specific, let us set $D\equiv[a,b]$, $a<b\in\mathbb{R}$, then an initial value problem on $D$ is defined as 
\begin{equation}
    y'(r)=f(r,y)\quad\text{with}\quad y(r=a)=y_a.
\end{equation}
Here we impose without loss of generality that the function $y$ should take the specific value $y_a$ at the beginning of the interval $[a,b]$. The evolution of $y$ is then governed by the differential equation. In this context, the condition at $r=a$ is often called the \textit{initial condition}. From the theoretical point of view, an initial value problem has good mathematical properties: the existence and uniqueness of the solution is guaranteed by the Cauchy-Lipschitz theorem \cite{Coddington1955}.

In physics, this type of problem arises in the field of dynamical systems. The variable $r$ is then often interpreted as a time variable. In this context, the ODE governs the time evolution of some physical quantities $(y^1,\dots,y^n)$ which are in a state $(y^1_a,\dots,y^n_a)$ at the initial time $r=a$.

There are several methods to solve an initial value problem numerically. In most cases, the integration domain first have to be discretized. For concreteness, let us consider a homogeneous discretisation of $D$ with $N+1$ nodes located at $r_i\in D$, $i\in\{0,\dots,N\}$,
\begin{equation}
\label{cststep}
    r_i=a+i\,\Delta r,\quad \Delta r=\frac{b-a}{N}\quad\Rightarrow\quad r_0=a,\; r_N=b.
\end{equation}
We attempt to calculate the values of $y$ at the nodes, $y_i\equiv y(r_i)$, given the initial value $y_0=y_a$. The most simple method to achieve this is called the \textit{explicit Euler method}. Each value $y_{i+1}$ is computed from $y_i$ by the recurrence relation
\begin{equation}
\label{euler}
    y_{i+1}=y_i+\Delta r\, f(r_i,y_i).
\end{equation}
Its derivation follows from the Taylor expansion of $y$ around $r_i$ up to the first order in $\Delta r$. Unfortunately, this method is very imprecise since it generates at each step errors of the order of $(\Delta r)^2$, leading after $N$ steps to a total accumulated error of the order of $\Delta r$. This would require in practice to use tiny step sizes. Therefore we choose to compute the successive values of $y_i$ by using the so-called \textit{Runge-Kutta 4} method:
\begin{equation}
\label{rk4}
    y_{i+1}=y_i+\frac{\Delta r}{6}(k_1+2k_2+2k_3+k_4),
\end{equation}
where the coefficient $k_{1,2,3,4}$ are given by
\begin{align*}
    k_1&=f\left(r_i,y_i\right),\\
    k_2&=f\left(r_i+\Delta r/2,y_i+\Delta r\,k_1/2\right),\\
    k_3&=f\left(r_i+\Delta r/2,y_i+\Delta r\,k_2/2\right),\\
    k_4&=f\left(r_i+\Delta r,y_i+\Delta r\,k_3\right).\numberthis{}
\end{align*}
At each step, the errors are of the order of $(\Delta r)^5$ so that the accumulated error is now of the order of $(\Delta r)^4$ ; this is why the Runge-Kutta 4 method is said to be of fourth-order. 

The performance of the method can be improved further by using an adaptive step size instead of the constant steps defined by the Eq.~\eqref{cststep}. Typically one can take big steps when the solution varies little, and small steps when it has big variations. There are several way to do this in practice, we refer for example to the Ref.~\cite{Press2007} for details.

\subsection{Boundary value problem}
\label{num_bvp}

A boundary value problem consists in a differential equation for which the conditions to be fulfilled by the unknown functions are specified on both boundaries of the domain $D$. Such a problem can be formulated as follows
\begin{equation}
    y'(r)=f(r,y)\quad\text{with}\quad C\left(y(a),y(b)\right)=0,
\end{equation}
where $C$ can be any vector-valued function, $C:\mathbb{R}^n\times\mathbb{R}^n\rightarrow\mathbb{R}^n$, encoding the \textit{boundary conditions}. The existence and uniqueness of the solution for this type of problem is in general not guaranteed as it was the case for initial value problems. 

From the numerical point of view, boundary value problems are also more complicated to solve since typically, the "initial" state $y(r=a)$ is not completely fixed and depends on some unknown parameters. The "missing" information resides in the conditions specified at the second boundary, for $y(r=b)$. For example, the boundary conditions can be such that the $m$ first components of $y$ at $r=a$ are fixed but not the remaining $n-m$ ones,
\begin{equation}
\label{bca}
    y^1(a)=\alpha_1,\;y^2(a)=\alpha_2,\;\dots,\;y^m(a)=\alpha_m.
\end{equation}
The problem then consists in finding the appropriate values of $y^{m+1}(a),y^{m+2}(a),\dots,y^{n}(a)$ such that on the second boundary one has,
\begin{equation}
\label{bcb}
    y^{m+1}(b)=\beta_1,\;y^{m+2}(b)=\beta_2,\;\dots,\;y^n(b)=\beta_{n-m}.
\end{equation}
One cannot use directly the Eqs.~\eqref{euler} or \eqref{rk4} to integrate the ODE in this case.

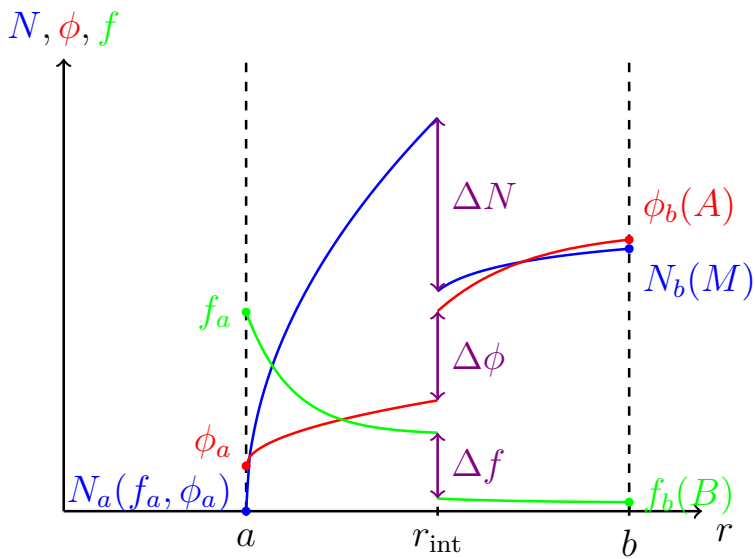
\begin{figure}
\centering
  \scalebox{1.2}{
\begin{tikzpicture}
\draw[thick,->] (-1,0) -- (6,0) node[anchor=north west] {$r$};
    \draw[thick,->] (-1,0) -- (-1,5) node[anchor=south] {${\color{blue}N},{\color{red}\phi},{\color{green}f}$}; 
    \draw[thick] (1,0.05) -- (1,-0.05) node[anchor=north] {$a$};
    \draw[thick] (5.2,0.05) -- (5.2,-0.05) node[anchor=north] {$b$};
    \draw[thick] (3.1,0.05) -- (3.1,-0.05) node[anchor=north] {$r_\text{int}$};
    \draw[thick,dashed] (1,0) -- (1,5);
    \draw[thick,dashed] (5.2,0) -- (5.2,5);
    
    \draw[thick,blue] plot[variable=\x,samples=60,smooth,domain=1:3.1] ({\x},{3.0*sqrt(\x-1)});
    \draw[thick,blue] plot[variable=\x,samples=60,smooth,domain=3.1:5.2] ({\x},{2.24956+sqrt(1-3.0/\x)});
    \draw[thick,blue,fill=blue] (1,0) circle (1pt) node[anchor=east,yshift=6pt] {$N_a(f_a,\phi_a)$};
    \draw[thick,blue,fill=blue] (5.2,2.9) circle (1pt) node[anchor=north west] {$N_b(M)$};
    
    \draw[thick,red] plot[variable=\x,samples=60,smooth,domain=1:3.1] ({\x},{0.5+0.5*sqrt(\x-1)});
    \draw[thick,red] plot[variable=\x,samples=60,smooth,domain=3.1:5.2] ({\x},{3.11033-20.0*exp(-\x)});
    \draw[thick,red,fill=red] (1,0.5) circle (1pt) node[anchor=south east,yshift=-2pt] {$\phi_a$};
    \draw[thick,red,fill=red] (5.2,3.0) circle (1pt) node[anchor=south west] {$\phi_b(A)$};
    
    \draw[thick,green] plot[variable=\x,samples=60,smooth,domain=1:3.1] ({\x},{0.846647+10.0*exp(-2.0*\x)});
    \draw[thick,green] plot[variable=\x,samples=60,smooth,domain=3.1:5.2] ({\x},{0.0944834+exp(-\x)});
    \draw[thick,green,fill=green] (1,2.2) circle (1pt) node[anchor=east] {$f_a$};
    \draw[thick,green,fill=green] (5.2,0.1) circle (1pt) node[anchor=west,yshift=2pt] {$f_b(B)$};
    
    \draw[thick,violet,<->] (3.1,4.34741) -- (3.1,2.42917) node[anchor=west,yshift=30pt] {$\Delta N$};
    \draw[thick,violet,<->] (3.1,2.20935) -- (3.1,1.22457) node[anchor=west,yshift=13pt] {$\Delta\phi$};
    \draw[thick,violet,<->] (3.1,0.866941) -- (3.1,0.139533) node[anchor=west,yshift=10pt] {$\Delta f$};
\end{tikzpicture}}
  \caption[Schematic representation of the resolution of a boundary value problem using the shooting method.]{Schematic representation of the resolution of a boundary value problem using the shooting method. The system of ODEs is of the form $N'(r)=\mathcal{D}_N(r,N,\phi,f)$, $\phi''(r)=\mathcal{D}_\phi(r,N,\phi,f)$, $f''(r)=\mathcal{D}_f(r,N,\phi,f)$ and can be rewritten as a system of 5 first order equations. The shooting parameters are $\{\phi_a,f_a,M,A,B\}$ and their values are obtained by resolving the matching conditions $\Delta N=0$, $\Delta\phi=0$, $\Delta f=0$, $\Delta\phi'=0$ and $\Delta f'=0$.}
  \label{shootfig}
\end{figure}

Among the various existing algorithms for dealing with boundary value problems \cite{Stoer2002,Press2007}, we use the \textit{shooting method}. We illustrate how this algorithm works in the Fig.~\ref{shootfig}. First, one splits the domain $D$ into two intervals,
\begin{equation}
    D_\text{I}=[a,r_\text{int}],\quad D_\text{II}=[r_\text{int},b],\quad\text{with}\quad a<r_\text{int}<b.
\end{equation}
Then, one chooses an initial guess for $y(a)$ and $y(b)$ which satisfy the boundary conditions. If the problem is well-posed, one should have
\begin{equation}
\label{initialguess}
    y(a)=y_a(c_0,\dots,c_n),\;\; y(b)=y_b(c_0,\dots,c_n)\;\;\text{such that}\;\; C(y_a,y_b)=0\;\;\forall c_k\in\mathbb{R}.
\end{equation}
The $c_k$'s are sometimes called the \textit{shooting parameters} and there should be as many as there are equations in the system, $k\in\{1,\dots,n\}$. Choosing an initial guess means assigning some numerical values to these parameters. The next step is to integrate the ODE on $D_\text{I}$ by using for example Eq.~\eqref{rk4}, starting from $r=a$ up to $r=r_\text{int}$ and obtain a value $y_\text{int,I}\equiv y(r_\text{int})$. Then, integrate the ODE for $r\in D_\text{II}$, starting from $r=b$ up to $r=r_\text{int}$ and obtain $y_\text{int,II}\equiv y(r_\text{int})$. At this stage, the values $y_\text{int,I}$ and $y_\text{int,II}$ would not agree, but since they are obtained from direct integration with the initial conditions \eqref{initialguess}, their difference depends on the shooting parameters, $y_\text{int,II}-y_\text{int,I}\equiv\Delta y(c_0,\dots,c_n)$. A global solution on $D=D_\text{I}\cup D_\text{II}$ is thus obtained only if
\begin{equation}
\label{matchcond}
    \Delta y(c_0,\dots,c_n)=0.
\end{equation}
This is formally an algebraic system of $n$ equations for the unknowns parameters $c_0,\dots,c_n$. The Eq.~\eqref{matchcond} is sometimes called the matching condition. Such a system can be solved by iteration using the \textit{Newton's method}. At each iteration, one has to integrate the ODE as described above, compute $\Delta y$, and adjust the values of the shooting parameters according to the Newton's formula (see the next subsection). The iterations are supposed to converge provided that the initial guess is not too far from the "true" solution. 

Summarizing, with the shooting method, a boundary value problem is transformed into two independent initial value problems so that we can use the Runge-Kutta 4 formula \eqref{rk4}. Then, the problem is reduced to an algebraic equation to solve which requires multiple iterations using the Newton's method. The drawback of this algorithm is its sensibility to the initial guess. Typically if the shooting parameters are initialized randomly, the iterations may not converge. It is also possible to encounter singularities that prevent integrating the ODEs up to $r=r_\text{int}$. Both of these issues can be avoided by using the \textit{multishooting method}. The idea is to generalize the shooting method by dividing the integration domain in several intervals, $D=D_\text{I}\cup D_\text{II}\cup D_\text{III}\cup\dots$. In addition to the original shooting parameters at the boundaries of $D$, we have new parameters to adjust which are the values of $y$ at each intersection between the intervals. The algebraic system to solve is thus bigger than the one obtained from the shooting method (it consists in more than $n$ equations). Therefore, the computation time may be increased, but the issues mentioned above should be avoided.

Finally, it is worth noting that the convergence of the (multi)shooting method towards a numerical solution does not prove the existence of the solution. For example, by satisfying the matching condition \eqref{matchcond}, the continuity of the numerical solution at $r=r_\text{int}$ is guaranteed, but not that of its derivatives. In practice, one can check the validity of a solution by varying the number of intervals used for the multishooting or their sizes. The numerical solution and its global physical quantities (for example, the mass) should not be affected by such changes.

\subsection{Newton's method}

In numerical analysis, it is very common that a differential system of equations is reduced to algebraic equations (for example, with the shooting method). For linear equations, finding the solution requires inverting a matrix (numerical methods for linear equations can be found in Ref.~\cite{Press2007}). Here we present briefly the Newton's method which is an algorithm for solving \textit{nonlinear} equations. Let us consider the following generic algebraic system of equations,
\begin{equation}
\label{alg_eq}
    F(x)=0,\quad x=(x^1,x^2,\dots,x^n)\in\mathbb{R}^n,
\end{equation}
where $F$ is a differentiable vector-valued function, $F:\mathbb{R}^n\rightarrow\mathbb{R}^n$, representing any system of $n$ equations for the $n$ real variables $x^1,\dots,x^n$. The strategy is to find an approximated solution by iterations starting from an initial guess $x_0$. The successive values $x_i$ are then computed by
\begin{equation}
    \label{newton}
    x_{i+1}=x_i-\alpha J^{-1}F(x_i),
\end{equation}
where $\alpha$ is a real parameter characterizing the step size and $J$ is the Jacobian matrix of $F$ whose coefficients are $J_{k\ell}\equiv\partial F^k/\partial x^\ell$. If the partial derivatives of the vector function $F$ are not known analytically (this is the case in Eq.~\eqref{matchcond} for the shooting method), one can use finite differences to compute them approximately,
\begin{equation}
    \frac{\partial F^k}{\partial x^\ell}\approx\frac{F^k(\dots,x^\ell+h,\dots)-F^k(\dots,x^\ell,\dots)}{h},\quad h>0.
\end{equation}

For a zero of $F$ with a multiplicity of 1, the convergence will be quadratic in the neighborhood of the solution. This means roughly that the number of correct digits of $x_i$ doubles at each step. One usually stops the algorithm when the difference $|x_{i+1}-x_i|$ is below a certain prescribed tolerance.

The drawback of the Newton's method is its sensibility to the initial guess used to start the iterations. If there is no unique solution to Eq.~\eqref{alg_eq}, finding the different solutions requires using different initial guesses and it may not be clear which solution the algorithm will converge to. In some cases, it may not converge at all. The convergence can be improved by choosing a suitable value for $\alpha$ at each step and we refer to the Ref.~\cite{Press2007} for more details.

\section{Partial Differential Equations}
\label{num_pde}

The theory of PDEs is a much richer topic than the study of ODEs. For example, one cannot transform a PDE of order $n$ into a system of $n$ first order equations which renders their classification more complicated. Here we will treat only the case of second order equations which are of great interest in physics. More specifically, in classical field theory, one usually encounters \textit{semilinear} PDEs. Such equations are linear only in the highest order derivatives and can be cast in the form
\begin{equation}
    \sum_{i=1}^n\sum_{j=1}^n a_{ij}\frac{\partial^2 u}{\partial x^i\partial x^j}+\text{lower-order terms}=0,
\end{equation}
where $u=u(x^1,\dots,x^n)$ is the unknown function and $a_{ij}=a_{ij}(x^1,\dots,x^n)$ are (real) coefficients. The second order PDEs are then classified according to the signature of the eigenvalues of the coefficient matrix $a_{ij}$:
\begin{itemize}
    \item\textit{elliptic}: the eigenvalues are all positive or all negative,
    \item\textit{parabolic}: the eigenvalues are all positive or all negative, except one that is vanishing,
    \item\textit{hyperbolic}: there is only one negative (resp. positive) eigenvalue and all the rest are positive (resp. negative).
\end{itemize}

The most basic examples of PDEs arising in physics are the \textit{wave equation} (hyperbolic), the \textit{heat equation} (parabolic) and the \textit{Laplace equation} (elliptic). The wave equation in one spatial dimension is
\begin{equation}
\label{waveeq}
    \frac{\partial^2u}{\partial t^2}-c^2\frac{\partial^2u}{\partial x^2}=0,
\end{equation}
where $x$ is the spatial coordinate, $t$ is the time coordinate\footnote{For hyperbolic or parabolic PDEs, one of the variables is naturally interpreted as the time variable} and $c$ is a parameter (typically interpreted as the wave velocity). The heat equation is
\begin{equation}
\label{heateq}
    \frac{\partial u}{\partial t}-\alpha\frac{\partial^2u}{\partial x^2}=0,
\end{equation}
where $\alpha$ is a parameter (the thermal diffusivity). Finally, the Laplace equation in two spatial dimensions is
\begin{equation}
\label{laplaceeq}
    \frac{\partial^2u}{\partial x^2}+\frac{\partial^2u}{\partial y^2}=0,
\end{equation}
where $x$, $y$ are spatial coordinates. 

The numerical methods for solving PDEs depends on the type of the equations. Here we shall describe only the \textit{finite element method} for elliptic equations. For a more detailed description of this method, we refer to the Refs.~\cite{Cook2007,Hutton2003,Reddy2005,allaire2019}.

\subsection{Weak formulation of PDEs}
\label{weak_app}

The equations \eqref{waveeq}-\eqref{laplaceeq} are the \textit{strong} formulations of the PDEs. The finite element method requires to convert the PDEs in their \textit{weak} (or \textit{variational}) formulations. This is an alternative way to formulate PDEs that involves integrals. We shall describe the procedure to obtain a weak form with an illustrative example.

Let us consider the Poisson equation on a domain $D\subseteq\mathbb{R}^n$ equipped with Dirichlet conditions on the boundary $\Gamma\equiv\partial D$,
\begin{align*}
    -\Delta u&=f\;\;\text{on}\;\;D,\\
    u&=0\;\;\text{on}\;\;\Gamma,\numberthis{}
    \label{poisson}
\end{align*}
where $\Delta$ is the Laplacian operator and $f$ is a given function on $D$. If $u$ solves the equations \eqref{poisson}, then for any function $v$ that satisfies also the Dirichlet boundary conditions we have
\begin{equation}
    -\int_D{v\,\Delta u\,dV}=\int_D{vf\,dV},
    \label{weak1}
\end{equation}
where $dV$ is the infinitesimal "volume" element on $D$. Conversely, if a function $u$ such that $u=0$ on $\Gamma$ solves Eq.~\eqref{weak1} for any $v$ then it solves also the original PDE \eqref{poisson}. This formulation is already a weak form of the Poisson equation but one can rewrite the left-hand side in a more symmetric way by using a Green's identity (multi-dimensional integration by parts):
\begin{equation}
    -\oint_\Gamma{v\,\nabla u\cdot n\,dS}+\int_D{\nabla u\cdot\nabla v\,dV}=\int_D{vf\,dV},
\end{equation}
where $\nabla$ denotes the gradient operator, $n$ is the outgoing unit vector normal to $\Gamma$ and $dS$ is the infinitesimal "surface" element. Since $v$ is assumed to vanish on the boundary, the surface integral vanishes and we obtain the weak formulation
\begin{equation}
    \int_D{\nabla u\cdot\nabla v\,dV}=\int_D{vf\,dV}.
    \label{weakpoisson}
\end{equation}
Up to some mathematical subtleties \cite{allaire2019}, solving the weak equation \eqref{weakpoisson} is equivalent to solve the original Poisson equation \eqref{poisson}. The function $v$ is usually called the \textit{test function}.

This procedure can be generalized for other types of boundary conditions (the conditions to be satisfied by the test function must be modified and the surface integral may not vanish \cite{allaire2019}) and for other PDEs. We can show that the weak formulation of any elliptic linear PDE is of the form
\begin{equation}
\label{genweak}
    a(u,v)=L(v),\quad\forall v\in V,
\end{equation}
where $V$ is a suitable space of functions, $a:V\times V\to\mathbb{R}$ is a bilinear application and $L:V\to\mathbb{R}$ is a linear application.

If the original problem involves several coupled PDEs, a weak formulation is obtained by introducing as many test functions as unknowns functions, applying the procedure described above for each equation, and then summing all the integral equations of the form \eqref{weakpoisson} together.

\subsection{Finite element method}

The idea of the finite element method is to transform a linear weak PDE into a linear algebraic system of equations. Nonlinear PDEs can also be solved numerically with this method, but it requires using an algorithm to solve nonlinear algebraic equations such as the Newton's method.

First of all, the integration domain must be discretized. In this thesis, we consider at most 2D problems (axisymmetric field configurations). The two spatial variables are typically a compactified radial coordinate $x\in[0,1]$ which maps spatial infinity to $x=1$ and a polar coordinate $\vartheta\in[0,\pi/2]$ so that the integration domain is a rectangle
\begin{equation}
    D=[0,1]\times[0,\pi/2].
    \label{domain}
\end{equation}
\begin{figure}
\centering
  \includegraphics[scale=0.35]{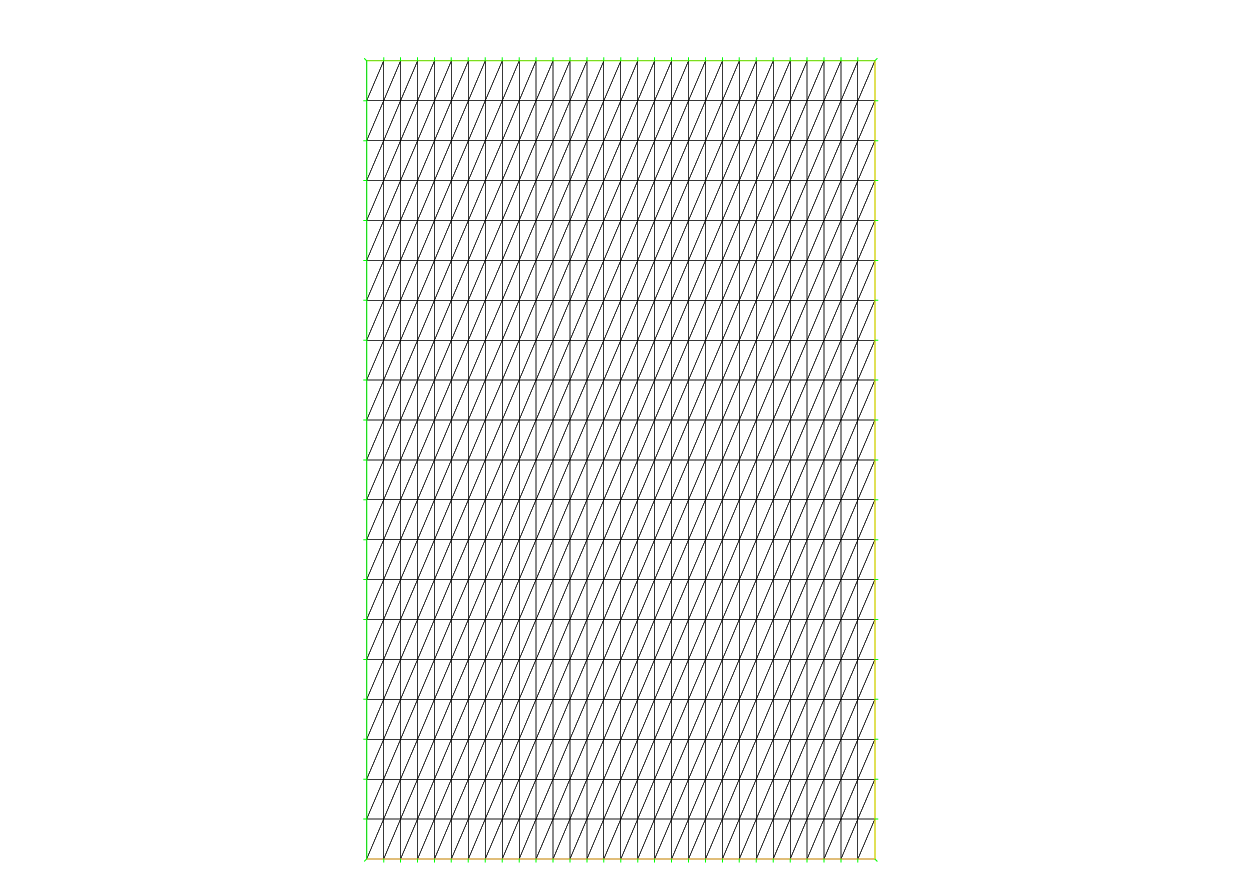}
  \includegraphics[scale=0.35]{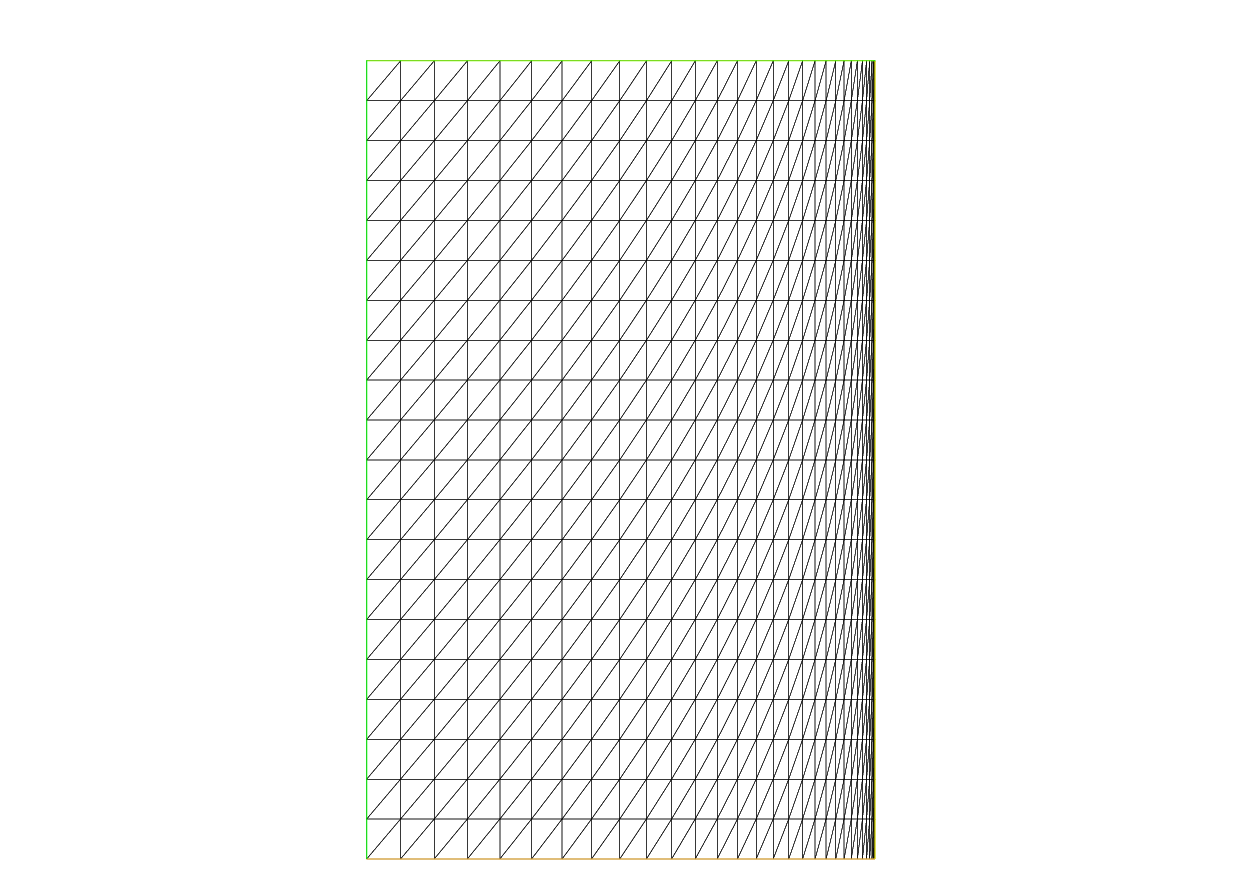}
  \includegraphics[scale=0.35]{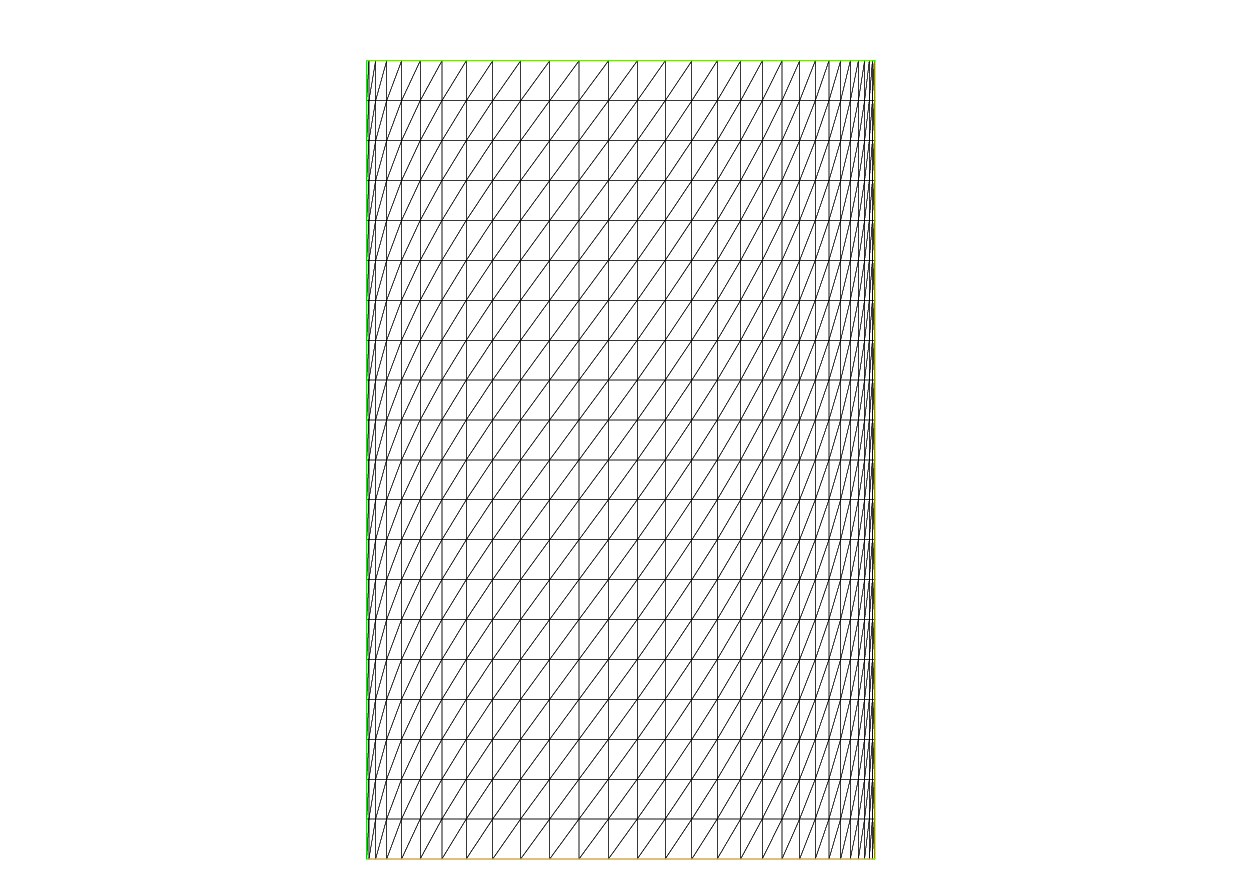}
  \includegraphics[scale=0.35]{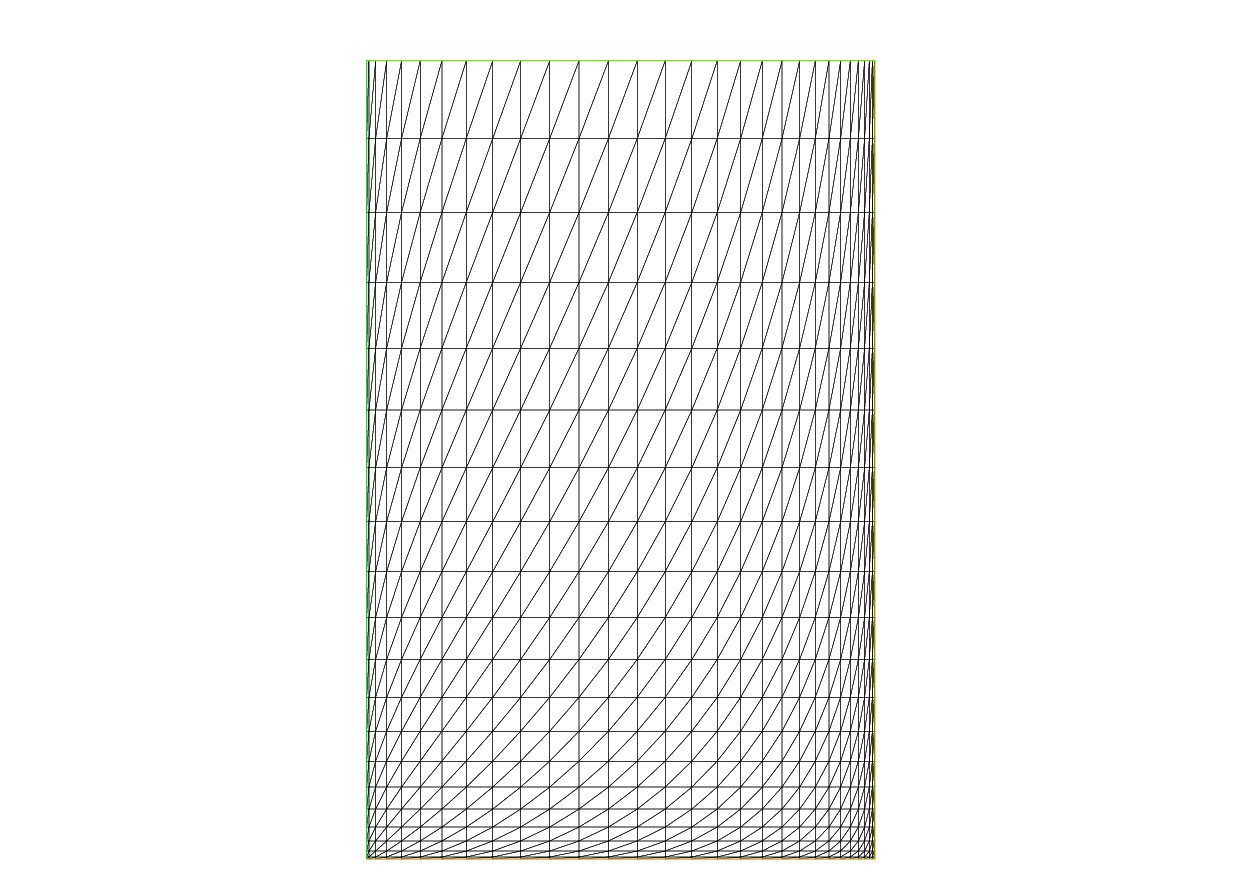}
  \caption[Examples of meshes generated by FreeFem on a rectangular domain.]{Examples of meshes generated by FreeFem on a rectangular domain. The distribution of triangles can be homogeneous (top left panel) or inhomogeneous (other panels).}
  \label{meshex}
\end{figure}
The most obvious way to discretize this domain is to consider a grid with rectangular cells covering $D$ (a 2D generalization of the line discretization presented in Eq.~\eqref{cststep}). However, many finite element softwares rather use triangular cells. The explicit construction of such a grid (or mesh) may seem complicated but the software we use, FreeFem \cite{MR3043640}, has its own internal mesher which automatically generates meshes over a given domain. To illustrate this, we present of Fig.~\ref{meshex} four different meshes covering the domain $D$ defined by Eq.~\eqref{domain}. The top left panel shows a homogeneous distribution of the triangles. Depending on the behavior of the field functions, the accuracy can be improved by choosing a higher density of triangles near the asymptotic boundary ($x=1$), near the origin/horizon ($x=0$) or near the symmetry axis ($\vartheta=0)$. This is shown on the other panels.

The next step is to choose a \textit{finite element space} on the mesh. It typically consists of a set of polynomial functions of order $k\geq 0$ with certain properties at the edges and nodes. A simple example of such a space is the set of piecewise linear functions. In one dimension, $D=[0,1]$, this finite element space is
\begin{equation}
        V_h=\left\{ u(x)=\sum_{i=0}^N{u_i\,\phi_i(x)}\text{  s.t.  }\phi_i(x_j)=\delta_{ij},\; \phi_i(x)\text{  is affine for } x\in\left]x_j,x_{j+1}\right[\right\},
\end{equation}
where $x_i=i/N$ are the nodes coordinates, $\phi_i(x)$ are the basis functions, and $u_i\in\mathbb{R}$ are the components of a function $u(x)\in V_h$ expanded in this basis. This clearly constitutes a finite dimension vector space. Of course all the functions defined on $D$ do not belong to $V_h$ because they live in an infinite dimensional function space. The finite element method approximates any of those functions by a finite expansion, $u_0\,\phi_0(x)+\dots+u_N\,\phi_N(x)\in V_h$.

\begin{figure}
\centering
  \includegraphics[scale=0.2]{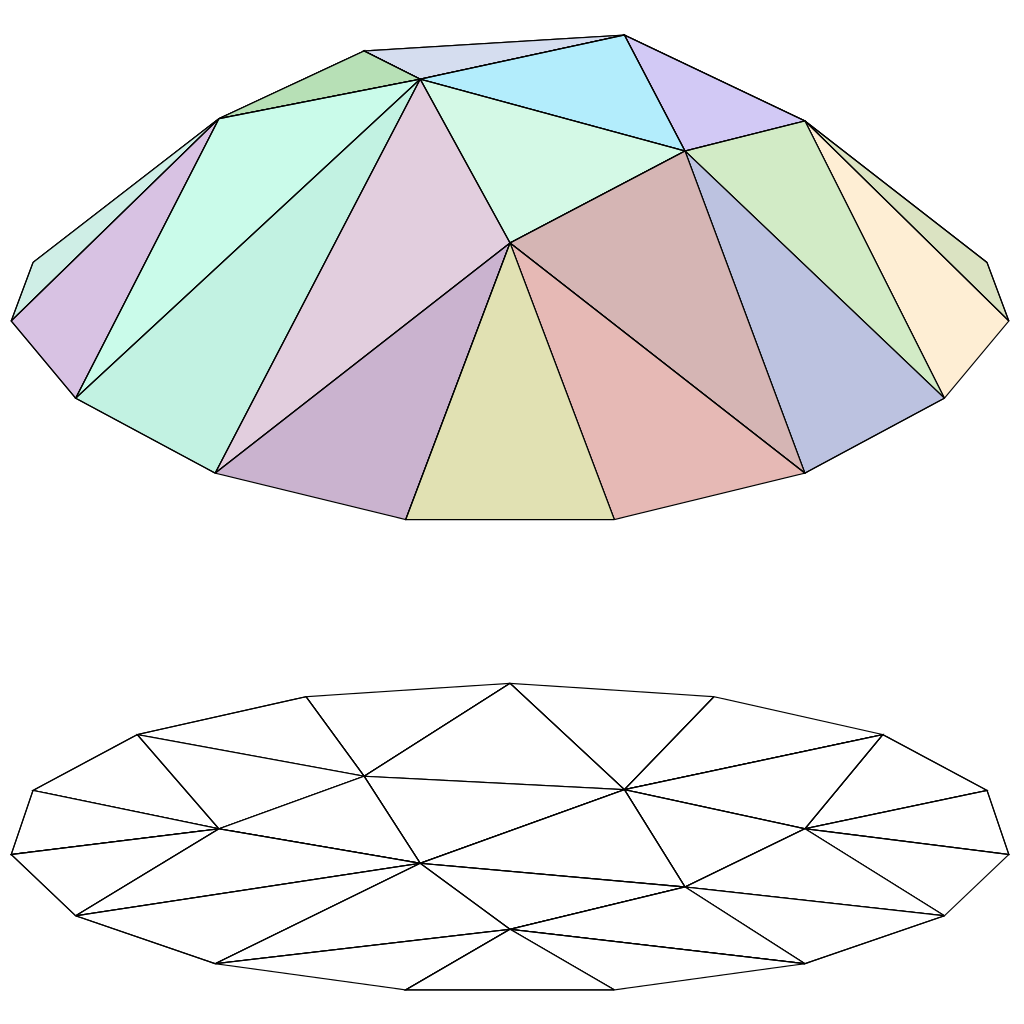}
  \caption{Example of a piecewise linear function in two dimensions.}
  \label{piecewise2d}
\end{figure}

The piecewise linear functions generalize in two dimensions, see Fig.~\ref{piecewise2d}. The corresponding basis functions are sometimes called the tent functions. In our work, we mainly use piecewise quadratic functions which give more accurate results.

\subsubsection{Linear PDEs}

We shall now describe how to find an approximated solution $u$ to a linear weak problem of the form \eqref{genweak}. For this, we require that the unknown function and the test function belong to the finite element space $V_h$. We expand them over the finite element basis,
\begin{equation}
    u=\sum_{i=1}^{M}{u_i\,\phi_i},\quad v=\sum_{i=1}^{M}{v_i\,\phi_i},
    \label{expbasis}
\end{equation}
where $M\equiv\text{dim}(V_h)$. Injecting this to \eqref{genweak}, the equation can be rearranged as
\begin{equation}
        \sum_j{v_j\left[\left(\sum_i u_i\,a(\phi_i,\phi_j)\right)-L(\phi_j)\right]}=0.
\end{equation}
Since this should hold for any test function, or in other word, for any coefficients $v_j\in\mathbb{R}$, the expression inside the square brackets has to be zero. Introducing the \textit{stiffness} matrix $A$ and a vector $B$ by
\begin{equation}
    A_{ij}\equiv a(\phi_i,\phi_j),\quad B_i\equiv L(\phi_i),
    \label{stiff}
\end{equation}
the problem reduces to the linear algebraic system,
\begin{equation}
    \sum_j{A_{ij}\,u_j}=B_i,
    \label{linsyst}
\end{equation}
for the $M$ unknowns $(u_1,\dots,u_M)$ which are sometimes called the \textit{degrees of freedom}. We emphasize that the basis functions $\phi_i$ are known so that all the components $A_{ij}$ and $B_i$ defined by Eq.~\eqref{stiff} can be computed. Finding the solution to \eqref{linsyst} thus requires the inversion of the stiffness matrix:
\begin{equation}
    u_i=\sum_j{(A^{-1})_{ij}\,B_j}.
\end{equation}
The approximated solution $u$ is expected to converge to the true solution of the original problem \eqref{genweak} when the dimension $M$ of the finite element space is increased. See Refs.~\cite{Cook2007,Hutton2003,Reddy2005,allaire2019} for details.

\subsubsection{Nonlinear PDEs}

Let us finally consider the case of nonlinear (elliptic) PDEs. Such equations can also be written in the generic form \eqref{genweak}, but the difference is that now the application $a(u,v)$ is nonlinear in $u$. To illustrate how to deal with the nonlinearities, we shall consider the following example:
\begin{align*}
    -\Delta u&=g(u)\;\;\text{on}\;\; D,\\
    u&=0\;\;\text{on}\;\;\Gamma,\numberthis{}
\end{align*}
where the function $g$ depends on the unknown function $u$ non-linearly. The corresponding weak form is 
\begin{equation}
    \int_D{\big(\nabla u\cdot\nabla v-g(u)\,v\big)\,dV}=0.
\label{weaknonlin}
\end{equation}
By expanding $u$ and $v$ over a finite element basis as in Eq.~\eqref{expbasis}, this problem can be rewritten as
\begin{equation}
    \sum_i{v_i\,F_i(u)}=0\quad\text{with}\quad F_i(u)=\int_D{\left(\nabla u\cdot\nabla\phi_i-g(u)\phi_i\right)dV}.
\label{wealnonlinbis}
\end{equation}
The nonlinear algebraic system to solve is thus
\begin{equation}
\label{nonlin_syst}
    F_i(u)=0,
\end{equation}
where the unknowns are the components of $u$ in the finite element basis, $(u_1,\dots,u_M)$.

We shall describe how to apply the Newton's method in this context. Introducing small variations $\delta u$, we differentiate the components $F_i(u)$ with respect to $u$,
\begin{equation}
    \delta F_i=\int_D{\left(\nabla\delta u\cdot\nabla\phi_i-g'(u)\,\delta u\,\phi_i\right)dV}.
\label{varweak}
\end{equation}
Then we expand the variations $\delta u$ over the finite element basis and rewrite $\delta F_i$ as
\begin{equation}
    \delta F_i=\sum_j{\delta u_j\,J_{ij}}\quad\text{with}\quad J_{ij}\equiv\int_D{\big(\nabla\phi_i\cdot\nabla\phi_j-g'(u)\,\phi_i\,\phi_j\big)\,dV}.
\end{equation}
Here $J_{ij}$ are the components of the Jacobian matrix of the nonlinear system \eqref{nonlin_syst}. We can now apply the Newton's formula \eqref{newton} to find a solution by iterations. The unknown $x$ entering this formula is here $(u_1,\dots,u_M)$ and the vector $F$ is defined by its components in Eq.~\eqref{wealnonlinbis}. Of course to start the iterations, an initial guess $(u_{1,0},\cdots,u_{M,0})$ should be provided. In practice, we choose a constant step size, $\alpha=1$, and adapt the initial guess to the physical situation.

	\newpage
	
	\addcontentsline{toc}{chapter}{Bibliography}
	

	\newpage
	
	\thispagestyle{empty}
	
	\mbox{}
	
	\newpage
	
	\newpage
    \thispagestyle{empty}

    \newgeometry{top=0.5cm,bottom=0.5cm,left=1.2cm,right=0.8cm}
    \hspace{-0.7cm}
	\includegraphics[width=0.3\textwidth]{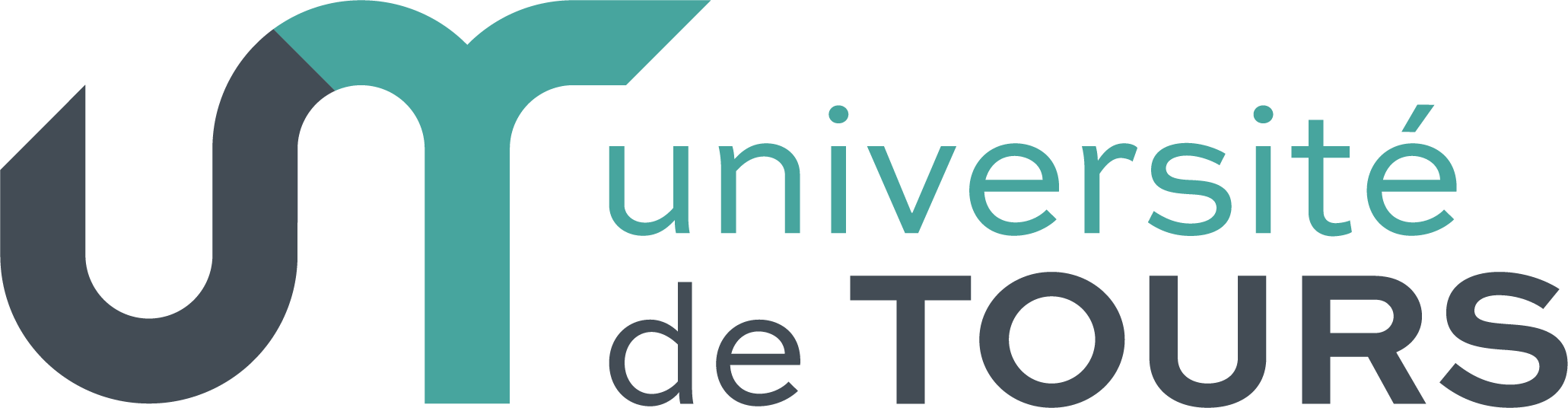}
    \hfill
    \includegraphics[width=0.22\textwidth]{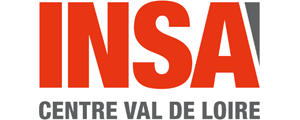}
    \hfill
	\includegraphics[width=0.2\textwidth]{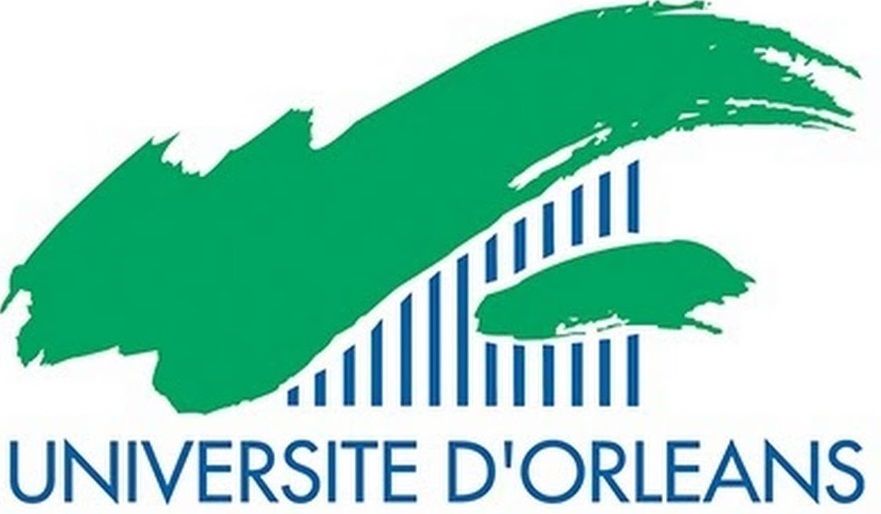}\\
    \vspace{-0.4cm}
    {\setlength{\parindent}{0cm}

    \begin{sffamily}

    \begin{center}
    \LARGE{Romain GERVALLE}\\ 
    \textbf{Trous noirs chevelus et autres objets compacts dans des théories de la gravité}
    \end{center}
	
    \vspace{0.2cm}

    {
    \fontsize{10pt}{10pt}\selectfont
    \fbox{
    \begin{minipage}{\textwidth}
    \textbf{Résumé :}\\
    Dans le cadre d’un espace-temps décrit par la relativité générale d’Einstein, sur lequel évolue uniquement le champ électromagnétique de Maxwell, les trous noirs stationnaires sont complètement caractérisés par leur masse, leur charge électrique ou magnétique, et leur moment cinétique : il s'agit de l'une des versions du théorème de calvitie. Pour autant, lorsque certaines des hypothèses de ce théorème sont omises, il a été établi qu'il cesse de s’appliquer. Cela conduit à l’émergence de trous noirs dits chevelus. Jusqu'à présent, les observations astronomiques ne permettent pas de détecter les «~cheveux~» des trous noirs. Cependant, avec le développement de détecteurs d'ondes gravitationnelles toujours plus précis, les trous noirs chevelus restent un sujet d'étude important en physique théorique. Dans cette thèse, nous considérons deux options permettant de s'affranchir du théorème de calvitie.

    \smallskip
    La première option consiste à décrire la métrique de l’espace-temps par une théorie de gravitation alternative. Nous étudierons la stabilité de trous noirs chevelus dans un espace-temps vide décrit par la théorie de la bigravité massive. Cette théorie est connue pour ses solutions cosmologiques permettant de décrire un Univers en expansion accélérée sans avoir besoin de recourir à la constante cosmologique. Nous montrerons que les trous noirs chevelus en bigravité, obtenus à l’aide de méthodes numériques, sont capables de représenter tant des trous noirs stellaires que des trous noirs supermassifs.

    \smallskip
    Une autre possibilité est de garder les équations d’Einstein, mais de considérer un contenu matériel autre que le champ électromagnétique de Maxwell. Nous choisirons pour cela les champs de la théorie électrofaible. Lorsque la gravitation est omise, cette théorie décrit des monopôles magnétiques de masse infinie. La relativité générale permet de les régulariser en masquant leur singularité coulombienne derrière un horizon des évènements. Les monopôles deviennent alors des trous noirs chargés magnétiquement, pouvant être chevelus. Ces trous noirs électrofaibles pourraient s'être formés lors de fluctuations primordiales, aux tout premiers instants de l’Univers. Après avoir étudié en détail la structure interne des monopôles en espace-temps plat, nous verrons comment leurs propriétés se généralisent au cas gravitationnel.

    \smallskip
    Lorsqu’un trou noir chevelu voit le rayon de son horizon se réduire à zéro, les champs externes restent, et la configuration ainsi obtenue est appelée un soliton. Nous étudierons pour terminer un cas particulier de soliton obtenu lorsqu’un champ scalaire complexe est couplé à la relativité générale. Nous construirons des chaînes de ces solitons, appelés étoiles à bosons. Les équations aux dérivées partielles sous-jacentes seront résolues à l’aide de la méthode des éléments finis : une approche originale et peu utilisée par la communauté de la relativité numérique.\\

    \textbf{Mots-clés :} trous noirs, gravité massive, méthodes numériques, monopôles magnétiques, théorie des champs classiques, relativité numérique.
    \end{minipage}
    }

    \vspace{0.7cm}

    \fbox{
    \begin{minipage}{\textwidth}
    \textbf{Abstract :}\\
    In the realm of spacetimes governed by Einstein's general relativity and containing only Maxwell's electromagnetic field, stationary black holes are fully characterized by their mass, electric or magnetic charge, and angular momentum -- a property encapsulated in a version of the no-hair theorem. However, the validity of this theorem is contingent on certain assumptions, and when these are relaxed, it has been established that the theorem does not always apply. This gives rise to the so-called hairy black holes. To date, astronomical observations have not provided concrete evidence of any type of black hole "hair". Nevertheless, the development of increasingly precise gravitational wave detectors has sparked renewed interest in hairy black holes. In this thesis, we delve into two approaches to circumvent the no-hair theorem.

    \smallskip
    The first option consists in describing the spacetime metric by an alternative theory of gravitation. We investigate the dynamical stability of hairy black holes in a vacuum spacetime described by the theory of massive bigravity. This theory is known for its cosmological solutions which can account for a self-accelerating expansion of the Universe without requiring the use of the cosmological constant. We show that hairy black holes in bigravity, which are obtained using numerical methods, can describe both stellar black holes and supermassive black holes.

    \smallskip
    Another approach is to keep Einstein's equations but to consider a different material content than Maxwell's electromagnetic field. For this we choose the fields of the electroweak theory. In the absence of gravitation, this theory describes magnetic monopoles with infinite mass. General relativity allows for their regularization by concealing their Coulombian singularity within an event horizon. As a result, monopoles become magnetically charged black holes which can exhibit a non-Abelian hair. These electroweak black holes might have formed during primordial fluctuations in the early Universe. After providing a detailed analysis of the internal structure of monopoles in flat space, we investigate how their properties generalize to the black hole case.

    \smallskip
    When the horizon radius of a hairy black hole shrinks to zero, only its external fields remain, giving rise to a configuration known as a soliton. Lastly, we study a particular example of soliton that arises when a complex scalar field is coupled to general relativity. We construct chains of these solitons, which are referred to as boson stars, by solving the underlying partial differential equations using the finite element method. This technique is not very common in the numerical relativity community and provides an alternative to the finite difference method.\\

    \textbf{Keywords :} black holes, massive gravity, numerical methods, magnetic monopoles, classical field theory, numerical relativity.
    \end{minipage}
    }

    \vfill
    }
    \end{sffamily}
    }

\end{document}